\documentclass[12pt,preprint]{aastex} \begin{document}
\def\fix#1{{#1}}
\def\fixed#1{{#1}}
\def\FUNNY#1{\null}
\def\dex#1{10$^{#1}$}
\def\tdex#1{$\times$10$^{#1}$}
\def\ltsim{<} 
\def\gtsim{>} 
\def\deg{$^\circ$}
\def\kms{km\,s$^{-1}$}
\def\cmm#1{cm$^{-#1}$}
\def\Msun{M$_\odot$}
\def\Msunpyr{M$_\odot$\,yr$^{-1}$}
\def\mA{m\AA}
\def\fu{erg\,\cmm2\,s$^{-1}$\,\AA$^{-1}$}
\def\e#1,#2{$\pm$#1$\pm$#2}
\def\E{$\pm$}
\def\ip{$\rho$}
\def\ref{\par}
\def\ms#1#2{\multicolumn{#1}{c}{#2}}
\def\ch#1{\colhead{#1}}
\def\Lya{Ly$\alpha$}  \def\Lyb{Ly$\beta$} \def\Lyg{Ly$\gamma$} \def\Lyd{Ly$\delta$} \def\Lye{Ly$\epsilon$} \def\Lyz{Ly$\zeta$} \def\Lyh{Ly$\eta$}
\def\Lyth{Ly$\theta$} \def\Lyi{Ly$\iota$} \def\Lyk{Ly$\kappa$} \def\Lyn{Ly$\nu$}    \def\Lyxi{Ly$\xi$}     \def\Lylim{Lylim}
\def\ionc#1#2#3{#1\,#3}
\def\HI{\ionc{H}{1}{I}} \def\HII{\ionc{H}{1}{II}} \def\H#1{H$_2$}
\def\OI{\ionc{O}{1}{I}} \def\OII{\ionc{O}{2}{II}}   \def\OIII{\ionc{O}{3}{III}} \def\OIV{\ionc{O}{4}{IV}} \def\OVI{\ionc{O}{6}{VI}} \def\OVII{\ionc{O}{7}{VII}}
\def\CI{\ionc{C}{1}{I}} \def\CII{\ionc{C}{2}{II}}   \def\CIII{\ionc{C}{3}{III}} \def\CIV{\ionc{C}{4}{IV}}
\def\NI{\ionc{N}{1}{I}} \def\NII{\ionc{N}{2}{II}}   \def\NIII{\ionc{N}{3}{III}} \def\NIV{\ionc{N}{4}{IV}} \def\NV{\ionc{N}{5}{V}}
\def\NeVIII{\ionc{Ne}{8}{VIII}}\def\NeIX{\ionc{Ne}{8}{IX}}
\def\MgI{\ionc{Mg}{1}{I}} \def\MgII{\ionc{Mg}{2}{II}}
\def\AlII{\ionc{Al}{2}{II}}
\def\SiII{\ionc{Si}{2}{II}} \def\SiIII{\ionc{Si}{3}{III}} \def\SiIV{\ionc{Si}{4}{IV}}
\def\SII{\ionc{S}{2}{II}} \def\SIII{\ionc{S}{3}{III}} \def\SVI{\ionc{S}{6}{VI}}
\def\ArI{\ionc{Ar}{1}{I}}
\def\FeII{\ionc{Fe}{2}{II}}
\def\lm{\,$\lambda$}
\def\llm{\,$\lambda\lambda$}
\def\OVIa{\OVI\lm1031.926}
\def\OVIb{\OVI\lm1037.617}
\def\OVIab{\OVI\llm1031.926, 1037.617}
\def\Wlim{$W_{\rm lim}$}
\def\Dv{$\Delta$$v$}
\def\Dva{$|$$\Delta$$v$$|$}
\def\Dz{$\Delta$$z$}
\def\Rgal{$D_{gal}$}
\def\Dgal{$D_{gal}$}
\def\vgal{$v_{gal}$}
\def\vabs{$v_{\rm abs}$}
\def\vlsr{$v_{\rm LSR}$}
\def\Lstar{L$_*$}
\def\vavg{$<$$v$$>$}
\def\vrb{\vgal, \Rgal, \ip}
\def\Hm{\,h$^{-1}_{71}$}
\def\FUSE{FUSE}
\def\HST{HST}
\def\FOS{{\it FOS}}
\def\STIS{{\it STIS}}
\def\GHRS{{\it GHRS}}
\def\COS{{\it COS}}
\def\NED{{\it NED}}
\def\Sintro{1}
\def\Sobs{2}
\def\SSabsdata{\Sobs.1}               \def\Tobs{1} \def\Tshift{2}
\def\SSidentify{\Sobs.2}              \def\Fokranges{1}
\def\SSmeasure{\Sobs.3}               \def\Fspectra{2}
\def\SSgaldata{\Sobs.4}               \def\FdiamvsMabs{3} \def\Tres{3} \def\Tion{4} \def\THI{5} \def\TOVI{6}

\def\Sabsresults{3}
\def\SScompare{\Sabsresults.1}
\def\SSOVIabs{\Sabsresults.2}         \def\TNOVI{7}
\def\SSlinewidth{\Sabsresults.2}      \def\Fdetwidth{4}
\def\SSdndz{\Sabsresults.3}           \def\Fdndz{5} \def\Tdndz{8}
\def\SSbevol{\Sabsresults.4}          \def\Fbevol{6}

\def\Sassocresults{4}
\def\SSassoccount{\Sassocresults.1}   \def\Tassoccount{9}
\def\SSnearest{\Sassocresults.2}      \def\Fnearest{7}
\def\SSassoc{\Sassocresults.3}        \def\Fassoc{8}
\def\SSvoid{\Sassocresults.4}
\def\SSvdiffhist{\Sassocresults.5}    \def\Fvdiffhist{9}
\def\SSwidthimp{\Sassocresults.6}     \def\Fwidthimp{10}
\def\SSewvsb{\Sassocresults.7}        \def\Fewvsb{11}

\def\Sgalresults{5}
\def\SSgaldens{\Sgalresults.1}        \def\Fewvsgaldens{12}
\def\SSgalaxycount{\Sgalresults.2}    \def\Tgalaxycount{10}
\def\SSdetfrac{\Sgalresults.3}
\def\SSdetfraccompare{\SSdetfrac.1}
\def\SSdetfracconstruct{\SSdetfrac.2} \def\TdetfracA{11} \def\TdetfracB{12} \def\Fdethistall{13} \def\Fdethistfield{14} \def\Fdethistgroup{15} \def\Fdethistgrpcnt{16} \def\Fdethistgrpcntcen{17}
\def\SSdetfraclyafield{\SSdetfrac.3}
\def\SSdetfraclyagroup{\SSdetfrac.4}
\def\SSdetfraclyb{\SSdetfrac.5}
\def\SSdetfracOVI{\SSdetfrac.6}
\def\SShalomap{\Sgalresults.4}        \def\Fhalomap{18} \def\Fimpvdiff{19}

\def\Sdisc{6}
\def\Sconcl{7}
\def\fox#1{\bf{#1}}

\def\captionokranges{%
Plot showing the velocity ranges where intergalactic \Lya, \Lyb, \OVIa\ and
\OVIb\ absorption may be obscured, for three different levels of interstellar
\H2\ absorption: none (top panel), medium ($J$=0 and $J$=1 lines only, middle
panel) and strong (bottom panel). The square boxes enclose an 80~\kms\ velocity
range around metal lines (with ion name given), 40~\kms\ around \H2\ lines (with
$J$-value given). Boxes with dotted borders are given for the geocoronal \OI*
lines, which are absent in orbital-night-only data.
}

\def\captionspectra{%
This figure shows the \Lya, \Lyb\ and \OVI\ spectra for all of the detected
intergalactic absorption lines. At the top of each panel, we give the AGN's
name, and within parentheses the galaxy with which the absorption is associated,
as well as the impact parameter of that galaxy. The flux scale has units of
\dex{-14}\,erg\,\cmm2\,s$^{-1}$\,\AA$^{-1}$. If orbital-night only data were
used, the line-identification includes ``(night)''. A velocity range of
800~\kms\ is shown, centered at the velocity of the absorber. Data are
represented by the histogram, while the fitted continuum (including the \H2\
model) is shown by the solid line. The dashed vertical line denotes the velocity
of the intergalactic absorber. Solid vertical lines give the velocities of
galaxies with impact parameter $<$1~Mpc, with the thick line identifying the
velocity of the galaxy associated with the absorber. At the bottom of each panel
interstellar Milky Way absorption lines are identified, with single numbers
given for the $J$-level of H$_2$ lines. An ``earth'' symbol is given at places
where geocoronal \Lyb\ or \OI* emission is present. The word ``flaw'' near
1043.2~\AA, indicates a flaw in the \FUSE\ detector, which makes any feature
near this wavelength suspect. Intergalactic absorption lines are identified
with the label raised relative to that identifying the Milky Way absorption. For
each intergalactic detection a ``+'' is added.  A ``$-$'' shows a non-detection.
A ''+'' or ''$-$'' in parentheses shows uncertain detections and upper limits
worse than 50~\mA. Finally, an ``x'' means that we cannot measure the line, most
often because of blending with Galactic absorption or geocoronal emission or
sometimes because there is no data.
}

\def\captiondiamvsMabs{%
\fix{Correlation between galaxy luminosity and galaxy diameter, for galaxies in
the RC3 for which both are given (excluding irregulars). The horizontal lines
correspond to galaxies with luminosity \Lstar, 0.5\,\Lstar, 0.25\,\Lstar\ and
0.1\,\Lstar. The vertical lines correspond to the corresponding diameters that
we use throughout the paper: 20.4, 14.6, 11.2 and 7.5~kpc. \fixed{The diagonal
line is a least squares fit (correlation coefficient 0.8) for $M$$<$$-$16 and
\Rgal$>$7.5~kpc, and has the functional form $M$=$-$5.78~log\,$R$$-$12,
corresponding to log\,$L$=2.31~log\,$R$$-$3.03}. {\it Selection criteria:} all
galaxies in the RC3 with both magnitude and diameter given.}
}

\def\captiondetwidth{%
Distributions of fitted linewidths for four different lines (\Lya, \Lyb, \OVIa\
and \OVIb), as indicated in the top left corner of each panel. $b$ values for
\Lya\ and \Lyb\ are listed in Table~\THI, those for \OVI\ in Table ~\TNOVI. The
top x-scale gives $b$-values, the bottom x-scale the full-width-at-half-maximum
(FWHM). \fix{{\it Selection criteria:} all lines that are not contaminated by
\H2\ absorption, the \FUSE\ detector flaw near 1043~\AA, or blended with other
absorption lines.} The dotted line in the \OVIa\ panel shows the distribution
found by Tripp et al.\ (2008), scaled to our sample size.
}

\def\captiondndz{%
Plots of the frequency of occurrence of \Lya, \Lyb\ and \OVI. Linear equivalent
width limit scale on the left, logarithmic scale on the right. Our results,
based on 102 \Lya, 34 \Lyb\ and 11 \OVI\ detections at $v$=400--5000~\kms, are
shown as the solid line with error bars. The stars in panels a and b show the
distribution of Penton et al.\ (2004), corrected for the difference in the way
that equivalent width limits are measured (see Sect.~\SSmeasure). In panels e
and f the closed stars show the $dN/dz$ distribution of Danforth \& Shull
(2005), the open stars that of Tripp et al.\ (2008) and the open plusses that of
Thom \& Chen (2008), with corrections for the difference in equivalent width
limit calculation when necessary. \fix{{\it Selection criteria:} all 133 \Lya,
\Lyb, \OVI\ lines at $v$=400--5000~\kms\ with $W$$>$$W_{\rm lim}$.}
}

\def\captionbevol{%
Parameters of the distribution of \Lya\ linewidths as function of redshift ($z$)
or lookback time. Top panels: fraction of lines with $b$$>$40~\kms. At $z$$<$1.0
the redshift interval is 0.1, at $z$$>$1.0 it is 0.2. The short horizontal lines
near the top of panel a show the redshift ranges of individual QSOs used to
derive the linewidth distribution. Bottom panels: horizontal lines show the
10th, 25th, 50th, 90th and 95th percentile of the distribution of $b$-values.
The points and vertical lines show the average and dispersion in each redshift
interval. \fix{{\it Selection criteria:} \Lya\ lines with $W$(rest)$<$500~\mA)
and implied optical depth $<$2.}
}

\def\captionnearest{%
Galaxy-galaxy and absorber-galaxy nearest-neighbor distributions. For a given
separation the curves give the fraction of galaxies closer to each other galaxy
or absorber. Galaxy-galaxy separations are calculated using directions and
estimated distances. Galaxy-absorber separations are 1.25 times the impact
parameter (see text). Black curves give galaxy-galaxy nearest-neighbor
distributions; solid black curves are given when comparing galaxies with the
same luminosity limit, dashed black curves when looking at the separation for
any galaxy from the reference galaxies. Colored curves give the nearest galaxies
to absorbers, red when the galaxy has $L$$>$\Lstar, blue when the galaxy has
$L$$>$0.1\,\Lstar, purple for all galaxies, magenta for the distributions given
by Penton et al.\ (2002). Solid colored curves include \Lya\ absorbers stronger
than some equivalent width limit, while dotted colored curves include only \Lya\
absorbers below some equivalent width. In panels a/b we compare our results to
those of Penton et al.\ (2002), in panels c/d all galaxies were included, in
panels e/f only field galaxies, and in panels g/h only group galaxies. \fix{{\it
Selection criteria:} $L$$>$\,0.1\,\Lstar\ (i.e.\ \Rgal$>$7.5~kpc) and
$v$$<$2500~\kms, or $L$$>$\,\Lstar\ (i.e.\ \Rgal$>$20.4~kpc) and
$v$$<$5000~\kms, classified as in the field or member of a group, and all \Lya\
lines with equivalent width above or below the given value.}
}

\def\captionassoc{%
For each absorber, this figure shows the impact parameter (horizontal axis) vs
the difference between the velocity of the absorption line and the systemic
velocity of the galaxies near it (vertical axis). The velocity is usually that
of \Lya\ but can be for \Lyb\ or \OVI\ if no \Lya\ data is available. The
absorber's velocity and the galaxy associated with it in Table~\Tres\ are given
on the second line of the label in the top left of each panel. The top right
label shows a panel number. Circles are for field galaxies, squares for galaxies
listed as part of a GH or LGG group. Filled symbols represent the galaxies that
are associated with the absorber. If the association is with a group, all group
galaxies are given a closed symbol. Open symbols with a plus in them are for
galaxies for which a non-detection is listed in Table~\Tres. If there is no
associated galaxy within 1~Mpc, a rightward pointing triangle is shown, whose
size and velocity placement are set by the nearest galaxy at \ip$>$1~Mpc. The
size of each symbol scales with the galaxy's diameter to the power 2/3.
}

\def\captionvdiffhist{%
Histograms of the difference in velocity between an absorber and the galaxy that
we associate with it. Top panels: \HI\ lines (\Lya\ or \Lyb), bottom panels:
\OVI\ lines. Left panels: only unambiguous associations are included (see
Sect.~\SSassoc). Right panels: all associations listed in Table~\Tres\ are
included. The smooth curves represent gaussians with the average velocity and
dispersion given in the labels.
}

\def\captionwidthimp{%
Plot of the linewidth of \Lya\ lines versus the impact parameter. Top panel: all
\Lya\ absorbers. Bottom panel: including only well-measured absorbers clearly
associated with field galaxies. Filled symbols are for reliably measured lines,
open symbols for problematic lines (e.g.\ saturated, noisy). Circles are given
for field galaxies, stars for group galaxies. Labels on the right side in the
bottom panel show the temperature corresponding to some FWHM/$b$-values for gas
at the given temperature. The histograms give the 10th, 50th and 90th percentile
of the distribution in 250~kpc (\ip$<$1~Mpc) or 500~kpc (\ip$>$2~Mpc) wide bins.
}

\def\captionewvsb{%
Scatter plot of the \Lya\ equivalent width against impact parameter. Left
panels: \fix{results for the absorbers listed in Table~\Tres\ (two absorbers
with equivalent width $>$1000~\mA\ are not shown in the top left panel).} The
symbol size scales as the 2/3rd power of the galaxy diameter. Right panels:
impact parameters to the nearest $L$$>$\Lstar\ galaxy. Top panels: linear scales,
bottom panels: logarithmic scales. The histograms in the top panels show the
10th, 50th and 90th percentiles of the distribution of equivalent width in 250
or 500~kpc wide impact parameter bins. In the bottom panels the dotted line
gives the relation between equivalent width and impact parameter claimed by Chen
et al.\ (2001), while the solid line is the relation given by Penton et al.\
(2002).
}

\def\captionewvsgaldens{%
Scatter plot of the total \Lya\ equivalent width within 500 or 1000~\kms\ from
an absorber versus the number density of galaxies with $L$$>$0.1~\Lstar\ in a
cylinder with radius 500, 1000 or 2000~kpc (see labels). Open circles show the
data for all detections in our sample, while filled stars are for the absorbers
in the sample of Bowen et al.\ (2002) and filled squares correspond to the
absorbers discussed by C\^ot\'e et al.\ (2005). In each panel, the column of
crosses corresponds to having no galaxy in the box, while the second column
gives the density with just one galaxy in each box. \fix{{\it Selection
criteria:} all \Lya\ lines; all galaxies fitting the criteria, where for
$L$$>$0.1\,\Lstar\ also $v$$<$2500~\kms\ and for $L$$>$\Lstar\ $v$$<$5000~\kms.}
}

\def\captiondethistall{%
Distribution of number of galaxies, number of detections and fraction of
detected galaxies as function of impact parameter. This includes every galaxy
with systemic velocity 400--5000~\kms\ and \ip$<$2~Mpc near any of the 76
sightlines in our sample, independent of brightness, size, completeness of the
galaxy survey near the sightline, or group membership. The histograms in the top
panels show the number of galaxies in impact parameter bins of 100~kpc. At
\ip$>$1~Mpc there are two histograms, with the lower one giving the actual
number and the higher values taking into account a correction for the
incompleteness of the \NED\ sample (see Sect.~\SSdetfracconstruct). The dotted
lines show the expected numbers for a sample of galaxies with random impact
parameters. The hatched areas in the top panels show the distribution of
detections. The bottom panels give the fraction of galaxies with which we
associate a detection; the thin vertical bars provide an estimate of the error
in the fraction, found from sqrt(\#detections). Finally, the numbers in the top
right corner give the total number of detections and galaxies in the plot.
\fix{{\it Selection criteria:} galaxies of any luminosity with
$v$=400--5000~\kms, all \Lya, \Lyb, \OVI\ lines found where the equivalent width
error is $<$100~\mA.}
}

\def\captiondethistfield{%
Same as Fig.~\Fdethistall. \fix{{\it Selection criteria:} galaxies with
\Rgal$>$7.5~kpc (equivalent to $L$$>$0.1\,\Lstar) and $v$=400--2500~\kms, and
not listed as a member of a galaxy group by Geller \& Huchra (1982, 1983) and/or
Garcia (1993). All \Lya, \Lyb, \OVI\ lines with $v$$<$2500~\kms\ found where the
equivalent width error is $<$100~\mA.}
}

\def\captiondethistgroup{%
Same as Fig.~\Fdethistall. \fix{{\it Selection criteria:} galaxies with
\Rgal$>$7.5~kpc (equivalent to $L$$>$0.1\,\Lstar) and $v$=400--2500~\kms, that
are listed as a member of a galaxy group by Geller \& Huchra (1982, 1983) and/or
Garcia (1993). All \Lya, \Lyb, \OVI\ lines with $v$$<$2500~\kms\ found where the
equivalent width error is $<$100~\mA.}
}

\def\captiondethistgrpcnt{%
Same as Fig.~\Fdethistall, except that we only count one detection or
non-detection for each galaxy group. \fix{The impact parameter, $\rho$, is that
to the group galaxy nearest the sightline. Only cases where $\rho$ is less than
half the diameter of the group are included.}
}

\def\captiondethistgrpcntcen{%
\fix{Same as Fig.~\Fdethistall, except that we only count one detection or
non-detection for each galaxy group, and unlike what is the case for
Fig.~\Fdethistgrpcnt, the impact parameter is that to the center of the group.}
}

\def\captionhalomap{%
Plot combining all sightline-galaxy associations. Each galaxy is rotated to have
the galaxy's major axis horizontal (possible for 261 of the 329 galaxies) (right
panels). For a small number this rotation is such that the approaching side is
on the left (the galaxies in the left panels). The sightline to the AGN is then
placed, using the following symbol code: colored symbols for detections of
either \Lya, \Lyb\ or \OVI, open symbols for non-detections. Stars indicate an
\OVI\ detection, circles are for \HI\ data with an \OVI\ upper limit, squares
when no \OVI\ data are available. An open plus is shown if there is just an
\OVI\ detection, a plus for just an \OVI\ upper limit, and a cross if we could
not check either of \HI\ and \OVI. Colors encode the difference in velocity
between the galaxy and the detection, being black if \Dv$<$20~\kms, yellow if
\Dv=20 to 50~\kms, orange if \Dv=50 to 100~\kms, red when \Dv$>$100~\kms,
light blue if \Dv=$-$50 to $-$20~\kms, dark blue if \Dv=$-$100 to $-$50~\kms,
and purple when \Dv$<$$-$100~\kms. The symbol size scales with the square root
of the equivalent width of the absorber. If just \Lyb\ is detected, its
equivalent width is scaled by a factor 3. Detections in panel a are for
3C\,232--NGC\,3067 (lightblue square), PG\,1259+593--UGC\,8146 (black star),
Ton\,S180--NGC\,247 (red star), Mrk\,771-UGC\,7697 (yellow circle),
PG\,0844+349-NGC\,2683 (dark blue star), PKS\,2155$-$304--ESO\,466-G32 (blue
circle), Ton\,S210-NGC\,253 (yellow star), Mrk\,335--NGC\,7817 (lightblue
circle), and MCG+10-16-111-NGC\,3556 (red square). Note, that toward
PKS\,2155$-$304 there are three lines, at 5105, 4990 and 5164~\kms; we only show
the strongest of these. The two non-detection with \ip$<$350~kpc are toward
Mrk\,110--NGC\,2841 and HE\,1228+0131--NGC\,4517. In panel c we have the lines
for HS\,1543+5921-SBS\,1543+593 (dark blue square), Mrk\,205-NGC\,4319 (dark
blue circle), PG\,0804+761-UGC\,4238 (black circle), Mrk\,876-NGC\,6140 (yellow
star), PG\,0953+414--NGC\,3104 (black star), Mrk\,290--NGC\,5983 (orange circle)
and HE\,0226$-$4110--NGC\,954 (purple star).
}

\def\captionimpvdiff{%
Histograms of the number of \Lya\ absorbers in 100~kpc wide bins of impact
parameter, separated by the difference in velocity between the absorbers and the
associated galaxy. Top panel: absorber velocity more than 20~\kms\ more negative
than that of the galaxy; bottom panel: absorber velocity more than 20~\kms\ more
positive than that of the galaxy; middle panel: absorber velocity within
20~\kms\ of that of the galaxy. The labels also give the average impact
parameter for all absorbers in the histogram.
}

\title{The Relationship Between Intergalactic \HI/\OVI\ and Nearby ($z$$<$0.017) Galaxies}
\author{B.P. Wakker\altaffilmark{1}, B.D. Savage\altaffilmark{1}}
\altaffiltext{1}{Department of Astronomy, University of Wisconsin, 475 N. Charter St, Madison, WI 53706.}

\begin{abstract}
We analyze intergalactic \HI\ and \OVI\ absorbers with $v$$<$5000~\kms\ in \HST\
and \FUSE\ spectra of 76 AGNs. The baryons traced by \HI/\OVI\ absorption are
clearly associated with the extended surroundings of galaxies; for impact
parameters $<$400~kpc they are $\sim$5 times more numerous as those inside the
galaxies. This large reservoir of matter likely plays a major role in galaxy
evolution. We tabulate the fraction of absorbers having a galaxy of a given
luminosity within a given impact parameter (\ip) and velocity difference (\Dv),
as well as the fraction of galaxies with an absorber closer than a given \ip\
and \Dv. We identify possible ``void absorbers'' (\ip$>$3~Mpc to the nearest
\Lstar\ galaxy), although at $v$$<$2500~\kms\ all absorbers are within 1.5~Mpc
of an $L$$>$0.1\,\Lstar\ galaxy. The absorber properties depend on \ip, but the
relations are not simple correlations. \fix{For four absorbers with
\ip=50--350~kpc from an edge-on galaxy with known orientation of its rotation,
we find no clear relation between absorber velocities and the rotation curve of
the underlying galaxy.} For \ip$<$350~kpc the covering factor of \Lya\ (\OVI)
around $L$$>$0.1~\Lstar\ galaxies is 100\% (70\%) for field galaxies and 65\%
(10\%) for group galaxies; 50\% of galaxy groups have associated \Lya. All \OVI\
absorbers occur within 550~kpc of an $L$$>$0.25\,\Lstar\ galaxy. \fix{The
properties of three of 14 \OVI\ absorbers are consistent with photoionization,
for five the evidence points to collisional ionization; the others are
ambiguous.} The fraction of broad \Lya\ lines increases from $z$=3 to $z$=0 and
with decreasing impact parameter, \fix{consistent with the idea that gas inside
$\sim$500~kpc from galaxies is heating up, although alternative explanations can
not be clearly excluded.}
\end{abstract}


\section{Introduction}
\par Intergalactic gas has been detected in optical spectra of QSOs since the
1970s. At ultraviolet wavelengths, the low redshift \HI\ Lyman $\alpha$ forest
of absorption lines was initially detected at moderate and low resolution with
the {\it Goddard High Resolution Spectrograph} (\GHRS) and the {\it Faint Object
Spectrograph} (FOS) aboard the Hubble Space Telescope (\HST) (Morris et al.\
1991; Bahcall et al.\ 1991). Data from the \HST\ QSO absorption-line Key Project
(Januzzi et al.\ 1998 and references therein), allowed a comparison between the
properties of the high- and low-redshift absorption lines and revealed evidence
for the evolution of the gas (Weymann et al.\ 1998). The realization that the
low column density portion of the low-redshift \HI\ absorption lines is tracing
very highly photoionized gas in the intergalactic medium (IGM) (containing
$\sim$30\% of the baryons at low $z$) followed from more extensive observational
studies (Penton et al.\ 2004), combined with theoretical insights about the very
large corrections required to convert measures of the observed \HI\ column
densities into total (\HI+\HII) column densities (Schaye 2001). More recent work
has revealed that the low $z$ \Lya\ forest includes both narrow and broad
absorption lines and that many of the broad lines are probably tracing gas
$\sim$3--10 times hotter than expected for gas in photoionization equilibrium
(Richter et al.\ 2004; Lehner et al.\ 2007). The detection of the relatively
high line density of intergalactic \OVI\ absorption at low redshift (Tripp et
al.\ 2000) provided additional information about the highly-ionized state of
some of the low redshift IGM. The most recent studies (Tripp et al.\ 2008;
Danforth \& Shull 2008; Thom \& Chen 2008) revealed that the \OVI\ is tracing a
complex mixture of highly photoionized and warm-hot collisionally ionized gas.
The combination of results from the narrow \Lya, broad \Lya\ and \OVI\
absorption line studies suggests that $\sim$40--50\% of the baryons at low $z$
probably resides in photoionized gas and the cooler part of the warm-hot IGM
predicted by cosmological hydrodynamical simulations (Cen \& Ostriker 1999;
Dav\'e et al.\ 2001). This implies that the baryonic content of the detected low
redshift IGM is $\sim$6 times larger than the baryonic content of galaxies,
which is estimated to be $\sim$8\% (Fukugita \& Peebles 2004). Four percent of
the baryons are in the hot plasmas found in galaxy clusters while the
hydrodynamical simulations suggest most of the remaining $\sim$50\% of the
baryons may reside in the hotter portions of the warm-hot IGM with
$T$$>$3\tdex5~K. The hotter gas has been detected through \NeVIII\ absorption
(Savage et al.\ 2005b; Narayanan et al.\ 2008). However, the reported X-ray
detections of \OVII\ absorption in the WHIM at $z$$>$0 toward Mrk\,421 (Nicastro
et al.\ 2005) are not supported by the independent studies of the X-ray spectrum
of this object (Kaastra et al.\ 2006; Rasmussen et al.\ 2007).    
\par At low redshift it is possible to study the relation between the
intergalactic gas and galaxies. In high column density \HI\ systems the very
strong \MgII\,$\lambda\lambda$2796.352, 2803.531 absorption lines are observable
from the ground at redshifts above about $z$=0.25, and with \HST\ at lower
redshifts. They have been associated with galaxies at impact parameters
$<$100~kpc (Bergeron 1991; Steidel et al.\ 1995). Steidel et al.\ (1995)
inferred that all $L$$>$0.05\,\Lstar\ galaxies are surrounded by spherical halos
with sizes on the order of 100~kpc, having 100\% covering fraction in \MgII, but
later work suggested that the \MgII\ halos are patchy, with covering fraction
about 50\% (Churchill et al.\ 2007; Kacprzak et al.\ 2008).
\par The first studies of the relation between galaxies and \Lya\ absorbers were
done near the 3C\,273.0 sightline (Morris et al.\ 1993). This study suggested
that \Lya\ clouds did not strongly associate with galaxies. Later, Lanzetta et
al.\ (1995), Tripp et al.\ (1998), Impey et al.\ (1999), Chen et al.\ (2001),
Bowen et al.\ (2002), Penton et al.\ (2002), C\^ot\'e et al.\ (2005), Aracil et
al.\ (2006) and Prochaska et al.\ (2006) studied this question. These authors
typically started with observations of 5--15 UV-bright AGNs, identified the
absorbers and then complemented this with a survey of the galaxies near the
sightline, where ``near'' typically means within 1\deg, or even within
10\arcmin. In a few cases the study started with a search for sightlines to
bright AGNs passing within about 200~kpc of a nearby galaxy (Bowen et al.\ 1996,
2002; C\^ot\'e et al.\ 2005). We discuss the parameters of these studies in
Sect.~\SSassoc, and compare the detailed results to ours in many of the
subsections of Sects.~\Sabsresults, \Sassocresults\ and \Sgalresults. Generally,
these authors concluded that \Lya\ absorbers at low impact parameters
($<$350~kpc or so) originate in galaxy halos, even though there are many \Lya\
absorbers far from galaxies, in the general intergalactic medium. Penton et al.\
(2002) in particular argued that about 20\% of the \Lya\ absorbers are ``void
absorbers'', which they defined as absorbers occuring more than 3~Mpc from the
nearest \Lstar\ galaxy.
\par For the low redshift intergalactic \OVI\ absorption, a number of papers
associated particular \OVI\ absorbers with galaxies (Danforth \& Shull 2008;
Tripp et al.\ 2008). Comparing the locations of \OVI\ absorbers with galaxy
catalogs, Stocke et al.\ (2006) found that most \OVI\ absorbers originate
relatively close to galaxies. Oppenheimer \& Dav\'e (2008) used theory to
conclude that \OVI\ absorbers consist of photoionized gas within 300~kpc from
$L$$>$0.1\,\Lstar\ galaxies. On the other hand Ganguly et al.\ (2008) used
modeling to predict that collisionally ionized \OVI\ should occur relatively far
from galaxies. Thus, the general relation between \OVI\ absorbers and galaxies
requires additional observational and theoretical work.
\par Theoretically (e.g.\ Oort 1970; Sommer-Larsen 2006; Fukugita \& Peebles
2006), galaxies are predicted to be surrounded by hot (\dex5--\dex6~K) coronae
with sizes on the order of several hundred kpc. These coronae may contain as
much or more matter as is present inside the galaxies. Some of the hot gas may
condense and rain down onto the galaxies, becoming visible as neutral
high-velocity clouds. UV absorption line studies show strong evidence for the
presence of 3\tdex5~K gas around the Milky Way (Sembach et al.\ 2003). The most
likely origin of this phase of the gas is in interfaces between cool
(\dex3--\dex4~K) condensations that are embedded in \dex6~K gas at distances of
a few to 100~kpc (Fox et al.\ 2005). Direct evidence for \dex6~K coronal gas is
ambiguous, however. Pedersen et al.\ (2006) claimed to have found the associated
X-ray emission for the case of NGC\,5746, but Yao et al.\ (2008) compared
Galactic \OVII\ and \NeIX\ X-ray absorption in several different sightlines and
concluded that the hot gas is confined to the Galactic thick disk only.
\par In this paper we analyze the relation between galaxies and absorbers by
concentrating on the nearest examples, i.e., galaxies and absorbers with
recession velocities \fix{above 400 and below 5000~\kms\ (with the exception of
three \OVI\ lines at $v$$<$400~\kms that are included in the tables and figures,
though not in statistical calculations).} At these velocities the galaxy sample
is basically complete down to 0.5~\,\Lstar. We also analyze a subsample, using
only absorbers with $v$$<$2500~\kms, where the galaxy sample is complete down to
0.1\,\Lstar. We look at \Lya\ (using 52 sightlines observed with \HST), and at
\Lyb\ and \OVI\ (using 63 sightlines observed with the {\it Far Ultraviolet
Spectroscopic Explorer}, \FUSE). Our galaxy sample combines the ``Third
Reference Catalogue of Galaxies'' (de Vaucouleurs et al.\ 1991) with all
galaxies within 5\deg\ of each AGN sightline listed in the {\it NASA
Extragalactic Database} (\NED: http://nedwww.caltech.ipac.edu). We also use the
group catalogues of Geller \& Huchra (1982, 1983) and Garcia (1993) to classify
galaxies as either field or group galaxies.
\par By concentrating on the very nearest galaxies and absorbers we can address
questions such as (1) What fraction of absorbers has a galaxy above a certain
luminosity near them? (2) What fraction of galaxies has associated \HI\ and/or
\OVI\ absorption, as function of impact parameter? (3) Do group and field
galaxies have the same or a different relation with the absorbers? (4) Is there
a relation between impact parameter and the parameters of the absorbers
(equivalent width, linewidth, difference in velocity with associated galaxy)?
(5) Are the properties of the lowest-redshift \OVI\ absorbers similar to those
at higher redshift, and if so, what does this imply for the association of \OVI\
with galaxies?
\par We describe our data and measurement methods in Sect.~\Sobs. In
Sects.~\Sabsresults, \Sassocresults\ and \Sgalresults\ we present our analyses.
In the first of these (Sect.~\Sabsresults), we study just the absorbers, without
reference to the galaxies near them. In Sect.~\Sassocresults\ we discuss the
statistics of the galaxies that can be found near the absorbers, as well as
individual galaxy-absorber associations. In Sect.~\Sgalresults\ we look at the
galaxies first and then determine the properties and statistics of the absorbers
found near them. In Sects.~\Sdisc\ and \Sconcl\ we discuss and summarize the
results.

\section{Observations}

\subsection{Absorption Line Data Origin}
\par The background targets were selected in the following manner. First, we
retrieved the data for the 421 extragalactic targets that were observed by the
{\it Far-Ultraviolet Spectroscopic Explorer} (\FUSE) (excluding stars in the
SMC, LMC, M31 and M33). In this sample, 106 targets have an S/N ratio per
resolution element $>$8 near 1031~\AA. The 53 of these that have recession
velocity $>$7500~\kms\ ($z$$>$0.025) were selected for the study in this paper.
Second, we searched the {\it Hubble Space Telescope} (\HST) archive for targets
observed with the {\it Goddard High Resolution Spectrograph} (\GHRS) or the {\it
Space Telescope Imaging Spectrograph} (\STIS) with spectrograph settings that
produce data with velocity resolution better than 30~\kms. There are 24 targets
with good \STIS-E140M data (S/N$>$5 per 6.5~\kms\ resolution element). For 20 of
these there is also good \FUSE\ data, while for four targets the \FUSE\ data
only has S/N$\sim$4 near 1031~\AA. There are 29 sightlines with good (S/N$>$5)
\STIS-G140M or \GHRS-G140M data, with a velocity resolution of 30 or 20~\kms,
respectively. For 15 of these \FUSE\ data with S/N$\sim$8 also exist, while for
7 targets the \FUSE\ data have low S/N. Eleven targets were only observed using
\STIS-G140M. Four final targets were added to the sample, even though they only
have \FUSE\ spectra with S/N$<$5, However, there are known galaxies with low
impact parameter ($<$150~kpc) and clear detections of \Lyb\ and/or \OVI\
absorption. These targets combine to make a sample size of 76.
\par We note that two targets are included that have redshifts below 7500~\kms.
ESO\,438-G09 has $v$=7200~\kms, while $v$(ESO\,185-IG13) is 5600~\kms. The first
of these is the only AGN observed with \STIS-G140M that has $v$$<$7500~\kms, and
there are absorbers at 1426 and 2215~\kms. So, rather than excluding it based on
its velocity, we decided to keep this target \fix{for some of the tables.
However, for the statistical analyses the low S/N ESO\,185$-$IG13 data were
automatically excluded, as we apply strict selection criteria.} ESO\,185-IG13 is
one of the targets with a low S/N \FUSE\ spectrum, but there is an absorber at
2635~\kms\ with low impact parameter (62~kpc). Since Tripp et al.\ (2008)
concluded that all absorbers with velocity differing by more than 2500~\kms\
from the redshift of the AGN are likely to be intergalactic, it is justified to
include ESO\,185-IG13 and ESO\,438-G09 in our sample.
\par The processing of the \FUSE\ data was described in detail by Wakker et al.\
(2003) and Wakker (2006), and therefore only a summary is given here. First, the
spectra were calibrated using v2.1 or v2.4 of the \FUSE\ calibration pipeline.
To correct for residual wavelength shifts, the central velocities of the Milky
Way interstellar lines were determined for each detector segment
(LiF1A/1B/2A/2B, SiC2A/2B) of each individual observation. The \FUSE\ segments
were then aligned with the LSR interstellar velocities implied by the
\STIS-E140M, or if no E140M data were available, with the LSR velocity of the
strongest component in the 21-cm \HI\ spectrum. For targets with a \STIS-E140M
spectrum, the interstellar reference velocity was determined by fitting all
Milky Way lines in that spectrum; the \STIS\ wavelength calibration is accurate
to about 1~\kms\ (Tripp et al.\ 2001, 2005). For sightlines with S/N$>$10 near
1031~\AA, the resulting shifts were given by Wakker (2006). Using these shifts,
LiF1A and LiF2B data are added together to produce the final spectrum for each
target. Although the data were aligned using an LSR velocity scale, we shifted
the spectrum to the heliocentric velocity scale to measure the intergalactic
absorption lines, as that is the convention for extragalactic studies
\par For \HST\ data, the calibrated fits files in the MAST archive were
retrieved. This is the only step needed, except for observations with the
\STIS-G140M grating and central wavelength 1222~\AA, where a 1-pixel shift seems
necessary. That conclusion is based on fourteen sightlines with good data and
relatively simple ISM absorption lines. For these sightlines the
\SII\lm1250.584, 1253.811 lines can be fit both in a G140M spectrum centered on
1222~\AA\ and in one centered on 1272~\AA, and sometimes also in an E140M
spectrum. To align the lines in the 1222~\AA-centered spectrum with those in the
other spectra, an average redward shift of 12~\kms\ is needed, which corresponds
to one pixel.
\par Finally, for observations with the \STIS-E140M echelle the MAST fits files
give the data for each of the 42 orders separately. These orders were combined
into a single spectrum by interpolating the photon counts and errors onto a
common grid, adding the photon counts (weighted by the rms at each pixel), and
converting back to a flux.
\par Using the final combined datasets, we measured the target flux and the S/N
ratio of each spectrum near 977, 1031 and 1238~\AA. Table \Tobs\ presents the
observation IDs, exposure times, fluxes and S/N ratios for all the targets in
the final sample. Table~\Tshift\ gives the \FUSE\ segment shifts for datasets
that were not included in Wakker (2006).
\def\FO{} \def\FF{\hbox to 0pt{$^a$\hss}} \def\FN{\hbox to 0pt{$^1$\hss}} \def\DN{\hbox to 0pt{$^2$\hss}}
\begin{deluxetable}{lrrrllrrrrrr}
\tablenum{1}
\tablewidth{0pt}
\tabletypesize{\scriptsize}
\tabcolsep=2pt
\tablecolumns{11}
\tablecaption{Target Exposure List}
\tablehead{%
\ch{object}&\ch{lon}      &\ch{lat}      &\ch{$z$}&\ch{type}&\ch{datasets$^1$}&\ch{T$_{\rm exp}^2$}&\ch{Flux$^3$}&\ch{S/N$^4$}&\ch{S/N$^4$}&\ch{S/N$^4$}\\
           &\ch{[$\circ$]}&\ch{[$\circ$]}&        &         &                 &\ch{[ks]}           &\ch{[f.u.]}  &\ch{1031}   &\ch{977}    &\ch{1238}   \\
\ch{(1)}&\ch{(2)}&\ch{(3)}&\ch{(4)}&\ch{(5)}&\ch{(6)}&\ch{(7)}&\ch{(8)}&\ch{(9)}&\ch{(10)}&\ch{(11)}
}\startdata
1H0419$-$577       & 266.99 & $-$42.00 & 0.1040 & Sey1    & D8080801, F0260101, F0260102 &  36.8 &      2.8 &    7 &    2 &    $-$ \\
1H0707$-$495       & 260.17 & $-$17.67 & 0.0411 & Sey1    & B1050101, B1050102, B1050103, E1190101 & 114.4 &      1.8 &   16 &    6 &    $-$ \\
1H0717+714       & 143.98 &  28.02 & 0.5000 & BLLac   & Z9071301, F0260302 &  70.0 &      3.2 &   15 &    6 &    $-$ \\
3C232            & 194.17 &  52.32 & 0.5305 & QSO     & O67002(Ga)       &  11.2 &      0.5 &    2 &    0 &   12 \\
3C249.1          & 130.39 &  38.55 & 0.3115 & QSO     & P1071601, P1071602, P1071603, S6010901 & 246.5 &      1.1 &   15 &    3 &    8 \\
                 &        &        &        &         & D1170101, D1170102, D1170103, U1027501 &       &     &     &     &     \\
                 &        &        &        &         & U1027502 &       &     &     &     &     \\
                 &        &        &        &         & O6E124$-$30(E)     &  68.8 &      1.1 &   15 &    3 &    8 \\
3C263            & 134.16 &  49.74 & 0.6460 & QSO     & E8480701, D8081701, G0440201, G0440202 & 220.1 &      1.2 &   14 &    3 &    $-$ \\
                 &        &        &        &         & G0440203, F0050101, F0050103, F0050104 &       &     &     &     &     \\
                 &        &        &        &         & F0050105 &       &     &     &     &     \\
3C273.0          & 289.95 &  64.36 & 0.1583 & QSO     & P1013501 &  43.2 &     26.9 &   31 &   19 &   27 \\
                 &        &        &        &         & O5D301(E)        &  18.7 &     26.9 &   31 &   19 &   27 \\
3C351.0          &  90.08 &  36.38 & 0.3719 & QSO     & O57901$-$04(E)     &  77.0 &      0.0 &    0 &   $-$0 &    8 \\
ESO141$-$G55       & 338.18 & $-$26.71 & 0.0360 & Sey1    & I9040104 &  40.6 &      5.2 &   16 &    8 &   20 \\
                 &        &        &        &         & Z3E702           &  15.6 &      5.2 &   16 &    8 &   20 \\
ESO185$-$IG13      & 343.64 & $-$29.37 & 0.0187 & HII     & Z9091401, G0200201, G0200203, G0200204 &  19.9 &      1.5 &    5 &    2 &    $-$ \\
ESO438$-$G09       & 277.55 &  29.36 & 0.0240 & Sey1.5  & O5EW06(Ga)       &  11.3 &      0.3 &    $-$ &    $-$ &   11 \\
Fairall 9         & 295.07 & $-$57.83 & 0.0470 & Sey1    & P1010601 &  33.9 &      1.5 &    6 &    3 &   19 \\
                 &        &        &        &         & Z3E704           &  14.4 &      1.5 &    6 &    3 &   19 \\
                 &        &        &        &         & Z26O02           &   8.1 &      1.5 &      &      &      \\
                 &        &        &        &         & Z3E704           &   6.9 &      1.5 &      &      &      \\
H1821+643        &  94.00 &  27.42 & 0.2844 & QSO     & P1016402, P1016405, C0950201, C0950202 & 276.5 &      3.0 &   29 &    8 &   13 \\
                 &        &        &        &         & O5E703$-$04(E)     &  50.9 &      3.0 &   29 &    8 &   13 \\
HE0226$-$4110      & 253.94 & $-$65.77 & 0.4950 & QSO     & P2071301, P1019101, P1019102, P1019103 & 206.9 &      2.7 &   25 &   11 &    9 \\
                 &        &        &        &         & P1019104, D0270101, D0270102, D0270103 &       &     &     &     &     \\
                 &        &        &        &         & O6E107$-$11(E)     &  43.8 &      2.7 &   25 &   11 &    9 \\
HE0340$-$2703      & 222.68 & $-$52.12 & 0.2830 & QSO     & O8EI03(Ga)       &   4.9 &      0.6 &    $-$ &    $-$ &   10 \\
HE1029$-$1401      & 259.33 &  36.52 & 0.0860 & QSO     & O4EC05(Ga)       &   4.1 &      5.1 &    $-$ &    $-$ &   28 \\
                 &        &        &        &         & O4EC05(Gb)       &   3.4 &      5.1 &      &      &      \\
HE1143$-$1810      & 281.85 &  41.71 & 0.0329 & Sey1    & P1071901 &   7.2 &      6.1 &    8 &    4 &    $-$ \\
HE1228+0131      & 291.26 &  63.66 & 0.1170 & QSO     & P1019001 &   4.0 &      4.9 &    5 &    2 &    5 \\
                 &        &        &        &         & O56A01$-$02(E)     &  27.2 &      4.9 &    5 &    2 &    5 \\
HS0624+6907      & 145.71 &  23.35 & 0.3700 & QSO     & P1071001, P1071002, S6011201, S6011202 & 113.5 &      1.0 &   11 &    3 &    7 \\
                 &        &        &        &         & O6E112$-$16(E)     &  62.0 &      1.0 &   11 &    3 &    7 \\
HS1543+5921      &  92.40 &  46.36 & 0.8070 & QSO     & O8MR01(Ga)       &  25.3 &      0.8 &    2 &    0 &   10 \\
                 &        &        &        &         & O8MR02$-$04(Gb)    &  28.8 &      0.8 &      &      &      \\
IRAS09149$-$6206   & 280.61 &  $-$9.20 & 0.0573 & Sey1    & A0020503, S7011002, S7011003, U1072201 & 100.0 &      1.7 &   13 &    2 &    $-$ \\
                 &        &        &        &         & U1072202, U1072203 &       &     &     &     &     \\
IRAS F22456$-$5125 & 338.51 & $-$56.63 & 0.1000 & Sey1    & Z9073901, Z9073902, E8481401 &  41.5 &      1.9 &    9 &    3 &    $-$ \\
MCG+10$-$16$-$111    & 144.21 &  55.08 & 0.0271 & Sey1    & O5EW02(Ga)       &  19.5 &      1.5 &    $-$ &    $-$ &   23 \\
MRC2251$-$178      &  46.20 & $-$61.33 & 0.0661 & QSO     & P1111010 &  51.4 &      2.1 &   12 &    3 &   30 \\
                 &        &        &        &         & O4EC03(Ga)       &   5.9 &      2.1 &   12 &    3 &   30 \\
                 &        &        &        &         & O4EC03(Gb)       &   4.6 &      2.1 &      &      &      \\
Mrk 9             & 158.36 &  28.75 & 0.0399 & Sey1.5  & P1071101, P1071102, P1071103, S6011601 &  52.4 &      2.3 &   13 &    2 &    $-$ \\
Mrk 106           & 161.14 &  42.88 & 0.1235 & Sey1    & C1490501 & 121.9 &      1.6 &   12 &    3 &    $-$ \\
Mrk 110           & 165.01 &  44.36 & 0.0353 & Sey1    & O4N352(Ga)       &   2.2 &      0.3 &    1 &    1 &   11 \\
Mrk 205           & 125.45 &  41.67 & 0.0708 & Sey1    & Q1060203, S6010801, D0540101, D0540102 & 223.8 &      1.2 &   17 &    6 &    8 \\
                 &        &        &        &         & D0540103, U1031102 &       &     &     &     &     \\
                 &        &        &        &         & O62Q03$-$05(E)     &  62.1 &      1.2 &   17 &    6 &    8 \\
Mrk 279           & 115.04 &  46.86 & 0.0305 & Sey1.5  & P1080303, P1080304, D1540101 & 181.8 &     10.2 &   45 &   21 &   32 \\
                 &        &        &        &         & O6JM01(E)        &  13.2 &     10.2 &   45 &   21 &   32 \\
                 &        &        &        &         & O8K101$-$05(E)     &  41.4 &     10.2 &      &      &      \\
Mrk 290           &  91.49 &  47.95 & 0.0296 & Sey1    & P1072901, D0760101, D0760102, E0840101 &  90.6 &      3.2 &   19 &    6 &   12 \\
                 &        &        &        &         & E0840102 &       &     &     &     &     \\
                 &        &        &        &         & Z3KH01           &   7.1 &      3.2 &   19 &    6 &   12 \\
Mrk 335           & 108.76 & $-$41.42 & 0.0258 & Sey1.2  & P1010203, P1010204 &  83.7 &      7.1 &   27 &   11 &   11 \\
                 &        &        &        &         & O8N504$-$05(E)     &  15.2 &      7.1 &   27 &   11 &   11 \\
Mrk 421           & 179.83 &  65.03 & 0.0300 & BLLac   & P1012901, Z0100101, Z0100102, Z0100103 &  83.7 &      9.6 &   30 &   14 &   23 \\
                 &        &        &        &         & Z2IA01           &  15.7 &      9.6 &   30 &   14 &   23 \\
Mrk 477           &  93.04 &  56.82 & 0.0378 & Gal     & D1180101 & 146.3 &      1.0 &   10 &    2 &    $-$ \\
Mrk 478           &  59.24 &  65.03 & 0.0791 & Sey1    & P1110909 &  14.0 &      3.5 &    8 &    5 &   37 \\
                 &        &        &        &         & O4EC14(Ga)       &   7.9 &      3.5 &    8 &    5 &   37 \\
                 &        &        &        &         & O4EC14(Gb)       &   6.3 &      3.5 &      &      &      \\
Mrk 501           &  63.60 &  38.86 & 0.0337 & BLLac   & P1073301, C0810101 &  29.8 &      3.1 &   10 &    3 &   11 \\
                 &        &        &        &         & Z1A652           &  31.3 &      3.1 &   10 &    3 &   11 \\
Mrk 509           &  35.97 & $-$29.86 & 0.0344 & Sey1.2  & X0170101, X0170102, P1080601 & 114.3 &      6.7 &   30 &    7 &   14 \\
Mrk 586           & 157.60 & $-$54.93 & 0.1553 & Sey1    & D0550101, D0550102 &  63.6 &      2.1 &   12 &    4 &    $-$ \\
Mrk 734           & 244.75 &  63.94 & 0.0502 & Sey1    & P1071702 &   4.9 &      4.0 &    6 &    3 &    $-$ \\
Mrk 771           & 269.44 &  81.74 & 0.0630 & Sey1    & P1072301 &   6.3 &      2.2 &    5 &    2 &   25 \\
                 &        &        &        &         & O4N305(Ga)       &   2.0 &      2.2 &    5 &    2 &   25 \\
                 &        &        &        &         & O4EC07(Ga)       &   5.8 &      2.2 &      &      &      \\
                 &        &        &        &         & O4EC07(Gb)       &   5.2 &      2.2 &      &      &      \\
Mrk 817           & 100.30 &  53.48 & 0.0315 & Sey1.5  & P1080403, P1080404 & 161.5 &      9.6 &   44 &   20 &   43 \\
                 &        &        &        &         & Z3E701           &  26.8 &      9.6 &   44 &   20 &   43 \\
Mrk 876           &  98.27 &  40.38 & 0.1290 & Sey1    & P1073101, D0280203 & 127.4 &      6.3 &   34 &   11 &   11 \\
                 &        &        &        &         & O8NN01$-$02(E)     &  29.2 &      6.3 &   34 &   11 &   11 \\
Mrk 926           &  64.09 & $-$58.76 & 0.0473 & Sey1.5  & O4EC12(Ga)       &   3.9 &      0.7 &    3 &    1 &    9 \\
                 &        &        &        &         & O4EC12(Gb)       &   3.8 &      0.7 &      &      &      \\
Mrk 1095          & 201.69 & $-$21.13 & 0.0323 & Sey1    & P1011201, P1011202, P1011203 &  55.8 &      2.0 &   11 &    4 &   18 \\
                 &        &        &        &         & Z3E706           &  14.9 &      2.0 &   11 &    4 &   18 \\
Mrk 1383          & 349.22 &  55.12 & 0.0865 & Sey1    & P1014801, P2670101 &  63.3 &      6.6 &   24 &   12 &    9 \\
                 &        &        &        &         & O8PG01$-$02(E)     &  19.2 &      6.6 &   24 &   12 &    9 \\
Mrk 1513          &  63.67 & $-$29.07 & 0.0630 & Sey1    & P1018301, P1018302, P1018303, P1018304 &  58.6 &      3.9 &   16 &    6 &   20 \\
                 &        &        &        &         & O4EC10(Ga)       &   7.3 &      3.9 &   16 &    6 &   20 \\
                 &        &        &        &         & O4EC10(Gb)       &   6.2 &      3.9 &      &      &      \\
MS0700.7+6338    & 152.47 &  25.63 & 0.1530 & Sey1    & P2072701, S6011501, D0550501, U1021403 & 181.9 &      1.6 &   16 &    4 &    $-$ \\
                 &        &        &        &         & U1021404 &       &     &     &     &     \\
NGC985           & 180.84 & $-$59.49 & 0.0431 & Sey1    & P1010903 &  50.6 &      3.4 &   11 &    5 &   29 \\
                 &        &        &        &         & O4EC11(Ga)       &   3.7 &      3.4 &   11 &    5 &   29 \\
                 &        &        &        &         & O4EC11(Gb)       &   3.8 &      3.4 &      &      &      \\
PG0804+761       & 138.28 &  31.03 & 0.1020 & QSO     & P1011901, P1011903, S6011001, S6011002 & 151.5 &      7.0 &   31 &   14 &   35 \\
                 &        &        &        &         & O4N301(Ga)       &   2.4 &      7.0 &   31 &   14 &   35 \\
                 &        &        &        &         & O4EC06(Ga)       &   4.9 &      7.0 &      &      &      \\
                 &        &        &        &         & O4EC06(Gb)       &   4.2 &      7.0 &      &      &      \\
PG0838+770       & 136.66 &  32.68 & 0.1310 & Sey1    & G0200104, G0200105, G0200106, G0200107 & 100.5 &      0.7 &    6 &    2 &    $-$ \\
PG0844+349       & 188.56 &  37.97 & 0.0640 & Sey1    & P1012002, D0280301, D0280302, D0280303 &  81.2 &      3.7 &   20 &    7 &    $-$ \\
                 &        &        &        &         & D0280304 &       &     &     &     &     \\
PG0953+414       & 179.79 &  51.71 & 0.2341 & QSO     & P1012201, P1012202 &  74.4 &      5.2 &   24 &   10 &   10 \\
                 &        &        &        &         & O63G01$-$04(E)     &  26.9 &      5.2 &   24 &   10 &   10 \\
PG1001+291       & 200.08 &  53.21 & 0.3297 & QSO     & P2073101 &  11.3 &      2.0 &    5 &    2 &    8 \\
                 &        &        &        &         & O6E117$-$23(E)     &  48.4 &      2.0 &    5 &    2 &    8 \\
PG1011$-$040       & 246.50 &  40.75 & 0.0580 & Sey1    & B0790101 &  85.3 &      2.6 &   17 &    7 &    $-$ \\
PG1049$-$005       & 252.28 &  49.88 & 0.3599 & QSO     & O4N303(Ga)       &   1.5 &      1.0 &    $-$ &    $-$ &    7 \\
PG1116+215       & 223.36 &  68.21 & 0.1765 & QSO     & P1013101, P1013102, P1013103, P1013104 &  76.7 &      5.7 &   25 &   12 &   12 \\
                 &        &        &        &         & P1013105 &       &     &     &     &     \\
                 &        &        &        &         & O5E701$-$02(E)     &  26.5 &      5.7 &   25 &   12 &   12 \\
PG1149$-$110       & 280.47 &  48.89 & 0.0490 & Sey     & O5EW05(Ga)       &   8.3 &      0.1 &    $-$ &    $-$ &    5 \\
PG1211+143       & 267.55 &  74.32 & 0.0804 & Sey1    & P1072001 &  52.2 &      5.2 &   17 &   10 &   18 \\
                 &        &        &        &         & O61Y01$-$08(E)     &  42.5 &      5.2 &   17 &   10 &   18 \\
PG1216+069       & 281.07 &  68.14 & 0.3313 & QSO     & P1072101 &  12.4 &      1.4 &    5 &    2 &    7 \\
                 &        &        &        &         & O6E131$-$39(E)     &  69.8 &      1.4 &    5 &    2 &    7 \\
PG1259+593       & 120.56 &  58.05 & 0.4778 & QSO     & P1080101, P1080102, P1080103, P1080104 & 553.8 &      1.8 &   37 &   14 &    8 \\
                 &        &        &        &         & P1080105, P1080106, P1080107, P1080108 &       &     &     &     &     \\
                 &        &        &        &         & P1080109, U1031801 &       &     &     &     &     \\
                 &        &        &        &         & O63G05$-$11(E)     &  95.8 &      1.8 &   37 &   14 &    8 \\
PG1302$-$102       & 308.59 &  52.16 & 0.2784 & QSO     & P1080201, P1080202, P1080203 & 145.9 &      1.6 &   16 &    5 &    4 \\
                 &        &        &        &         & O5BU61$-$02(E)     &  22.1 &      1.6 &   16 &    5 &    4 \\
PG1341+258       &  28.71 &  78.15 & 0.0870 & QSO     & O5EW01(Ga)       &   8.1 &      0.9 &    $-$ &    $-$ &   16 \\
PG1351+640       & 111.89 &  52.02 & 0.0882 & Sey1    & P1072501, S6010701 & 118.4 &      1.6 &   16 &    4 &   13 \\
                 &        &        &        &         & O4EC54(Ga)       &   8.5 &      1.6 &   16 &    4 &   13 \\
                 &        &        &        &         & O4EC54(Gb)       &   6.3 &      1.6 &      &      &      \\
PG1444+407       &  69.90 &  62.72 & 0.2673 & QSO     & P1072701 &  10.0 &      1.8 &    4 &    3 &    9 \\
                 &        &        &        &         & O6E101$-$06(E)     &  48.6 &      1.8 &    4 &    3 &    9 \\
PG1553+113       &  21.91 &  43.96 & 0.3600 & BLLac   & E5260501, E5260502, E5260503 &  47.1 &      2.7 &   12 &    4 &    $-$ \\
PG1626+554       &  84.51 &  42.19 & 0.1330 & Sey1    & C0370101 &  90.9 &      1.5 &   14 &    0 &    $-$ \\
PHL1811          &  47.47 & $-$44.81 & 0.1920 & QSO     & P2071101, P1081001, P1081002, P1081003 &  75.7 &      4.9 &   17 &    0 &   11 \\
                 &        &        &        &         & O8D901$-$04(E)     &  33.9 &      4.9 &   17 &    0 &   11 \\
PKS0405$-$12       & 204.93 & $-$41.76 & 0.5726 & QSO     & B0870101, D1030101, D1030102 & 140.3 &      2.2 &   19 &    7 &    8 \\
                 &        &        &        &         & O55S01$-$02(E)     &  27.2 &      2.2 &   19 &    7 &    8 \\
PKS0558$-$504      & 257.96 & $-$28.57 & 0.1370 & QSO     & P1011504, C1490601 &  93.0 &      3.3 &   19 &    9 &    $-$ \\
PKS2005$-$489      & 350.37 & $-$32.60 & 0.0710 & BLLac   & P1073801, C1490301, C1490302 &  48.7 &      5.0 &   18 &    8 &   57 \\
                 &        &        &        &         & O4EC09(Ga)       &   6.1 &      5.0 &   18 &    8 &   57 \\
                 &        &        &        &         & O4EC09(Gb)       &   5.4 &      5.0 &      &      &      \\
PKS2155$-$304      &  17.73 & $-$52.25 & 0.1160 & BLLac   & P1080701, P1080705, P1080703 & 121.5 &     12.8 &   31 &   12 &   12 \\
                 &        &        &        &         & O5BY01$-$02(E)     &  28.5 &     12.8 &   31 &   12 &   12 \\
RX J0048.3+3941  & 122.28 & $-$23.18 & 0.1340 & QSO     & D1310101, D1310102, D1310103, D1310104 & 191.3 &      1.4 &   14 &    4 &    $-$ \\
                 &        &        &        &         & D1310105, D1310106, D1310107 &       &     &     &     &     \\
RX J0100.4$-$5113  & 299.48 & $-$65.84 & 0.0620 & Sey1    & D8060301, E8970201 &  23.0 &      3.1 &   10 &    3 &   14 \\
                 &        &        &        &         & O8P802(Ga)       &   2.3 &      3.1 &   10 &    3 &   14 \\
                 &        &        &        &         & O8P802(Gb)       &   1.1 &      3.1 &      &      &      \\
RX J1830.3+7312  & 104.04 &  27.40 & 0.1230 & Sey1    & G0200302 &  25.4 &      3.5 &    6 &    0 &   25 \\
                 &        &        &        &         & O5EW09(Ga)       &   5.8 &      3.5 &    6 &    0 &   25 \\
Ton S180         & 139.00 & $-$85.07 & 0.0620 & Sey1.2  & P1010502, D0280101 &  26.9 &      6.3 &   13 &    6 &   32 \\
Ton S210         & 224.97 & $-$83.16 & 0.1160 & QSO     & P1070301, P1070302 &  52.7 &      6.2 &   20 &    8 &    4 \\
                 &        &        &        &         & O6L001$-$02(E)     &  17.3 &      6.2 &   20 &    8 &    4 \\
VIIZw118         & 151.36 &  25.99 & 0.0797 & Sey1    & P1011604, P1011605, P1011606, S6011301 & 129.8 &      2.0 &   18 &    6 &   19 \\
                 &        &        &        &         & O4EC13(Ga)       &   9.5 &      2.0 &   18 &    6 &   19 \\
\enddata
\tablecomments{
1: This column identifies the \FUSE\ and \HST\ datasets; \FUSE\ datasets consist of an 8-character code giving the observing program, object id, and exposure number;
   \HST\ datasets list a 4-character program id, followed by a 2-digit object id and a 3-digit observation id, which we omit;
   for \HST\ \STIS\ observations an extra identifier between parentheses shows whether the data were taken using the G140M grating centered at 1222~\AA\ (Ga) or 1272~\AA\ (Gb) or with the E140M echelle
   (E); datasets starting with ``Z'' were obtained with the \GHRS.
2: Exposure time in kiloseconds, given separately for each \FUSE\ exposure (which corresponds to several orbits); only a single value is given for each multi-orbit \HST\ exposure.
3: Flux in units of 10$^{-14}$\,erg\,cm$^{-2}$\,s$^{-1}$\,\AA$^{-1}$; for \FUSE\ datasets this is the flux at 1031~\AA, while for targets with only \HST\ datasets it is the flux at 1238~\AA.
4: Signal-to-noise ratio per resolution element at 1031, 977 and 1238~\AA; for \FUSE\ data the resolution element is 20~\kms, while for \HST\ data it depends on the
   instrument and grating: 20~\kms\ for \GHRS\ spectra, 30~\kms\ for \STIS\ grating (G140M) exposures, and 6.5~\kms\ for \STIS\ echelle (E140M) data.
}
\end{deluxetable}
\begin{deluxetable}{llrrrrrrrr}
\tablenum{2}
\tablewidth{0pt}
\tabletypesize{\footnotesize}
\tabcolsep=3pt
\tablecolumns{10}
\tablecaption{\FUSE\ Velocity Shifts for Exposures not in Wakker (2006)}
\tablehead{%
\ch{object}      & \ch{dataset} & \ch{Cal}  & \ch{T$_{\rm exp}$}& \multicolumn{6}{c}{shifts} \\
                 &              & \ch{FUSE} & \ch{[ks]} & \ch{LiF1A} & \ch{LiF1B} & \ch{LiF2A} & \ch{LiF2B} & \ch{SiC2A} & \ch{SiC2B} \\
\ch{(1)}&\ch{(2)}&\ch{(3)}&\ch{(4)}&\ch{(5)}&\ch{(6)}&\ch{(7)}&\ch{(8)}&\ch{(9)}&\ch{(10)}
}\startdata
1H\,0419$-$577     & D8080801     &    2.4    &    4.7    &     $-$22    &     $-$22    &     $-$22    &     $-$22    &     $-$48    &     $-$48    \\
                 & F0260101     &    2.4    &   16.9    &            &            &      16    &      27    &       0    &            \\
                 & F0260102     &    2.4    &   15.2    &            &            &      16    &      27    &       0    &            \\
1H\,0707$-$495     & E1190101     &    2.4    &   48.2    &     $-$19    &     $-$19    &     $-$19    &     $-$19    &     $-$13    &     $-$13    \\
3C\,249.1        & U1027501     &    2.4    &   13.2    &      14    &      14    &            &      58    &       0    &            \\
                 & U1027502     &    2.4    &   16.9    &      14    &      14    &            &      58    &       0    &            \\
3C\,263          & E8480701     &    2.4    &    7.1    &       9    &       9    &      42    &      42    &      42    &      42    \\
                 & D8081701     &    2.4    &    3.3    &       9    &       9    &      42    &      42    &      42    &      42    \\
                 & G0440201     &    2.4    &   14.7    &       9    &       9    &      42    &      42    &      42    &      42    \\
                 & G0440202     &    2.4    &   14.7    &       9    &       9    &      42    &      42    &      42    &      42    \\
                 & G0440203     &    2.4    &   20.1    &       9    &       9    &      42    &      42    &      42    &      42    \\
                 & F0050101     &    2.4    &   79.2    &      25    &      25    &      54    &      54    &      42    &      42    \\
                 & F0050103     &    2.4    &   17.3    &            &            &      54    &      54    &      42    &      42    \\
                 & F0050104     &    2.4    &   40.1    &            &            &      54    &      54    &      42    &      42    \\
                 & F0050105     &    2.4    &   20.7    &            &            &      54    &      54    &      42    &      42    \\
ESO\,185$-$IG13    & Z9091401     &    2.4    &   12.8    &       0    &        0   &       0    &       0    &       0    &       0    \\
                 & G0200201     &    3.2    &    2.1    &       0    &        0   &       0    &       0    &            &       0    \\ 
                 & G0200203     &    3.2    &    2.6    &       0    &        0   &            &            &            &            \\ 
                 & G0200204     &    3.2    &    2.4    &       0    &        0   &       0    &       0    &            &       0    \\ 
HE\,1143$-$1810    & P1071901     &    2.4    &    7.3    &       5    &        5   &       5    &       5    &      $-$2    &      $-$2    \\
HE\,1228+0131    & P1019001     &    2.4    &    4.0    &      17    &       17   &       6    &       6    &      10    &      10    \\
IRAS\,09149$-$6206 & A0020503     &    2.4    &   13.1    &      $-$6    &       $-$6   &     $-$19    &      $-$6    &            &            \\
                 & S7011002     &    2.4    &   13.6    &      $-$6    &       $-$6   &     $-$12    &            &            &            \\
                 & S7011003     &    2.4    &   14.1    &     $-$12    &      $-$12   &     $-$12    &            &            &            \\
                 & U1072202     &    2.4    &   39.1    &       7    &        7   &      35    &      43    &            &            \\
                 & U1072203     &    2.4    &   16.2    &     $-$12    &      $-$12   &      35    &      43    &      $-$5    &      $-$5    \\
IRAS\,F22456$-$5125& Z9073901     &    2.4    &    5.7    &     $-$11    &      $-$11   &     $-$11    &     $-$11    &      20    &      20    \\
                 & Z9073902     &    2.4    &   31.6    &     $-$11    &      $-$11   &     $-$11    &     $-$11    &      10    &      10    \\
                 & E8481401     &    2.4    &    4.2    &      $-$1    &       10   &      10    &      $-$1    &      10    &      10    \\
Mrk\,205         & U1031102     &    3.2    &   15.9    &       7    &        7   &      58    &      58    &            &            \\
Mrk\,478         & P1110909     &    2.1    &   14.0    &      34    &       34   &      34    &      34    &      34    &      34    \\
Mrk\,734         & P1071702     &    2.4    &    4.9    &      $-$4    &       04   &      $-$9    &      $-$4    &     $-$31    &     $-$31    \\
MS\,0700.7+6338  & U1021403     &    2.4    &   23.1    &      17    &       17   &      42    &      42    &       0    &       0    \\
                 & U1021404     &    2.4    &   54.3    &      17    &       17   &      42    &      42    &       0    &       0    \\
PG\,0838+770     & G0200101     &    2.4    &    5.9    &      29    &       29   &      67    &      67    &      29    &      29    \\
                 & G0200104     &    2.4    &   31.9    &      29    &       29   &      67    &      67    &      29    &      29    \\
                 & G0200105     &    2.4    &   28.8    &      29    &       29   &      67    &      67    &      29    &      29    \\
                 & G0200106     &    2.4    &   20.7    &      29    &       29   &      67    &      67    &      29    &      29    \\
                 & G0200107     &    2.4    &   19.0    &      29    &       29   &      67    &      67    &      29    &      29    \\
PG\,1001+291     & P2073101     &    2.4    &   11.3    &      36    &       36   &      36    &      36    &      36    &      36    \\
PG\,1444+407     & P1072701     &    2.4    &   10.0    &      32    &       32   &      32    &      32    &      32    &      32    \\
RX\,J0048.3+3941 & D1310101     &    2.4    &   37.6    &      27    &       27   &      31    &      31    &      46    &      46    \\
                 & D1310102     &    2.4    &   32.9    &      27    &       27   &      31    &      31    &      46    &      46    \\
                 & D1310103     &    2.4    &    6.6    &      27    &       25   &      31    &      31    &      46    &      46    \\
                 & D1310104     &    2.4    &   39.3    &      25    &       25   &      13    &      13    &      34    &      34    \\
                 & D1310105     &    2.4    &   24.4    &      25    &       25   &      13    &      13    &      34    &      34    \\
                 & D1310106     &    2.4    &   26.6    &      25    &       25   &      13    &      13    &      34    &      34    \\
                 & D1310107     &    2.4    &   23.9    &      25    &       25   &      13    &      13    &      34    &      34    \\
RX\,J0100.4$-$5113 & D8060301     &    2.4    &    5.8    &       0    &        0   &       0    &       0    &       0    &       0    \\
                 & E8970201     &    2.4    &   17.2    &       0    &        0   &       0    &       0    &       0    &       0    \\
RX\,J1830.3+7312 & G0200302     &    2.4    &   25.4    &            &            &      30    &      30    &            &            \\
\enddata
\end{deluxetable}

\subsection{Absorption Line Identification}
\par For each background target, we began by identifying all intrinsic and
intergalactic hydrogen and metal lines, looking for redshifted Lyman series
lines and high-ionization lines such as those of \OVI, \OIII, \NV, \NIV, \NIII,
\CIV, \CIII, \SiIII, \SVI, as well as low-ionization lines, if appropriate. We
also identified all high- and low-ionization Milky Way and high-velocity cloud
(HVC) ionic and molecular absorption lines. For each background target, we
modeled all \H2\ absorption lines, using the method presented by Wakker (2006).
A further complication is the contamination by the geocoronal \OI* emission
lines near \Lyb, \Lyg, \OVIa\, \OVIb\ and \CIII\lm977.020. The geocoronal lines
are absent in \FUSE\ data taken during the orbital night, but for many
sightlines the orbital-night-only data have much lower S/N ratios.
\begin{figure}\plotfiddle{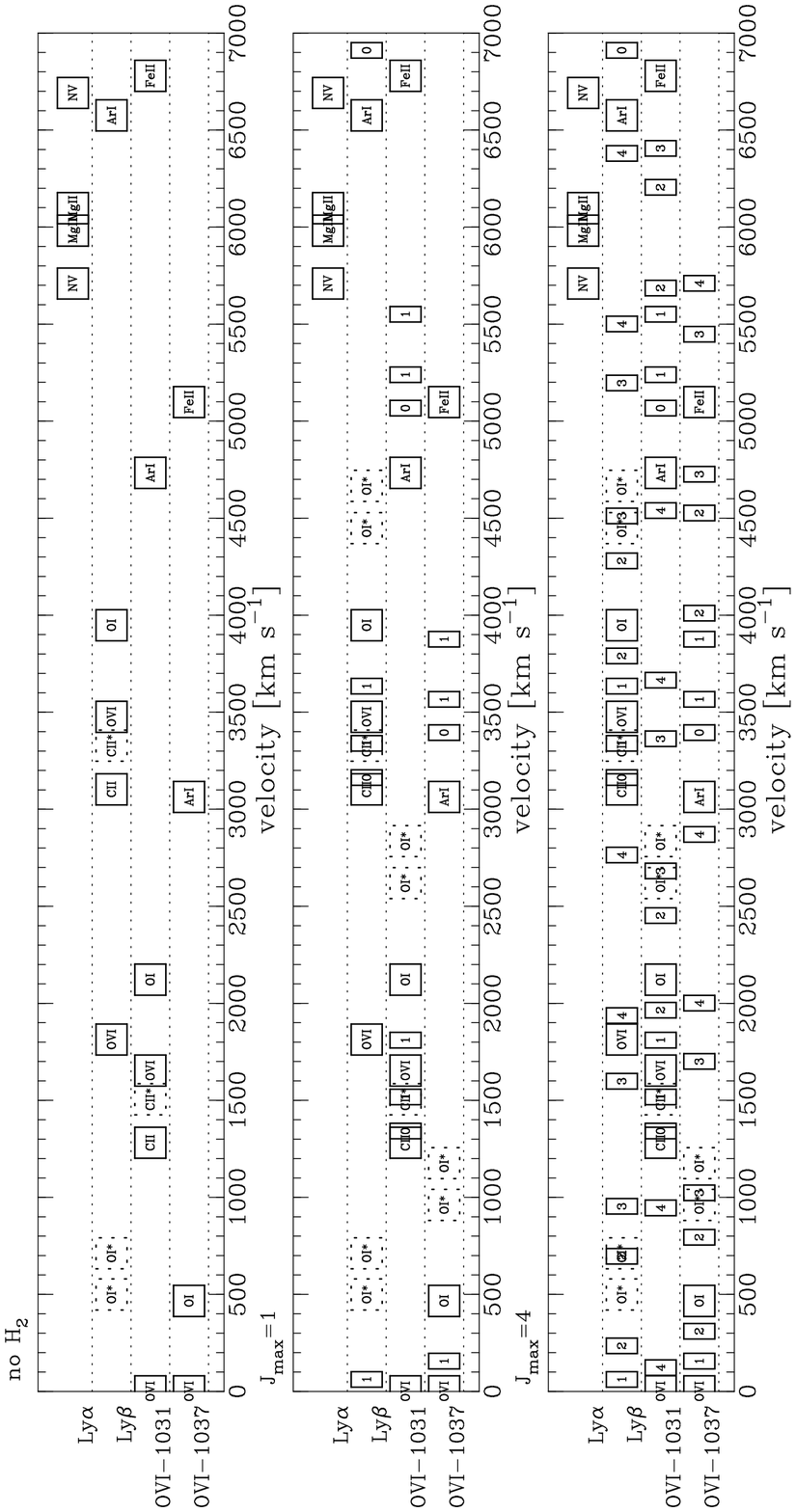}{0in}{270}{250}{480}{0}{-350}\figurenum{1}\caption{\captionokranges}\end{figure}
\par Figure~\Fokranges\ shows the velocities of the interstellar lines relative
to \Lya, \Lyb\ and the two \OVI\ lines. The three panels are for three different
levels of contamination by Galactic molecular hydrogen (\H2): no \H2\ with good
night-only data, median contamination with just $J$=0 and 1 \H2\ lines, and
strong \H2\ (lines up to $J$=4 are seen). A box that is 80~\kms\ wide is drawn
around each ISM line. This is a typical absorption width, though it will be
different in detail for each sightline.
\par If there are no \H2\ lines, both \OVI\ lines are visible for velocities of
about 500 to 1200~\kms\ and above 2300~\kms\ (except for a few regions
contaminated by interstellar \ArI\lm1048.220 and \FeII\lm1063.176). More
typical, however, is the situation in the middle panel, where some H$_2$ is
present, but useful orbital-night-only data do exist. In that case, \Lyb\ can be
seen over most of the velocity range between 500 and 7000~\kms, except between
about 3000 and 3700~\kms, where \CII\lm1036.337, \OVIb\ and \OI\lm1039.230
interfere. The most-easily visible \OVI\ line alternates between the 1031 and
the 1037 line, with \OVIa\ mostly uncontaminated in the velocity ranges
500--1200~\kms\ and 3000--7000~\kms, while the \OVIb\ line is clear from
1300--3000~\kms\ and above 4000~\kms. The bottom panel represents the worst
case. It is clear that in this case intergalactic lines will only be visible if
they occur at just the right velocity.
\par With the intrinsic, interstellar and geocoronal contamination in mind, we
checked whether any of the identified intergalactic absorbers occur near the
systemic velocity of each galaxy with impact parameter \ip$<$1~Mpc. If we did
not find intergalactic absorption, we measured equivalent width upper limits and
noted these. If we did find intergalactic absorption having a small velocity
difference with a galaxy, we measured the line(s) and decided whether or not to
associate the galaxy with the absorber. These associations are discussed in more
detail in Sect.~\SSassoc. Between the lines from interstellar ions, interstellar
\H2, intrinsic lines and intergalactic absorption-line systems there are only a
few marginally significant features that have remained unidentified in the
\FUSE\ data. They do not match known \Lya\ lines, nor can they be \Lyb\ as no
corresponding \Lya\ is seen.

\subsection{Absorption Line Measurements}
\par For each background target, we determined a continuum for each absorption
line that we looked at by fitting a low-order (up to 4th) polynomial through
line-free regions, using the method described by Sembach \& Savage (1992). For
most sightlines the velocity range of the fit is about 5000~\kms, but for
complex spectra, the continuum fit is made in several pieces. In the case of
\Lya, we fitted the continuum through the parts of the spectrum outside the
damping wings of the Galactic \Lya\ line, and then we modeled the ISM \Lya\
Voigt profile. The detailed results of this effort will be reported in a future
paper (Brown et al., in preparation).
\par To measure the absorption lines, we integrated over the velocity range
where absorption is apparent, using the fitted continua as a reference. We also
determined an error estimate, which consists of two parts, a statistical and a
systematic error, listed separately in Cols.~11--14 of Table~\Tres. The
determination of these errors was described in detail by Wakker et al.\ (2003).
The statistical error combines the random noise in the data with the uncertainty
associated with the placement of the continuum. The systematic error combines
the uncertainty associated with the choice for the minimum and maximum velocity
of the absorption (i.e.\ the change in measured equivalent width when changing
the velocity range by \E5~\kms) with the fixed-pattern noise (6~\mA\ for \FUSE\
data, 1.2~\mA\ for \STIS-G140M and 0.3~\mA\ for \STIS-E140M observations). In
addition to measuring the equivalent width by straight integration, the central
velocity and line width of the absorption line are estimated by fitting a
gaussian to the observed absorption profile.
\par We also note non-detections associated with galaxies (see next subsection).
For these we determined a 3$\sigma$ upper limit on the equivalent width as three
times the quadrature sum of the statistical and fixed-pattern noise error
obtained when integrating over 60~\kms\ around the systemic velocity of the
galaxy. The 60~\kms\ figure corresponds to three resolution elements for \FUSE\
spectra, two for \STIS-G140M data, and nine for \STIS-E140M. It is based on the
median line width of detected features, discussed in Sect.~\SSlinewidth.
Previous authors used different widths for this estimate. Tripp et al.\ (2008)
did integrate over the width of a typical detected line, but used 15 \STIS-E140M
pixels, corresponding to about 55~\kms. On the other hand, Penton et al.\
(2000a,b, 2002, 2004) and Danforth \& Shull (2005, 2008), determined 4$\sigma$
detection limits by integrating over a single resolution element, i.e.\ 20~\kms\
for \FUSE, 40~\kms\ for \STIS-G140M, 6.5~\kms\ for \STIS-E140M data. This leads
to equivalent width limits that are a factor $\sqrt{60/20}*3/4$ $\sim$1.3
smaller for \FUSE\ spectra, and a factor 2.3 smaller for \STIS-E140M data. Since
the narrowest absorption lines have an FWHM of about 25~\kms, integrating over a
single resolution element yields detection limits that are generally optimistic,
and most appropriate only for the small fraction of narrow lines. Thom \& Chen
(2008) used yet another method: they derived 3$\sigma$ limits assuming lines
have $b$=10~\kms\ (FWHM 25~\kms), which gives values a factor $\sqrt{60/25}$=1.5
lower than ours. When comparing results between these different papers, we have
to correct for the differences in these definitions.
\par Figure~\Fspectra\ presents the \Lya, \Lyb, \OVIa\ and \OVIb\ spectra for
all detected intergalactic systems. For each of these we list the galaxy that we
associate with the detection. We discuss each \Lya, \Lyb\ and \OVI\ absorber and
the galaxy association separately for each sightline in the Appendix.
Figure~\Fspectra\ shows a total of 133 intergalactic absorption line systems,
including 115 with \Lya, 40 with \Lyb, 13 with \OVIa, and 5 with \OVIb. \HI\
(i.e.\ \Lya\ {\it or} \Lyb\ is seen in 129 systems), \OVI\ (i.e.\ either line)
in 14 systems. For two systems (at 3073~\kms\ toward Mrk\,290 and at 260~\kms\
toward Ton\,S180) we only list the two \OVI\ lines, while we identify two lines
as uncorroborated \OVIa\ -- at 3109~\kms\ toward PG\,1302$-$102 and at 288~\kms\
toward Ton\,S210.
\begin{figure}\plotfiddle{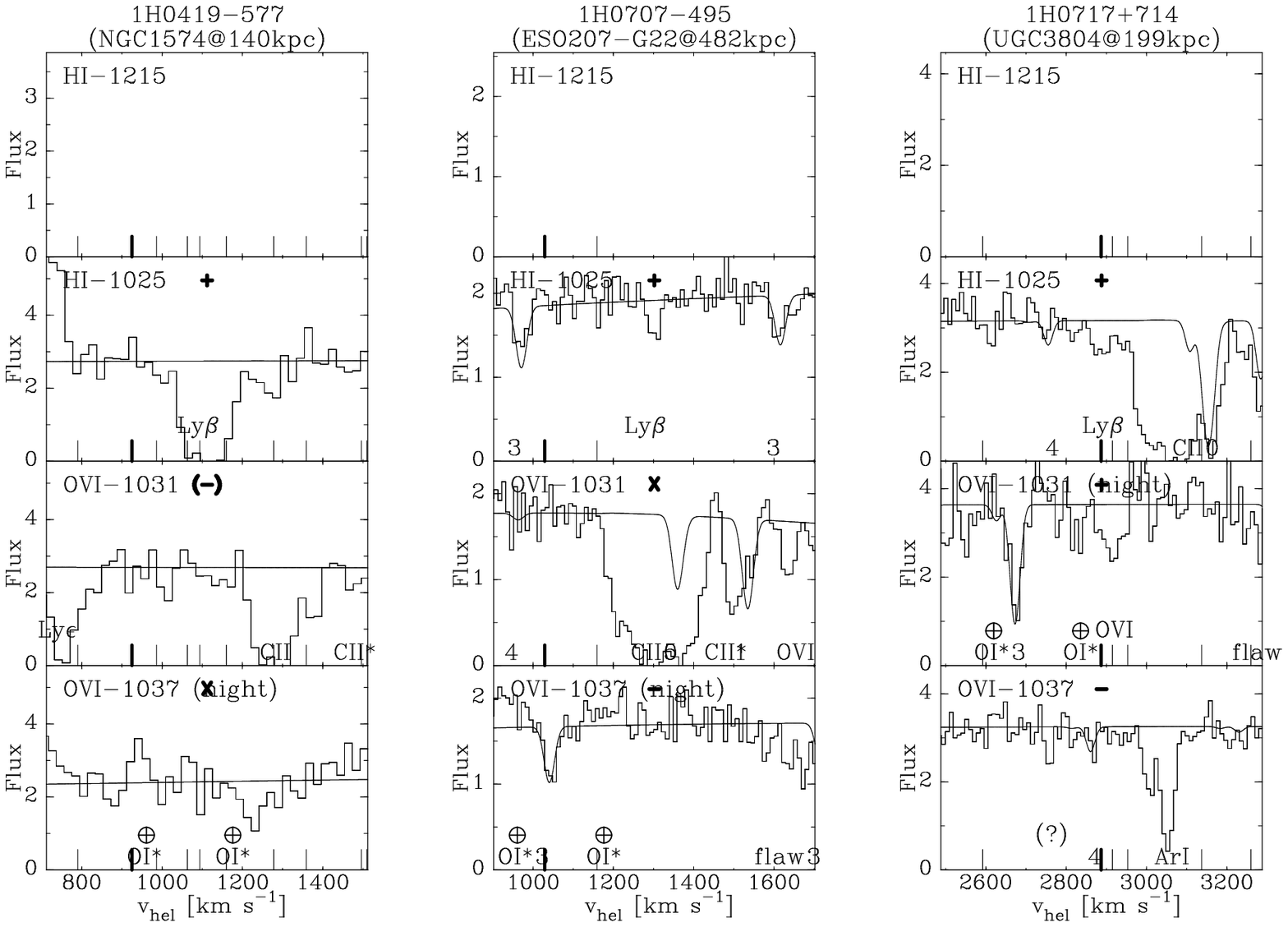}{0in}{0}{440}{285}{0}{0}\figurenum{2}\caption{\captionspectra}\end{figure}
\begin{figure}\plotfiddle{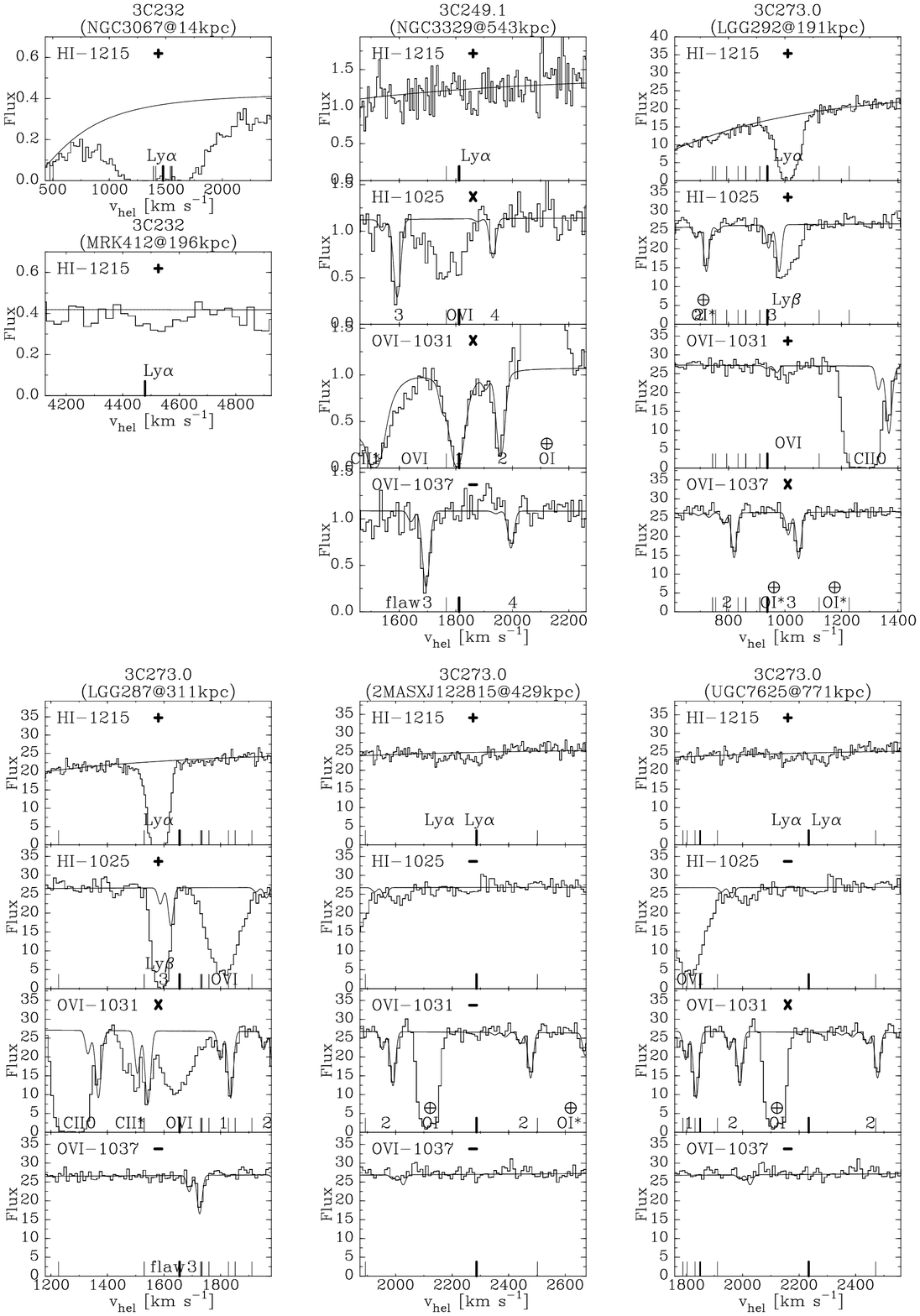}{0in}{0}{440}{570}{0}{0}\figurenum{2}\caption{Continued.}\end{figure}
\begin{figure}\plotfiddle{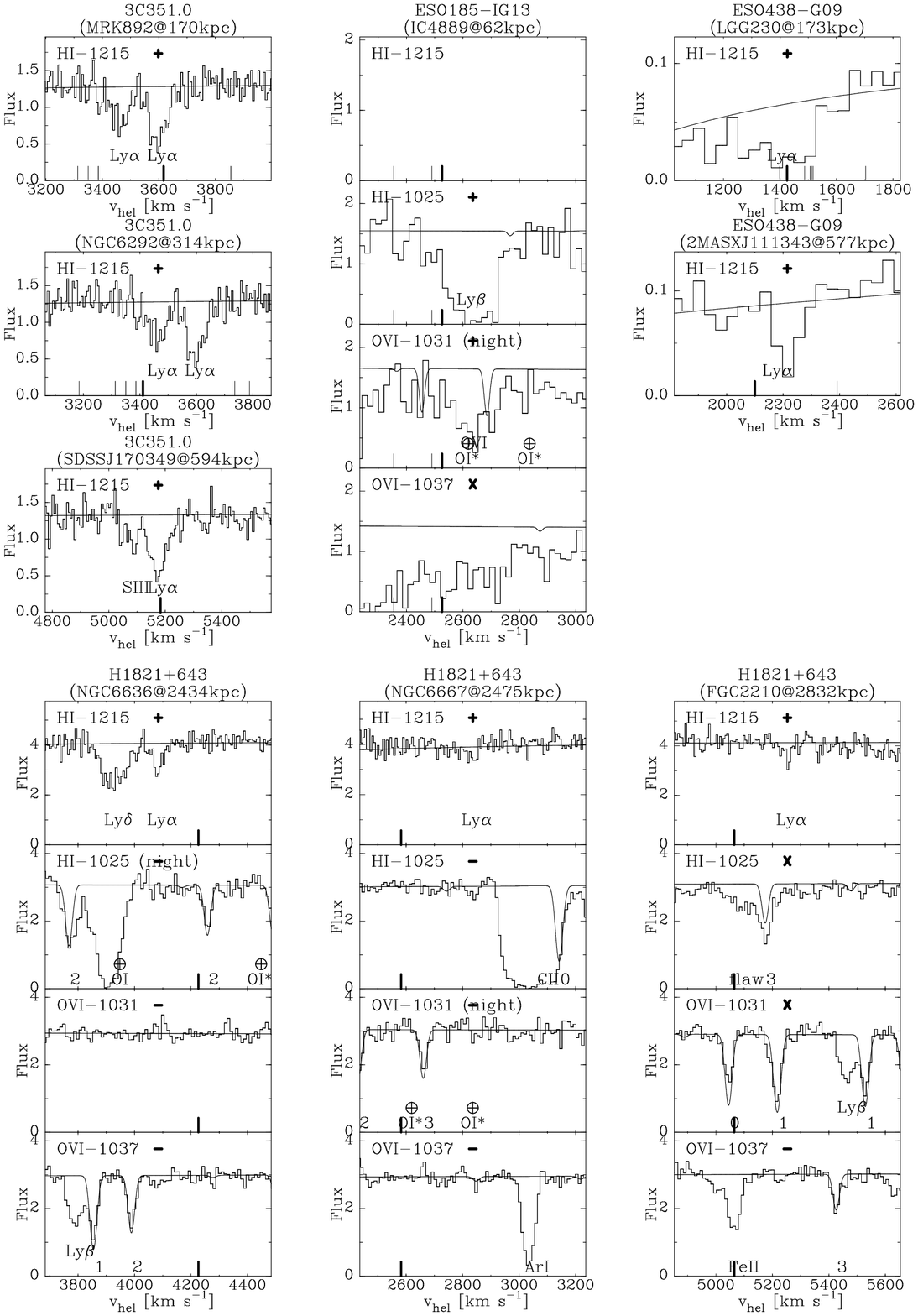}{0in}{0}{440}{570}{0}{0}\figurenum{2}\caption{Continued.}\end{figure}
\begin{figure}\plotfiddle{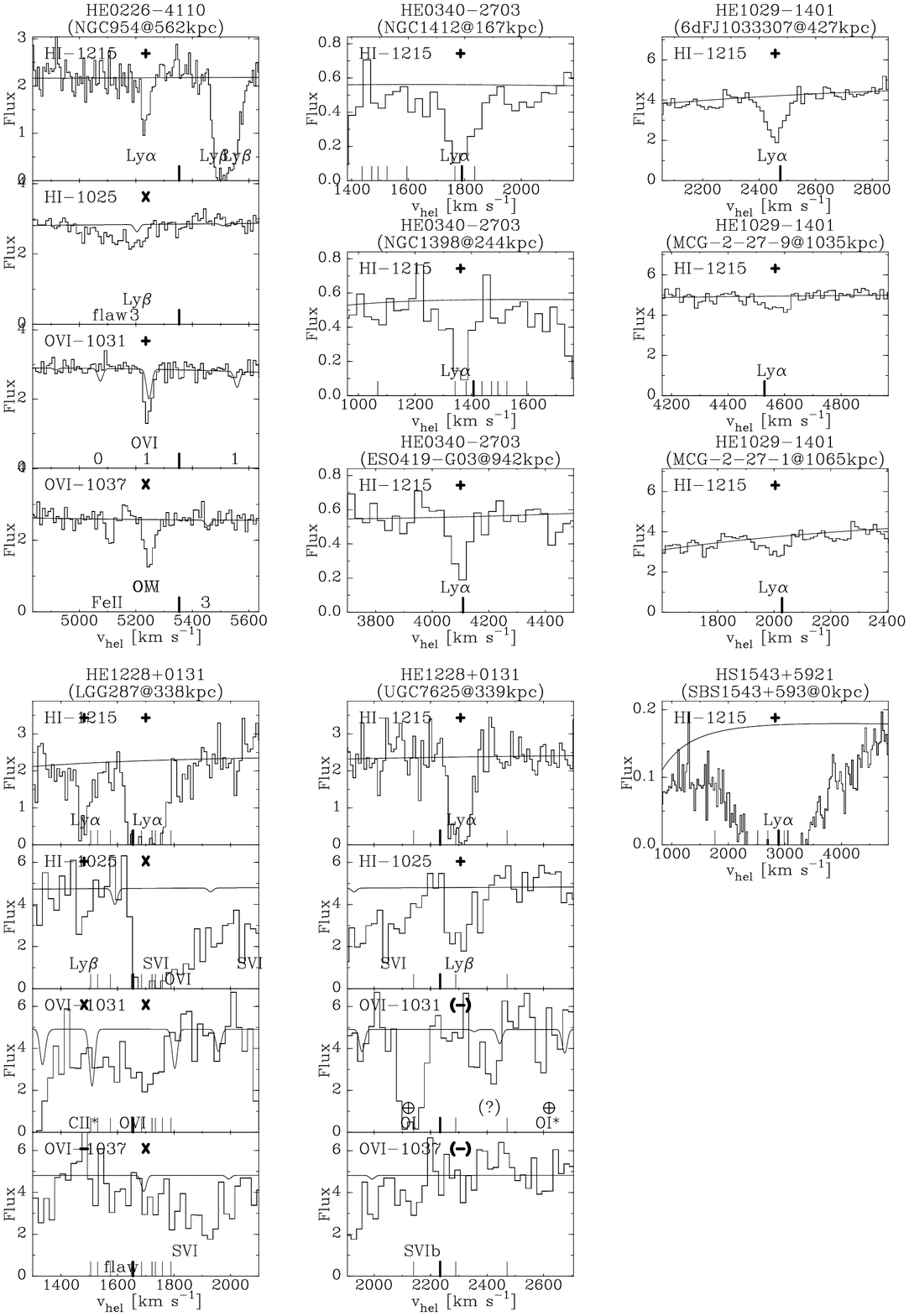}{0in}{0}{440}{570}{0}{0}\figurenum{2}\caption{Continued.}\end{figure}
\begin{figure}\plotfiddle{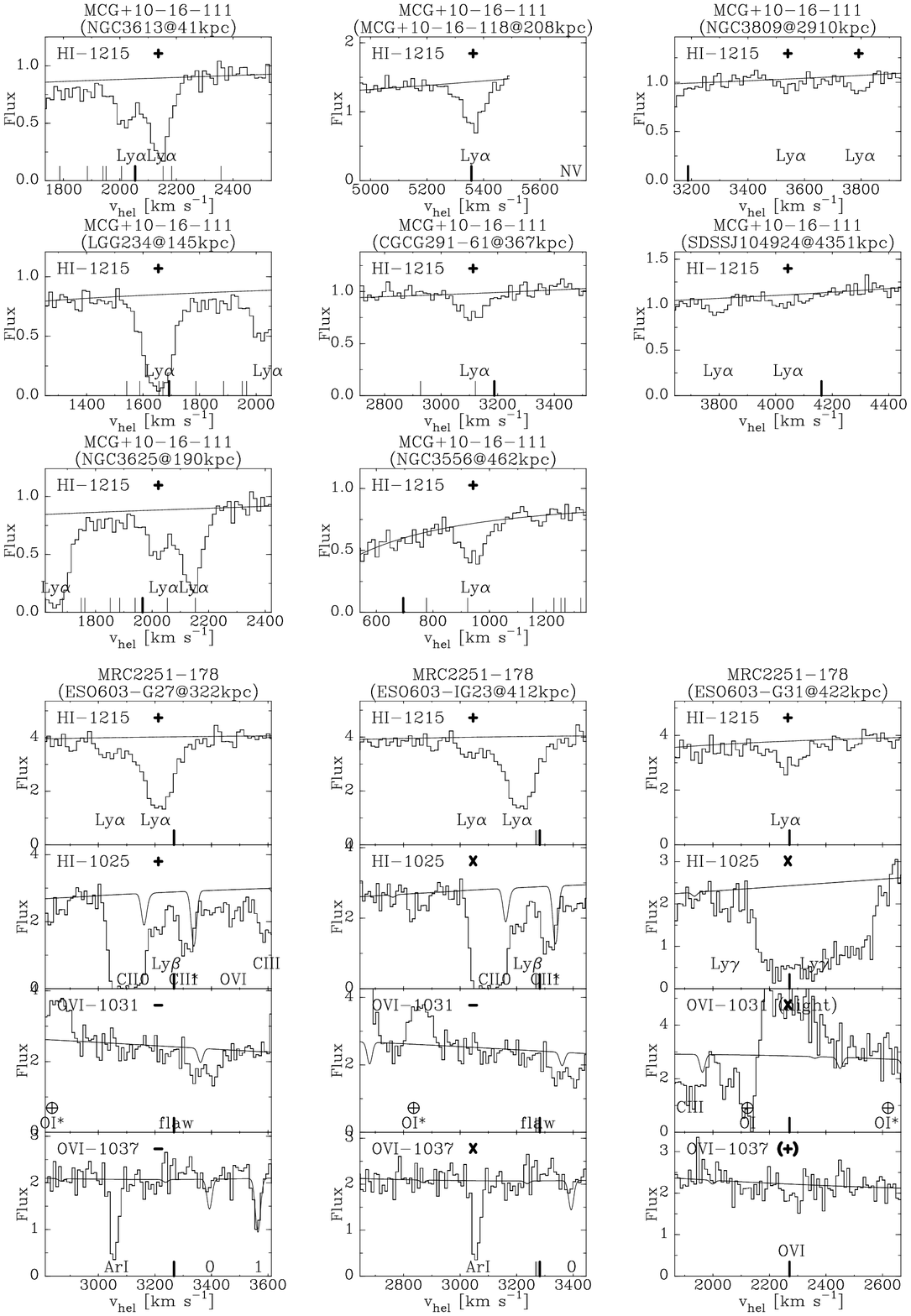}{0in}{0}{440}{570}{0}{0}\figurenum{2}\caption{Continued.}\end{figure}
\begin{figure}\plotfiddle{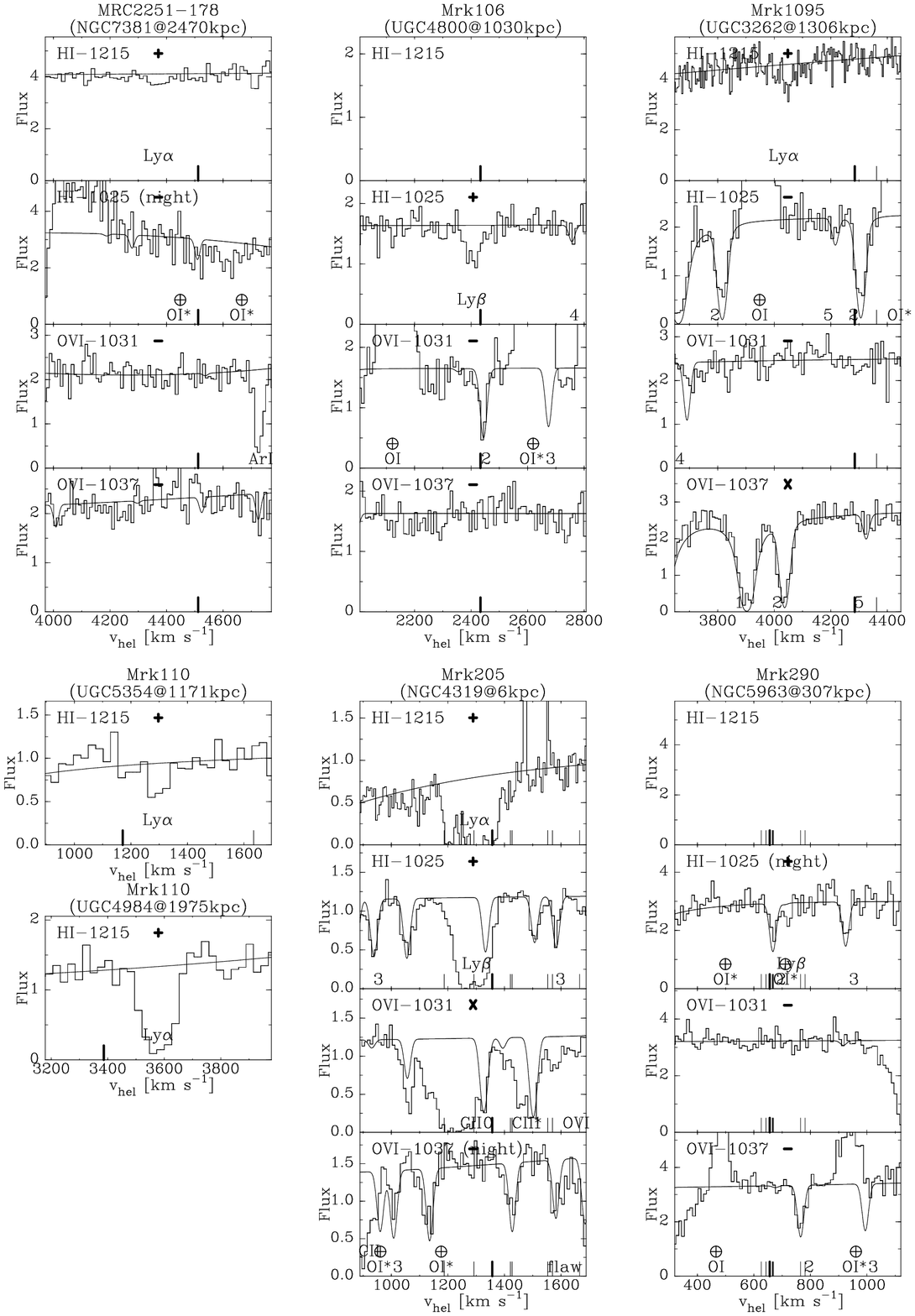}{0in}{0}{440}{570}{0}{0}\figurenum{2}\caption{Continued.}\end{figure}
\begin{figure}\plotfiddle{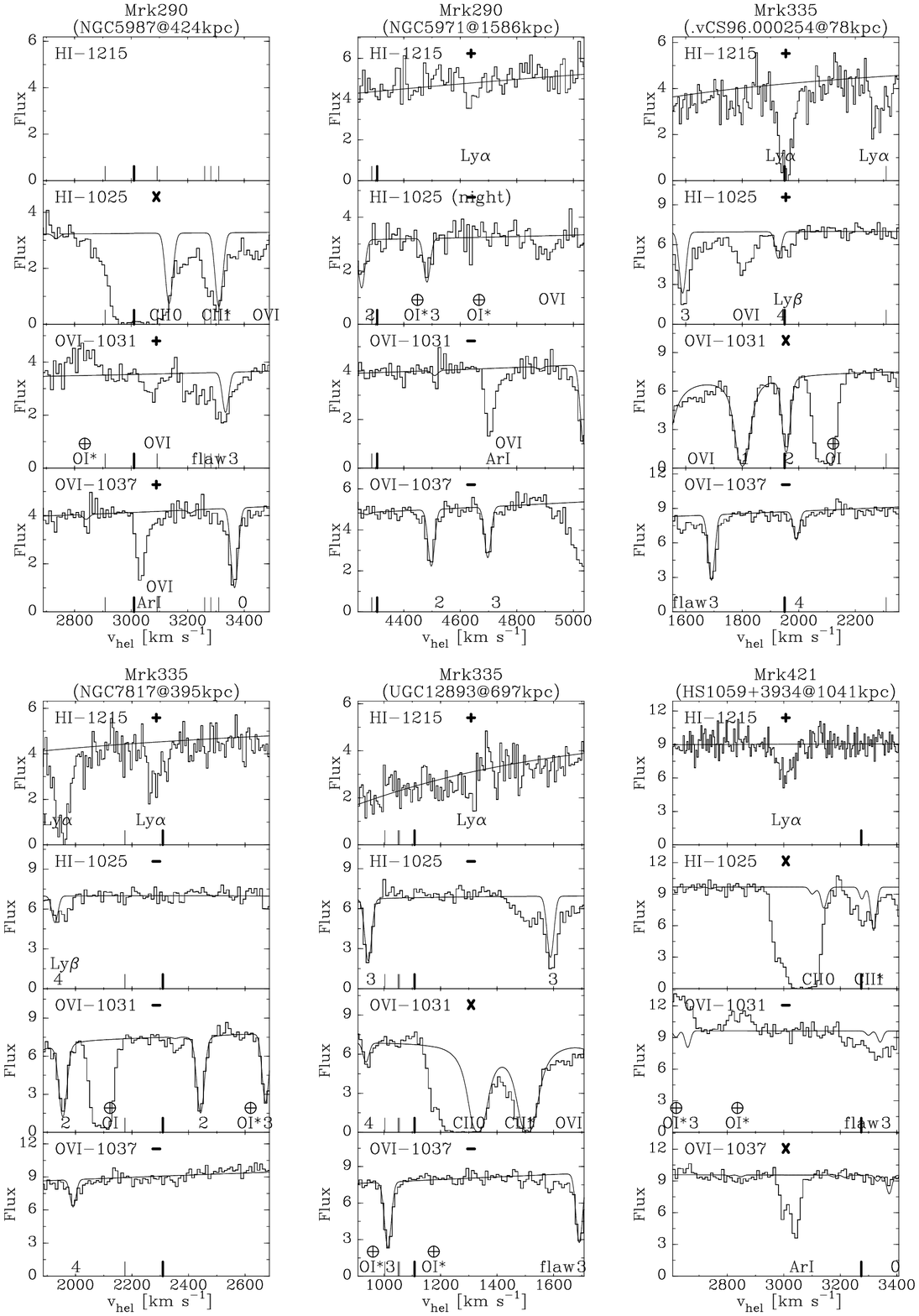}{0in}{0}{440}{570}{0}{0}\figurenum{2}\caption{Continued.}\end{figure}
\begin{figure}\plotfiddle{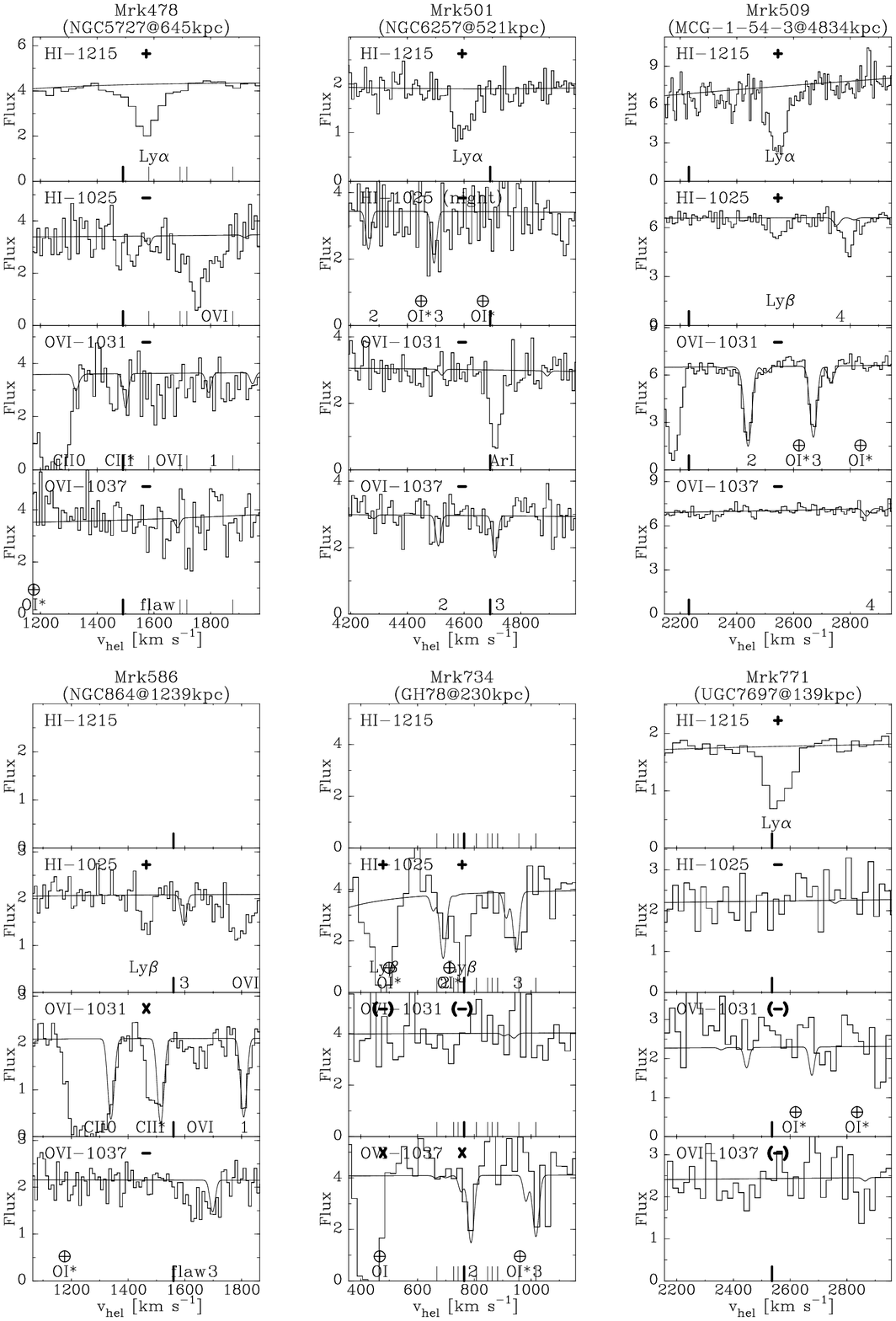}{0in}{0}{440}{570}{0}{0}\figurenum{2}\caption{Continued.}\end{figure}
\begin{figure}\plotfiddle{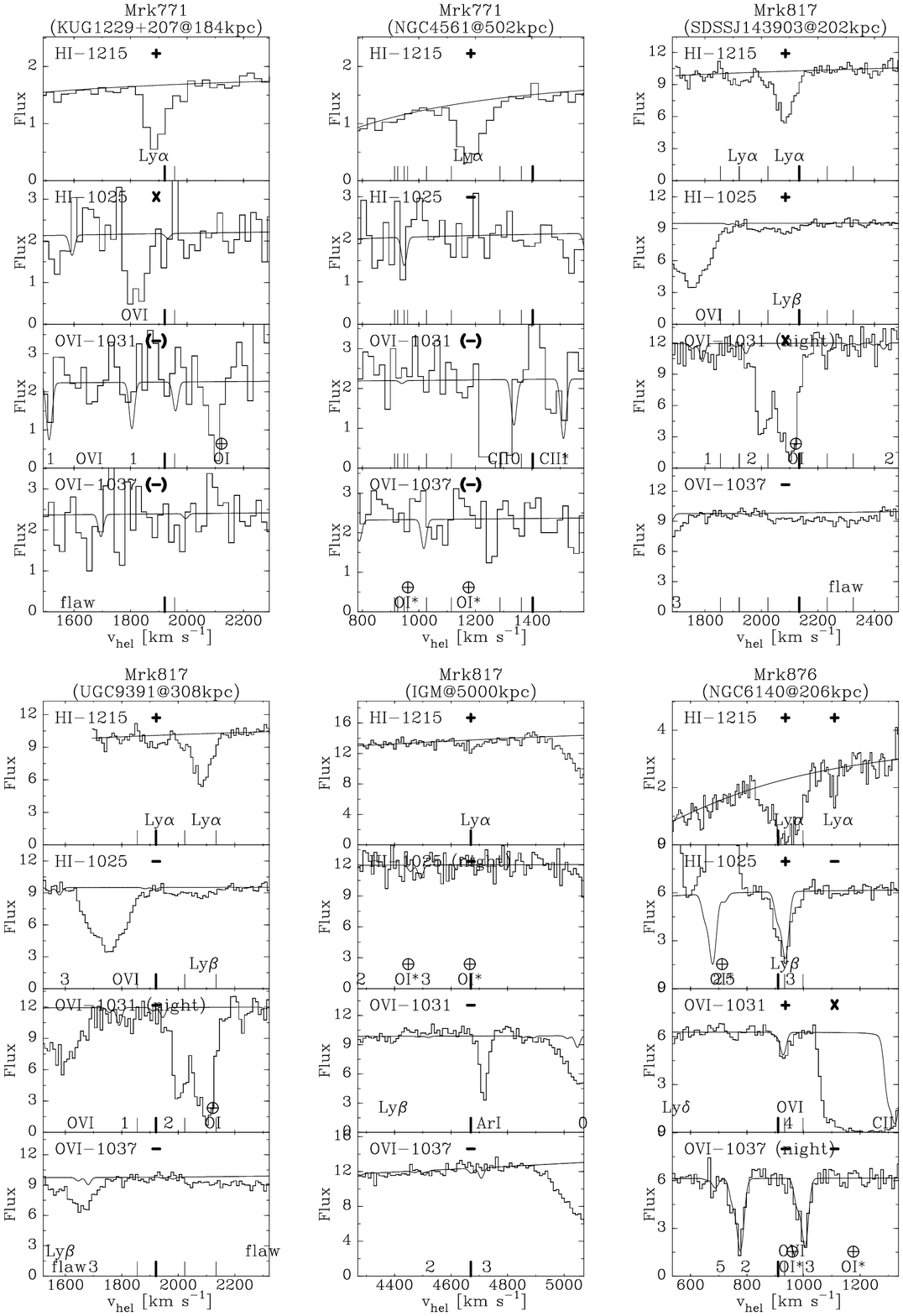}{0in}{0}{440}{570}{0}{0}\figurenum{2}\caption{Continued.}\end{figure}
\begin{figure}\plotfiddle{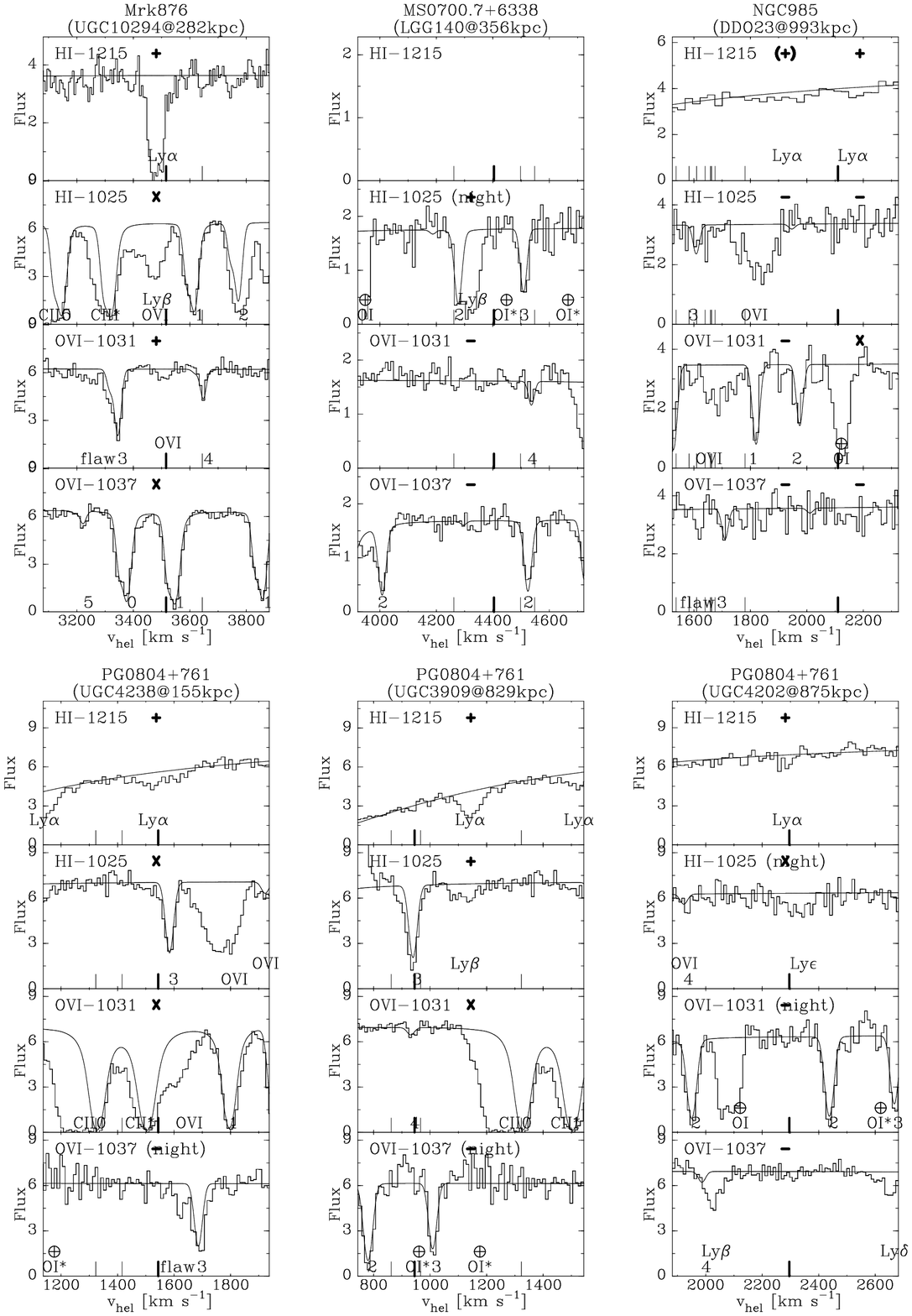}{0in}{0}{440}{570}{0}{0}\figurenum{2}\caption{Continued.}\end{figure}
\begin{figure}\plotfiddle{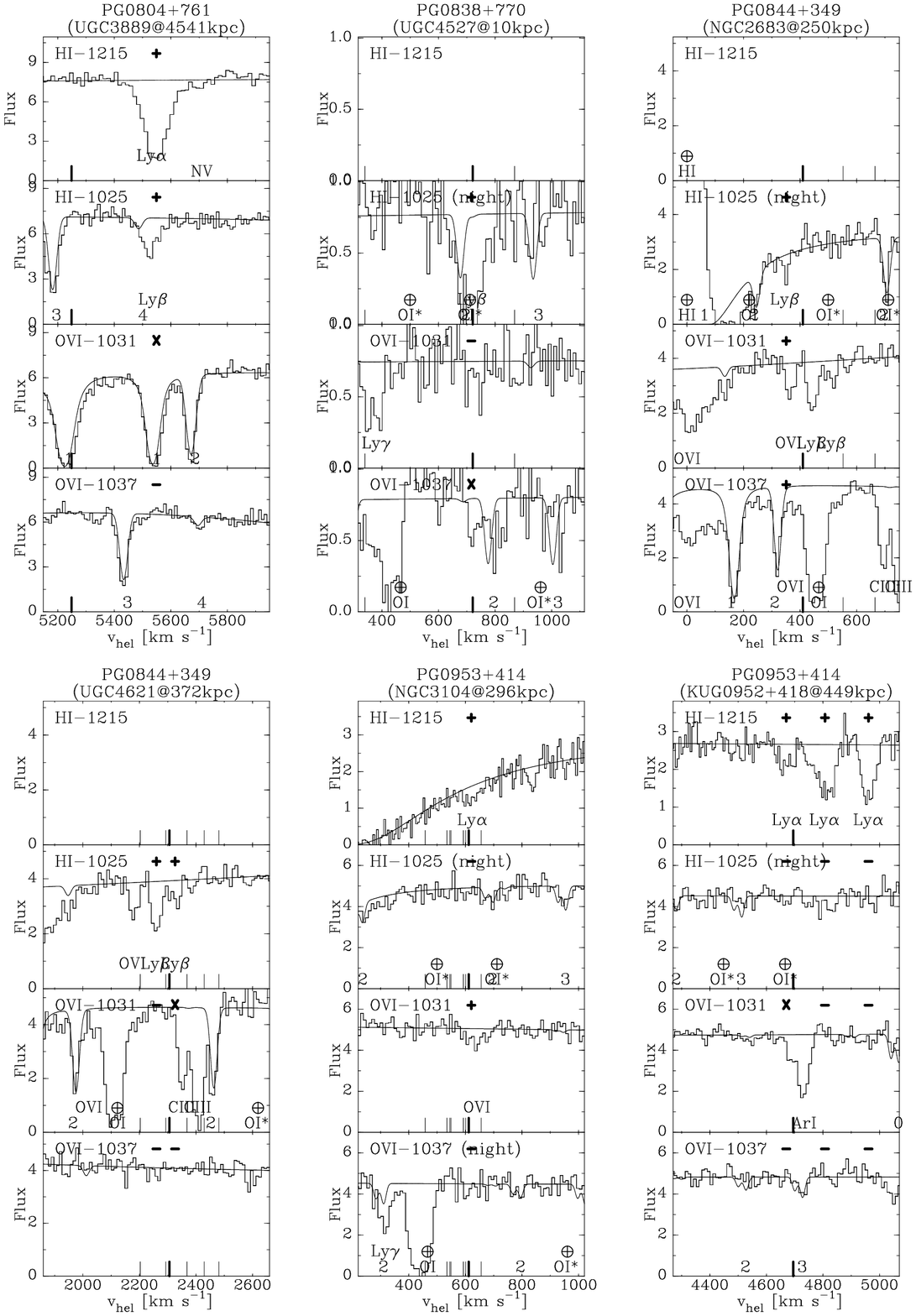}{0in}{0}{440}{570}{0}{0}\figurenum{2}\caption{Continued.}\end{figure}
\begin{figure}\plotfiddle{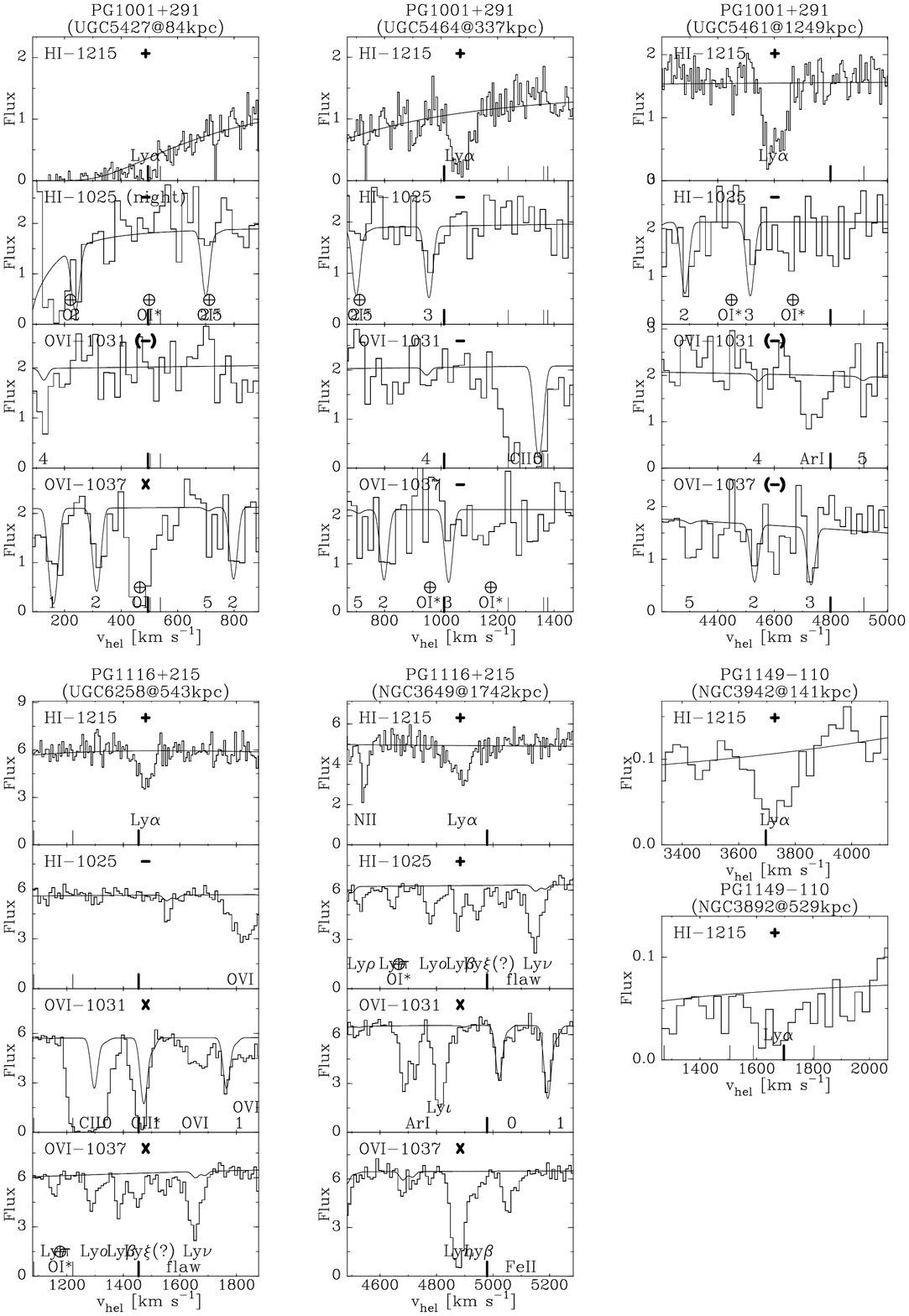}{0in}{0}{440}{570}{0}{0}\figurenum{2}\caption{Continued.}\end{figure}
\begin{figure}\plotfiddle{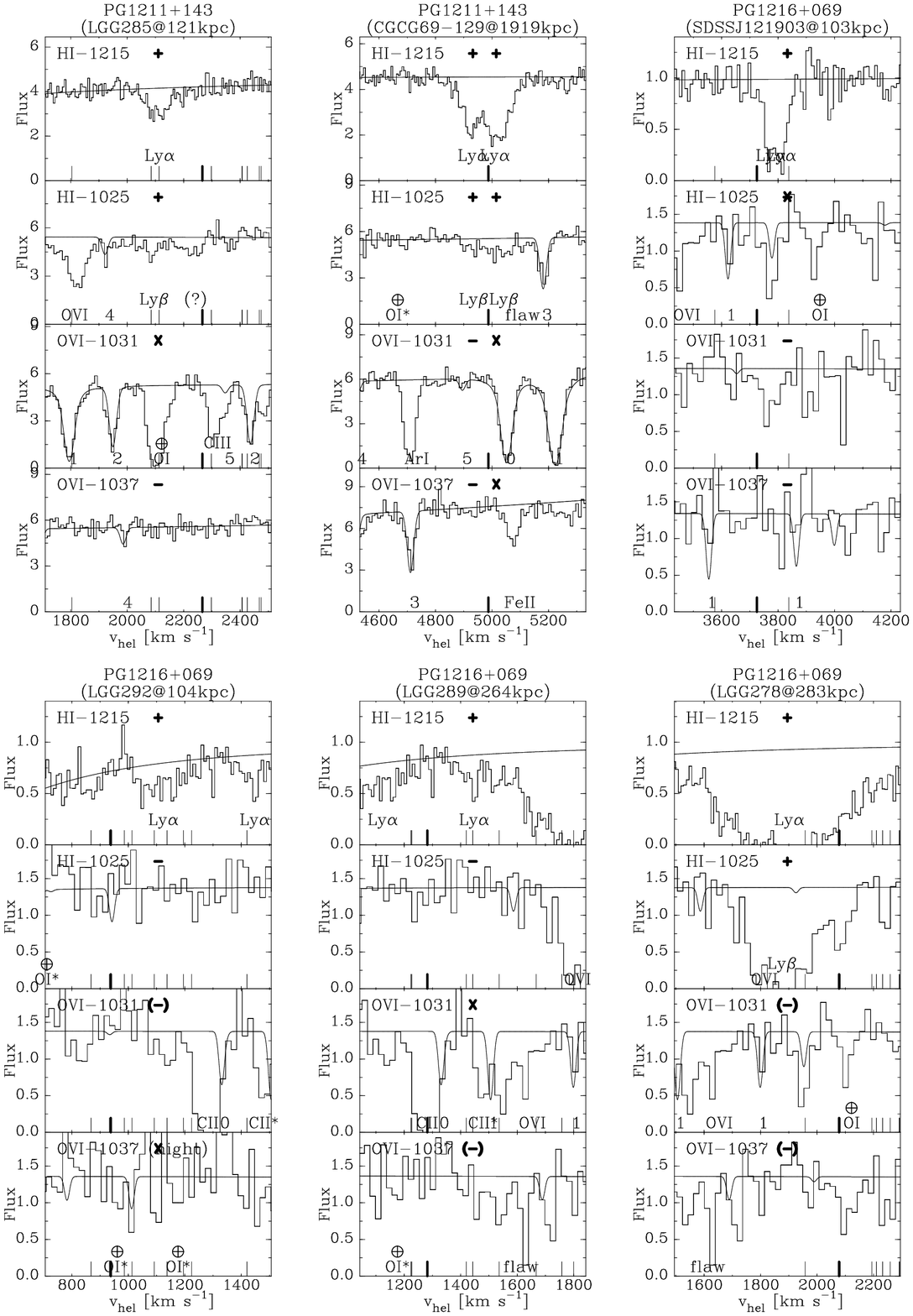}{0in}{0}{440}{570}{0}{0}\figurenum{2}\caption{Continued.}\end{figure}
\begin{figure}\plotfiddle{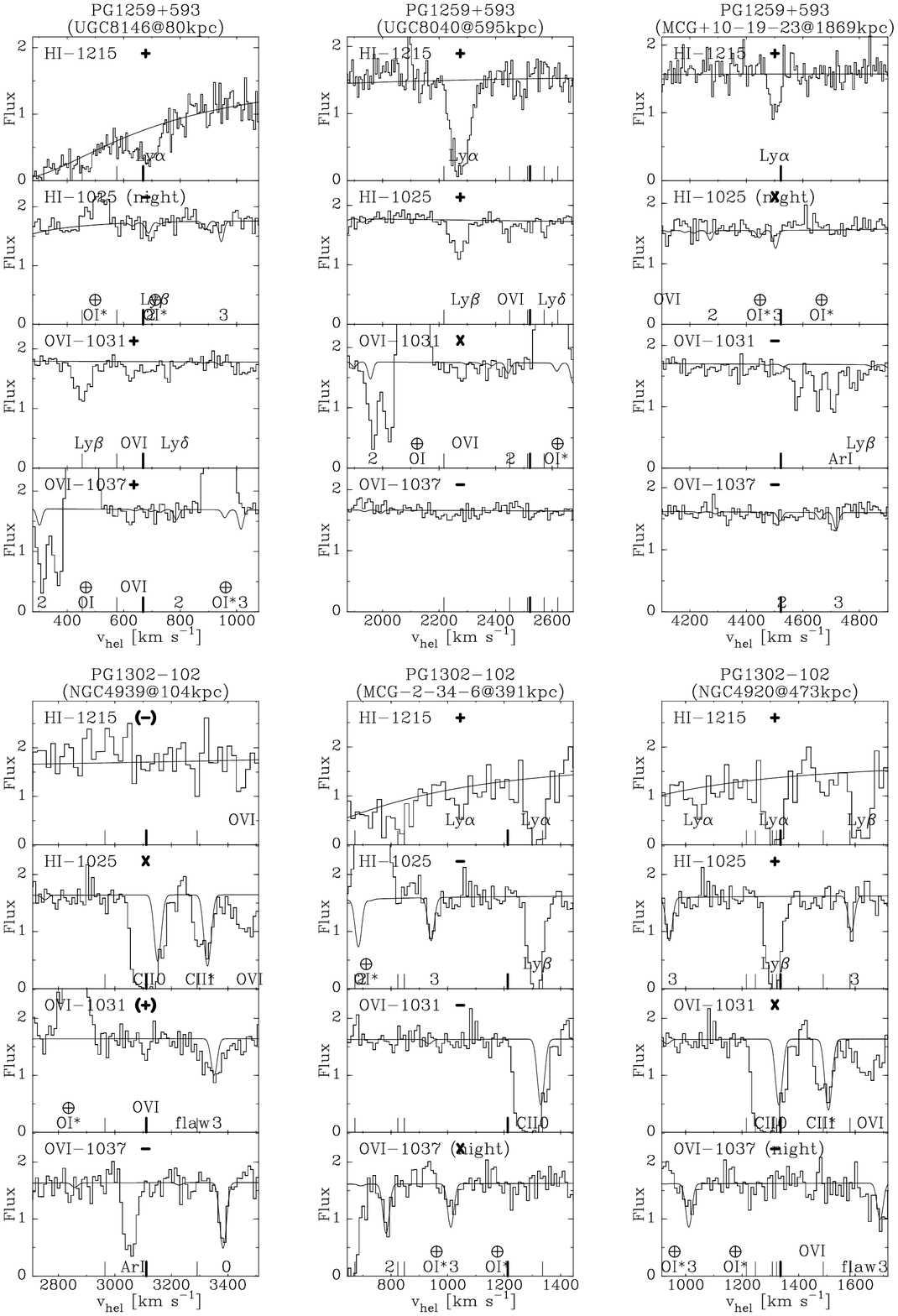}{0in}{0}{440}{570}{0}{0}\figurenum{2}\caption{Continued.}\end{figure}
\begin{figure}\plotfiddle{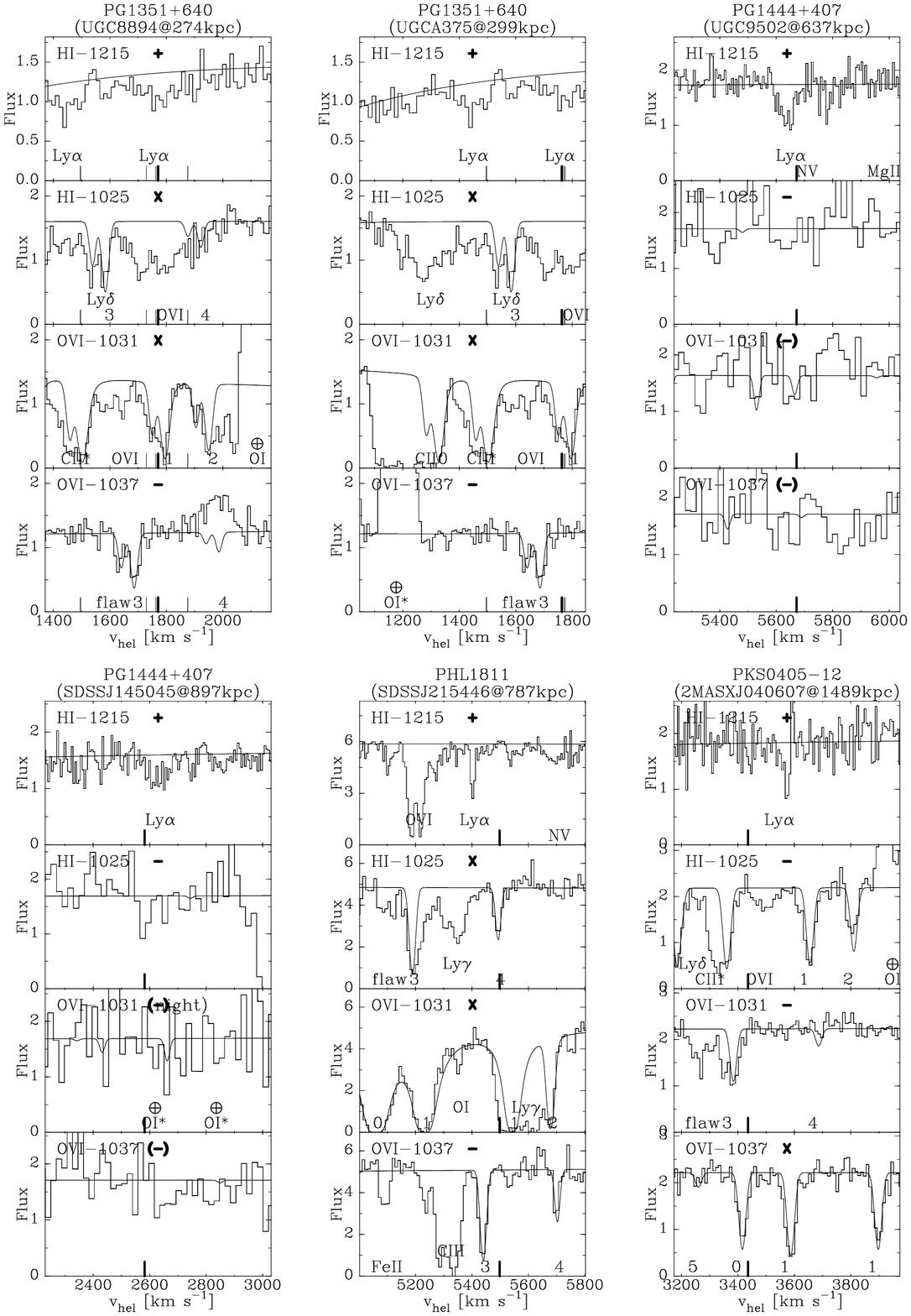}{0in}{0}{440}{570}{0}{0}\figurenum{2}\caption{Continued.}\end{figure}
\begin{figure}\plotfiddle{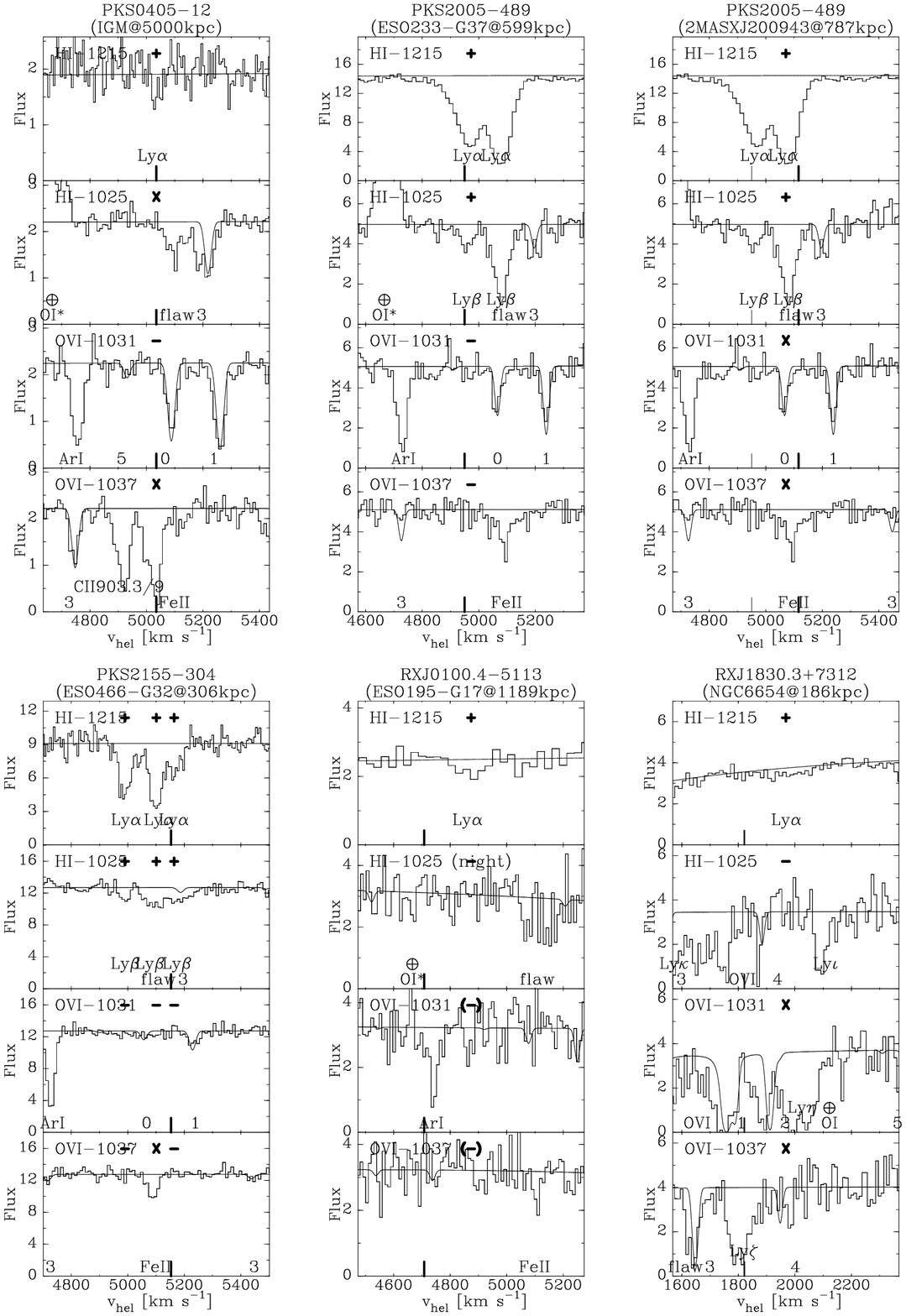}{0in}{0}{440}{570}{0}{0}\figurenum{2}\caption{Continued.}\end{figure}
\begin{figure}\plotfiddle{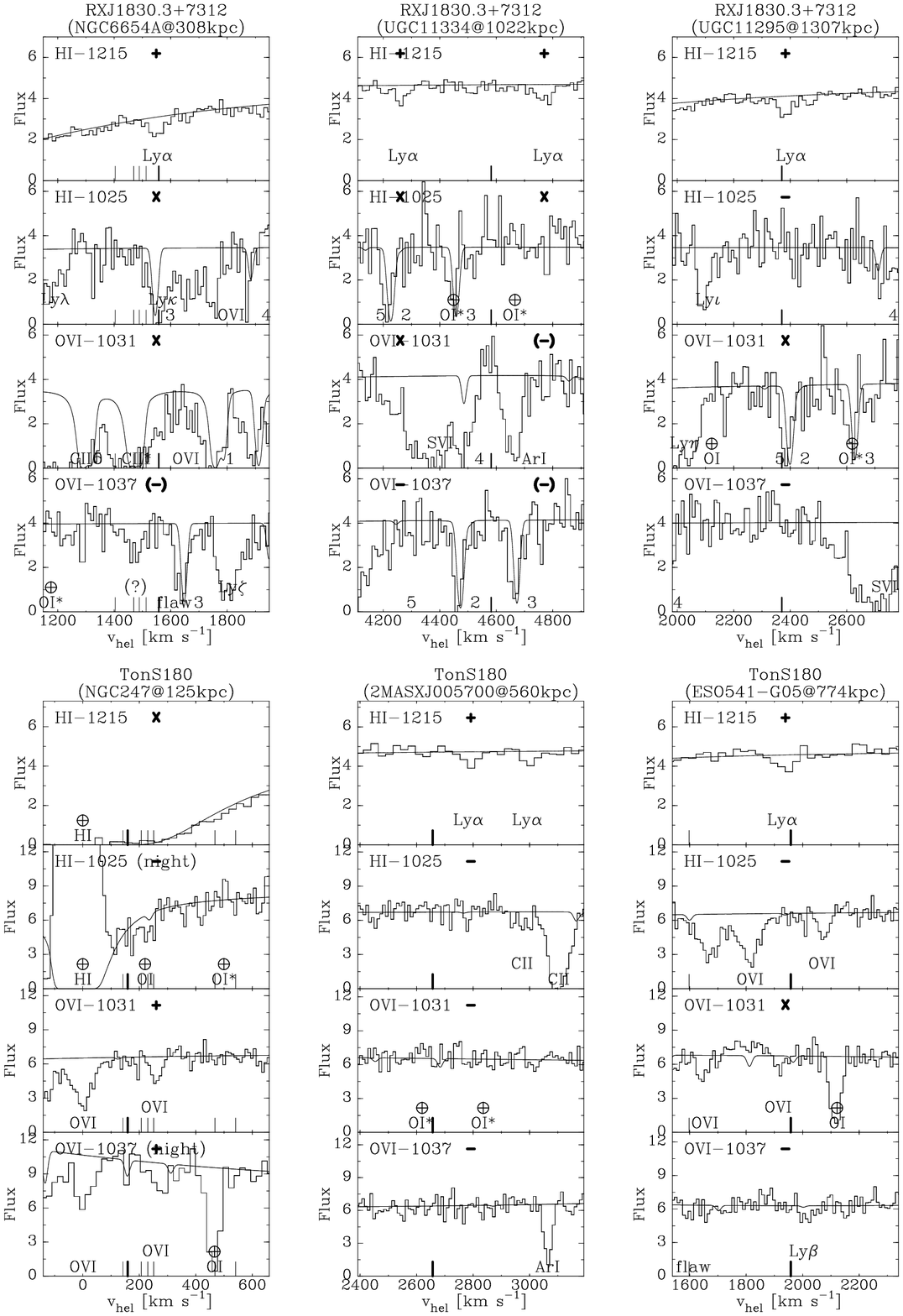}{0in}{0}{440}{570}{0}{0}\figurenum{2}\caption{Continued.}\end{figure}
\begin{figure}\plotfiddle{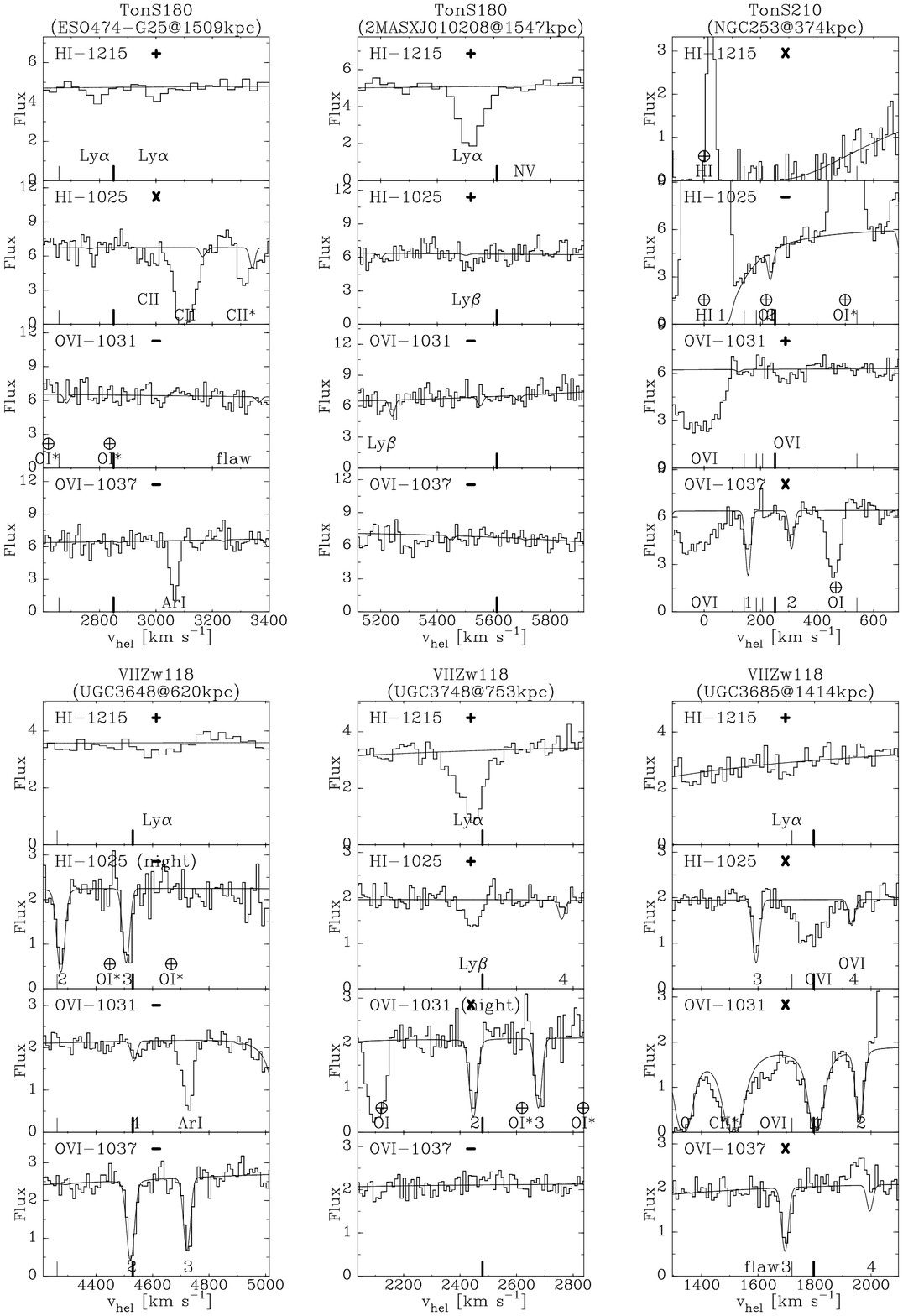}{0in}{0}{440}{570}{0}{0}\figurenum{2}\caption{Continued.}\end{figure}

\subsection{Galaxy Data Origin}
\par For each target, we searched for nearby galaxies with low impact parameter
(\ip), as described below. We started with the ``Third Reference Catalogue of
Galaxies'' (de Vaucouleurs et al.\ 1991). This catalogue gives names and
positions for 23011 galaxies, velocities for 16689 galaxies and angular
diameters for 21605 galaxies. A distance estimate was made for each of these
galaxies, using the following method.
\par Distances for Local Group galaxies were taken from Mateo (1998). For about
100 galaxies individually determined distances were taken from Sandage \&
Tammann (1975) and Freedman et al.\ (2001), though the distances of Sandage \&
Tammann (1975) were corrected to correspond to a Hubble constant of
71~\kms\,Mpc$^{-1}$, rather than the value of 50~\kms\,Mpc$^{-1}$ used in the
original paper. Next, the group catalogues of Geller \& Huchra (1982, 1983) and
Garcia (1993) were used. In these papers, galaxy groups were defined using
automated algorithms. In the remainder of this paper, these groups will be
identified by their GH (Geller \& Huchra 1983) or LGG (Garcia 1993) number (LGG
stands for ``Lyon Galaxy Groups''). These papers listed the galaxies that are
considered members of each of the 176 GH or 486 LGG groups. The GH catalogue
only includes galaxies at declinations $>$$-$3\deg, while the LGG catalogue
covers the whole sky. There is overlap between the two catalogues, so that many
GH groups have an LGG counterpart. However, rarely is the exact same set of
galaxies used to define a group, mostly because the LGG catalogue is newer, so
Garcia used more galaxies with known radial velocities, but also because he used
slightly stricter criteria. In crowded areas, GH groups are often split into two
or even more groups in the LGG list. We add two new groups to the ones listed in
these two papers (see notes to 3C\,351.0 and RX\,J1830.3+7312), as there are
clear concentrations of galaxies, but no GH or LGG group, probably because not
all radial velocities were measured before 1993.
\par For each galaxy listed as an LGG or GH group member, the galaxy's
heliocentric velocity was found from the RC3. An average velocity was then
determined for each group using the velocities of its members. This was
converted into a distance estimate after correcting for a Virgocentric flow
(following Geller \& Huchra 1982), which corresponds to a velocity of 300~\kms\
in the direction R.A.=186\fdg7833, Dec=12\fdg9333. All galaxies in a group were
then assigned the distance corresponding to this corrected average group
velocity, using a Hubble constant of 71~\kms\,Mpc$^{-1}$. This procedure yields
a distance for $\sim$4800 RC3 galaxies. In the remainder of the paper, we
usually do not explicitly add the Hubble constant scale to each distance, but we
use $H_o$=71~\kms\ throughout.
\par The remainder of the galaxies in the RC3 are not considered a member of a
GH or LGG group, so each galaxy's individual velocity was corrected for a
Virgocentric flow to obtain a distance estimate. For the $\sim$5000 RC3 galaxies
without listed velocity measurement, a velocity of 2000 (usually) or 4000~\kms\
(for UGC galaxies) was assumed to continue the calculation (but see below).
Finally, a distance of 2.8~Mpc was assigned to any galaxy for which the
calculation gave distance smaller than 2.8~Mpc. None of these very nearby
galaxies are included in the final sample of galaxies with small impact
parameter.
\par \fixed{As we were completing the study described in this paper, Tully et
al.\ (2008) published a paper giving the most recent distance estimates for 1791
galaxies, derived from a variety of methods. We compared these distances to our
earlier estimates and found that Tully et al.\ (2008) give values for 40 of the
galaxies in our final sample. For many of the galaxies several different methods
were used, and the different estimates usually agree (but not always). For 26 of
the 39 overlapping galaxies our estimate is within 30\% of the Tully et al.\
(2008) estimate. For the remaining 13 our value and the value given by Tully et
al.\ (2008) differ by $>$30\%. In five cases the differences can sometimes be
attributed to the fact that the galaxy was assigned to a GH group when it
apparently should not have been (NGC\,1398, NGC\,2841, NGC\,3067, NGC\,3692,
NGC\,5963). In the other eight cases (NGC\,1533, NGC\,2671, NGC\,4802,
NGC\,4939, NGC\,5727, NGC\,5981, UGC\,10014, IC\,2763) the Tully et al.\ (2008)
value just differs from the value suggested by \vgal/\Dgal.}
\par The RC3 galaxy sample is more or less complete down to a B magnitude of
15.5, and angular diameter larger than about 1 arcmin. To complement this
catalogue with fainter, smaller, lower surface brightness, and more recently
discovered galaxies, we searched the {\it NASA Extragalactic Database} (\NED:
http://nedwww.caltech.ipac.edu) for all galaxies with $v$$<$7000~\kms\ lying
within 5 degrees (the maximum distance allowed by \NED) of each target. This
yields $\sim$6000 galaxies with impact parameter $<$3~Mpc, including about 250
NGC/UGC and 100 IC galaxies that were not listed in the RC3, as well as
$\sim$1600 low-redshift galaxies discovered by the {\it Sloan Digital Sky
Survey} (SDSS), $\sim$800 from the {\it Two-Micron All-Sky Survey} (2MASS), and
between 100 and 300 each from the {\it 2-degree Field} (2dF) survey, the {\it
Kiso Ultraviolet Galaxy} (KUG) survey, and the ESO, CGCG, MCG and VCC
catalogues.
\par We estimated a distance for each of the additional galaxies found in \NED\
using the same procedure described above for RC3 galaxies. Since an angular
distance of 5 degrees corresponds to a linear distance $<$1~Mpc for galaxies
with distance $>$11.5~Mpc, or velocity $>$800~\kms, it is unlikely that any
target-galaxy pairs with impact parameter $<$1~Mpc have been missed. Of course,
galaxies for which \NED\ does not include a velocity are still not included in
this sample.
\par Next, we calculated the angular distance between the extragalactic UV
source and each RC3 and \NED\ galaxy, and converted this into an impact
parameter by multiplying with the galaxy's distance. We kept all galaxies with
impact parameter $<$1~Mpc and systemic velocity between 400 and 6000~\kms.
However, for some of the statistical work described below, we used a maximum
velocity of 5000~\kms\ (or even 2500~\kms\ in some cases), as the galaxy sample
becomes too incomplete at higher velocities. This velocity corresponds to a
distance of 5000/(H$_o$=71) = 70.4~Mpc, or a distance modulus of 34.2. Since
$M_B$=$-$19.57 for an \Lstar\ galaxy (Marzke et al.\ 1994), an \Lstar\ galaxy
with $v$=5000~\kms\ has $m_B$=14.7. This means that for this velocity limit the
RC3 should be more or less complete for galaxies brighter than 0.5\,\Lstar. For
velocities $<$3700~\kms, the sample is complete above $L$$>$0.25\,\Lstar, and
for velocities $<$2500~\kms\ above $L$$>$0.1\,\Lstar.
\begin{figure}\plotfiddle{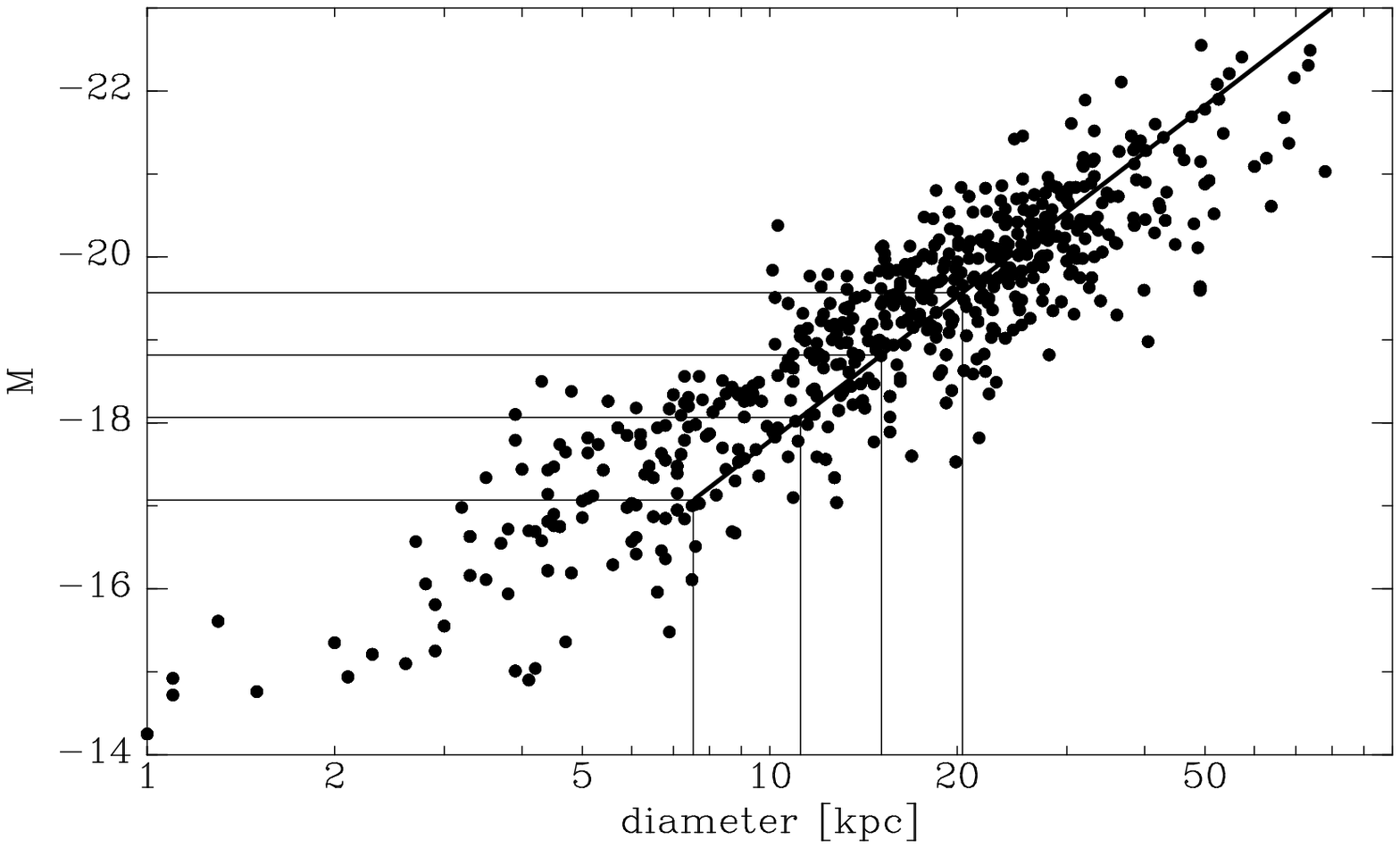}{0in}{0}{450}{350}{0}{-150}\figurenum{3}\caption{\captiondiamvsMabs}\end{figure}
\par For several of the analyses we do in Sects.~\Sassocresults\ and
\Sgalresults\ we would like to apply a galaxy luminosity selection criterion.
However, the RC3 includes a magnitude for only 17\% of the listed galaxies,
while the angular size at 25th magnitude surface brightness is given for 94\%.
The angular diameters are converted to linear diameter using the galaxy's
distance. We correlated linear diameter (\Rgal) and absolute magnitude when both
are known and we use a galaxy's diameter as a proxy for brightness in the
remainder of the paper. \fix{This correlation is shown in Fig.~\FdiamvsMabs,
using the galaxies in the RC3 with both diameter and absolute magnitude given.}
The correlation yields a diameter \Rgal=20.4~kpc for an \Lstar\ galaxy,
\Rgal=14.6~kpc for 0.5~\Lstar, \Rgal=11.2~kpc for 0.25\,\Lstar, and
\Rgal=7.5~kpc for a 0.1~\Lstar\ galaxy. Using this conversion, we find 1366
galaxies in the RC3 with \Rgal$>$7.5~kpc (equivalent to\ $L$$>$0.1\,\Lstar), and
23 additional ones from \NED. The RC3 has 1033 galaxies with \Rgal$>$14.6~kpc
($L$$>$0.5\,\Lstar), and we add 13 from the \NED\ sample. For \Rgal$>$20.4
($L$$>$\Lstar), the RC3 includes 588 galaxies, \NED\ adding 4 more.
\par\fix{For a given diameter limit, 80--90\% of the galaxies fit the
corresponding luminosity criterion. Using the diameter criterion will include
about 5\% of galaxies that are fainter than the corresponding luminosity, by up
to about 1 magnitude. On the other hand, about 10\% of the galaxies that are
brighter than a given limit are excluded by the diameter criterion. This is a
small price to pay for the ability to classify $>$95\% of the galaxies instead
of $<$20\%.}
\par We could instead have tried to extract magnitudes from \NED\ for individual
galaxies, but the magnitudes that \NED\ gives are on many different systems,
including, but not limited to, $B_T$ as defined for the RC3, $UBV$, SDSS $ugriz$
and photographic. Trying to regularize these would give as much uncertainty as
using diameters as a proxy. The only proper way to determine consistent galaxy
luminosities would be to measure all galaxies on the same system, such as that
used for a portion of the RC3.
\par As mentioned earlier, we assumed a value of 2000 or 4000~\kms\ for galaxies
for which the RC3 does not provide a velocity. In a few cases this resulted in
small impact parameters. We then checked whether \NED\ gives a velocity for the
galaxy. If so, we redid the impact parameter calculation. In the end there are
quite a few galaxies whose velocity remains unknown, but which may have impact
parameter $<$1~Mpc to a background target. Most of these galaxies are smaller
and less luminous and have similar impact parameter as the galaxy that we
associate with the absorber. However, for 13 absorbers there are a total of 16
galaxies with unknown velocity that could have significantly lower impact
parameter than the one we list in our results table (see Appendix notes for
1H\,0707$-$495, 3C\,249.1, Mrk\,509, Mrk\,1095, PG\,0804+761, PG\,1211+143,
PG\,1302$-$102, PG\,1444+407, RX\,J0100.4$-$5113 and VII\,Zw\,118). In three
cases the galaxy with the lowest known impact parameter is at \ip$>$1~Mpc, but
the galaxy with unknown velocity could have \ip$<$1~Mpc (see notes for
PG\,1211+143, RX\,J0100.4$-$5113 and VIII\,Zw\,118).

\par Using the sample of galaxies, we associated absorption lines with some of
these, as we discuss in more detail in Sect.~\SSassoc. For the other galaxies,
we derived upper limits on the equivalent widths of \Lya, \Lyb\ and \OVI\
absorption. Table~\Tres\ lists all the galaxies with \vgal$<$6000~\kms\ that lie
within 1~Mpc of a sightline to a background target \fix{(sorted on impact
parameter),} and for which the ratio of impact parameter to galaxy diameter is
$<$125. We justify the latter criterion in Sect.~\SSassoc. All individual
associations in Table~\Tres\ are discussed in detail in the Appendix, on a
sightline-by-sightline basis. In Table~\Tion\ we separately list the detections
for lines other than \Lya, \Lyb\ or \OVI.
\par\fix{We have made the full galaxy table available in the on-line version of
this paper. This includes {\it all} galaxies near each sightline. This table is
inhomogeneous, however. Near some sightlines deep surveys for dwarf galaxies
exist. Near others special searches were done. For some sightlines (especially)
in the Virgo cluster region we did not check the diameter in \NED\ for some of
the additional galaxies with impact parameter $<$1~Mpc. Although these
additional galaxies come from searches for dwarfs, it is possible (but unlikely)
that there are cases where a galaxy should have been listed in Table~\Tres\ but
was not. The on-line table should therefore be treated with care. It is only
reliably complete for statistical analyses for large galaxies.} \par In
Tables~\THI\ and \TOVI\ we summarize all the detected \Lya, \Lyb\ and \OVI\
lines, and compare our measurements with previously published values.
\def\nd#1{{$\lbrace$#1$\rbrace$}}
\def\h{}
\def\NC{$^1$}
\def\V#1{[#1\hbox to 0pt{]$^2$\hss}}
\def\OR{$^3$}
\def\mcp#1+#2 {#1$^4$}
\def\mcpo#1+#2 {#1$^{3,4}$}
\def\mcpl#1+#2 {#1$^{3,5}$}
\def\mcn#1$-$#2 {#1$^4$}
\def\ORLI{$^{4,5}$}
\def\LI{$^5$}
\begin{deluxetable}{lllrrrrccccccc}
\tablenum{3}
\tablewidth{0pt}
\tabletypesize{\scriptsize}
\tabcolsep=3pt
\tablecolumns{13}
\voffset0cm
\rotate
\voffset3cm
\tablecaption{\HI\ and \OVI\ results\NC}
\tablehead{%
\ch{Target}    &\ch{Galaxy}   &\ch{Group}    &\ch{\vgal}   &\ch{$D$}   &\ch{\Rgal}    &\ch{$\rho$}    &\ch{Q}    &\ch{HI}   & \ch{OVI}   & \ch{\Lya}   & \ch{\Lyb}   & \ch{\OVI$-$1031}   & \ch{\OVI$-$1037}   \\
\ch{}          &\ch{}         &\ch{}         &\ch{[\kms]}  &\ch{[Mpc]} &\ch{[kpc]}    &\ch{[kpc]}     &\ch{}     &\ch{}     & \ch{}      & \ch{[\mA]}  & \ch{[\mA]}  & \ch{[\mA]}       & \ch{[\mA]}       \\
\ch{(1)}&\ch{(2)}&\ch{(3)}&\ch{(4)}&\ch{(5)}&\ch{(6)}&\ch{(7)}&\ch{(8)}&\ch{(9)}&\ch{(10)}&\ch{(11)}&\ch{(12)}&\ch{(13)}
}\startdata
HS1543+5921   &SBS1543+593\OR &GH158  & 2889 & 42.7 &  7.8 &  0.3 &!& + &    & $>$5000@2829    &               &                 &              \\
Mrk205        &NGC4319\OR     &GH107  & 1357 & 27.7 & 23.8 &    6 &!& + &  $-$ & 801\e13,4@1289& 477\e14,8@1283\h& \nd{\CII}       &  $<$38         \\
PG0838+770    &UGC4527        &       &  721 & 11.8 &  4.7 &   10 &!& + &  $-$ &               & 330\e53,13@716\h&  $<$54            &  \nd{\OI}    \\
3C232         &NGC3067\OR     &GH50   & 1476 & 23.9 & 17.0 &   14 &!& + &    & $>$5000@1434    &               &                 &              \\
MCG+10$-$16$-$111 &NGC3613        &LGG232 & 2054 & 31.2 & 35.3 &   41 &2& + &    & 301\e11,7@2136&               &                 &              \\
MRK205        &NGC4291        &LGG284 & 1757 & 27.7 & 15.3 &   51 &!& $-$ &  $-$ &  $<$29          &  \nd{\OVI}    &  $<$22            &  \nd{\OI}    \\
ESO185$-$IG13   &IC4889         &       & 2526 & 34.5 & 28.9 &   62 &!& + &  + &               & 641\e52,14@2635& 335\e68,15@2627\h& \nd{\Lyb}\\
Mrk110        &MCG+09$-$16$-$10   &       &  457 &  9,1 &  1.7 &   66 &!& $-$ &    &  $<$54          &               &                 &              \\
\enddata
\end{deluxetable}
\def\NC{$^1$}
\begin{deluxetable}{lllrrlrr}
\tablenum{4}
\tablewidth{0pt}
\tabletypesize{\scriptsize}
\tabcolsep=3pt
\tablecolumns{8}
\tablecaption{Detections of other ionic lines\NC}
\tablehead{%
\ch{Target}    &\ch{Galaxy}          &\ch{Group}    &\ch{\vgal}    &\ch{\ip}    &\ch{ion}    &\ch{Ion EW@$v$}    &\ch{\HI/\OVI?}    \\
\ch{}          &\ch{}                &\ch{}         &\ch{[\kms]}   &\ch{[kpc]}  &            &\ch{[\mA@\kms]}    &                  \\
\ch{(1)}&\ch{(2)}&\ch{(3)}&\ch{(4)}&\ch{(5)}&\ch{(6)}&\ch{(7)}&\ch{(8)}
}\startdata
Mrk205         &NGC4319              &GH107         & 1357         &     6      & \CIII       &  293\e 33,18@1273 & +\Lya,+\Lyb,$-$OVI \\
               &                     &              &              &            & \SiIII      &  283\e 10, 5@1293 & "                \\
               &                     &              &              &            & \CIV\lm1548 &  282\e 19, 2@1280 & "                \\
               &                     &              &              &            & \CIV\lm1550 &  165\e 17, 3@1290 & "                \\
3C232          &NGC3067              &GH50          & 1476         &    14      & \SiIII      & 1043\e 21,20@1493 & +\Lya            \\
               &                     &              &              &            & \CIV\lm1548 & 1081\e 39,22@1462 & "                \\
               &                     &              &              &            & \CIV\lm1550 &  761\e 44, 5@1462 & "                \\
               &                     &              &              &            & \NV\lm1238  &  233\e 35, 5@1377 & "                \\
\enddata
\tablecomments{%
1: Description of columns:
Column~1: AGN name;
Column~2: Associated galaxy;
Column~3: Galaxy group if galaxy is in group;
Column~4: Heliocentric systemic velocity of galaxy;
Column~5: Impact parameter;
Column~6: Ion detected;
Column~7: Measurement result, giving ``equivalent width $\pm$ statistical error $\pm$ systematic error @ fitted velocity'';
Column~8: Entries show whether \Lya, \Lyb\ and/or \OVI\ were also detected (+) or not ($-$); if line info is missing there is no data for that ion.
\ \ \ \ \ \ \ \
2: Full name is 2MASX J200943.13$-$481105.2.
}
\end{deluxetable}
\def\Nref{1} \def\NDSRS{2} \def\Nwidth{3} \def\Nflag{4} \def\Nnone{5}
\def\D{$^\NDSRS$}
\def\NCIIA{6}
\def\NTripp{7}
\def\NHEtwtw{8}
\def\NMCG{9}
\def\NMRC{10}
\def\NLybHt{11}
\def\NPGoeof{12}
\def\NPGthfo{13}
\def\NPHL{14}
\def\NPKStoof{15}
\def\NPKStoff{16}
\def\NVII{17}
\begin{deluxetable}{lrrrlrrrcccclrc}
\tablenum{5}
\tablewidth{0pt}
\tabletypesize{\scriptsize}
\tabcolsep=3pt
\tablecolumns{15}
\rotate
\voffset2.1cm
\tablecaption{\Lya\ and \Lyb\ detections and comparison with values in the literature}
\tablehead{%
\ch{Object}       &\ms4{Literature Values}                                      &\ms5{This paper}                                                                               &\ms2{Flags$^\Nflag$}&\ch{Galaxy}&\ch{\ip}  &\ch{Note}\\
                  &\ch{$v$(\Lya)}&\ch{EW(\Lya)}&\ch{EW(\Lyb)}&\ch{Ref.$^\Nref$} &\ch{$v$(\Lya)}&\ch{EW(\Lya)}&\ch{EW(\Lyb)} &\ch{$b$(\Lya)$^\Nwidth$} &\ch{$b$(\Lyb)$^\Nwidth$} &\ch{\Lya}&\ch{\Lyb} &           &          &         \\
                  &\ch{[\kms]}   &\ch{[\mA]}   &\ch{[\mA]}   &                  &\ch{[\kms]}   &\ch{[\mA]}   &\ch{[\mA]}    &\ch{[\kms]}              &\ch{[\kms]}              &         &          &           &\ch{[kpc]}&         \\
\ch{(1)}&\ch{(2)}&\ch{(3)}&\ch{(4)}&\ch{(5)}&\ch{(6)}&\ch{(7)}&\ch{(8)}&\ch{(9)}&\ch{(10)}&\ch{(11)}&\ch{(12)}&\ch{(13)}&\ch{(14)}&\ch{(15)}
}\startdata
1H\,0419$-$577      &      &         &         &          & 1112 &             & 463\E25\E15 &    & 79 &   & + & NGC1574        &  140 &          \\ 
1H\,0707$-$495      &      &         &         &          & 1302 &             &  26\E 7\E 8 &    & 19 &   & + & ESO207$-$G22     &  482 &          \\ 
1H\,0717+714      &      &         &         &          & 2888 &             &  41\E 7\E 8 &    & 32 &   & + & UGC3804        &  199 & \NCIIA   \\ 
3C\,232           & 1428 &6400\E230&         & Ke+2005  & 1434 &    DLA      &             &  $-$ &    &   &   & NGC3067        &   14 &          \\ 
                  &      &         &         &          & 4526 & 153\E30\E 4 &             & 94 &    & + &   & Mrk412         &  196 &          \\ 
3C\,249.1         &      &         &         &          & 1861 &  40\E 9\E 3 &             & 29 &    & + &   & NGC3329        &  543 &          \\ 
                  &      &         &         &          & 6672 & 245\E15\E 3 &  39\E13\E 3 & 58 & 59 & + &   & CGCG351$-$57     &  858 &          \\ 
3C\,273.0         & 1015 & 369\E36 &         & PSS2000\D& 1010 & 394\E 7\E 1 & 120\E 4\E10 & 53 & 44 &   &   & LGG292         & $>$191 &          \\
\enddata
\end{deluxetable}
\def\Nref{1} \def\NEW{$^2$} \def\Nqual{3} \def\Ndflag{4}
\def\NHE{5}\def\NMRC{6}\def\NHALF{7} \def\NMrkfse{8} \def\NMRKess{9} \def\NPGoeof{10} \def\NPGonft{11} \def\NPGooos{12} \def\NPGotoo{13} \def\NPGtwfn{14} \def\NPGotot{15} \def\NPHL{16} \def\NPKtoof{17}
\begin{deluxetable}{lrrrlrrrcclrc}
\rotate
\voffset2cm
\tablenum{6}
\tablewidth{0pt}
\tabletypesize{\scriptsize}
\tabcolsep=3pt
\tablecolumns{13}
\tablecaption{\OVI\ detections and comparison with values in the literature}
\tablehead{%
\ch{Object}       &\ms4{Literature Values}                                      &\ms3{This paper}                           &\ch{Qual$^\Nqual$}&\ch{Flag$^\Ndflag$}
                                                                                                                                  &\ch{Galaxy}&\ch{\ip}  &\ch{Note}\\
                  &\ch{$v$(\OVI)}&\ch{EW(1031)}&\ch{EW(1037)}&\ch{Ref.$^\Nref$} &\ch{$v$(\OVI)}&\ch{EW(1031)}&\ch{EW(1037)} &     &     &           &          &         \\
                  &\ch{[\kms]}   &\ch{[\mA]}   &\ch{[\mA]}   &                  &\ch{[\kms]}   &\ch{[\mA]}   &\ch{[\mA]}    &     &     &           &\ch{[kpc]}&         \\
\ch{(1)}&\ch{(2)}&\ch{(3)}&\ch{(4)}&\ch{(5)}&\ch{(6)}&\ch{(7)}&\ch{(8)}&\ch{(9)}&\ch{(10)}&\ch{(11)}&\ch{(12)}&\ch{(13)}
}\startdata
1H\,0717+714      &      &         &         &          & 2914 & 66\E15\E 9 &$<$23         &?&+& UGC3804          & 199 &         \\ 
3C\,273.0         & 1013 &  26\E10 & $<$30     & Se+2001  & 1008 & 21\E 3\E 7 &            &!& & LGG292           &$>$191 &         \\ 
                  & 1015 &  35\E 6 &         & DSRS2006 &      &    "       &            & & & "                &     &         \\ 
                  & 1001 &  31\E 7 &         & Tr+2008  &      &    "       &            & & & "                &     &         \\
                  & 1589 &         &  17\E10 & Se+2001  & 1580 &            &$<$10         & &$-$& LGG287           &$>$311 &         \\ 
                  & 1586 &         &  $<$9\NEW & DSRS2006 &      &            &  "         & & & "                &     &         \\
ESO\,185$-$IG13     &      &         &         &          & 2627 &335\E68\E15 &            &!&+& IC4889           &  62 &         \\ 
HE\,0226$-$4110     & 5234 &  40\E10 &         & Le+2006  & 5240 & 41\E 6\E 4 &            &+& & NGC954           & 562 & \NHE    \\
\enddata
\end{deluxetable}

\section{Absorption Line Results}
\par In this section we first compare our line identifications listed in
Tables~\Tres, \Tion, \THI\ and \TOVI\ with those claimed in previous work
(Sect.~\SScompare).  We then discuss a number of analyses that can be done from
the perspective of the \Lya/\Lyb/OVI\ lines, without reference to the galaxies
near the sightlines, specifically the distribution of linewidths
(Sect.~\SSlinewidth), the frequency of absorbers ($dN/dz$, Sect.~\SSdndz), and
the evolution of linewidths over time (Sect.~\SSbevol). Since the width and
$dN/dz$ distributions and their implications have been presented and discussed
previously (Penton et al.\ 2004; Lehner et al.\ 2007 for \Lya\ and Tripp et al.\
2008 for \OVI), and since our distributions are similar, we defer to those
papers for a more detailed discussion of the implications.

\subsection{Comparison with Previous Work}
\par A number of previous authors have published measurements for very low
redshift \Lya\ and \OVI\ absorbers at $z$$<$0.017. Among these are Bowen et al.\
(1996, 2002), Penton et al.\ (2000a, 2000b, 2002, 2004), C\^ot\'e et al.\
(2005), Danforth et al.\ (2006) and Danforth \& Shull (2008). Where Bowen et
al.\ and C\^ot\'e et al.\ concentrated on looking for absorption associated with
nearby galaxies (with sample sizes of 7 and 5 sightlines, respectively), Penton
et al.\ looked at the statistics of all \Lya\ absorbers found in their 30
sightlines and Danforth \& Shull (2008) studied the \OVI\ absorption in the
Penton et al.\ sample.
\par Our sample includes the sightlines from these papers, but we also include
30 new sightlines for which low-redshift data have not yet been published. Of
the 129 \HI\ systems that we list (see Tables~\Tres\ and \THI) Bowen et al.\
(1996, 2002) previously published values for 23, C\^ot\'e et al.\ (2005) for 5,
and Penton et al.\ (2000a, b, 2002, 2004) for 48. A few other papers also
presented systems in some sightlines, but 45 systems (33\% of the sample, 29
\Lya\ lines and 36 \Lyb\ lines) are presented here for the first time.
\par We do not confirm 20 of the previously published \HI\ systems. These are
indicated by a ``$-$'' in Column 8 or 9 in Table~\THI\ and they are discussed
individually in the Appendix. Nine of these disagreements are for systems that
are listed as $<$3$\sigma$ detections in the various Penton et al.\ papers
(toward HE\,1029$-$1401, HE\,1228+0131, Mrk\,335, PKS\,2005$-$489 and
PKS\,2155$-$304), while three are due to different interpretations of some of
the spectra in Bowen et al.\ (2002) (toward MCG+10-16-111, PG\,1216+069 and
PG\,1341+258). Of the remaining six, four were listed as $<$4$\sigma$ detections
by the respective authors (toward HE\,0226$-$4110, HE\,1228+0131, HS\,0624+607
and PKS\,0405$-$12), while two toward PHL\,1811 were misidentified as redshifted
\OVI\ lines. So, on balance, the only systems with whose reality we do not agree
were at the limit of detection.
\par For \OVI\ the literature comparison shows a much worse situation. Sembach
et al.\ (2001), Richter et al.\ (2004), and Lehner et al.\ (2006) previously
published \OVI\ absorptions toward 3C\,273.0, PG\,1259+593 and HE\,0226$-$4110,
respectively. We agree with all of their low-$z$ detections. Danforth et al.\
(2006) previously systematically analyzed the low-redshift \OVI\ absorbers. We
agree with only three of the six relevant detections that they listed (toward
3C\,273.0, PG\,1259+593 and Mrk\,876). We also find three new \OVI\ lines in
sightlines that they studied (toward MRC\,2251$-$178, Mrk\,876 and
PG\,0953+414). In addition, we find seven new \OVI\ absorbers in other
sightlines (1H\,0717+714, ESO\,185-IG13, Mrk\,290, PG\,0844+349, PG\,1302$-$102,
Ton\,S180 and Ton\,S210), making for a total of fourteen positively identified
\OVI\ absorbers at $v$$<$6000~\kms. All these systems are commented on in
Sect.~\SSOVIabs\ and in the Appendix.
\par Danforth et al.\ (2006) and Danforth \& Shull (2008) listed \OVI\
detections and non-detections for all \Lya\ systems in their sightline sample,
but we only compared to our results for the 33 systems at $v$$<$6000~\kms. As
mentioned above, we agree with three of their detections, and do not confirm
three other detections. We further agree with 20 of their non-detections, but we
consider seven non-detections to have been given in error (see below). I.e., we
agree with 23 of 33 (70\%) of their results. Since we find three new detections
in their sightlines, the detection statistics are not affected much, even though
the samples overlap for only three of the fourteen positive identifications.
\par The three absorption lines for which do not agree with the Danforth et al.\
(2006) claim that they are redshifted \OVI\ are the following. a) A 29\E16~\mA\
line at 3205~\kms\ toward MRC\,2251$-$178, complementing strong \Lya\ and \Lyb;
as can be seen in Fig.~\Fspectra, there is no convincing evidence for \OVI\
absorption at this velocity, so we set an upper limit of 27~\mA. b) \OVI\ at
1147~\kms\ toward PG\,0804+761; this feature is more likely to be
\CII\lm1037.337\ in the Galactic high-velocity cloud complex~A (see Appendix).
c) A 45\E14~\mA\ \OVIb\ absorber at 2130~\kms\ toward PG\,1211+143;
Fig.~\Fspectra\ clearly shows no evidence for a feature here and we set an upper
limit of 19~\mA.
\par For seven non-detections listed by Danforth et al.\ (2006), we instead
argue for a detection in three cases (see above). For four others (toward
PG\,1116+215, PHL\,1811 and PKS\,2005$-$489) there are other absorption lines
(\Lyg, \Lyh, \Lyxi, \H2) at the velocity of \OVI\ corresponding to a \Lya\
detection. So, in practice it is not possible to set a useful upper limit.

\subsection{Discussion of \OVI\ Absorbers}
\par In this subsection we summarize the \OVI\ absorbers in our sample. The
spectra can be seen in Fig.~\Fspectra. We derive column densities using the
apparent optical depth method (Sembach \& Savage 1992). We also refer to line
widths, which were measured by fitting a gaussian to the apparent optical depth
profile and correcting the result for instrumental broadening (see also
Sect.~\SSlinewidth). The data for the \OVI\ absorbers are collected in
Table~\TNOVI, listing the equivalent widths, linewidths and column densities for
the \OVI\ lines, as well as other lines detected in each system.
\smallskip
\par {\it 1H\,0717+714 at 2915~\kms}
\par In the direction toward 1H\,0717+714 there is absorption at 1041.960~\AA\
that is best interpreted as \OVIa\ at a velocity of 2915~\kms\ (reported here
for the first time). The feature is located just longward of geocoronal
\OI*\lm1041.688 emission, and we measure it using the orbital-night only data,
although it is visible in the combined data. The line is 66$\pm$15~\mA, where
the 3$\sigma$ detection limit for a 60~\kms\ wide line is 12~\mA. In the
combined day+night data the detection limit is 7~\mA, so the feature is clearly
significant. The corresponding \OVIb\ line is not seen, with a detection limit
of 24~\mA. The expected value is 33$\pm$8$\pm$8~\mA, using the statistical error
near the \OVIb\ line, and a systematic error based on the \OVIa\ line, but any
\OVIb\ absorption would also be contaminated by \H2 L(5-0) P(4)\lm1047.550. 
\par The feature at 1035.603~\AA\ is very likely to be the corresponding \Lyb\
absorption, although its apparent velocity differs by 27~\kms\ from that of the
\OVI\ line. This feature is unlikely to be interstellar \CII\lm1036.337 at
$-$207~\kms\ originating in an ionized envelope of the high-velocity cloud
complex~A, observed about one degree away. Assuming that this HVC has a
metallicity of 0.1 solar, the logarithmic carbon abundance would be $-$4.61, and
if the feature is \CII, the implied total hydrogen column density would be
$\sim$2\tdex{18}\,\cmm2. However, there is no evidence for \HI\ absorption at
$-$207~\kms\ in the higher Lyman lines. Thus, intergalactic \Lyb\ at 2888~\kms\
is the best interpretation for the feature at 1035.603~\AA. The spectrum near
\CII\ also shows a feature centered at $-$150~\kms, and this does have
corresponding \HI\ absorption in the Lyman lines, with a column density
compatible with the possible 1--2\tdex{18}\,\cmm2\ seen in the 21-cm spectrum.
It is likely that this component originates in the outskirts of complex~A.
\def\NC{$^1$}
\begin{deluxetable}{llrrrr}
\tablenum{7}
\tablewidth{0pt}
\tabletypesize{\scriptsize}
\tabcolsep=3pt
\tablecolumns{6}
\tablecaption{Intergalactic \OVI\ systems at $z$$<$0.017\NC}
\tablehead{%
\ch{Object}   &\ch{Line}    &\ch{$v$}    & \ch{$W$}    & \ch{$b$}    & \ch{log $N$}      \\
              &             &\ch{[\kms]} & \ch{[\mA]}  & \ch{[\kms]} & \ch{[\cmm2]}      \\
\ch{(1)}&\ch{(2)}&\ch{(3)}&\ch{(4)}&\ch{(5)}&\ch{(6)}
}\startdata
1H\,0717+714  & \Lyb        &     2888   &  41\E 7\E 8 &    29\E4    & 13.86\E0.08\E0.06 \\ 
1H\,0717+714  & \OVI\lm1031 &     2915   &  66\E15\E 9 &    32\E2    & 13.81\E0.17\E0.04 \\
1H\,0717+714  & \OVI\lm1037 &            & $<$23         &             &$<$13.87             \\
3C\,273.0     & \Lya        &     1010   & 394\E 7\E1  &    53\E3    & 14.27\E0.08\E0.01 \\ 
3C\,273.0     & \Lyb        &     1013   & 120\E 4\E10 &    44\E2    & 14.32\E0.02\E0.02 \\
3C\,273.0     & \OVI\lm1031 &     1008   &  21\E 3\E 7 &    34\E6    & 13.24\E0.15\E0.15 \\
ESO\,185$-$IG13 & \Lyb        &     2635   & 641\E52\E14 &    88\E3    &$>$15.35             \\ 
ESO\,185$-$IG13 & \OVI\lm1031 &     2627   & 335\E67\E15 &    87\E7    & 14.63\E0.15\E0.01 \\
ESO\,185$-$IG13 & \CIII\lm977 &     2625   & 470\E113\E42&   108\E16   & 14.18\E0.35\E0.19 \\
HE\,0226$-$4110 & \Lya        &     5235   &  60\E 7\E 2 &    17\E3    & 13.17\E0.06\E0.01 \\ 
HE\,0226$-$4110 & \OVI\lm1031 &     5240   &  41\E 6\E 4 &    17\E3    & 13.57\E0.08\E0.02 \\
HE\,0226$-$4110 & \CIII\lm977 &     5259   &  23\E 9\E 3 &    19\E4    & 12.64\E0.23\E0.15 \\
HE\,0226$-$4110 & \CIV\lm1548 &     5232   &  39\E11\E2  &     7\E4    & 13.03\E0.20\E0.14 \\
MRC\,2251$-$178 & \Lya        &     2265   & 133\E12\E 3 &    64\E4    & 13.43\E0.05\E0.01 \\ 
MRC\,2251$-$178 & \OVI\lm1037 &     2283   &  40\E12\E 9 &    35\E5    & 13.87\E0.25\E0.08 \\
Mrk\,290      & \OVI\lm1031 &     3073   &  49\E 8\E 7 &    34\E4    & 13.63\E0.08\E0.06 \\ 
Mrk\,290      & \OVI\lm1037 &     3073   &  20\E 8\E 7 &    31\E3    & 13.46\E0.14\E0.17 \\
Mrk\,876      & \Lya        &      936   & 476\E14\E 3 &    77\E2    & 14.27\E0.08\E0.01 \\ 
Mrk\,876      & \Lyb        &      933   &  79\E 6\E33 &    75\E2    & 14.14\E0.04\E0.03 \\
Mrk\,876      & \OVI\lm1031 &      945   &  17\E 4\E 8 &    29\E3    & 13.18\E0.13\E0.16 \\
Mrk\,876      & \OVI\lm1037 &            & $<$16         &             &$<$13.13             \\
Mrk\,876      & \SiIII\lm1206&     912   &  42\E 9\E 1 &    10\E4    & 12.34\E0.15\E0.11 \\
Mrk\,876      & \Lya        &     3481   & 267\E12\E 3 &    37\E3    &$>$14.06             \\ 
Mrk\,876      & \OVI\lm1031 &     3508   &  18\E 4\E 7 &    29\E4    & 13.18\E0.15\E0.16 \\
PG\,0844+349  & \Lyb        &      351   &  25\E 8\E 2 &     6\E2    & 13.66\E0.09\E0.03 \\ 
PG\,0844+349  & \OVI\lm1031 &      365   &  37\E 5\E 2 &    28\E2    & 13.57\E0.08\E0.02 \\
PG\,0844+349  & \OVI\lm1037 &      365   &  27\E 5\E 7 &    ..       & 13.74\E0.09\E0.04 \\
PG\,0953+414  & \Lya        &      621   &  70\E 9\E 3 &    41\E3    & 13.19\E0.06\E0.01 \\ 
PG\,0953+414  & \Lyb        &            & $<$23         &             &$<$13.52             \\
PG\,0953+414  & \OVI\lm1031 &      637   &  39\E 8\E 7 &    53\E2    & 13.57\E0.11\E0.07 \\
PG\,1259+593  & \Lya        &      678   & 231\E 9\E 3 &    62\E3    & 13.81\E0.02\E0.01 \\ 
PG\,1259+593  & \Lyb        &            & $<$15         &             &$<$13.26             \\
PG\,1259+593  & \OVI\lm1031 &      627   &  25\E 5\E 2 &    24\E4    & 13.33\E0.11\E0.03 \\
PG\,1259+593  & \OVI\lm1037 &      622   &  13\E 3\E 2 &    15\E3    & 13.41\E0.12\E0.06 \\
PG\,1302$-$102  & \Lya        &            & $<$72         &             &$<$13.21             \\
PG\,1302$-$102  & \OVI\lm1031 &     3109   &  19\E 6\E 8 &    13\E5    & 13.23\E0.15\E0.15 \\
PG\,1302$-$102  & \OVI\lm1037 &            & $<$23         &             &$<$13.34             \\
Ton\,S180     & \Lyb        &            & $<$37         &             &$<$13.76             \\ 
Ton\,S180     & \OVI\lm1031 &      260   &  51\E10\E 8 &    24\E4    & 13.68\E0.09\E0.05 \\
Ton\,S180     & \OVI\lm1037 &      285   &  34\E13\E 8 &    33\E5    & 13.75\E0.24\E0.09 \\
Ton\,S180     & \CIII\lm977 &      279   &  60\E28\E 7 &    41\E8    & 13.10\E0.26\E0.16 \\
Ton\,S210     & \Lyb        &            & $<$20         &             &$<$13.45             \\ 
Ton\,S210     & \OVI\lm1031 &      288   &  25\E 8\E 7 &    41\E6    & 13.38\E0.13\E0.10 \\
\enddata
\tablecomments{%
1: Column (1) gives the target name. Column (2) gives the \OVI\ lines and other
lines in the system that are detected or, in the cases of \Lya, \Lyb, \OVI,
where it is possible to set an upper limit. Columns (3), (5) and (6) give the
central (heliocentric) velocity, linewidth and logarithmic column density
derived from a gaussian fit to the apparent optical depth profile (see Sembach
\& Savage 1992). Column (4) gives the equivalent width, \fix{with upper limits
determined by integrating over a 60~\kms\ wide window centered on the central
velocity of the detected ion.}
}
\end{deluxetable}
\par The galaxy nearest the sightline is UGC\,3804, at \vgal=2887~\kms,
\ip=199~kpc, with diameter 22.8~kpc. It is a member of the LGG\,141 group, but
most of the group galaxies have impact parameters $>$650~kpc.
\par Nominally, the velocity of \Lyb\ and \OVIa\ differ by 27~\kms, but the S/N
in the \OVI\ spectrum is too low to know for sure. The linewidths of the \Lyb\
and \OVIa\ absorption are very similar (29 vs 32~\kms). Thus, the velocity
offset between \HI\ and \OVI\ suggests collisional ionization, while the
similarity in the linewidths suggests photoionization (although the \OVI\ line
would be relatively wide). \fix{However, we conclude that the measurements are
too noisy to properly distinguish between the different ionization origins.}
\smallskip
\par {\it 3C\,273.0 at 1010~\kms}
\par This \OVIa\ absorption is relatively clear, and it was previously reported
by Sembach et al.\ (2001), Danforth \& Shull (2005) and Tripp et al.\ (2008).
The strong \Lya\ and \Lyb\ lines originate in LGG\,292, one of the groups near
the Virgo cluster. The nearest group galaxy is MCG0-32-16 (\vgal=1105~\kms,
\Rgal=6.3~kpc), with impact parameter 191~kpc. A number of smaller galaxies have
\ip=200--500~kpc. As can be seen in Fig.~\Fspectra, the \OVIb\ line is strongly
contaminated by \H2\ L(5-0) R(3)\lm1041.158 absorption. Although the \Lya\ line
is very strong, the \HI\ column densities derived from the \Lya\ and \Lyb\ lines
match. The \OVI\ line is only slightly narrower than the \HI\ lines
($b$=34~\kms\ vs $b$$\sim$50~\kms). \fix{Nominally this implies a temperature of
8.5\tdex4~K and $b$(turbulent)$\sim$30~\kms.} As the \HI\ and \OVI\ velocities
are close (1010 vs 1008~\kms), the system is likely to be mostly photoionized
gas. although the temperature would be very high for gas in photoionization
equilibrium.
\smallskip
\par {\it ESO\,185-IG13 at 2627~\kms}
\par The \Lyb\ line in this spectrum is so strong that it is saturated.
Similarly, the \OVIa\ at 2627~\kms\ is very strong, so even though we can only
use orbital-night-only data, and even though the spectrum is rather noisy, the
detection is clear. The \OVIb\ line, unfortunately, is hidden by the intrinsic
\Lyb\ absorption, but it would probably be visible in a spectrum with higher S/N
ratio. The sightline to ESO\,185-IG13 is the seventh closest to another galaxy
in our sample. IC\,4889 (\vgal=2526~\kms, \Rgal=28.9~kpc) has impact parameter
62~kpc. \fix{The \Lyb\ and \OVI\ lines are very broad (88 vs 87~\kms) and close
in velocity. However, the saturation of the \Lyb\ line limits the value of the
line width comparison. The \OVI\ could be either photoionized or collisionally
ionized.}
\smallskip
\par {\it HE\,0226$-$4110 at 5240~\kms}
\par Near a velocity of 5240~\kms\ there is a set of narrow lines that is likely
to contain intergalactic \Lya, \Lyb\ and \OVIa. These were previously reported
by Lehner et al.\ (2006) and Tripp et al.\ (2008), but were not listed by
Danforth \& Shull (2005, 2008). The \Lya\ line is clear, though rather narrow
($b$=17~\kms), which sets an upper limit on the gas temperature of 6500~K. The
\Lyb\ line is confused by the \FUSE\ detector flaw near 1043~\AA. So although
there appears to be a feature where \Lyb\ is expected, it can't be reliably
measured. At the velocity of the \OVIb\ line is a \H2\ line (L(4-0)
R(1)\lm1049.960), but the \H2\ model that is based on all \H2\ lines combined
shows that the feature is too strong. The best explanation is additional
intergalactic \OVI\ at 5240~\kms. After correcting for the \H2\ line, the
linewidth and velocity of this feature matches that of \Lya, strongly suggesting
that the gas is photoionized. Unfortunately, the corresponding \OVIb\ is
contaminated by intergalactic \OIV\lm787.711 at $z$=0.34035 (see Lehner et al.\
2006).
\par This sightline passes 562~kpc from NGC\,954 (\vgal=5353~\kms,
\Rgal=33.0~kpc), a member of a group of galaxies with $v$$\sim$5000~\kms\ that
is not included in Garcia et al.\ (1993). The impact parameter is the largest
for any of the \OVI\ absorbers in our sample, but the galaxy surveys in this
part of the sky are not very deep, so there is a good chance that an
$L$$>$0.1~\Lstar\ galaxy with lower impact parameter can be found.
\smallskip
\par {\it MRC\,2251$-$178 at 2283~\kms}
\par In this sightline a clear \Lya\ feature at 2265~\kms\ appears to be matched
by an \OVIb\ line at 2283~\kms, although this feature has a significance of only
about 3.5$\sigma$. Both the \Lya\ and the \OVI\ line may have two (matching)
components. The \OVIa\ line is contaminated by strong geocoronal emission, even
in the orbital-night-only data. This is therefore one of the most uncertain
\OVI\ detections. There is a suggestion that both the \Lya\ and the \OVI\
absorptions have two components, but the S/N ratio of the data is not good
enough to definitively decide this. If the \OVI\ line is real, it can be
associated with ESO\,603-G31 (\vgal=2271~\kms, \Rgal=9.1~kpc) at impact
parameter 422~kpc. Danforth \& Shull (2005) did not report this feature, nor did
they give a lower limit. \fix{If we assume that the structure inside the line is
real, photoionization is the most likely explanation for this, since the lines
would be narrow. If instead the structure is noise (quite likely), then we
conclude that the width of the \HI\ line ($b$=64$\pm$4~\kms) is about twice that
of the \OVI\ line ($b$=35$\pm$5~\kms), which would imply $T$$\sim$1.8\tdex5~K
and $b$(turbulent)$\sim$30~\kms. Thus the gas in this system is probably
collisionally ionized.}
\smallskip
\par {\it Mrk\,290 at 3073~\kms}
\par In this sightline, a clearly significant (49$\pm$8~\mA) feature at
1042.504~\AA\ is best explained as intergalactic \OVI\ absorption at 3073~\kms.
There are no interstellar lines at this wavelength. The feature cannot be \Lyb\
at 5163~\kms, as there is no matching, stronger, \Lya\ line in the \GHRS\
spectrum. The \Lya\ line corresponding to the \OVI\ has not been observed, but a
\COS\ spectrum is planned; detecting \Lya\ would confirm the \OVI\
interpretation. The corresponding \Lyb\ line is obscured by saturated Galactic
\CII\ absorption. There does appear to be a matching \OVIb\ absorption. Although
this is blended on one side with Galactic \ArI\lm1048.220, no other Galactic
interstellar absorption line has a wing on the positive-velocity side. Moreover,
if we measure just the positive-velocity side of this feature, its strength and
width match the values expected from the \OVIa\ absorption. On balance, this
\OVI\ absorber appears secure. \fix{However, there is not enough information to
determine the origin of the ionization of \OVI.}
\par Several galaxies with \vgal$\sim$3100~\kms\ have impact parameters of
300--600~kpc. In Table~\Tres\ we associate the absorber with NGC\,5987
(\ip=424~kpc), as it is by far the largest (\vgal=3010~\kms, \Rgal=51.7~kpc).
\smallskip
\par {\it Mrk\,876 at 945~\kms}
\par Toward Mrk\,876 there is a feature at 1035.179~\AA\ that we interpret as a
blend of \H2\ L(6-0) P(4) at 1035.783~\AA\ and \OVIa\ redshifted to a velocity
of 945~\kms. As can be seen in Fig.~\Fspectra, the \H2\ model does not exactly
match this feature, unlike what is the case for the other \H2\ $J$=4 lines (see
e.g.\ the Mrk\,876--UGC\,10294 panel in Fig.~\Fspectra). After removal of the
\H2\ absorption, a 4$\sigma$ absorption remains, which we interpret as
intergalactic \OVIa, because there also is \Lya\ and \Lyb\ absorption at its
velocity; these lines were previously reported by Shull et al.\ (2000) and
C\^ot\'e et al.\ (2005). Finally, there is a 42$\pm$9~\mA\ feature at
1210.170~\AA\ that is most likely \SiIII\ at 912~\kms.
\par The \Lya\ line is broad ($b$=75~\kms) and at a velocity of 936~\kms. The
S/N ratio of the \STIS-E140M spectrum is insufficient to determine whether this
line is a single or multi-component absorber, although the apparent optical
depth profile looks compatible with a single component. The corresponding \Lyb\
absorption is strongly contaminated by two-component \H2\ L(6-0) R(3)\lm1028.985
absorption, but after correcting for this contamination, the implied \HI\ column
density matches that of the \Lya\ line to within the errors.
\par Although the width of the \OVI\ line is difficult to measure, it is less
than half that of the \Lya\ line ($b$=29~\kms). \fix{The difference in the \HI\
and \OVI\ linewidths implies $T$=3\tdex5~K and $b$(turbulent)=23~\kms, implying
the origin of the ionization is almostly certainly collisional ionization.}
\par A large (\Rgal=27.1~kpc) isolated galaxy (NGC\,6140, \vgal=910~\kms) lies
only 206~kpc from the sightline, and it is the only candidate galaxy to
associate with the absorber.
\smallskip
\par {\it Mrk\,876 at 3508~\kms}
\par The strong \Lya\ line at 3481~\kms\ (previously reported by Danforth \&
Shull 2005) is matched by \Lyb\ absorption, although the latter is blended with
Galactic \OVIb, and no equivalent width can be measured. The presence of \Lyb\
is shown by the fact that the apparent optical depth profiles of Galactic \OVIa\
and \OVIb\ do not match, the only sightline in the \FUSE\ sample for which this
is the case (see Wakker et al.\ 2003). A 4.5$\sigma$ (17$\pm$4~\mA) feature at
1044.001~\AA\ is best explained as a corresponding \OVIa\ line at 3501~\kms,
even though it is offset in velocity from \Lya\ by 27~\kms. Danforth \& Shull
(2005) reported an upper limit of 16~\mA\ at this velocity. There is only one
galaxy that can clearly be associated with the absorption: UGC\,10294
(\vgal=3516~kpc, \Rgal=27.6~kpc, \ip=282~kpc). The widths of the \Lya\ and
\OVIa\ lines are similar (37 vs 29~\kms), which would support the idea that this
system consists of photoionized gas. However, such an origin does not explain
the 27~\kms\ velocity offset, so collisional ionization is more likely.
\smallskip
\par {\it PG\,0844+349 at 365~\kms}
\par The \FUSE\ spectrum of this target has several features that are difficult
to interpret, but which we decided are \Lyb, \OVIa\ and \OVIb\ at
$\sim$360~\kms. Unfortunately, this velocity is such that any \Lya\ line would
be hidden in the Galactic \Lya\ line. In the \OVIa\ part of the spectrum there
are three clear features. We interpret the two at the most positive velocities
as \Lyb\ at 2260 and 2326~\kms, especially since there is a strong feature in
the low-resolution \FOS\ spectrum near these velocities. Normally, we would be
inclined to identify the third feature also as \Lyb, although this line is only
a 3$\sigma$ detection, and appears to be extremely narrow. Furthermore, the
spectrum lies below the continuum at the wavelengths where \OVIb\ is expected,
even though this falls between a \H2\ line (L(5-0) R(2)\lm1038.689) and Galactic
\OI\lm1039.230. \fix{The low quality of the \Lyb\ measurement makes it difficult
to determine the origin of the ionization of \OVI\ even though the \OVI\ line
appears to be broader than the \Lyb\ line and there is a velocity offset.}
\smallskip
\par {\it PG\,0953+414 at 637~\kms}
\par Toward this target there are two clear features that can be interpreted as
\Lya\ at 621~\kms\ and \OVIa\ at 637~\kms, which have not been reported before.
Both of these features are very significant (70$\pm$9 and 39$\pm$8~\mA), and
intergalactic \OVI\ is the most likely interpretation for the feature at
1034.129~\AA. This system can be associated with NGC\,3104 (\vgal=612~\kms,
\Rgal=11.5~kpc, \ip=296~kpc), which is the only galaxy with $v$$\sim$600~\kms\
and \ip$<$600~kpc. The widths of the \Lya\ and \OVIa\ lines are \fix{difficult
to measure because the \HI\ line is noisy and in the wing of Galactic \Lya.}
They are similar, though large ($b$=41$\pm$3 and 53$\pm$2~\kms). Since the
velocities differ by 16~\kms, this \OVI\ system may originate in a collisionally
ionized gas.
\smallskip
\par {\it PG\,1259+593 at 627~\kms}
\par In the spectrum of this target there is a set of aligned features. The
\Lya\ line at 678~\kms\ is clear, and fairly broad. Tripp et al.\ (2008)
interpreted this as a two-component system, probably basing this on the possibly
double \OVI\ line. However, the noise in the \Lya\ spectrum is too high to be
certain. Therefore, we list this as a single system. The expected corresponding
\Lyb\ line is weak, and, if present, would be hidden by \H2\ L(6-0)
P(2)\lm1028.104. At velocities of 627 and 622~\kms\ there is a matching set of 
\OVI\ features, whose equivalent widths are in the ratio 2:1. We interpret this
set of absorbers as an intergalactic system associated with UGC\,8146   
(\vgal=669~\kms, \Rgal=12.6~kpc, \ip=80~kpc), even though there is a dwarf   
galaxy (SDSS\,J130206.46+584142.8, \Rgal=1.8~kpc) with slightly lower impact 
parameter (72~kpc). These features were previously reported by Richter et al.\
(2004), C\^ot\'e et al.\ (2005), Danforth et al.\ (2006) and Tripp et al.\
(2008).
\par \fix{The widths of the two \OVI\ lines appear to differ ($b$=24$\pm$4 and
15$\pm$3~\kms), but this can be attributed to the noisiness of the data. The
average of these is about a factor three smaller than the width of the \Lya\
line ($b$=62$\pm$3~\kms). However, mostly because the \HI\ and \OVI\ velocities
differ by 51~\kms, collisional ionization is the more likely explanation for the
origin of this system.}
\smallskip
\par {\it PG\,1302$-$102 at 3109~\kms}
\par The possible \OVIa\ absorber at 3109~\kms\ that we list toward
PG\,1302$-$102 is the least certain detection in our sample. \Lyb\ is hidden by
Galactic \CII\ absorption, \OVIb\ would be too weak and blends with
\ArI\lm1048.220. For \Lya\ only an upper limit of 72~\mA\ can be set. However,
since two of the seven \Lya\ lines in systems with \OVI\ have $W$$\sim$70~\mA,
the \Lya\ non-detection is not problematic. The \OVIa\ line itself is not
strong, but appears clear (see Fig.~\Fspectra). The line is narrow
($b$=13~\kms), so if this system is real, it probably consists of photoionized
material, \fix{as turbulent broadening usually is at least this large.}
\par The galaxy NGC\,4939 (\vgal=3111~\kms, \Rgal=24.6~kpc) lies only 104~kpc
from the sightline. It is the only galaxy with velocity near 3109~\kms\ with
\ip$<$900~kpc, and thus the only candidate for associating with the absorber.
\fix{The small impact parameter to this galaxy is one of the arguments we use to
interpret the feature as a real line, as we find all other $L$$>$0.1\,\Lstar\
field galaxies with impact parameter $<$350~kpc have associated \Lya\
absorption.}
\smallskip
\par {\it Ton\,S180 at 260~\kms}
\par In this sightline there is a clear pair of \OVI\ lines, centered at
260~\kms\ (first reported by Wakker et al.\ 2003). The separately derived column
densities are compatible with each other. However, no \HI\ is detected, as \Lyb\
is blended with geocoronal \OI*\lm1027.431, and any \Lya\ would be hidden by the
Galactic \Lya\  absorption. On the other hand, there is a possible \CIII\ line
centered at 279~\kms\ that goes with the \OVI\ detection. On balance, the
identification of the \OVI\ lines appears secure. In spite of their relatively
low velocity (below 400~\kms), we identify the lines as intergalactic, rather
than as originating in the Milky Way halo, since no other Galactic high-velocity
gas with high positive velocities is known in the part of the sky where
Ton\,S180 is located, and because NGC\,247 (\vgal=159~\kms, \Rgal=15.7~kpc) has
an impact parameter of only 125~kpc. \fix{There is not enough information to
determine the origin of the ionization of \OVI.}
\smallskip
\par {\it Ton\,S210 at 288~\kms}
\par The feature at 1032.933~\AA\ is interpreted as probable \OVIa\ at 288~\kms,
even though there is no corroborating \Lya\ (hidden in Galactic \Lya), \Lyb\
(confused by geocoronal \OI*\lm1027.431) or \OVIb\ (blended with \H2\ L(5-0)
R(2)\lm1038.689). However, there are no other intergalactic absorption systems
that produce e.g.\ \Lyb\ or metal lines at this wavelength, so \OVIa\ is the
most likely interpretation. If so, this line is associated with NGC\,253
(\vgal=251~\kms, \Rgal=20.7~kpc, \ip=374~kpc). However, the velocity of this
absorber is low enough that another possibility is that the gas is located in
the Local Group. \fix{There is not enough information to constrain the origin of
the \OVI\ absorption.}
\smallskip
\par In their analysis of 78 \OVI\ absorbers, Tripp et al.\ (2008) plotted the
\OVI/\HI\ column density ratio vs $N$(\HI) and found a strong correlation
between the two. If we do this for the 11 systems in Table~\TNOVI\ with both
\HI\ and \OVI\ results, we find that our systems follow the same relation, with
$N$(\HI) ranging from 13.17 to 14.27, which falls in the middle of the range
found by Tripp et al.\ (2008). Further, for the most part, the linewidths of our
\OVI/\HI\ absorbers follow the distribution found by Tripp et al.\ (2008) (see
Sect.~\SSlinewidth). This suggests that the conclusions of Tripp et al.\ (2008)
concerning the nature of the low-redshift \OVI\ absorbers are also valid for
these 11 systems.
\par\fix{As discussed above, we have a reasonably good handle on the origin of
the ionization in 8 of the 14 systems. Based on the linewidth and velocity
offset comparison, we conclude that collisional ionization is likely for five
absorbers (toward MRC\,2251$-$178, Mrk\,876 (both systems), PG\,0953+414 and
PG\,1259+593), while for three photioioniation is more likely (toward 3C\,273.0,
HE\,0226$-$4110 and PG\,1302$-$102). The origin of the \OVI\ ionization remains
undetermined in 6 of the 14 systems.}
\par As can be seen from the cases above, 10 of the 14 \OVI\ absorbers originate
within 550~kpc from an \Lstar\ galaxy, and the other four originate within
450~kpc of an 0.1\,\Lstar\ galaxy. \fix{We therefore agree with Stocke et al.\
(2006)} that {\it in general, low-redshift intergalactic \OVI\ only originates
within 500~kpc from bright galaxies.}

\subsection{Distribution of \Lya, \Lyb\ and \OVI\ Linewidths}
\begin{figure}\plotfiddle{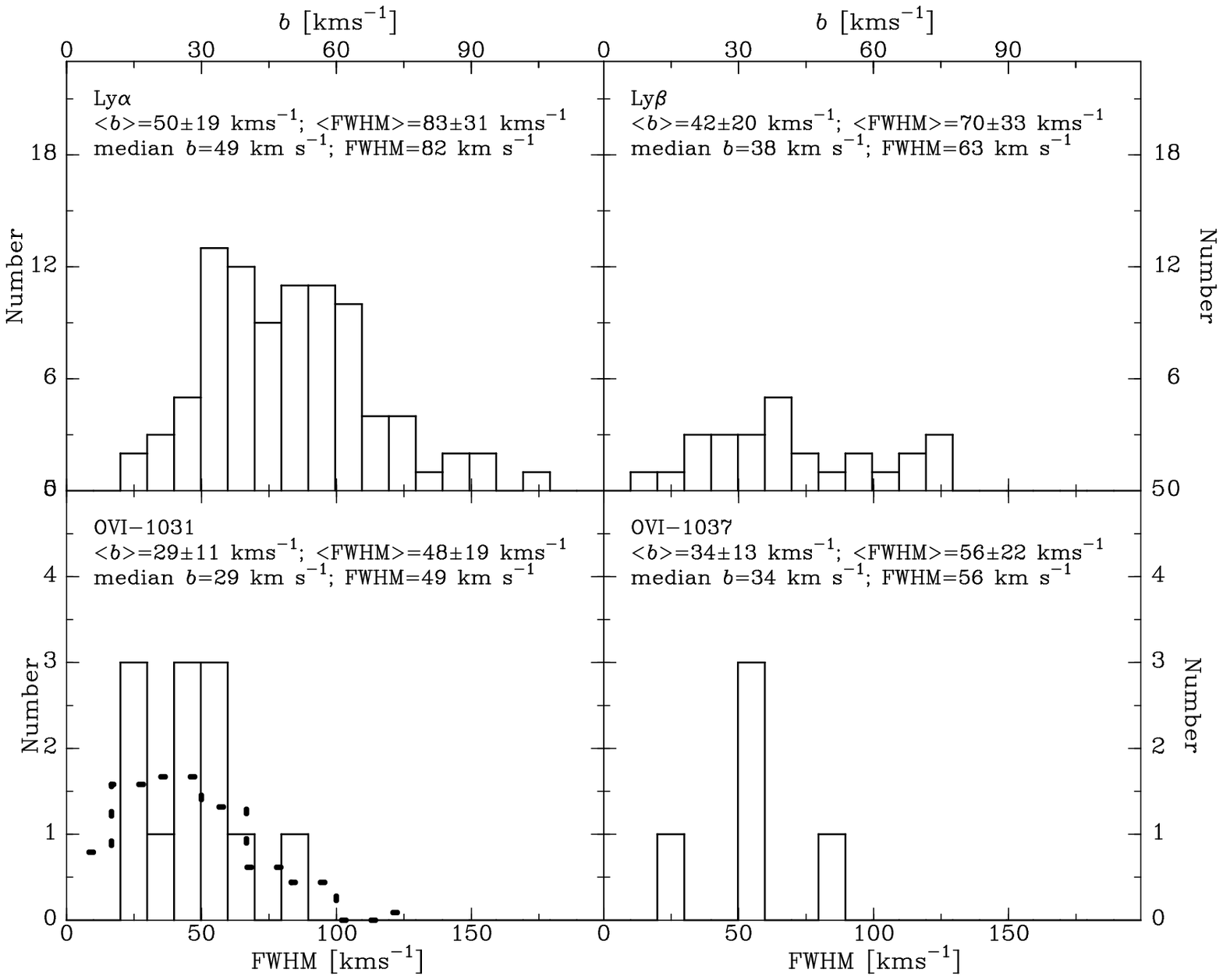}{0in}{0}{450}{350}{0}{-150}\figurenum{4}\caption{\captiondetwidth}\end{figure}
\par Figure~\Fdetwidth\ shows the distributions of fitted linewidths, separately
for each of four lines -- \Lya, \Lyb, \OVIa\ and \OVIb. The widths have been
corrected for instrumental broadening, although this makes only a slight
difference, as the resolution is 6.5~\kms\ for \STIS-E140M data and
$\sim$30 and 20~\kms\ for \STIS-G140M and \FUSE\ data, respectively, while the
full width at half maximum (FWHM) of the lines typically is much larger. We
removed saturated \Lya\ lines, some weak lines with low significance and lines
in noisy spectra, as well as lines observed with the \GHRS, as its line-spread
function has broad wings. We also removed lines that are blended with \H2\ or
with the \FUSE\ detector flaw or which are otherwise problematic (such as
multi-component lines).
\par For \Lya\ we find $<$$b$$>$=50~\kms, median 49~\kms\ and dispersion
19~\kms, where Penton et al.\ (2000b) measured $<$$b$$>$=38~\kms, median
35~\kms\ and dispersion 16~\kms, while Danforth \& Shull (2008) give a median of
28~\kms\ and dispersion 16~\kms. Although 47 of our 115 \Lya\ lines overlap with
the Penton et al.\ (2000b) sample, the discussion in the next subsection
suggests that the difference in the medians may be significant and that the
typical linewidth at $z$=0--0.017 may be larger than at $z$=0.017--0.2.
\par A new result is our derivation of these values for \Lyb: $<$$b$$>$=42~\kms,
median 38~\kms\ and dispersion 20~\kms, i.e.\ on average the \Lyb\ lines seem to
be slightly narrower than the \Lya\ lines. This may indicate that some of the
\Lya\ lines that we thought were well-measured are still affected by saturation
and other effects.
\par For \OVI, $<$$b$$>$=29~kms, median 29~\kms\ and dispersion 11~\kms\
(excluding the noisy, apparently broad detection associated with IC\,4889).
Thus, the average/median is smaller for \OVI\ than for \HI. As discussed in
Sect.~\SSOVIabs, we find individual cases where $b$(\OVI)$<$$b$(\HI), but also
cases where the linewidths are similar. Tripp et al.\ (2008) showed the
linewidth distribution of 74 \OVI\ absorbers that were found in a sample of 16
AGNs with redshifts up to 0.5. We show their distribution by the dotted line in
Fig.~\Fdetwidth\ (scaling to our sample size of 14). When compared to the 14
absorbers at $z$$<$0.017 that form our sample, it is clear that the two
distributions are similar.

\subsection{$dN/dz$ -- Frequency of Intergalactic \Lya, \Lyb\ and \OVI\ Absorption}
\begin{deluxetable}{ccccc}
\tablenum{8}
\tablewidth{0pt}
\tabletypesize{\footnotesize}
\tablecolumns{5}
\tablecaption{$dN$(OVI)/$dz$ values$^1$}
\tablehead{%
\ch{item}      &\ch{This paper}&\ch{Danforth \& Shull (2005)$^2$}&\ch{Tripp et al.\ (2008)}&\ch{Thom \& Chen (2008)}\\
\ch{(1)}&\ch{(2)}&\ch{(3)}&\ch{(4)}&\ch{(5)}
}\startdata
path@\Wlim=50~\mA & 0.4         & 2.0        & 2.8                 & 2.4            \\
\# sightlines     & 76          & 31         & 16                  & 16             \\
\Wlim\ scale$^3$  & 1.0         & 1.3        & 1.0                 & 1.5            \\
\Wlim=15~\mA      &             & 19\E3 (38) &                     &                \\
\Wlim=20~\mA      & 50\E22 (10) &            &                     &                \\
\Wlim=30~\mA      & 16\E9 (7)   & 17\E3 (35) & 15.6\E2.9\E2.4 (41) & 10.4\E2.2 (22) \\
\Wlim=50~\mA      &  8\E5 (3)   &  9\E2 (19) &                     &  6.7\E1.7 (16) \\
\Wlim=70~\mA      &  2\E3 (1)   &            &  8.8\E2.1\E1.7 (27) &                \\
\Wlim=100~\mA     &             &  3\E2 (6)  &  4.5\E1.5\E1.2 (14) &                \\
\Wlim=200~\mA     &             &            &  2.2\E1.2\E0.8 (6)  &                \\
\Wlim=300~\mA     &             &            &  0.9\E1.0\E0.5 (3)  &                \\
\enddata
\tablecomments{%
1: Entries gives the published values for low-redshift $dN$(\OVI)/$dz$, followed
by the number of detections (in parentheses) at the given equivalent width
limit. 
2: Danforth \& Shull (2008) show a plot of $dN/dz$ based on more data, but they
do not give a table. However, the distribution in their plot follows the numbers
in Danforth \& Shull (2005).
3: The different studies used different ways to define the equivalent width
limit at a given wavelength. The listed values are those reported in the
original publications. However, to properly compare the $dN/dz$ values the
equivalent width limits need to be scaled with the values in row three. E.g.,
$dN/dz$ (Danforth \& Shull at 30~\mA) should be compared to $dN/dz$ (this paper
at 40~\mA).
}
\end{deluxetable}
\par
\begin{figure}\plotfiddle{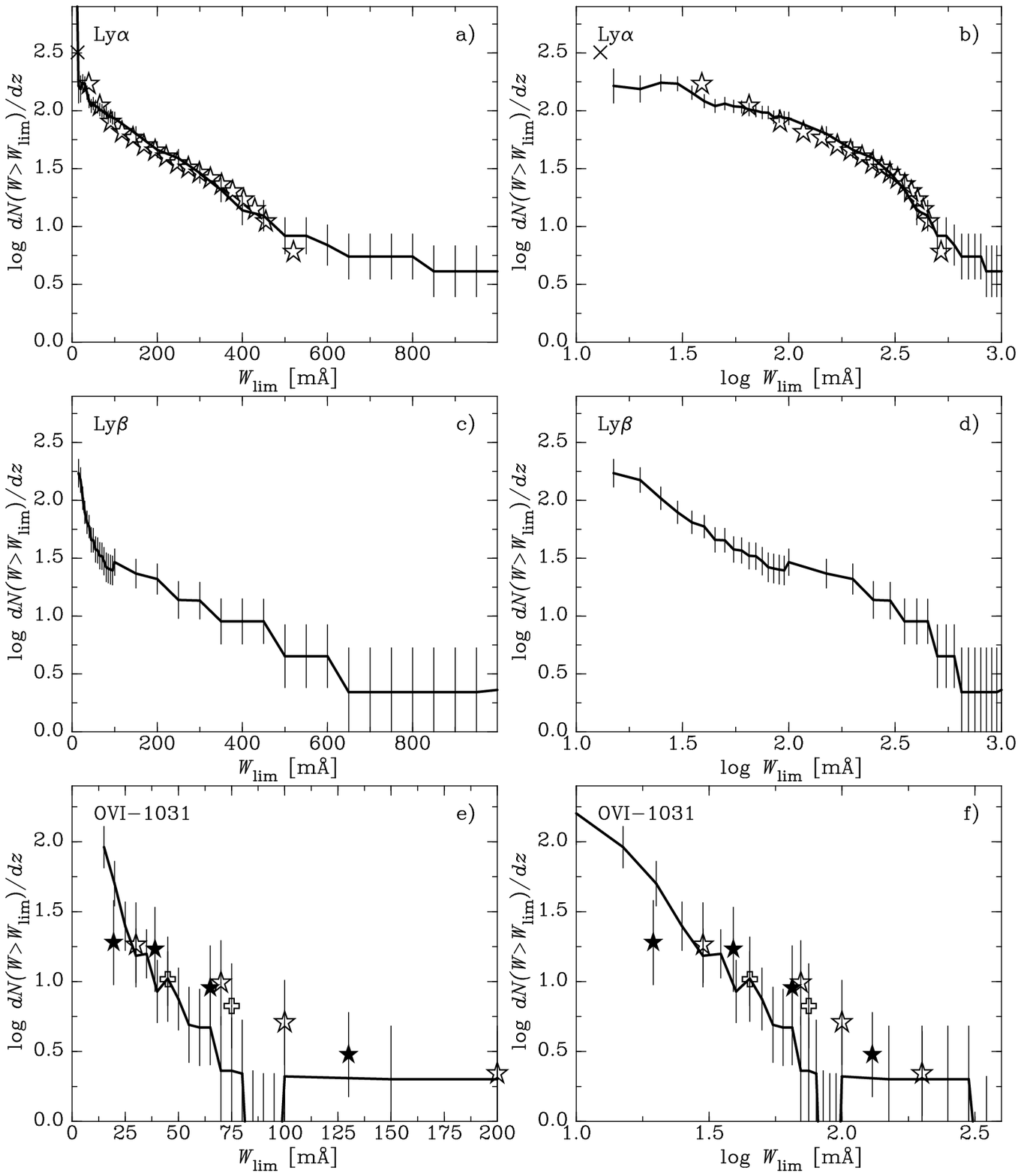}{0in}{0}{400}{400}{0}{-150}\figurenum{5}\caption{\captiondndz}\end{figure}
\par In Fig.~\Fdndz\ we present plots for the frequency with which \Lya, \Lyb\
and \OVIa\ occur ($dN/dz$), as function of the limiting equivalent width. This
is a fundamental quantity that can be predicted by theoretical models (e.g., Cen
et al.\ 2001; Fang \& Bryan 2001; Furlanetto et al.\ 2005; Cen \& Fang 2006), as
well as one that can be converted to an estimate of the baryon density
associated with the ion. Penton et al.\ (2004) and Tripp et al.\ (2008)
presented their results for $dN/dz$ for \Lya\ and \OVIa\ absorbers in the low
redshift universe, which in their case meant $z$$<$0.5 in 15 and 16 sightlines,
respectively.
\par Estimating $dN/dz$ as function of limiting equivalent width \Wlim\ requires
counting the number of absorbers with equivalent width $W$ larger than \Wlim\
and dividing by the total path \Dz\ over which these absorbers could have been
detected. In practice, this calculation is not trivial, because the limiting
equivalent width varies widely between the different sightlines in our sample,
and even within a single spectrum (e.g., if there are strong emission lies
intrinsic to the background AGN). We thus calculate the 3$\sigma$ equivalent
width upper limit for a 60~\kms\ wide absorber as function of wavelength for
each of \Lya, \Lyb\ and \OVIa. For a given \Wlim\ we then find the total
velocity range between 400 and 5000~\kms\ over which lines with $W$$>$\Wlim\
could have been detected. We subtract the ranges where Galactic or
higher-redshift intergalactic lines block the possible detection of
intergalactic lines. This includes both ionic and molecular low- and
high-velocity Galactic absorption. We also estimate an error for the pathlength,
taking into account the fuzzy edges of the Galactic absorption lines (50~\kms\
per sightline for \Lya, 200~\kms\ per sightline for \Lyb\ and \OVIa).  In our 76
sightlines, in the velocity range $v$=400--5000~\kms\ ($z$=0.0013--0.0167), and
for \Wlim$>$100~\mA, we build up \Dz=0.68 for \Lya, \Dz=0.41 for \Lyb\ and
\Dz=0.48 for \OVIa, This can be compared to the values \Dz=1.1 for \Lya\ in
Penton et al.\ (2004) and \Dz=3.2 for \OVIa\ in Tripp et al.\ (2008). For lower
\Wlim, the redshift path decreases \fix{(\Dz=0.51, 0.33, 0.40 for \Lya, \Lyb,
\OVIa, respectively at \Wlim=50\,mA)}, until it hits zero at \Wlim$\sim$20~\mA.
\par The resulting $dN/dz$ distributions are shown in Fig.~\Fdndz, with a
log-log scale in the right column and a linear scale in the left column. The
error estimate for each $dN/dz$ combines sqrt(\# detections) with the estimated
error in the path; the latter is only important at the lowest values of \Wlim.
\par In Figs.~\Fdndz a, b, the connected points with error bars show our data,
i.e.\ log\,$dN$(\Lya)/$dz$ for velocities $<$5000~\kms\ ($z$$<$0.0167). The
stars show the results from Penton et al.\ (2004), which were taken from the
plot in their paper, although we multiplied \Wlim\ by a factor 1.3 to account
for the difference in the way the equivalent width limits are calculated (see
Sect.~\SSmeasure). It is clear that the two distributions are basically
identical, implying that the somewhat higher redshifts \Lya\ absorbers at
$v$=5000--20000~\kms\ in the Penton et al.\ (2004) sample should have the same
relationship to galaxies as we find below in Sects.~\Sassocresults\ and
\Sgalresults.
\par We make three additional remarks. First, Penton et al.\ (2004) showed a
\Lya\ point at \Wlim=10~\mA, even though only two of their detections are weaker
than 20~\mA, and there is almost no redshift path at \Wlim$<$20~\mA\
(\Dz$<$0.02); this point is shown by the crosses in Figs.~\Fdndz a, b. Second,
in the log\,$dN/dz$ vs log\,\Wlim\ plot (Fig.~\Fdndz b), our \Lya\ distribution
appears to show a slight turnover for the weakest absorbers (\Wlim$<$25~\mA).
This is most likely an indication that we may have missed some weak lines. Only
with more sensitive data (such as will be provided by COS) will it be possible
to extend the measurements to $W$$<$20~\mA. Third, we find relatively more
strong ($W$$>$500~\mA) \Lya\ absorbers per unit redshift than did Penton et al.\
(2004). They find two (both toward PG\,1211+143, at $v$$>$5000~\kms) for
\Dz=1.1, while we find six (toward 3C\,232, HS\,1543+5921, PG\,1216+069,
Mrk\,205, HE\,1228+0131 and ESO\,438-G09) for \Dz=0.7. None of these six
sightlines was in the Penton et al.\ (2004) sample. The net result of this is
that the dropoff in $dN/dz$ at high equivalent width limits is not as steep as
found by Penton et al.\ (2004) found. We can also compare to the results of
Weymann et al.\ (1998), who showed $dN/dW$ for strong lines found in the \FOS\
QSO Absorption Line Key Project (\Dz$\sim$30), though not $dN/dz$ as function of
\Wlim. However, above some (high) value of the equivalent widths, the redshift
path will stay the same, so that the shape of $dN/dW$ will be the same as that
of $dN/dz$. Their $dN/dW$ is a power-law between $W$=300 and 900~\mA. Our sample
shows an apparent rise in the number of absorbers at \Wlim$>$500~\mA.
\fixed{This difference suggests that the results at high equivalent widths are
affected by object selection bias, especially since some sightlines were
a-priori selected to pass close to or even through nearby galaxies. In
particular, these are 3C\,232, ESO\,438-G09, HS\,1543+5921, Mrk\,205 and
MCG+10-16-111, which yield four of the six strongest lines.}
\par Figures~\Fdndz c,d shows $dN$(\Lyb)/$dz$, which has not been shown in
previous papers. For a given equivalent width limit, there are 2--3 times fewer
\Lyb\ lines per unit redshift than \Lya\ lines. Since the ratio of the optical
depths of \Lya\ and \Lyb\ is 5.3, this means that more \Lyb\ lines are
detected than would naively be expected. We used the measured linewidths
(corrected for instrumental broadening) to convert the equivalent widths to
optical depths, and checked the ratio of \Lya\ and \Lyb\ in the 26 systems in
which both lines are detected. Both lines are unsaturated, not contaminated and
well-measured in just eight systems (at 3212~\kms\ toward MRC\,2251$-$178,
1954~\kms\ toward Mrk\,335, 2545~\kms\ toward Mrk\,509, 2085~\kms\ toward
Mrk\,817, 1144~\kms\ toward PG\,0804+761, 4932~\kms\ toward PG\,1211+143,
2275~\kms\ toward PG\,1259+593 and 5519~\kms\ toward Ton\,S180). In each of
these systems the optical depth ratio is found to be near the expected value of
5.3 (within the errors). In the other 18 systems the derived optical depth ratio
ranges from 0.70 to 7.0, with an average of 3.2. For the cases with unsaturated
\Lya\ lines, the \Lyb\ lines tend to be in multi-component systems or in noisy
spectra, making the measurements uncertain. When \Lya\ is saturated, the
equivalent width does not increase as fast with increasing column density as is
the case for \Lyb, resulting in a ratio of \Lya\ and \Lyb\ equivalent widths
that is smaller than the ratio of optical depths.
\par Table~\Tdndz\ and Figs.~\Fdndz e, f present our $dN$(\OVI)/$dz$ results and
compare them to the three previous studies of low-redshift $dN$(\OVI)/$dz$:
Danforth \& Shull (2005; 40 systems at $z$$<$0.15), Tripp et al.\ (2008; 91
systems at $z$=0.15--0.5) and Thom \& Chen (2008; 27 systems at $z$=0.15--0.5).
We note that Tripp et al.\ (2008) discussed a difference between \OVI\
components and \OVI\ systems, since some of their \OVI\ detections come in
groups; we compare to their results for systems.
\par Before discussing the similarities and differences between the four
studies, we note that different conventions are used to calculate the detection
limits (see Sect.~\SSmeasure). To repeat: Danforth \& Shull (2005) defined
\Wlim\ as a 4$\sigma$ non-detection of a line one resolution element (20~\kms)
wide; Tripp et al.\ (2008) defined \Wlim\ as three times the error for a
15-pixel (55~\kms) wide line; Thom \& Chen (2008) defined \Wlim\ as a 3$\sigma$
non-detection for a line with $b$=10~\kms. We define \Wlim\ as the 3$\sigma$
error over a 60~\kms\ interval. As Fig.~\Fdetwidth\ shows, all detected \OVI\
lines have FWHM$>$30~\kms, i.e., the integration range will typically be
60~\kms\ or more. Thus, using a narrower integration interval to define the
detection limit underestimates that limit. A proper comparison between the
different studies thus requires increasing the equivalent width limits of
Danforth \& Shull (2005) by a factor $\sqrt{60/20}*3/4$$\sim$1.3 and those of
Thom \& Chen (2008) by a factor $\sqrt{60/(1.66*10)}$=1.5.
\par With these corrections, we can see in Figs.~\Fdndz e,f that the different
studies agree that $dN$(\OVI)/$dz$ at a 30~\mA\ equivalent width limit is about
17. However, we find fewer strong systems ($W$$>$70~mA) than expected from the
number of \OVI\ systems seen at $z$=0.1--0.5. The Tripp et al.\ (2008) study has
27 systems over a path \Dz=3.1 for \Wlim=70~\mA. For our path \Dz=0.43 we would
thus expect to find four systems. We find just one, although we would have
counted the detection toward ESO\,185-IG13 if we had had better data. Similarly,
for \Wlim=100~\mA\ we expect two detections, but find one (toward
ESO\,185-IG13). Since we are working with a relatively small total pathlength
and small number statistics, the discrepancy is not problematic.
\par Figures~\Fdndz e,f also show that our high value for $dN$(\OVI)/$dz$ at
\Wlim=20~\mA\ (50\E22) is compatible with extrapolating the other studies. We
derive this from a redshift path \Dz=0.1195 built up in the 17 \FUSE\ sightlines
that have S/N$>$20, and which yield six detections with $W$(\OVI)$>$20~\mA. The
{\it Cosmic Origins Spectrograph} (COS) will provide an increase in the typical
S/N at $\lambda$$>$1200~\AA\ from about 10 to about 30--50, giving a typical
detection limit of about 10--15~mA. Compared to the Tripp et al.\ (2008) survey,
we thus expect to see an increase in the number of detected \OVI\ absorbers per
unit redshift interval in the COS data of a factor about two.
\par The discussion above leads us to the conclusion that: {\it To compare the
different published analyses of the distribution of the number of absorption
lines between $z$=0 and $z$=0.5 as function of limiting equivalent width, it is
necessary to take into account the different definitions of the detection limits
that were used. After correcting, we find that the distributions for \Lya\ and
\OVI\ found in different studies do agree for equivalent width limits between 20
and 40~\mA.}

\subsection{Time Evolution of \Lya\ Linewidths}
\begin{figure}\plotfiddle{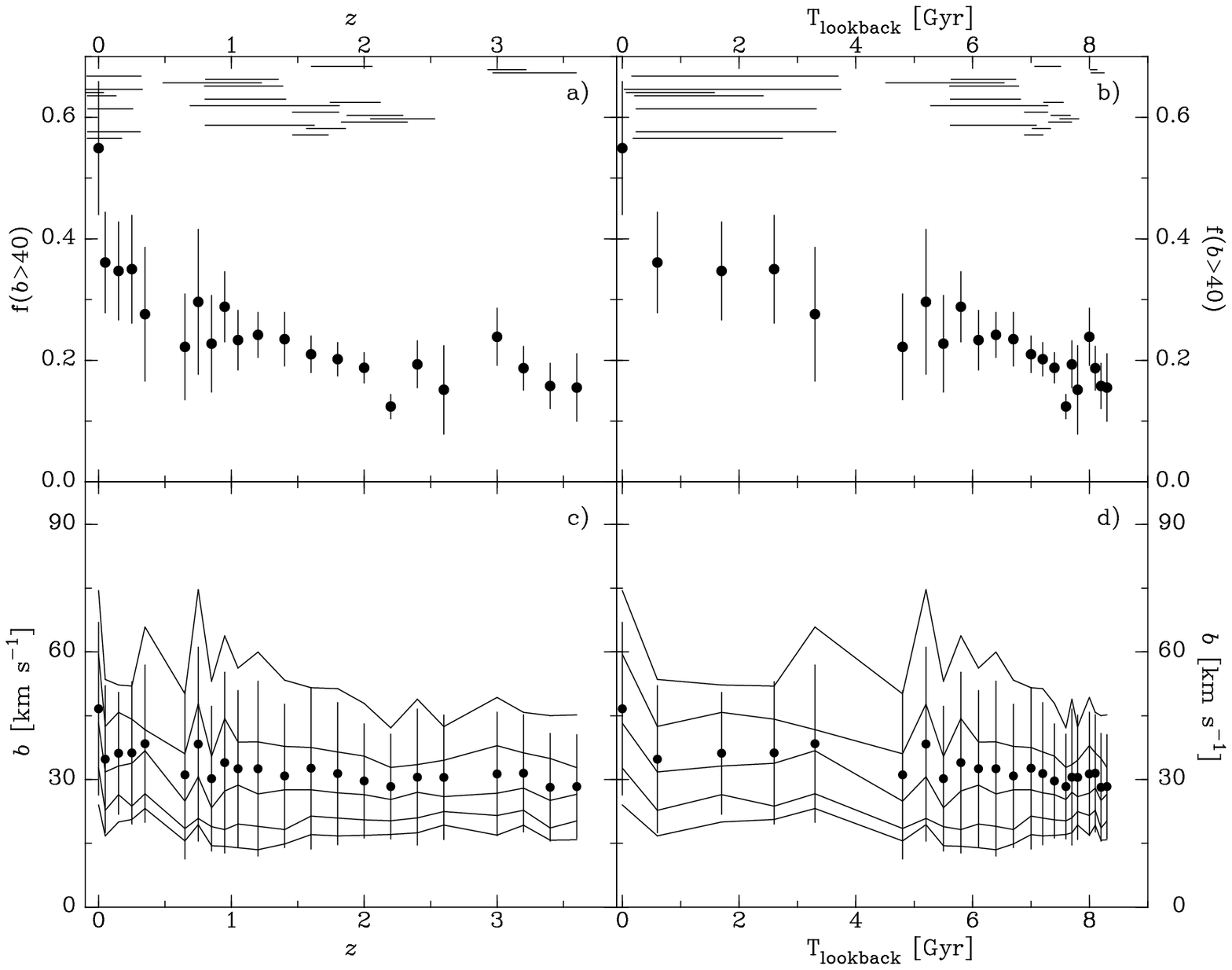}{0in}{0}{450}{350}{0}{-150}\figurenum{6}\caption{\captionbevol}\end{figure}
\par Lehner et al.\ (2007) did a detailed study of the distribution of 316 \Lya\
lines toward seven QSOs at $z$$<$0.5 with high-quality \STIS-E140M spectra. They
compared their sample to those of Kim et al.\ (2001, 2002; 2315 \Lya\ lines at
$z$=1.5--3.6) and Janknecht et al.\ (2006; 1325 \Lya\ lines at $z$=0.6--1.9),
and plotted the relative number of broad ($b$$>$40~\kms) absorption lines in
different redshift intervals (0.5 units wide). If the broadening is thermal,
$b$=40~\kms\ corresponds to a temperature of \dex5~K. Lehner et al.\ (2007)
found that the fraction of broad \Lya\ lines increases with decreasing redshift,
suggesting that the intergalactic medium may be heating up over time. We revisit
this analysis and add our measurements at $z$=0--0.017.
\par\fix{We note that Kim et al.\ (2001) gave the S/N ratio of the optical
(VLT/UVES) data as ``typically 40--50''. The mid-UV spectra used by Janknecht
et al.\ (2006) have S/N ratios of 10--15 per resolution element. The seven
high-quality low-redshift spectra of Lehner et al.\ (2007) also have S/N about
10--15 per resolution element. Thus, the quality of the several UV datasets is
comparable, while that of the higher-redshift optical data is better.}
\par For each of the detected \Lya\ lines in the three samples referenced above,
values were given for the fitted column density and linewidth. We convert these
to equivalent widths ($W$) and optical depths ($\tau$) and then look at the
distribution of linewidths, using redshift intervals 0.1 wide for $z$$<$1, and
0.2 wide for $z$$>$1, excluding lines with $W$(rest)$>$500~\mA\ and $\tau$$>$2.
These criteria remove saturated lines, for which the fitted $b$-value is not a
good measure of the intrinsic width. We also checked the Lehner et al.\ (2007)
sample in more detail and further remove lines that are clear blends, as well as
ones that are hard to discern. \fix{For our sample we only include the 225
well-measured lines in their sample of 316.}
\par Figure~\Fbevol\ shows some of the resulting parameters of the linewidth
distribution, as function of both redshift and lookback time. In Figs.~\Fbevol
a,b we give the fraction of lines with $b$$>$40~\kms\ (the parameter shown by
Lehner et al.\ 2006), with errors based on Poisson statistics. Figures~\Fbevol
c,d shows curves that represent the 10th, 25th, 50th, 75th and 90th percentile
of the distribution, as well as points and bars giving the average and
dispersion. We also include the redshift ranges of the individual QSO spectra
(line segments at the top of panels a and b). This shows that most of the
irregularities in the distributions are associated with the edges of the
spectra. For instance, for the $z$=2.3--2.7 interval almost all \Lya\ lines come
from the spectrum of HE\,1347$-$2457, but at $z$$<$2.36 other QSOs contribute.
Similarly, the depression at $z$=1.4 happens at the break between the optical
sample of Kim et al.\ (2001, 2002) and the \STIS-E230M sample of Janknecht et
al.\ (2006). Further, the wild fluctuations at $z$=0.6--0.9 occur because these
bins are based on a relatively noisy spectrum (S/N$\sim$10) of a single QSO,
PG\,1634+706. Finally, the $z$=0.45 bin uses data from the long-wavelength edges
of the \STIS\ spectra, where the noise increases and variations in the detector
sensitivity may result in fluctuations. So, lines that are narrower or broader
than average may become more difficult to discern. Obviously, a more thorough
analysis of each of these spectra is required before one can conclude that the
fluctuations in the distribution of $b$ are real or artifacts.
\par With these caveats, clear trendlines can be seen. The 10th percentile of
the distribution stays more or less constant at 16~\kms, while from $z$=3.5 to
$z$$\sim$0 the 50th percentile (i.e.\ the median) increases from 25 to
$\sim$35~\kms, and the 90th percentile increases from 40 to $\sim$60~\kms.
Similarly, the fraction of lines wider than $b$=40~\kms\ increases from 10\% at
$z$=3.5 to 55\% at $z$=0. Fig.~\Fbevol b suggests a mostly continuous increase
in the fraction of wide lines as the universe evolves. Because of the apparent
problems with instrumental breaks, it is not possible to derive the precise
manner in which the linewidths have increased over time. The data support a
linear increase over time. Note, however, that our sample (at $z$=0) has a
substantially larger fraction of wide lines than the lowest redshift point in
the Lehner et al.\ (2007) sample (at $z$=0.05). This may be caused by the
differing particular set of sightlines in the two samples.
\par The largest fraction of wide lines and the largest average linewidth occur
at $z$=0. In spite of the large fluctuations and large errors in $f$($b$$>$40),
the difference between the points at $z$=0 and $z$=0.1--0.4 is likely
significant. To confirm this, it will be necessary to reassess all linewidth
measurements, using the same method at all redshifts. \fix{We note that compared
to Lehner et al.\ (2007), we only used the reliable lines in their sample, i.e.,
we excluded lines with large optical depth or blends. This did not fundamentally
change the conclusion derived in that paper.}
\par To summarize, we conclude from Fig.~\Fbevol\ that {\it the widths of the
\Lya\ lines have increased over cosmic time, that the $z$=0 sample has the
largest fraction of wide lines, and that the distribution function of the
linewidths is the widest at the present time.}

\section{Galaxies Associated with Absorbers}
\par In making the associations between absorbers and galaxies near the
sightline, some previous authors (such as Bowen et al.\ 1996, 2002) implicitly
used the assumption that \Lya\ absorbers are associated with the halos of the
galaxies near them (a reasonable assumption when concentrating on sightlines
with low impact parameters to galaxies). However, other authors argued that most
absorbers are intergalactic. For instance, Impey et al.\ (1999) compared their
absorber sample with galaxy surveys in the Virgo region, and concluded that
galaxy halos are not responsible for the absorbers. Penton et al.\ (2002)
concurred, except that they allowed that strong absorbers with small impact
parameters may originate in halos. A number of other authors analyzed imaging
surveys of galaxies up to $z$$\sim$0.2 near several sightlines (e.g.\ Tripp et
al.\ 1998; Sembach et al.\ 2004; Tumlinson et al.\ 2005; Prochaska et al.\
2006). However, the field of view is usually small, the luminosity limit is not
constant and most faint galaxies are not found, so the results have not been
unambiguous. The advantage of our approach of looking at the very nearest
galaxies is that the galaxy content of the survey volume is relatively uniformly
well known, at least down to a given luminosity limit (0.5\,\Lstar\ at
5000~\kms, 0.1\,\Lstar\ at 2500~\kms).
\par In this section we present several analyses of the \Lya/\Lyb/OVI\
detections that look at the galaxies near each of the absorbers. For the first
two analyses we do not make a-priori assumptions about which galaxy a particular
absorption line might be associated with, by just looking at the probability of
finding galaxies of a given luminosity near the absorber as function of impact
parameter and absorber-galaxy velocity difference (Sect.~\SSassoccount). We then
compare the distribution of galaxy-absorber impact parameters to that of
galaxy-galaxy separations (Sect.~\SSnearest). In these two subsections we
establish that absorbers do associate with galaxies. Therefore, we continue with
a description of the particular associations that we make, which are listed in
Table~\Tres\ (Sect.~\SSassoc). This leads to a discussion of the issue of ``void
absorbers'' (Sect.~\SSvoid). Having established the individual galaxy-absorber
associations, we look at the distribution of velocity differences
$v$(abs)$-$v(gal) (Sect.~\SSvdiffhist), at the distribution of absorber
linewidths as function of impact parameter (Sect.~\SSwidthimp), and at the
relation between absorber equivalent width and impact parameter
(Sect.~\SSewvsb).

\subsection{The Fraction of Absorbers Having a Galaxy of Given $L$ Within \ip, \Dv}
\begin{deluxetable}{llrrrrr}
\tablenum{9}
\tablewidth{0pt}
\tabletypesize{\footnotesize}
\tablecolumns{7}
\tablecaption{Fraction of \Lya\ absorbers having a galaxy within \ip\ and \Dv$^1$}
\tablehead{%
\ch{\ip}   &\ch{\Dv}    &\ch{all} &\ch{$L$$>$0.1\,\Lstar\phantom{0}} &\ch{$L$$>$0.25\,\Lstar} &\ch{$L$$>$0.5\,\Lstar\phantom{0}} &\ch{$L$$>$\Lstar\phantom{1.00}} \\
\ch{[kpc]} &\ch{[\kms]} &         &                                  &                        &                                  &                                \\
\ch{(1)}&\ch{(2)}&\ch{(3)}&\ch{(4)}&\ch{(5)}&\ch{(6)}&\ch{(7)}
}\startdata
$<$200 & $<$200 & 28; 21\%& 20; 15\%& 18; 13\%& 14; 10\%& 8; 6\%\\
$<$200 & $<$400 & 30; 22\%& 23; 17\%& 20; 15\%& 15; 11\%& 9; 6\%\\
$<$200 & $<$1000 & 31; 23\%& 24; 18\%& 21; 15\%& 15; 11\%& 9; 6\%\\
\hline
$<$400 & $<$200 & 60; 45\%& 49; 36\%& 44; 33\%& 34; 25\%& 23; 17\%\\
$<$400 & $<$400 & 63; 47\%& 52; 39\%& 47; 35\%& 37; 27\%& 27; 20\%\\
$<$400 & $<$1000 & 63; 47\%& 52; 39\%& 47; 35\%& 37; 27\%& 27; 20\%\\
\hline
$<$1000 & $<$200 & 93; 69\%& 78; 58\%& 67; 50\%& 52; 39\%& 36; 27\%\\
$<$1000 & $<$400 & 100; 75\%& 83; 62\%& 70; 52\%& 54; 40\%& 39; 29\%\\
$<$1000 & $<$1000 & 100; 75\%& 83; 62\%& 70; 52\%& 54; 40\%& 39; 29\%\\
\hline
$<$2000 & $<$200 & 111; 83\%& 95; 71\%& 80; 60\%& 63; 47\%& 44; 33\%\\
$<$2000 & $<$400 & 122; 91\%& 103; 77\%& 86; 64\%& 68; 51\%& 47; 35\%\\
$<$2000 & $<$1000 & 122; 91\%& 103; 77\%& 86; 64\%& 68; 51\%& 47; 35\%\\
\hline
$<$3000 & $<$200 & 113; 84\%& 97; 72\%& 82; 61\%& 64; 48\%& 44; 33\%\\
$<$3000 & $<$400 & 128; 96\%& 109; 81\%& 92; 69\%& 71; 53\%& 49; 36\%\\
$<$3000 & $<$1000 & 128; 96\%& 109; 81\%& 92; 69\%& 71; 53\%& 49; 36\%\\
\hline
\enddata
\tablecomments{%
1: This table gives number and percentage of \Lya\ absorbers associated with
galaxies. E.g. on the fourth line in Col.\ (7), there are 23 (or 17\%) absorbers
for which it is possible to find a galaxy with luminosity $>$\Lstar\ that lies
within 400~kpc and whose systemic velocity differs by less than 200~\kms\ from
that of the absorber. For this count no difference is made between group and
field galaxies. The total number of absorbers is 133.
}
\end{deluxetable}
\par In Table~\Tassoccount\ we summarize the fraction of \Lya\ absorbers for
which we can find a galaxy of a given luminosity within some impact parameter
and differing in velocity by less than a given value. Note that we do the
opposite analysis (i.e.\ we ask whether an absorber can be found near a
particular galaxy) in Sect.~\SSgalaxycount. To construct Table~\Tassoccount, we
find the nearest galaxy that fits the impact parameter, velocity difference and
luminosity criterion. The table shows that for the great majority (128 of 133,
or 96\%) of intergalactic absorbers a galaxy can be found with impact parameter
\ip$<$3~Mpc and velocity difference \Dv$<$400~\kms. As the criteria are made
more strict, the fraction decreases. If there is a galaxy within a given impact
parameter limit, most absorbers have velocities that are within 200~\kms\ from
that galaxy, with generally just a few having \Dv=200--400~\kms. The fraction of
absorbers decreases by a factor of about 2.2 as the luminosity limit of the
nearest galaxy is increased from 0.1~\Lstar\ to \Lstar. Further, for any given
impact parameter and velocity difference limit, about 15--20\% of the absorbers
do not have a galaxy brighter than 0.1~\Lstar\ as the nearest galaxy. Only 23
systems (17\%) occur within what would generally be called the halo of a
luminous galaxy, i.e.\ within 400~kpc and 200~\kms\ of a galaxy with
$L$$>$\Lstar, and only 8 of these originate in the inner part of these halos
(\ip$<$200~kpc).
\par We can also determine the median impact parameter for finding a galaxy
brighter than a given luminosity and with \Dv$<$400~\kms\ to an absorber. We
find that 50\% of the absorbers with $v$$<$2500~\kms\ has a galaxy with
$L$$>$0.1~\Lstar\ within 370~kpc. The median impact parameter for an absorber
with $v$$<$3700~\kms\ to an $L$$>$0.25~\Lstar\ galaxy is 390~kpc, while for
absorbers with $v$$<$5000~\kms\ it is 730~kpc to an $L$$>$0.5~\Lstar\ galaxy and
870~kpc to an $L$$>$\Lstar\ galaxy.
\par Based on Table~\Tassoccount\ we conclude that {\it for almost all \Lya\
absorbers there is a galaxy with \ip$<$3~Mpc having \Dv$<$400~\kms, and most
(81\%) of these are brighter than 0.1\,\Lstar. On the other hand, only a small
fraction ($\sim$20\%) occurs within the halos of $L$$>$\Lstar\ galaxies.}

\subsection{Distribution of Nearest-Neighbor Separations}
\begin{figure}\plotfiddle{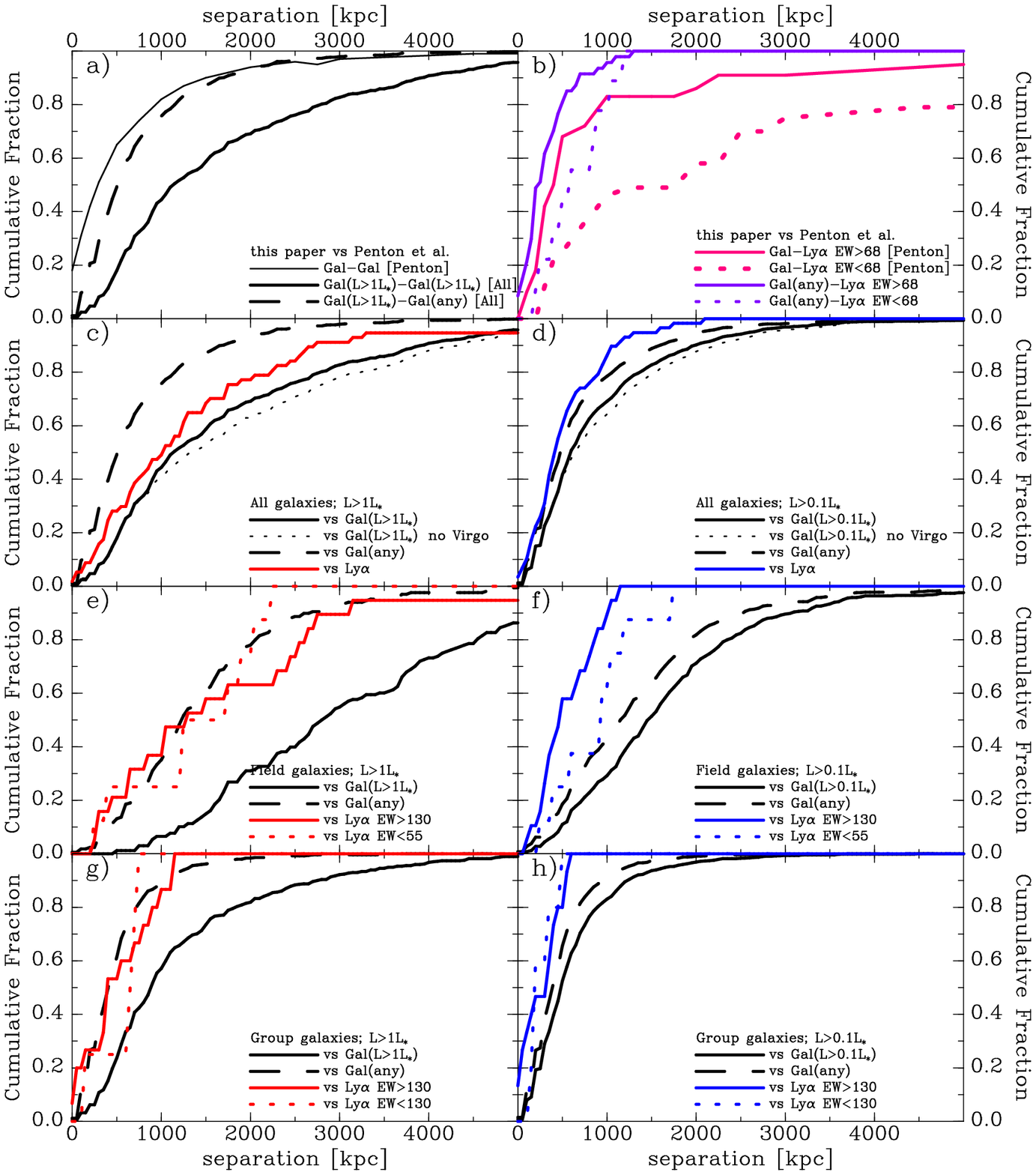}{0in}{0}{350}{350}{0}{-150}\figurenum{7}\caption{\captionnearest}\end{figure}
\par A different way of exploring the relation between absorbers and galaxies is
to compare the 3-D separation between galaxies to the 3-D separation between
absorbers and galaxies. This was previously done by Penton et al.\ (2002) and
Stocke et al.\ (2006). Penton et al.\ (2002) used relatively complete galaxy
surveys down to at least \Lstar\ near fifteen of their sightlines, while Stocke
et al.\ (2006) used a combination of galaxy surveys, containing \dex6 galaxies
with known redshifts. In their Fig.~3 (Penton et al.\ 2002) and Fig.~1 (Stocke
et al.\ 2006), these authors presented the cumulative nearest-neighbor galaxy
distribution, i.e., the fraction of galaxies whose nearest neighbor is separated
by less than a given value. We first analyze our data in the manner presented by
these authors and then add some new twists.
\par To calculate our version of the nearest-neighbor distributions, we start
with the direction and distance of every galaxy with $L$$>$\Lstar\ and
\vgal$<$4500~\kms\ and find the nearest $L$$>$\Lstar\ galaxy with
\vgal$<$5000~\kms. We use a 4500~\kms\ limit for the base galaxy to avoid edge
effects, since otherwise we might miss the nearest galaxy if it has
\vgal$>$5000~\kms. A 500~\kms\ velocity difference corresponds to about 10~Mpc,
which is larger than all but a few of the largest separations that are found.
For a given impact parameter we then find the fraction of \Lstar\ galaxies whose
nearest neighbor \Lstar\ galaxy is closer. We also do this calculation using
only $L$$>$0.1\,\Lstar\ galaxies with $v$$<$2000~\kms. Comparing these
distributions to the nearest \Lstar\ or 0.1\,\Lstar\ neighbor of an absorber, we
can determine whether or not absorbers associate more with galaxies of a given
luminosity than the galaxies associate with each other.
\par We further look at the nearest neighbor galaxy of any luminosity to each
$L$$>$\,\Lstar\ or $>$0.1\,\Lstar\ galaxy. This comparison allows a comparison
of the typical impact parameter of an absorber to the typical separation between
dwarf galaxies and bright galaxies, even though the dwarf galaxy sample is
incomplete.
\par Since the galaxy-galaxy nearest-neighbor separations were calculated in
3-dimensional space, while the absorber-galaxy impact parameters are
2-dimensional projections, we assume that the gas cloud with impact parameter
\ip\ can lie up to \ip\ kpc in front or behind the projection plane. Assuming a
random placing along this line, the 3-D separation between an absorber and a
galaxy will on average be a factor 1.25 times the impact parameter. All
absorber-galaxy separations in Fig.~\Fnearest\ are thus increased by this factor
over the raw impact parameters.
\par In Fig.~\Fnearest a, b we compare the Penton et al.\ (2002) curves to our
data, using the same luminosity and equivalent width criteria. The thin solid
line in panel a shows their galaxy-galaxy distribution. The Stocke et al.\
(2006) curves are the same as those of Penton et al.\ (2002). The thick solid
black line gives our version of the distribution of the nearest \Lstar\ neighbor
for \Lstar\ galaxies. The dashed black line gives the distribution for the
nearest neighbor of any luminosity to \Lstar\ galaxies, using only galaxies with
\vgal$<$2500~\kms. The faint galaxies include many Virgo dwarfs, and many dwarf
galaxies that lie near well-studied bright galaxies or near some of the AGN
sightlines. The comparison between the dashed black and thin black curve in
Fig.~\Fnearest a shows that the nearest neighbor function for this inhomogeneous
sample is quite similar to that found by Penton et al.\ (2002). We conclude that
Penton et al.\ (2002) showed the nearest-neighbor of any luminosity to each
\Lstar\ galaxy, although their text and figure caption would suggest that they
showed the distribution of the separation between \Lstar\ galaxies. Stocke et
al.\ (2006) explicitly addressed this issue and mention that they used both
approaches.
\par The Penton et al.\ (2002) galaxy-absorber distributions are shown in
Fig.~\Fnearest b by the magenta lines, separately for strong ($W$$>$68~\mA) and
weak ($W$$<$68~\mA) lines. Comparing the galaxy-galaxy and galaxy-absorber
distributions, Penton et al.\ (2002) as well as Stocke et al.\ (2006) concluded
that the typical nearest neighbor galaxy to a strong absorber is about as close
as that galaxy's nearest neighbor, while weak lines occur much further from
galaxies. This would imply that absorbers are {\it not} generally associated
with galaxy halos, but instead occur in intergalactic filaments. As presented,
their analysis is correct, but we find that there is more to the story, and that
the conclusion needs to be modified.
\par The purple lines in Fig.~\Fnearest b give our version of the
galaxy-absorber nearest neighbor distribution for \Lya\ absorbers with
equivalent width $>$68~\mA\ and $<$68~\mA. It is clear that the Penton et al.\
(2002) curves have a kink at around 1.5~Mpc impact parameter, whereas ours are
more smooth across this impact parameter range. That is, we find a continuous
distribution of the number of absorbers as function of impact parameter, whereas
Penton et al.\ (2002) found a deficit of absorbers with \ip$\sim$1.5~Mpc. The
purple curve is also significantly higher than the Penton et al.\ (2002)
galaxy-\Lya\ curve for impact parameters above 500~kpc, which means that we can
find galaxies with \ip=500--1500~kpc near some absorbers, where Penton et al.\
(2002) found relatively fewer.
\par In Fig.~\Fnearest c, d, we present our nearest-neighbor distributions,
using the $L$$>$\Lstar\ sample in Fig.~\Fnearest c, the $L$$>$0.1\,\Lstar\
sample in Fig.~\Fnearest d. We also show the galaxy-galaxy distributions that
are obtained when excluding galaxies within 30\deg\ from the Virgo cluster (thin
black lines). These distributions are compared to the galaxy-absorber
distributions -- red for the distance between absorbers and $L$$>$\Lstar\
galaxies, blue for $L$$>$0.1\,\Lstar\ galaxies. We conclude that:
\par\noindent (1) For 50\% of $L$$>$\Lstar\ galaxies, the nearest \Lstar\ galaxy
is at a separation $<$1.2~Mpc, while the nearest \Lstar\ galaxy to an absorber
has a median impact parameter 1.1~Mpc. The nearest $L$$>$0.1\,\Lstar\ galaxy on
average lies 600~kpc from another $L$$>$0.1\,\Lstar\ galaxy, while for \Lya\
lines such a galaxy is on average found within 450~kpc. We note that Stocke et
al.\ (2006) quoted values of 1.8~Mpc and 250~kpc for the median separations
between \Lstar\ and 0.1\,\Lstar\ galaxies. We have not identified the origin of
this discrepancy, but it may partly be due to the inhomogeneity in the magnitude
systems used to define the different galaxy catalogues that were used
\par\noindent (2) The Virgo cluster has a significant effect on the median
separation between galaxies. Excluding it leads to a median nearest \Lstar\
neighbor for an \Lstar\ galaxy of 1.4~Mpc.
\par\noindent (3) The nearest \Lstar\ galaxy to an absorber has a median
separation of 1.1~Mpc, while the nearest 0.1\,\Lstar\ galaxy lies at 450~kpc.
The strongest lines ($W$$>$68~\mA) have a median distance to \Lstar/0.1\,\Lstar\
galaxies of 900/400~kpc, while the median for weak lines ($W$$<$68~\mA) is
1750/950~kpc. Stocke et al.\ (2006) quoted 1130/445~kpc for strong lines and
2150/1455~kpc for weak lines. The large difference in median separations for
strong and weak lines led Stocke et al.\ (2006) to conclude that strong lines
correlate more strongly with galaxies than weak lines. Our data support this,
though we also conclude that the difference may not be as pronounced as Stocke
et al.\ (2006) found.
\par There is a big problem with the galaxy-galaxy distributions in
Fig.~\Fnearest c, d, however. This is that in galaxy groups the nearest
neighbor will typically be closer than the nearest neighbor to a field galaxy.
Gaseous halos of galaxies in groups may touch and possibly merge, or
alternatively, the group environment may destroy them. Halos around field
galaxies are much more likely to form a single structure.
\par To deal with this problem, we separate the absorber and galaxy samples into
absorbers/galaxies in the field and in groups. We count a galaxy as a group
galaxy if it was identified as such in the GH+LGG group sample that we discussed
previously (Sect.~\SSgaldata), or if it falls within a group's outline on the
sky and within the velocity range of the group galaxies used to define the
group. For absorbers, we overlaid the sightlines on maps of the distribution of
group galaxies, and we counted absorbers as associated with a group if the
sightline passes through or close to the edge of a group. This is the case for
19 of the systems. The other 96 absorbers occur in the field.
\par Some have argued that all galaxies really are members of groups. One could
also argue that the GH/LGG definitions of groups is arbitrary. Nevertheless,
using these defined groups will separate the galaxies into a set of galaxies
located in regions of high or low galaxy density, and thus the general
conclusion remains valid.
\par Following Penton et al.\ (2002) and Stocke et al.\ (2006), we divide the
absorber sample into strong and weak lines. However, rather than using 68~\mA,
we set the dividing line at 130~\mA, as this is the median for our sample.
\par The curves in Fig.~\Fnearest c, d would seem to indicate that the typical
impact parameter of a \Lya\ absorber to a galaxy is not much smaller than the
typical distance between galaxies. However, we use Fig.~\Fnearest e--h to show
that this is because field and group galaxies were averaged. In Figs.~\Fnearest
e--h we can see the following:
\par\noindent (1) The median distance between \Lstar\ field galaxies is 2.9~Mpc
(see solid black line in Fig.~\Fnearest e) while the median separation between a
field \Lya\ absorber and the nearest \Lstar\ galaxy is 1.5~Mpc (thick red line).
Conversely, for only 14\% of \Lstar\ galaxies in the field is the nearest
\Lstar\ galaxy closer than 1.5~Mpc.
\par\noindent (2) The dashed black line in Fig.~\Fnearest e shows that half of
the \Lstar\ field galaxies have a smaller galaxy within about 1.2~Mpc. This is
comparable to the typical distance between an absorber and an \Lstar\ galaxy.
That is, in the field the typical separation between an \Lstar\ galaxy and
another, smaller, galaxy is similar to the typical separation between it and an
absorber. \fix{A KS-test can test whether the differences are statistically
significant. Comparing the distributions of the 396 galaxies and that of the 39
strong (EW$>$130~\mA) or 21 weak (EW$<$55~\mA) absorbers, the hypothesis that
they differ is rejected only at the 50\% (for strong lines) or 75\% (for weak
line) confidence level.}
\par\noindent (3) On average, field \Lya\ absorbers occur significantly closer
to the nearest 0.1~\,\Lstar\ field galaxy (median impact parameter 600~kpc) than
these galaxies are to each other (median impact parameter 1.5~Mpc) -- compare
the solid black line with the blue lines in Fig.~\Fnearest f. \fix{The KS-test
rejects the hypothesis that the distributions are the same at the $>$99.9\%
confidence level.}
\par\noindent (4) The typical separation between an absorber and the nearest
field $L$$>$0.1\,\Lstar\ galaxy (median 600~kpc) is smaller than the typical
separation between the nearest galaxy of any luminosity to an $L$$>$0.1\,\Lstar\
galaxy (1.1~Mpc) -- compare blue and dashed black lines in Fig.~\Fnearest f.
\par\noindent (5) When looking at the separation between $L$$>$0.1\,\Lstar\
field galaxies and absorbers, the strong absorbers occur on average closer to
those galaxies than the weak absorbers -- compare solid and dotted blue lines in
Fig.~\Fnearest f. \fix{The KS-test rejects the hypothesis that the distributions
for the strong (EW$>$130~\mA) and weak (EW$<$55~\mA) lines are the same at the
99.3\% level.}
\par\noindent (6) In galaxy groups the median distance between \Lstar\ galaxies
is 900~kpc (solid black line in Fig.~\Fnearest g), between 0.1\,\Lstar\ galaxies
it is 500~kpc (solid black line in Fig.~\Fnearest h).
\par\noindent (7) For half of the \Lya\ absorbers that originate in groups it is
possible to find an \Lstar\ or 0.1\,\Lstar\ galaxy within 550 or 350~kpc,
respectively (solid colored lines in Figs.~\Fnearest g, h). \fix{The KS-test
rejects the hypothesis that the distribution of the 1761 $L$$>$0.1\,\Lstar\
group galaxies is the same as that of the 13 strong or 19 weak absorbers with
$>$98.5\% confidence. On the other hand, there is no difference between strong
and weak absorbers in this case: the KS-test rejects the hypothesis that they
are the same only at the 64\% confidence level, i.e., the difference is a
0.5$\sigma$ effect.} Thus, as is the case for field absorbers, these \Lya\
absorbers typically occur closer to galaxies than the galaxies do to each other,
though the contrast is not as pronounced.
\par\noindent (8) Comparing Fig.~\Fnearest c--h shows that when combining the
field and group sample, the averaging of the galaxy-galaxy nearest-neighbor
distributions and the galaxy-absorber nearest-neighbor distributions works out
such that the two curves are close together in the combined sample, because
group galaxies are closer together than absorbers are to field galaxies.
\par From these comparisons we conclude that {\it \Lya\ absorbers are associated
with $L$$>$0.1\,\Lstar\ galaxies, both in the field and in groups. Reaching this
conclusion requires looking at group and field galaxies and absorbers
separately.}
\par As a final note, we mention that an alternative way of comparing the
galaxy-galaxy and galaxy-absorber distributions is to use the two-point
correlation function. This was done by Wilman et al.\ (2007), who used the 381
absorbers from the \HST\ QSO Absorption Line Key Project and a sample of 685
galaxies at $z$$<$1 to calculate the two-point correlation function for a number
of absorber column density intervals as function of impact parameter and
line-of-sight separation. They find that the absorber-galaxy two-point
correlation function is more concentrated toward small separations than the
galaxy-galaxy function, and also that the absorber-galaxy function extends out
to beyond 1~Mpc. This analysis thus supports our contention that absorbers
associate more with galaxies than that the galaxies associate with each other.

\subsection{Associating \Lya/\Lyb/\OVI\ with Individual Galaxies}
\begin{figure}\plotfiddle{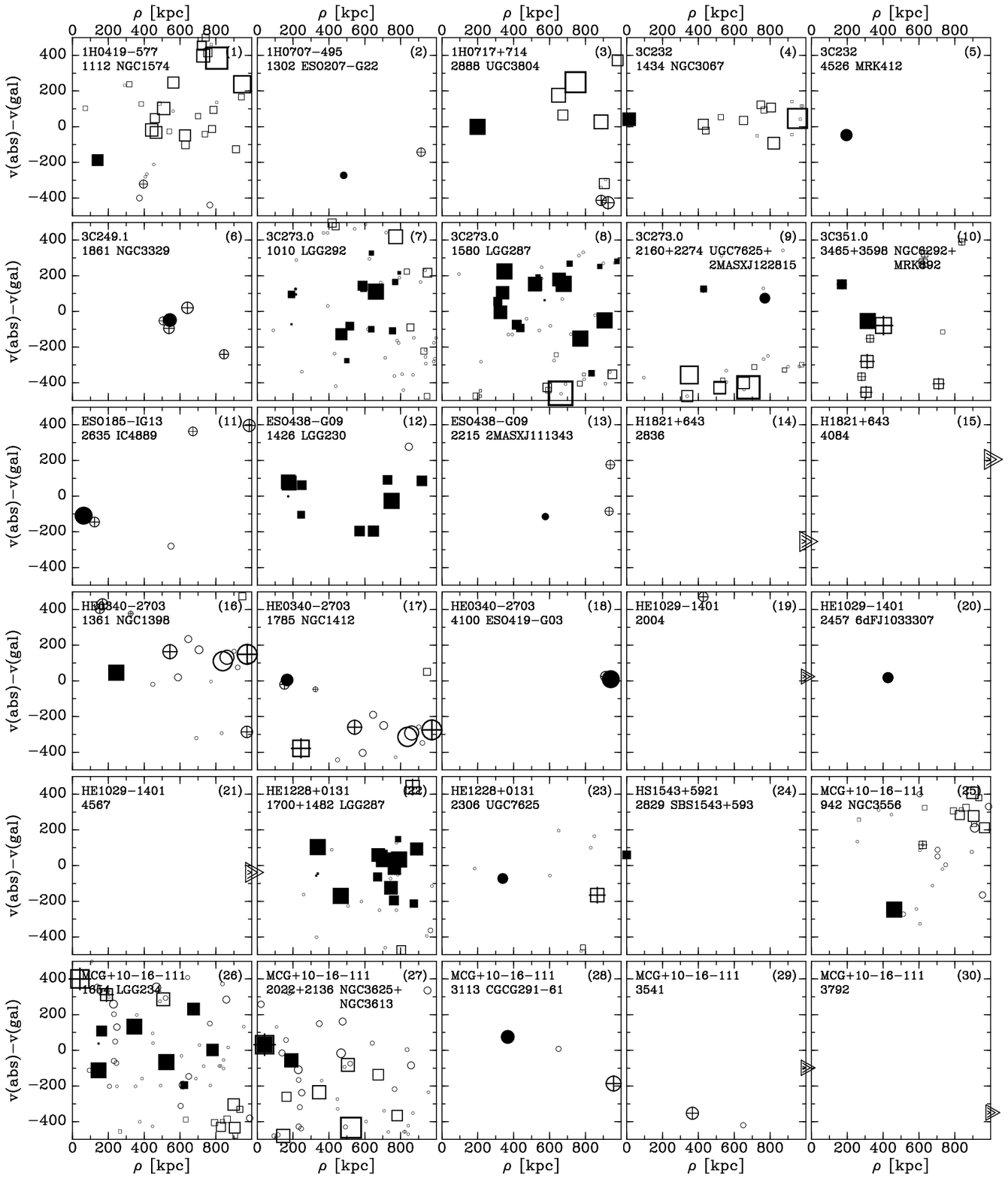}{0in}{0}{400}{400}{0}{-150}\figurenum{8}\caption{\captionassoc}\end{figure}
\begin{figure}\plotfiddle{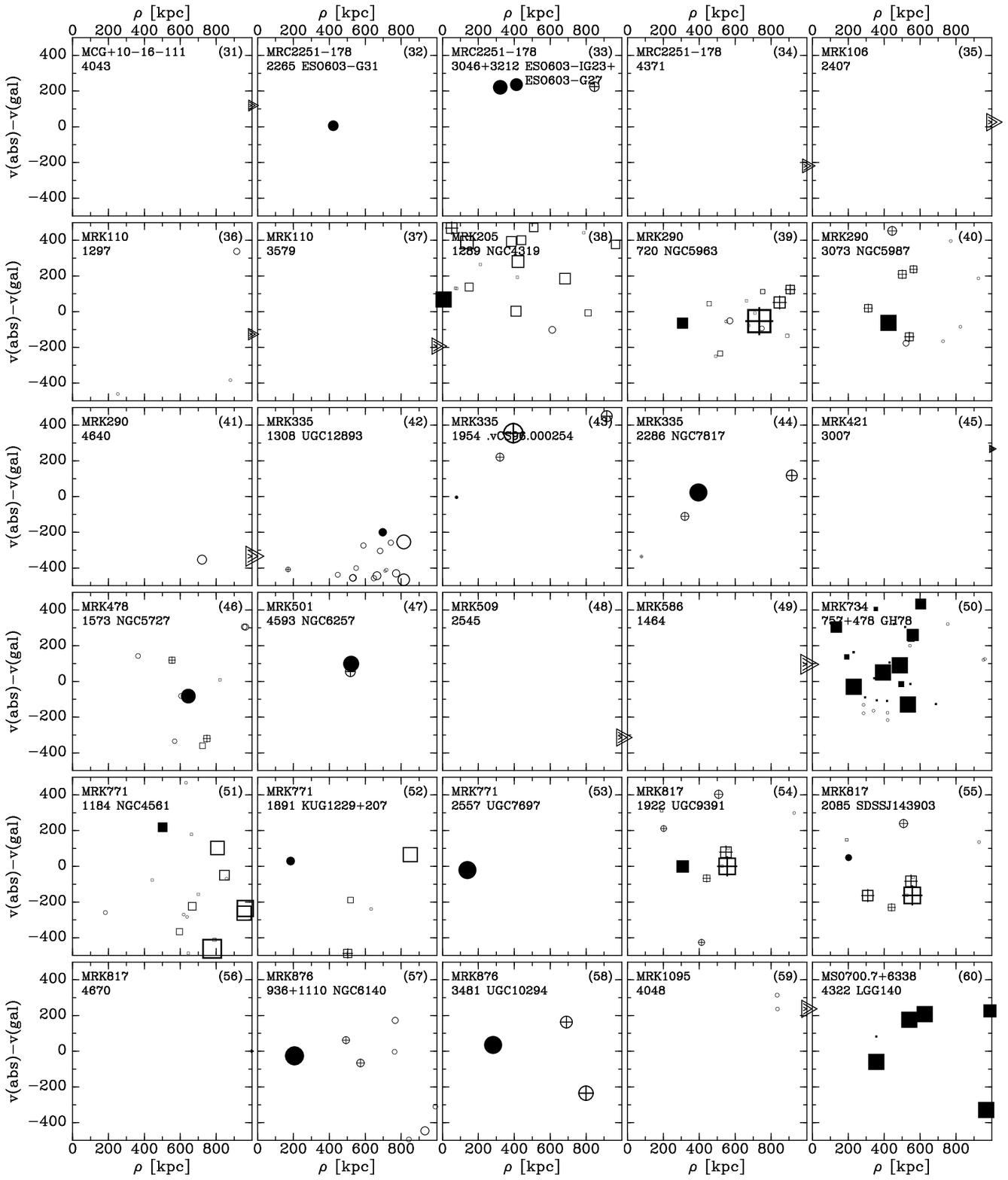}{0in}{0}{400}{400}{0}{-150}\figurenum{8}\caption{Continued.}\end{figure}
\begin{figure}\plotfiddle{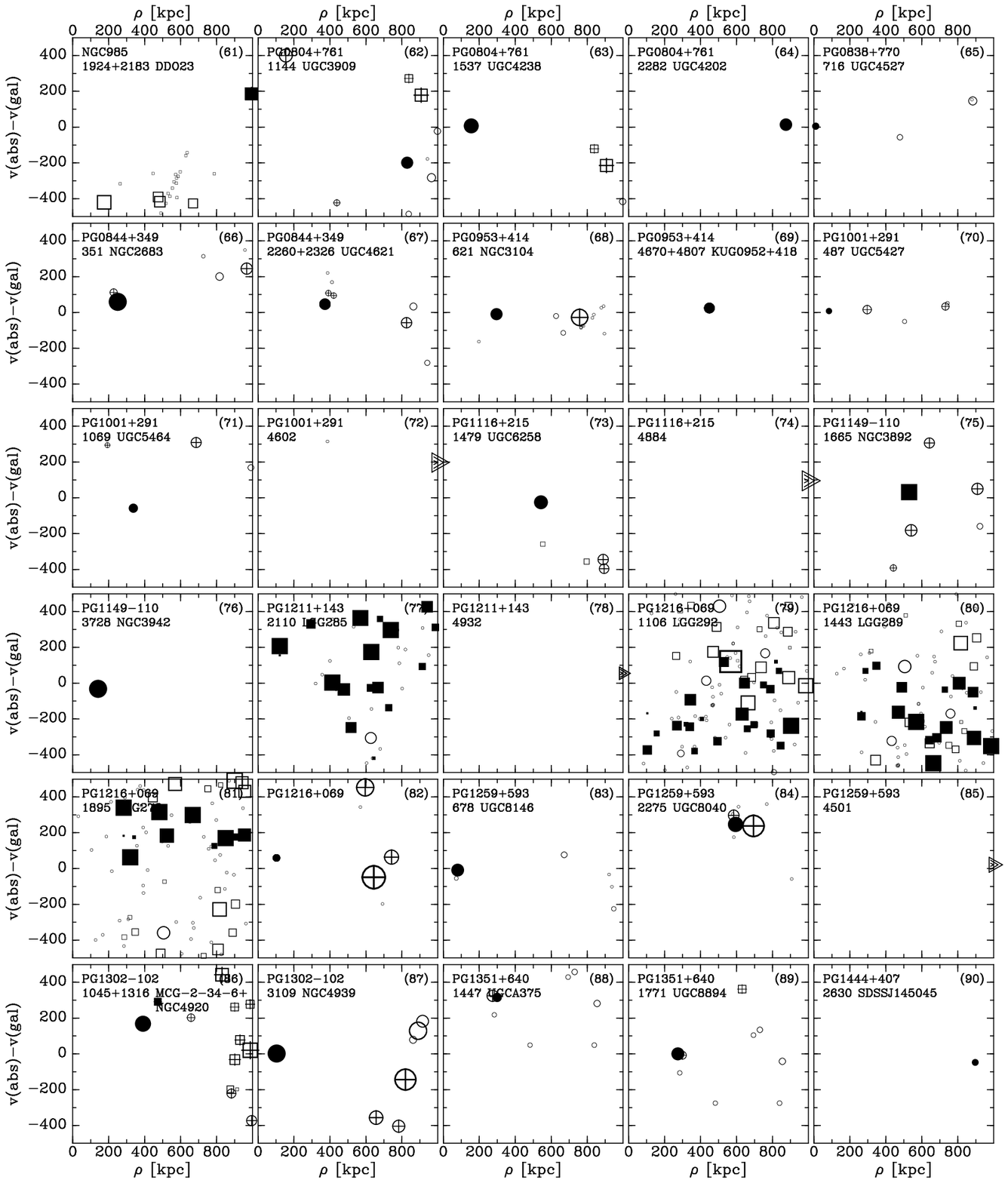}{0in}{0}{400}{400}{0}{-150}\figurenum{8}\caption{Continued.}\end{figure}
\begin{figure}\plotfiddle{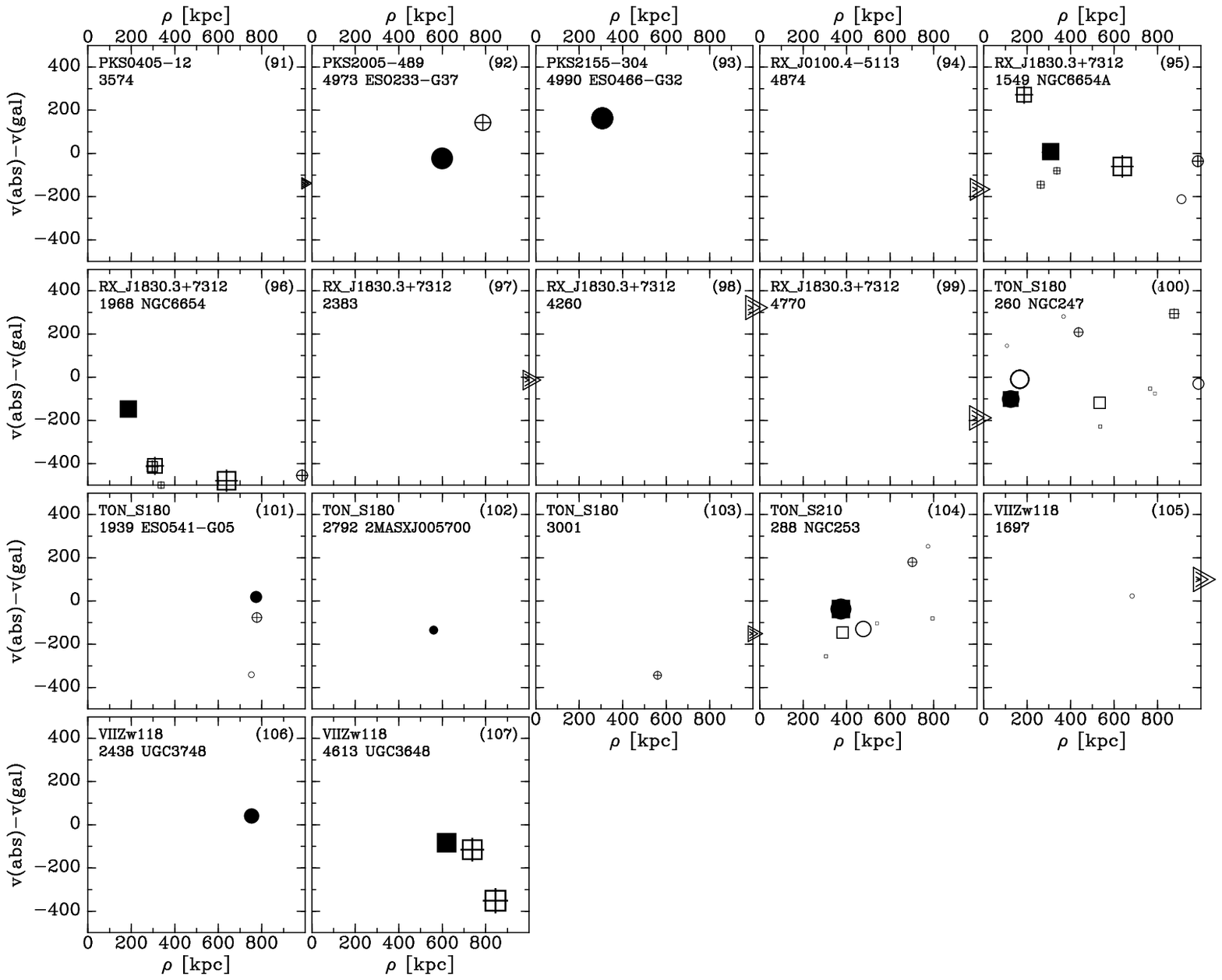}{0in}{0}{400}{300}{0}{-150}\figurenum{8}\caption{Continued.}\end{figure}
\par Now that we have established that absorbers associate with galaxies, we
discuss the general properties of the associations that we make to construct
Table~\Tres. But first we summarize the different a-priori criteria that were
used in previous work. These papers used either $H_0$=100~\kms\ or
$H_0$=71~\kms. We rescale all distances and impact parameters to the latter, but
when discussing results from other papers we include a factor \Hm=71/$H_0$ as a
reminder.
\par Lanzetta et al.\ (1995) identified 46 galaxies at $z$=0.07--0.55 having
impact parameter \ip=25--500\Hm~kpc in fields toward six QSOs observed with the
{\it Faint Object Spectrograph} (FOS) on HST. These data allowed a search for
absorbers with equivalent width $>$150~\mA. They found associated absorbers
(\Dva$<$500~\kms) near 20 galaxies.
\par Bowen et al.\ (1996) analyzed absorptions near 38 galaxies at $z$=0--0.08
with \ip=60--700\Hm~kpc using FOS spectra of 10 QSOs. Nine were found to have
associated \Lya.
\par Tripp et al.\ (1998) studied the relation between galaxies and \Lya\
absorbers in two QSO sightlines (H\,1821+643 and PG\,1116+215). They found 17
galaxy-absorber pairs with \ip$<$1~Mpc and \Dv$<$350~\kms. Within \ip$<$600~kpc
and \Dv$<$1000~\kms\ they found a galaxy for 100\% of the absorbers. They
further found that the weakest \Lya\ absorbers ($W$$<$100~\mA) associate with
galaxies sightly less than the strong absorbers.
\par Impey et al.\ (1999) discussed ten low-resolution QSO spectra observed with
the \GHRS. All sightlines were in the general direction of the Virgo cluster.
Eleven of the intergalactic \Lya\ lines in their sample were correlated against
a sample of galaxies complete to $M_B$=$-$16 (0.04\,\Lstar). From this
comparison, they concluded that absorbers are not preferentially associated with
galaxy halos, looking at impact parameters $<$2~Mpc and velocity differences
$<$300~\kms.
\par Bowen et al.\ (2002) obtained STIS-G140M spectra of seven AGN sightlines
passing close to (\ip$<$700\Hm~kpc) eight nearby galaxies, deciding in favor of
an association between \Lya\ absorption and galaxy if the velocity difference
was less than about 500~\kms. They considered the complications that occur when
associating an absorber with the galaxy nearest in velocity. For systems with
multiple absorbers this can mean that a weak absorber is associated with the
galaxy with low impact parameter, while a strong absorber is associated with a
smaller galaxy with higher impact parameter. In their sample this happens for
the sightlines to ESO\,438-G09, MCG+10-16-111 and RX\,J1830.3+7312. As can be
seen from the discussions in the Appendix, these three are among the more
problematic sightlines, and in this respect Bowen et al.\ (2002) got unlucky
with their sample.
\par Penton et al.\ (2002) studied 15 sightlines and listed the nearest three
galaxies to each \Lya\ absorber, where they used a ``retarded Hubble flow''
model to convert velocity differences into spatial distances. That is, they
assumed that $D$(gal--abs)=\ip\ when \Dva$<$300~\kms\ and add (in quadrature) a
line-of-sight radial distance as (\Dva$-$300)/H$_o$. Stocke et al.\ (2006) used
the same method, but extended the velocity range to $\pm$500~\kms. The problem
with this method is that it weighs the impact parameter on the sky much more
than the velocity difference. Further, any galaxy with \Dv$>$370~\kms\ is
a-priori assumed to be more than 1~Mpc away from the absorber, while
\Dv=440~\kms\ already implies 2~Mpc. We therefore decided not to introduce this
complication in our analyses.
\par Prochaska et al.\ (2006) looked for galaxies with $R$ magnitude $<$20 near
metal-line absorbers toward PKS\,0405$-$12. They claimed to be complete down to
$L$=0.1\Lstar\ for \ip$<$1~Mpc at $z$=0.1 and concluded that the \OVI\ absorbers
toward this sightline arise in individual halos, galaxy groups, filamentary
structure as well as voids, i.e.\ in a variety of environments.
\par Clearly, a variety of criteria have been used in the literature. To
determine the ``best'' criteria more data are needed, as well as an approach
that is not a-priori biased to assuming a maximum impact parameter or maximum
velocity difference. Combined with the relative completeness of the galaxy
sample at velocities $<$5000~\kms, our sample of sightlines and absorbers is
just large enough that we do not have to choose the association criteria
a-priori.
\par In Fig.~\Fassoc\ we graphically show the justification for each association
that we make between an absorption line and a galaxy. We generally associate the
absorption line with the galaxy with the smallest combination of \ip\ and
\Dv=$v$(gal)$-$$v$(abs), where a velocity difference of a few hundred \kms\ is
considered equivalent to an impact parameter of about a Mpc. However, sometimes
we prefer a larger galaxy over a dwarf galaxy with slighly lower \ip\ but higher
\ip/\Rgal.

\par For 33\% (44 of 133) of the systems listed in Table~\Tres\ an unambiguous
association can be made with a galaxy, and in 42 of these the velocity
difference is $<$200~\kms. For instance toward 1H\,0717+714, we detect \Lyb\ at
2892~\kms. There is one large group galaxy (UGC\,3804 in LGG\,141,
\Rgal=22.8~kpc) with impact parameter 199~kpc, shown by the filled square in
Fig.~\Fassoc. A few other group galaxies lie further away (\ip$>$650~kpc) and
are shown by the open squares.
\par For a few sightlines (Mrk\,876, PG\,0844+349, PG\,0953+414, PG\,1216+069,
PKS\,2155$-$304), there are two or three \Lya\ detections listed in Table~\Tres\
as associated with one galaxy. This happens when there are no other galaxies
with which the absorption line could reasonably be associated.
\par For another few sightlines (3C\,273.0, 3C\,351.0, MCG+10-11-116,
MRC\,2251$-$178, PG\,1302$-$102, PKS\,2005$-$489) two \HI\ absorption features
are seen with similar velocity, and they can be associated with two galaxies
that have fairly similar impact parameter. Then both galaxies are listed in
Table~\Tres\ as having associated absorption, with the associations made to
minimize the total velocity difference. In these cases, it is possible to argue
that one galaxy should be associated with both detections, or that the
association should be reversed. However, this will generally make no difference
for doing statistical analyses. In Fig.~\Fassoc\ just one panel is shown in
these cases, rather than a separate panel for each absorption line.
\par Toward 3C\,249.1, HE\,0340$-$2703 (twice), Mrk\,478, Mrk\,501, Mrk\,771,
Mrk\,817 (twice), PG\,1001+291, PG\,1259+593, Ton\,S180 and VII\,Zw\,118 there
are multiple galaxies with similar impact parameter and similar velocity
difference, but only one intergalactic absorption feature. We then list the
largest galaxy in Table~\Tres, and list non-detections for the rest. In such
cases the listed galaxy may or may not be the proper one to chose.
\par Toward 3C\,273, ESO\,438-G09, HE\,1228+0131, MCG+10$-$16$-$111, Mrk\,734,
MS\,0700.7+6338, PG\,1211+143 and PG\,1216+069 the sightline passes through a
dense group of galaxies. In their spectra we find a total of 13 systems, for
which we don't list a particular galaxy in Table~\Tres\ as the associated
galaxy, but instead we list the group itself, using the smallest impact
parameter to a member of the group.
\par The cases summarized above add up to 93 of the 133 absorbing systems. For
almost all of these a well-justified association can be made between the
absorption and a galaxy with low ($<$1~Mpc) impact parameter. And if the
association is not clear-cut, there are galaxies with similar impact parameter
to choose from. This leaves 40 absorbers without clear-cut association with a
galaxy. Seven of these (toward 1H\,0149$-$577, Mrk\,335 (2 lines), NGC\,985 (2
lines), PG\,0804+761 and PG\,1351+643) are ambiguous. For instance, toward
NGC\,985 the only known galaxy with small velocity difference has large impact
parameter, while galaxies with lower impact parameter would have unusually high
velocity difference. Or, toward 1H\,0419$-$577 the candidates are a medium-sized
galaxy at \ip=140~kpc and another galaxy half its size at \ip=70~kpc. All of
these cases are discussed in detail in the Appendix.
\par Finally, in the directions to H\,1821+643, HE1029$-$1401,
MCG+10$-$11$-$116, MRC\,2251$-$178, Mrk\,106, Mrk\,110, Mrk\,290, Mrk\,421,
Mrk\,509, Mrk\,586, Mrk\,817, Mrk\,1095, PG\,0804+761, PG\,1001+291,
PG\,1116+215, PG\,1211+143, PG\,1259+593, PKS\,0405$-$12, RX\,J0100.4$-$5113,
RX\,J1830.3+7312, Ton\,S180 and VII\,Zw\,118 there are a total of 33
intergalactic \HI\ lines at a velocity for which we cannot find a galaxy with
impact parameter less than 1~Mpc. These cases are collected at the end of
Table~\Tres. In these sightlines, the search for associated galaxies was
extended to impact parameters of 5~Mpc. We can find a galaxy within 2~Mpc for 22
absorbers, and within 3~Mpc for 28. In all but two cases it is possible to find
a galaxy within 5~Mpc, and this is the galaxy listed in Table~\Tres. However,
since we do not list all the non-detections for galaxies with impact parameter
between 1 and 5~Mpc, these detections are treated slightly differently in the
remainder of the paper.
\par Among all the associations there are 14 cases where \OVI\ is seen. Twelve
of these fall into the clear-association category, one is in a multi-component
system (toward Mrk\,876), and one is associated with a group (toward 3C\,273).
\par Of the 128 associations between an intergalactic \HI\ absorber and a galaxy
listed in Table~\Tres, just 15 have \Dva$>$200~\kms. Seven of these occur for
galaxies with \ip=400--1000~kpc and seven have the nearest galaxy at
\ip$>$1~Mpc. The remaining one is the line at 1447~\kms\ (\Dv=316~\kms) toward
PG\,1351+640, which is ambiguously associated with UGCA\,375, at
\vgal=1763~\kms\ (see Sect.~\SSassoc). There are only two other cases for which
the nearest galaxy with \ip$<$3~Mpc also has \Dva$>$300~\kms. These are the line
at 4638~\kms\ toward Mrk\,290, for which the nearest galaxy with similar
velocity is NGC\,5971 (\ip=1586~kpc, \Dv=332~\kms), and the line at 3792~\kms\
toward MCG+10-16-111, with \Dv=349~\kms\ and the nearest galaxy NGC\,3809 at
\ip=2910~kpc.
\par In Table~\Tres\ we do not list galaxies with \ip$<$1~Mpc for which
\ip/\Rgal$>$125, where \Rgal\ is the galaxy's diameter. This criterion allows us
to avoid listing non-detections for many dwarf galaxies that are close to a
larger galaxy with similar impact parameter. We also avoid listing many dwarf
galaxies whose impact parameter is much larger than that of a large galaxy with
which the absorption can be associated. However, by looking at the ratio of
impact parameter to diameter (rather than just the diameter) we do allow small
galaxies with small impact parameter to appear in the table of results. The
criterion is based on the fact that we found that \ip/\Rgal$<$125 for all but
one likely associations that have \ip$<$1~Mpc. Even when the nearest galaxy to
an absorber is at \ip$>$1~Mpc, \ip/\Rgal\ is $<$125 for half the cases. In fact,
for half the associations \ip/\Rgal\ is $<$23 and for 90\% it is $<$80. Note,
however, that for many of the analyses presented in Sects.~\Sassocresults\ and
\Sgalresults\ we include {\it all} galaxies with impact parameter $<$1~Mpc, even
if they are not listed in Table~\Tres.
\par Summarizing the results of our attempts at associating absorbers with
galaxies we conclude that: {\it For the majority (100 of 133, 75\%) of
absorption lines, it is possible to find a galaxy within 1~Mpc and 400~\kms. For
about half of these (54) there is just one galaxy that is a likely candidate.
For most of the other half, there either are two galaxies equally likely to be
associated with the absorber (12 cases) or there are two absorption lines and
two likely galaxies (12 cases). In a small number of cases (13) the absorption
occurs in a group of galaxies and no unambiguous choice can be made. Ambiguities
also exist for a small fraction (7 cases) of absorbers outside groups.}

\subsection{\Lya\ Absorbers in Voids}
\par A question that is complementary to asking whether \Lya\ absorbers
associate with galaxies is whether there are any \Lya\ absorbers that lie in
voids. This question was studied most clearly by Penton et al.\ (2002), who
defined a ``void absorber'' as a \Lya\ line for which there is no galaxy within
3~Mpc. They based their 3~Mpc void size on the contention that it is the median
distance from a random point in the universe to the nearest \Lstar\ galaxy. As
Fig.~\Fnearest c shows, this is true if one only looks at galaxies with
$L$$>$\Lstar. A 3~Mpc definition for the separation from the nearest \Lstar\
galaxy would imply that half of the \Lstar\ galaxies lie in voids.
\par Taking a nearest-neighbor distance of 3~Mpc as the limit for a void
absorber, we can find a galaxy within that distance and with \Dva$<$400~\kms\
for 99 of the 102 \Lya\ lines at $v$$<$5000~\kms. This would imply a 3$\pm$2\%
fraction of void absorbers. In particular, these would be the lines at
2545~\kms\ toward Mrk\,509 (nearest galaxy at \ip=4834~kpc), at 4043~\kms\
toward MCG+10-16-111 (nearest galaxy at \ip=4351~kpc), and at 4670~\kms\ toward
Mrk\,817 (nearest known galaxy at \ip=8739~kpc). The fact that all three lines
are at $v$$>$4000~\kms\ suggests that maybe the galaxy surveys near these
sightlines were not deep enough to find all galaxies. We note that near Mrk\,509
there are five galaxies with unknown velocity (and $L$$\sim$0.25\,\Lstar) whose
impact parameter would be between 1.5 and 2.5~Mpc, if their velocity were about
2500~\kms. In that case the detection would no longer be considered a ``void
detection'' (see Appendix for more details). Similarly, MCG+10-17-2A could have
\ip=1.4~Mpc from MCG+10-16-111 if that galaxy's velocity is about 4043~\kms.
Thus, it is quite possible that there are galaxies within 2~Mpc of the two best
current candidates for ``void absorbers'' 
\par If we impose that the nearest galaxy has to be brighter than 0.1\,\Lstar,
there are two more candidate void absorbers: the line at 3574~\kms\ toward
PKS\,0405$-$12 (nearest $L$$>$0.1\,\Lstar\ galaxy at \ip=4492~kpc; Table~\Tres\
lists 2MASX\,J04060761$-$1023272 with \Rgal=5.7~kpc, \ip=1489~kpc) and the line
at 3007~\kms\ toward Mrk\,421 (nearest $>$0.1\,\Lstar\ galaxy at \ip=4840~kpc;
nearest galaxy HS\,1059+3934 with \Rgal=3.8~kpc, \ip=1041~kpc).
\par In principle, some of the candidate void absorbers could instead represent
gas expanding at very-high-velocity from the AGN. For Mrk\,421 this expansion
velocity would be 5993~\kms, for Mrk\,509 it would be 7767~\kms, and for
Mrk\,817 it is 4760~\kms. These would be comparatively large expansion
velocities, however. Tripp et al.\ (2008) concluded that absorbers with velocity
differing more than 2500~\kms\ from that of the AGN are almost certain to be
intergalactic, rather than associated with the AGN. Thus, the absorbers toward
Mrk\,421. Mrk\,509 and Mrk\,817 are unlikely to be associated with the AGNs.
\par The foregoing discussion does not use a complete sample of galaxies,
however. If we do restrict ourselves to complete samples, we find that there may
be no absorbers that can be classified as a ``void absorber''. At velocities
below 2500~\kms, where our galaxy sample is complete down to $L$$>$0.1\,\Lstar,
we can find an $L$$>$0.1\,\Lstar\ galaxy within \ip$<$1.4~Mpc for {\it all} of
the 58 absorption lines with $v$$<$2500~\kms. We can find an $L$$>$0.25\,\Lstar\
galaxy for 71 of the 75 absorbers with $v$$<$3700~\kms, and an
$L$$>$0.5\,\Lstar\ galaxy for 92 of the 102 absorbers with $v$$<$5000~\kms.
Finally, there are 17 absorbers among the 102 with $v$$<$5000~\kms\ for which we
cannot find an $L$$>$\Lstar\ galaxy within 3~Mpc. This fraction of 17/102 or
17$\pm$4\% is the equivalent of the value of 22$\pm$8 that Penton et al.\ (2002)
quoted for the fraction of absorbers in voids -- they looked at four sightlines
to compare the locations of $L$$>$\Lstar\ galaxies relative to absorbers at
$v$$<$20000~\kms. We note that 12 of our 17 void absorbers have $v$$>$3000~\kms.
So, it is conceivable that galaxies were still missed near the sightlines with
candidate void absorbers.
\par Looking only at the 58 absorbers with $v$$<$2500~\kms, we find an
$L$$>$0.1\,\Lstar\ galaxy within 1.4~Mpc for each of these, as well as a
$>$0.25\,\Lstar\ galaxy within 2.2~Mpc, and a $>$0.5\,\Lstar\ galaxy within
2.5~Mpc. Finally, for 54 of the 58 absorbers there is an $L$$>$\Lstar\ galaxy
within 2.6~Mpc. Thus, using a 3~Mpc limit, there are {\it no} void absorbers
when looking for galaxies brighter than 0.5\,\Lstar, and there is a 7$\pm$3\%
fraction when looking for \Lstar\ galaxies. This would suggest that there are
{\it no} void absorbers if the galaxy sample is sufficiently complete at a
luminosity limit better than 0.5\,\Lstar.
\par McLin et al.\ (2002) looked in more detail near seven of the void absorbers
claimed by Penton et al.\ (2002) (four at $v$$<$5000~\kms), determining
redshifts of galaxies down to $\sim$0.1\,\Lstar\ lying within
20\arcmin--40\arcmin\ of the sightlines. They did not find any galaxies with
\ip$<$250~kpc. However, for three of the four at $v$$<$5000~\kms\ we find an
$L$$>$0.3\,\Lstar\ galaxy within 1.1~Mpc (\Lya\ at 2426~\kms\ toward
VII\,Zw\,118, UGC\,3748 at 753~kpc; \Lya\ at 1979~\kms\ toward HE\,1029$-$1401,
MCG$-$2-27-1 at 1065~kpc; \Lya\ at 2548~\kms\ toward Mrk\,509, MCG$-$54-3 at
2231~kpc). Only for the 3007~\kms\ line toward Mrk\,509 is the nearest
$L$$>$0.1\,\Lstar\ galaxy (UGC\,6383) at large impact parameter (4840~kpc). All
these galaxies have angular separation larger than 1 degree, so the
20\arcmin--40\arcmin\ field used by McLin et al.\ (2002) may have been too small
to find all the relevant galaxies.
\par Based on these results, we conclude that {\it it depends on the luminosity
limit and completeness of the galaxy sample whether an absorber is called a
``void absorber'' or not.} Further, {\it we find that there may not be any void
absorbers (i.e.\ absorbers occuring more than 3~Mpc from the nearest galaxy) if
a luminosity limit of 0.1\,\Lstar\ is used. In fact, we find a $>$0.1\,\Lstar\
galaxy within 1.5~Mpc of each of 58 absorbers with $v$$<$2500~\kms, and a
$>$0.5~\Lstar\ galaxy within 2.5~Mpc of each absorber with $v$$<$2500~\kms.}
Finally, {\it we find that just 10\% of the absorbers with $v$$<$5000~\kms\ lies
more than 3~Mpc from the nearest 0.5~\Lstar\ galaxy. Finally, where Penton et
al.\ (2002) found a fraction of 22$\pm$8\% for absorbers more than 3~Mpc from
the nearest \Lstar\ galaxy, we find a fraction of 17$\pm$4\% for absorbers with
$v$$<$5000~\kms, but just 7$\pm$3\% when only looking at absorbers with
$v$$<$2500~\kms.}

\subsection{Velocity Difference Between \Lya/\OVI\ Absorbers and Associated Galaxies}
\begin{figure}\plotfiddle{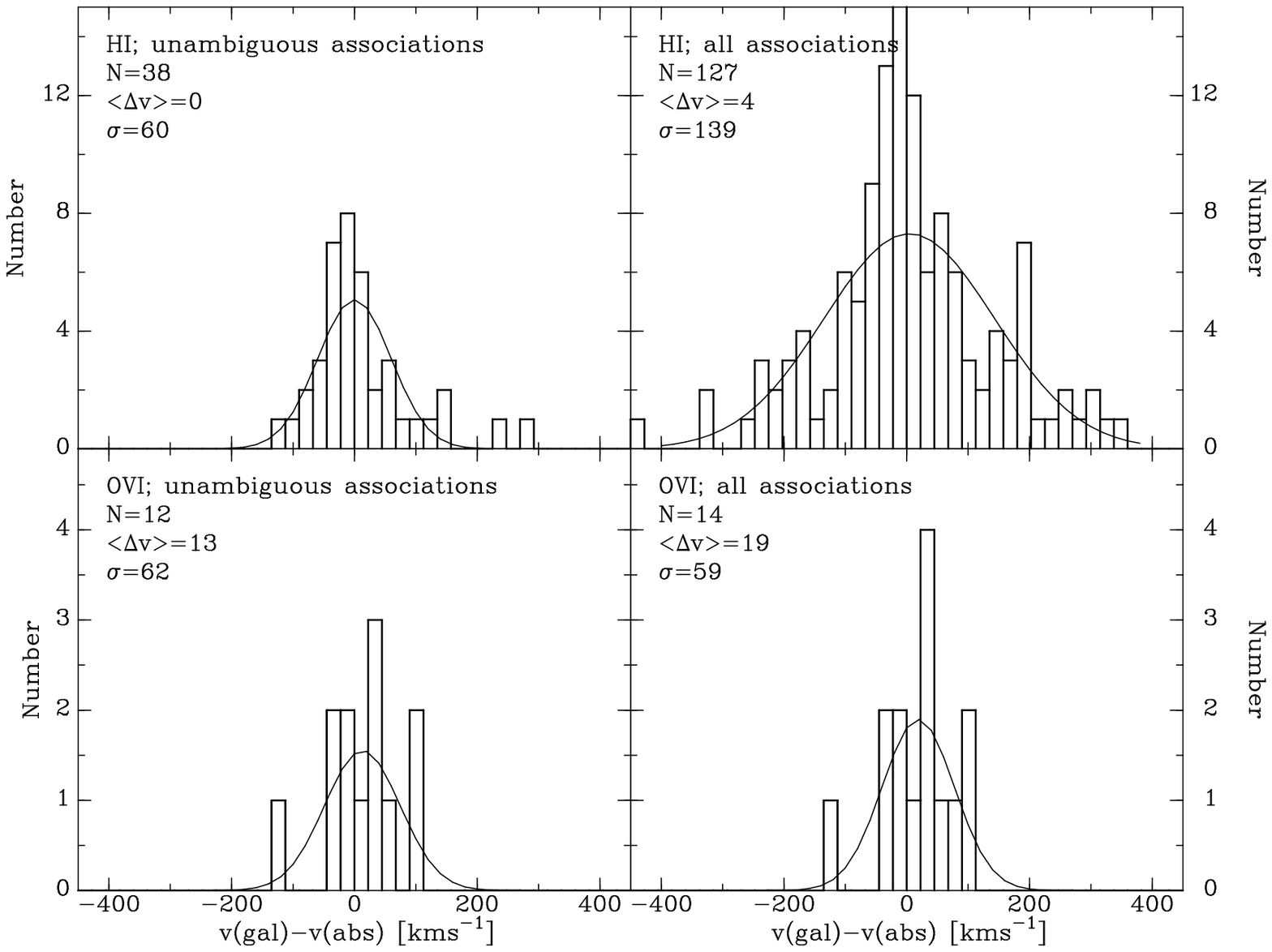}{0in}{0}{450}{350}{0}{-150}\figurenum{9}\caption{\captionvdiffhist}\end{figure}
\par To determine (a-posteriori) the best velocity difference criterion for
associating an absorber with a galaxy, we show in Fig.~\Fvdiffhist\ the
distribution of the difference between the velocity of the absorber (either
\Lya\ or \Lyb\ in the top panels, \OVI\ in the bottom panels) and the associated
galaxy (as listed in Table~\Tres, and as discussed in the Appendix and
Sect.~\SSassoc). In the left panels we show the distribution when selecting only
the unambiguous cases with \ip$<$1~Mpc, while on the right all associations are
included. Clearly, the distribution of unambiguous cases is basically
symmetrical around \Dv=0; excluding the two outliers, the dispersion is 60~\kms,
with a range from $-$118 to +147~\kms. The full distribution (right panels)
includes the cases with multiple absorbers associated with one galaxy, cases
with one or two lines and two galaxies with similar impact parameter and
velocity, detections associated with a group and detections with the nearest
galaxy at \ip$>$1~Mpc.
\par As we limited ourselves to the part of the universe where most galaxies
have previously been found, our sample of galaxies is more complete than that
used in previous studies. Therefore, we draw the conclusion that {\it for
unambiguous associations (i.e.\ just a few galaxies are known within 1~Mpc and
400~\kms, just one or two of which are bright), the difference in velocity
between the intergalactic absorption and the galaxy's systemic velocity
ranges from $-$118 to 147~\kms, with a dispersion of 60~\kms. \fix{This suggests
that in general it is OK to use a limit of \Dv$<$400~\kms\ for associating an
absorber with a galaxy, although if there are no galaxies with low impact
parameter and low velocity difference in individual cases larger velocity
differences might exist.}}

\subsection{\Lya\ Linewidths versus Impact Parameter}
\begin{figure}\plotfiddle{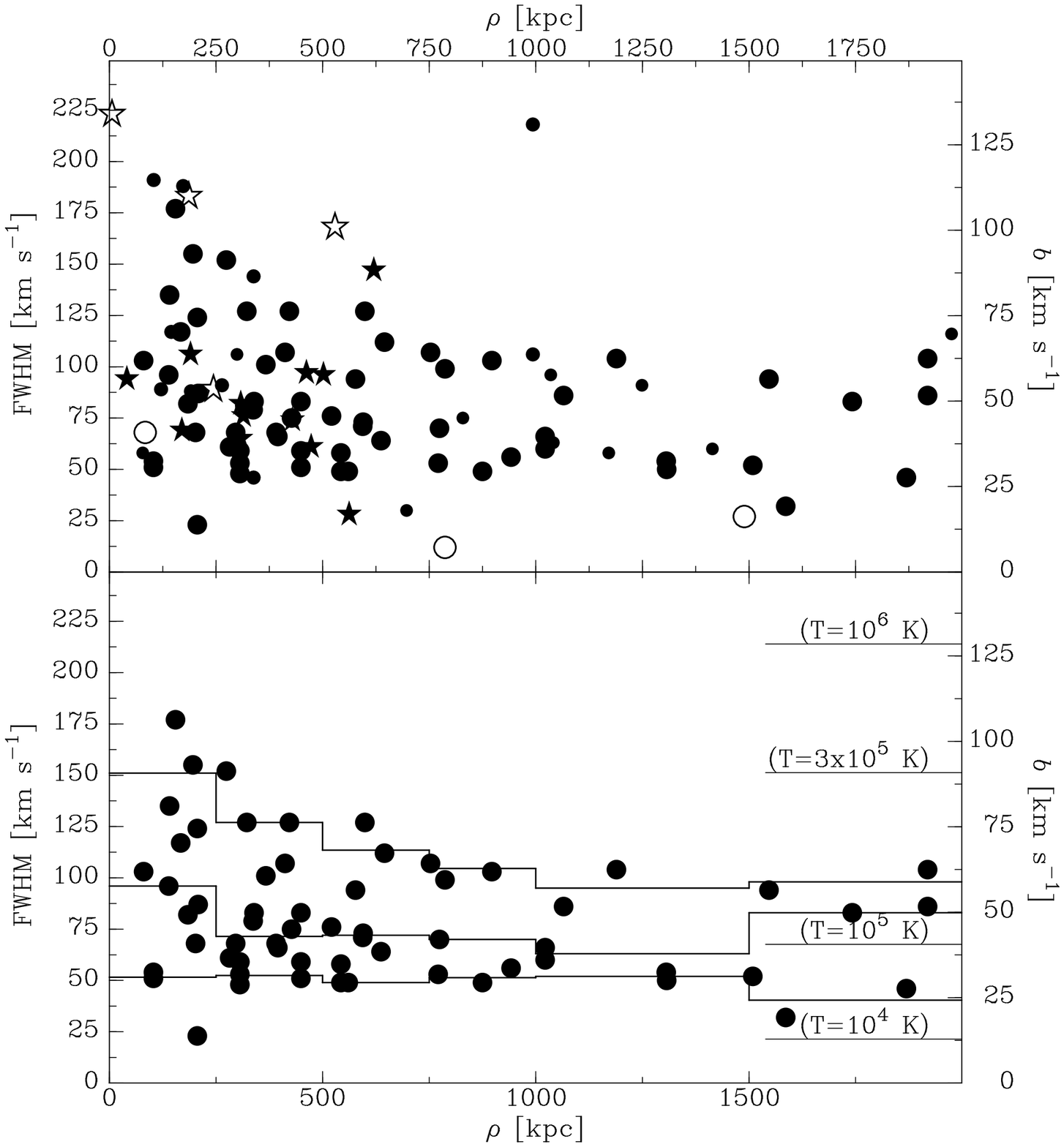}{0in}{0}{450}{450}{0}{-150}\figurenum{10}\caption{\captionwidthimp}\end{figure}
\par In Fig.~\Fwidthimp\ we look at the distribution of fitted line widths of
\Lya\ versus impact parameter. In the top panel we show the scatter plot for all
lines. Absorbers that we associate with groups are shown by stars, field
absorbers by circles. Closed symbols correspond to reliable absorbers, open
circles to unreliable ones. Unreliable lines are those that are saturated as
well as lines for which the width measurement is questionable, either because
the spectrum is too noisy, or the line is too shallow. Finally, small symbols
are for lines where the association with a galaxy is ambiguous, i.e.\ the
assigned impact parameter is unreliable. In the bottom panel we include only
reliable field absorbers with a clear association with a galaxy (see discussion
in Sect.~\SSassoc).
\par The outlier at (206~kpc, 21~\kms) in both panels is the secondary \Lya\
line at 1109~\kms\ seen toward Mrk\,876. The outlier at (993~kpc, 219~\kms) in
the top panel corresponds to the line at 1924~\kms\ seen toward NGC\,985. This
is a weak line, which may have multiple components. Also, the association we
list with DDO\,23 is one of the most ambiguous in the sample. In fact, Bowen et
al.\ (2002) associated it with NGC\,988, which has impact parameter 175~kpc, but
in that case $v$(\Lya)$-$$v$(gal) would be 419~\kms, larger than for any other
absorber-galaxy association. Associating the line with the LGG\,71 group is also
problematic, since the highest velocity for any group galaxy is 1665~\kms\
(\ip=446~kpc). If the line is indeed a single broad line, its location in
Fig.~\Fwidthimp\ would suggest that there should be galaxy with low impact
parameter, which could be NGC\,988. A \COS\ spectrum of this target could
resolve these issues.
\par Both panels of Fig.~\Fwidthimp\ shows a clear pattern: at all impact
parameters there is a large spread in measured linewidths, but there is a trend
for the maximum linewidth to increase with decreasing impact parameter. The
histograms in the bottom panel show the 10th, 50th and 90th percentile of the
distribution of linewidths. At impact parameters above 200~kpc, the 50th
percentile (i.e.\ the median) is 75--80~\kms, but for the 0--200~kpc bin it is
105~\kms. Also, the upper envelope of the distribution shows a clear trend of
increasing maximum linewidth with decreasing impact parameter.
\par On the right side of the bottom panel of Fig.~\Fwidthimp\ we also show
horizontal lines at FWHMs of 21, 68, 151 and 214~\kms, corresponding to the
thermal width of \HI\ at temperatures of \dex4, \dex5, 3\tdex5, and \dex6~K.
However, the width of the \HI\ absorption line is not necessarily thermal.
\fix{It is often assumed that \Lya\ absorbers consist of photoionized gas and
such modeling gives consistent results (e.g.\ Penton et al.\ 2004).} Also, some
lines may have multiple components. Alternatively, the gas may be highly
turbulent. However, a turbulent width of 100~\kms\ is an order of magnitude
higher than typical values seen inside galaxy disks. If the gas is orbiting a
galaxy, but stretched over many kpc, the projection of its orbital velocity may
change from the front to the back side, broadening the observed profile. For
cloud densities in the gas typical of that of clouds in the Milky Way corona
($\sim$\dex{-3}\,\cmm2, Fox et al.\ 2005), a cloud with total hydrogen column
density $\sim$\dex{20}~\cmm2\ would be about 30~kpc deep. Orbiting with
200~\kms\ at 100~kpc, the changing velocity projection then introduces a
velocity gradient of about 8~\kms, much smaller than the observed 50~\kms\
increase in the maximum velocity width.
\par\fix{A final possibility is that large velocity gradients were introduced
for the gas in the extended halos by tidal effects. Near the Milky Way, the
Magellanic Stream provides evidence for such processes, and in the Galactic
Standard of Rest reference system it has an apparent velocity gradient of
400~\kms\ over 180\deg. Sightlines through the stream result in profiles that
when fit by a single-gaussian can have FWHM up to $\sim$80~\kms. This is much
smaller than the absorber linewidths at low impact parameter. However, we cannot
exclude that the apparent linewidths will be larger if the sightline passes
through a tidal feature along its long axis. On the other hand, it is unlikely
that the axis of the feature and the sightline line up exactly in many cases. On
balance, tidal stretching may be one of the causes of the line broadening, but
it is unlikely to be the explanation for {\it all} of the lines that are broader
than 150~\kms.}
\par At present, the S/N ratios of spectra with candidate broad lines are too
low to analyze the detailed shapes of the profiles and decide whether they can
be described by a single component gaussian. The installation of \COS\ on \HST\
will allow observations with sufficiently high S/N ratio to check this. With
this caveat in mind, however, Fig.~\Fwidthimp\ suggests that {\it there is an
increase in the temperature of the gas within a few hundred kpc of galaxies.
Such behavior inside gravitational wells is predicted in hydrodynamical
simulations of structure formation in the universe (Cen \& Ostriker 1999; Dav\'e
et al.\ 2001; Cen \& Fang 2006).}

\subsection{\Lya\ Equivalent Width versus Impact Parameter}
\begin{figure}\plotfiddle{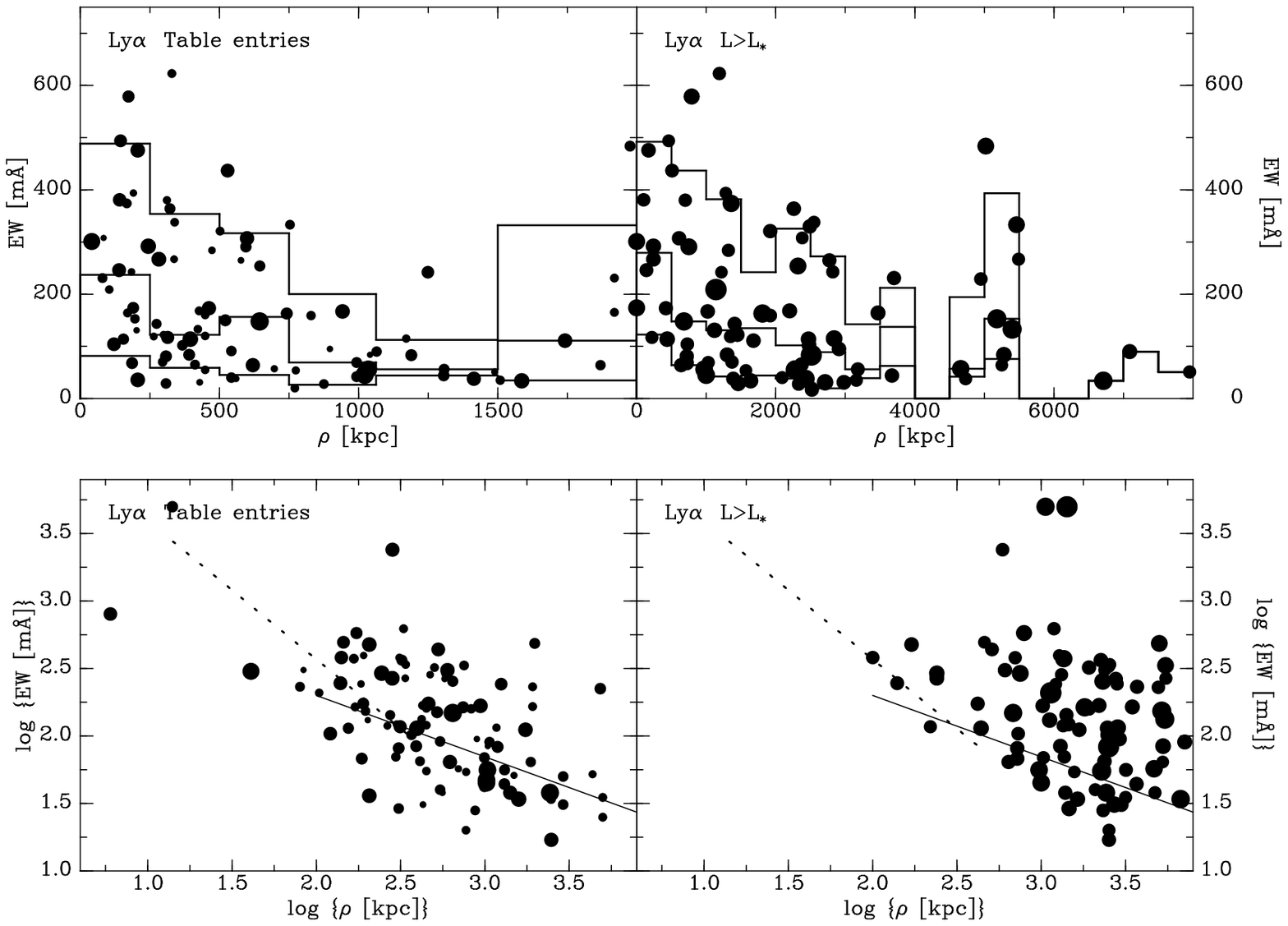}{0in}{0}{450}{350}{0}{-150}\figurenum{11}\caption{\captionewvsb}\end{figure}
\par The question of an anti-correlation between the equivalent width of \Lya\
absorbers and the impact parameter to the nearest galaxy has been the subject of
much discussion and disagreement in the literature. Lanzetta et al.\ (1995)
reported an anti-correlation between equivalent width ($W$) and \ip, but they
only had 7 detections with \ip$<$100\Hm~kpc. Tripp et al.\ (1998) plotted nine
absorbers at \ip$<$600\Hm~kpc, and these do show an anti-correlation between $W$
and \ip. Chen et al.\ (2001) claimed a tight anti-correlation between $W$ and
\ip\ for their 34 associations with \ip$<$200\Hm~kpc. In particular, they
claimed that log$W$= $-$(0.96$\pm$0.11)~log\ip\,+\,constant. However, also they
showed plots of the difference between log\,$W$ predicted by this equation and
the actual log\,$W$, and these reveal that the residuals have a large spread of
$\pm$0.5 dex. Chen et al.\ (2001) further claimed that the anti-correlation
improves if the luminosity of the associated galaxy is taken into account. Impey
et al.\ (1999) combined earlier studies with their sample of 139 \Lya\ lines,
and saw a more mixed picture, concluding that at high impact parameters there is
no anti-correlation between $W$ and \ip, but at \ip$<$200\Hm~kpc there is a
trend of finding higher $W$ at lower \ip. The well-defined sample of 6 clear
associations with \ip$<$200\Hm~kpc of Bowen et al.\ (2002) also lead them to
conclude that log\,$W$ and log\,\ip\ do anti-correlate. Penton et al.\ (2002)
studied this once more, using their sample of 81 low-redshift absorbers and
concluded that at \ip$>$50\Hm~kpc the $W$ vs.\ \ip\ plot is a scatter plot. At
lower \ip\ they did not find a strong anti-correlation, but the lowest measured
$W$ is 500~\mA, so that viewed over a wide range of \ip\ there seems to be some
relation.
\par We show our results in Fig.~\Fewvsb, using linear scales in the top two
panels, logarithmic scales in the bottom two. In the studies listed above only
logarithmic scales were used for $W$ and \ip, but as we discuss below, this
obscures the real concluson that can be drawn from the scatter plot. In the left
panels of Fig.~\Fewvsb\ we use the impact parameters and equivalent widths given
in Table~\Tres. In the plot on the right we check to see what happens if we were
only able to find the nearest $L$$>$\Lstar\ galaxy to an absorber.
\par For comparing with the literature results (bottom panels), we include the
relations given by Chen et al.\ (2001) (dotted line) and Penton et al.\ (2002)
(solid line). This shows that our impact parameters are typically much larger
than those of Chen et al.\ (2001). The distribution of points at \ip$>$100~kpc
looks like a scatter plot, except when looking at the nearest $L$$>$\Lstar\
galaxy, in which case there seems to be an impact parameter dependent lower
envelope, i.e.\ $W$$<$100~\mA\ only occurs at \ip$>$0.5~Mpc, $W$$<$50~\mA\ at
\ip$>$1~Mpc and $W$$<$25~\mA\ at \ip$>$2.5~Mpc from the nearest \Lstar\ galaxy.
A similar pattern holds when looking at the impact parameter relative to
$L$$>$0.1\,\Lstar\ galaxies, for which $W$$<$100~\mA\ occurs at \ip$>$150~kpc,
$W$$<$50~\mA\ at \ip$>$300~kpc and $W$$<$25~\mA\ at \ip$>$750~kpc.
\par On the other hand, strong lines occur mostly in the neighborhood of
galaxies, with just two of the ten $>$400~\mA\ lines occurring at impact
parameters \ip$>$350~kpc (at 1665~\kms\ toward PG\,1149$-$110, \ip=529~kpc, and
at 3579~\kms\ toward Mrk\,110, \ip=1975~\kms). At the $>$300~\mA\ level 19 of 23
lines (82\%) occur within 600~kpc of a galaxy, and 16 of 23 (70\%) have
\ip$<$350~kpc. This includes 6 of the 10 detections that we associate with a
group rather than a single galaxy. We can look at this in reverse, and note that
for about 10\% of the strong lines the nearest galaxy has impact parameter
$>$1~Mpc (5 of 38 with $W$$>$200~\mA, 2 of 23 with $W$$>$300~\mA, and 1 of 10
with $W$$>$400~\mA).
\par The top panels of Fig.~\Fewvsb\ show that using a logarithmic scale for the
equivalent width hides an important facet of the relation between $W$ and \ip.
The bins in the top panels give the 10th, 50th and 90th percentile of the
distribution of $W$ in a 250~kpc (at \ip$<$1~Mpc) or 500~kpc (at \ip$>$1000~kpc)
wide interval. The 10th percentile remains more or less constant with impact
parameter (except for the lowest impact parameters to the nearest \Lstar\
galaxy), while the 90th percentile strongly anti-correlates with \ip, especially
at \ip$<$1500~kpc. Note that the top panels do not show the four strongest
lines, three of which occur inside a galaxy (\ip=0.3~kpc toward HS\,1543+5921,
\ip=6~kpc toward Mrk\,205, \ip=14~kpc toward 3C\,232 and the sub-DLA at
1895~\kms\ toward PG\,1216+069). It is clear that at any impact parameter, the
distribution of equivalent widths of \Lya\ absorbers is wide, but the maximum
equivalent width is larger at smaller impact parameters.
\par We checked whether there was any difference in the distribution when
selecting only field or group galaxies, or when selecting only galaxies with
diameters in a given range, or when comparing galaxies with \vgal$<$2500 vs.\
galaxies with \vgal$<$5000~\kms. In all but one cases the 10th, 50th and 90th
percentile bins are basically identical. The lone exception is that all the
lines stronger than 100~\mA\ at \ip$>$1500~kpc occur at velocities above
2500~\kms. Since the galaxy sample is less complete for the fainter galaxies, it
is possible that a better search for galaxies could turn up an $L$$<$0.1~\Lstar\
galaxy with lower impact parameter.
\par From Fig.~\Fewvsb\ we derive the following conclusions: {\it At any impact
parameter there is a wide range in \Lya\ equivalent widths, but the strongest
lines are stronger at lower impact parameter. Also: 80\% of strong
($W$$>$300~\mA) lines occur within 600~kpc of a galaxy, while 70\% originate
within 350~kpc. On the other hand, weak lines only occur far from \Lstar\
galaxies, and the weaker the line, the larger the minimum impact parameter;
specifically all \Lya\ lines with $W$$<$25~\mA\ have \ip$>$750/2500~kpc to the
nearest 0.1/1.0\,\Lstar\ galaxy, while all \Lya\ lines with $W$$<$50~\mA\ have
\ip$>$300/1000~kpc to the nearest $L$$>$0.1/1.0\Lstar\ galaxy.}

\section{Absorbers Near Galaxies}
This section presents a number of analyses from the perspective of the galaxies.
By restricting ourselves to very low redshift, we can use general surveys and
catalogs to produce a fairly complete galaxy sample. Looking at many sightlines,
we can thus study the detection fraction as function of impact parameter using
different selection criteria. First we look at the amount of intergalactic gas
as function of the density of galaxies near the absorbers (Sect.~\SSgaldens). In
Sect.~\SSgalaxycount\ we tabulate the fraction of galaxies having an absorber
within a given impact parameter and velocity, while in Sect.~\SSdetfrac\ we show
histograms of the detection fraction as function of impact parameter, using
several criteria to select galaxies. We also combine all the sightlines to
create a synthetic map of a galaxy's gaseous envelope (Sect.~\SShalomap).

\subsection{\Lya\ Equivalent Width versus Galaxy Density}
\begin{figure}\plotfiddle{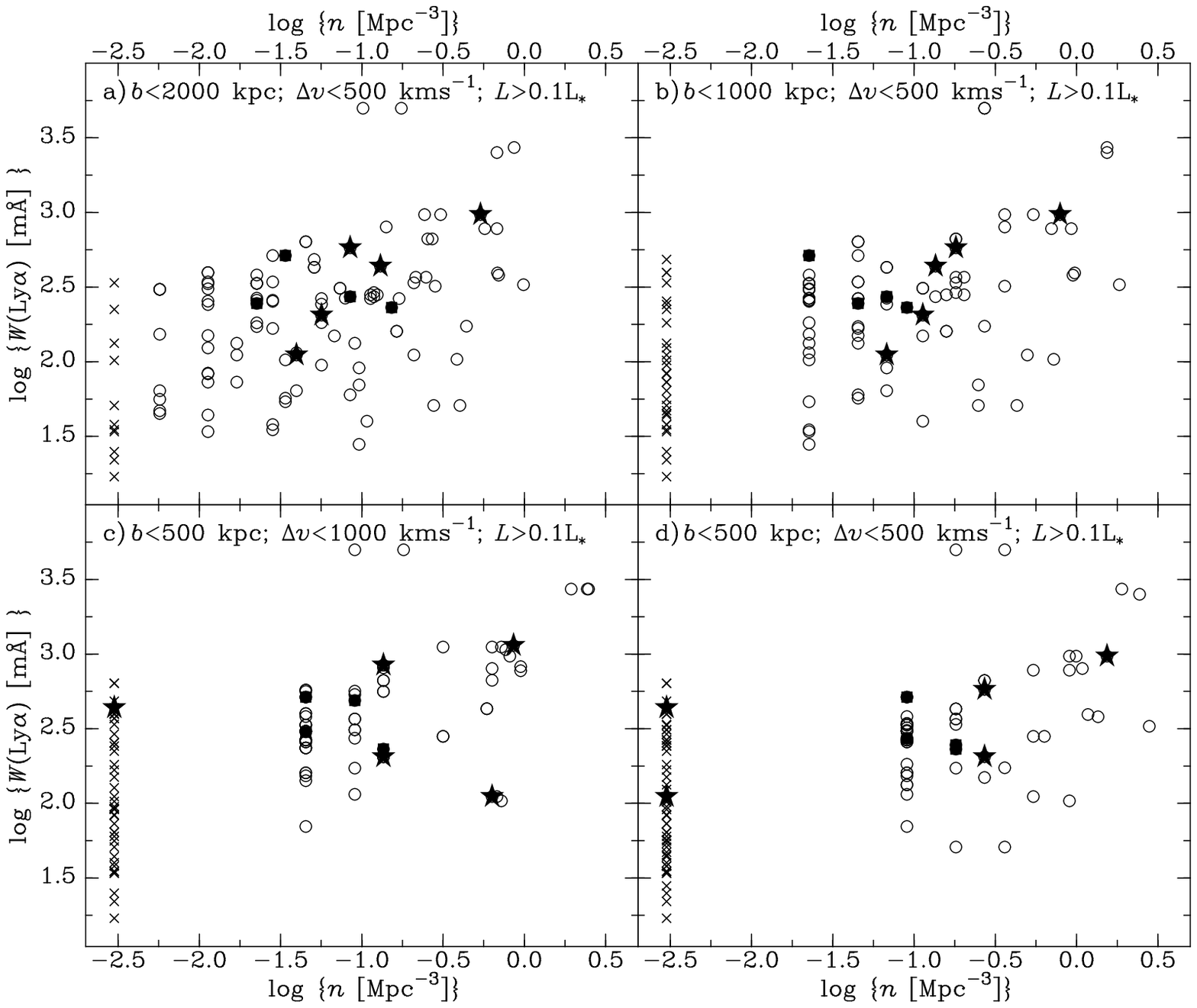}{0in}{0}{450}{400}{0}{-150}\figurenum{12}\caption{\captionewvsgaldens}\end{figure}
\par A way of looking at the relation between galaxies and absorbers is to look
at the density of galaxies in the neighborhood of an absorber. Bowen et al.\
(2002) did this for the six sightlines in their sample. They calculated the
number of galaxies in a cylindrical volume with impact parameter $<$2~Mpc and
velocity within $\pm$500~\kms\ from the absorber. Since the Hubble constant is
71~\kms\,Mpc$^{-1}$, the depth of such a cylinder may be larger than its width.
However, galaxies usually have large peculiar velocities and thus \Dv/H$_o$ is
not a good measure of the differential distance of the galaxy along the line of
sight. Bowen et al.\ (2002) adopted 500~\kms, arguing that this is near the
maximum observed for groups. They also made the point that one should compare
the galaxy density to the total equivalent width (or column density) of the
absorbers in a velocity interval -- if there are multiple absorbers, taking each
absorber separately would underestimate the density of intergalactic gas. With
this method and criteria, Bowen et al.\ (2002) found that there was a very tight
correlation between the galaxy density and the total \Lya\ equivalent width for
the six main absorbing systems in their sample. C\^ot\'e et al.\ (2005) added
four more systems, and although their plot of log\,$W$ vs log\,$n$ showed larger
scatter, they thought that they confirmed the conclusion of Bowen et al.\
(2002).
\par In Fig.~\Fewvsgaldens\ we show our results, for several choices for the
maximum impact parameter and velocity difference between absorber and galaxies.
For each absorber, we first add the equivalent width of it and all other
absorbers within \Dv$<$500 or $<$1000~\kms. We then count the number of galaxies
with $L$$>$0.1~\Lstar\ in each cylinder and divide by its volume. In each of the
four panels we show the systems in Bowen et al.'s (2002) study as filled stars
and the systems of C\^ot\'e et al.\ (2005) as filled squares. We looked at this
scatter plot for cylinders with radius 0.2, 0.5, 1, 2 and 3~Mpc and velocity
range 200, 400, 500, 1000 and 3000~\kms, as well as separately for absorbers
with $v$$<$2500~\kms\ and $v$$>$2500~\kms\ and using different luminosity limits
for the galaxies. All of these choices lead to the same conclusions.
\par Figure~\Fewvsgaldens a presents the plot using the criteria used by Bowen
et al.\ (2002) (\ip$<$2~Mpc, \Dv$<$500~\kms). The filled stars then fall on a
line \fix{(correlation coefficient 0.88)}, and the filled squares scatter near
this line. However, it is clear that this must be an artifact of the particular
set of systems that Bowen et al.\ (2002) and C\^ot\'e et al.\ (2005) observed.
The scatter plot does suggest that there is a tendency for an increase of the
maximum total equivalent width with increasing galaxy density \fix{(correlation
coefficient 0.95)}, but there clearly is no strong general correlation \fix{(the
correlation coefficient for all data is 0.40).} If we look at a somewhat smaller
box (1~Mpc by 500~\kms, panel b), we come to the same conclusion. If we look
closer to the absorbers, using a 500~kpc by 1000~\kms\ box, we find that there
may a rough correlation between the amount of absorbing intergalactic gas and
the galaxy density (panel c). But looking even closer (panel d), the properties
of the scatter plot revert to those of the larger boxes.
\par In Fig.~\Fewvsgaldens c a number of outliers are located at log\,$n$=
$-$0.2, log\,$W$=2.05. Two of these correspond to the detections at 1924 and
2183~\kms\ toward NGC\,985. As discussed in Sect.~\SSassoc\ and in the notes to
NGC\,985, and as can be seen in Fig.~\Fassoc, this is an absorber that is very
difficult to associate with a galaxy. The galaxies in the nearby group LGG\,71
range in velocity from 1145 to 1781~\kms, and the sightline passes 175~kpc from
the edge of the group. The derived galaxy density is thus sensitive to the
parameters of the cylinder. For \Dv=1000~\kms\ many group galaxies are included,
but for \Dv=500~\kms\ they are not, so that the point falls on what appeared to
be a linear relation between log\,$W$ and log\,$n$. The third outlier
corresponds to the line at 2110~\kms\ seen toward PG\,1211+143, which is normal
in every other aspect.
\par From Fig.~\Fewvsgaldens\ we conclude that {\it there is no correlation
between the total \Lya\ equivalent and the galaxy density near the absorber,
although the maximum \Lya\ equivalent width may depend on the galaxy density. At
all galaxy densities there is a wide range in the observed \Lya\ equivalent
widths.}

\subsection{The Fraction of Galaxies of Given $L$ Having a \Lya\ Absorber Within \ip, \Dv}
\par In Table~\Tgalaxycount\ we look at the fraction of galaxies that have an
associated \Lya\ absorber stronger than either 50 or 300~\mA\ within a given
impact parameter and velocity difference. This table is the complement to
Table~\Tassoccount. To construct Table~\Tgalaxycount\ we first find for each
galaxy with \vgal$<$2500~\kms\ the impact parameter to an AGN sightline, and we
check the 3$\sigma$ \Lya\ equivalent width error at the velocity of the galaxy.
Then we note whether or not there is an intergalactic line within a given impact
parameter and velocity difference. We use a velocity limit of 2500~\kms, rather
than 5000~\kms\ because then our galaxy sample is basically complete down to
0.1\,\Lstar, and then we can compare the fractions across different
luminosities.
\par Each entry in the table consists of three parts. The middle part is the
number of galaxies with equivalent width limit $<$50~\mA\ (first group of 15
rows) or $<$300~\mA\ (second group of 15 rows), impact parameter less than the
number in Col.~1, and luminosity larger than the limits given in Cols. 3--6. The
first part of the entry is the number of these galaxies for which there is a
\Lya\ or \Lyb\ absorption line with $W$$>$50 or $>$300~\mA\ whose velocity
differs from the galaxy's systemic velocity by less than the number in Col.~2.
The third part of each entry is the ratio of these two numbers, converted to a
percentage. So, for example, there are 135 galaxies with \vgal$<$2500~\kms\ and
$L$$>$0.1\,\Lstar\ within 1~Mpc of an AGN sightline, in whose spectrum the
3$\sigma$ equivalent width limit is $<$50~\mA\ at the velocity of the galaxy.
For 68 (50\%) of these, the AGN spectrum shows a $>$50~\mA\ \Lya\ or \Lyb\ line
with velocity within 400~\kms\ of the systemic velocity of the galaxies.
\par We summarize the conclusions that can be drawn from this table below. We
also looked at the fractions separately for field and group galaxies and even
for the four sightlines going through or near the Virgo cluster (3C\,273.0,
HE\,1228+0131, PG\,1211+143 and PG\,1216+069). The fractions are basically the
same, however, typically differing by less than 10\%.
\par From Table~\Tgalaxycount\ we can draw the following conclusions: {\it (1)
At low impact parameter ($<$400~kpc), it is possible to find a $>$50~\mA\ \Lya\
pabsorber within 1000~\kms\ for all galaxies, and within 400~\kms\ for the
majority ($\sim$80\%) of them. (2) Detecting a $>$300~\mA\ \Lya\ absorber within
400~\kms\ is possible for $\sim$40--50\% of galaxies with \ip$<$400~kpc and for
$\sim$45\% of galaxies with \ip$<$200~kpc. (3) The fraction of galaxies with a
\Lya\ line within a given velocity difference does not depend on their
luminosity. (4) At higher impact parameters the fraction of galaxies with an
associated \Lya\ line decreases.}
\begin{deluxetable}{llrrrrr}
\tablenum{10}
\tablewidth{0pt}
\tabletypesize{\footnotesize}
\tablecolumns{6}
\tablecaption{Fraction of galaxies with $v$$<$2500~\kms\ and a \Lya\ absorber within \ip\ and \Dv$^1$}
\tablehead{%
\ch{\ip}   &\ch{\Dv}    &\ch{$L$$>$0.1\,\Lstar\phantom{0}} &\ch{$L$$>$0.25\,\Lstar} &\ch{$L$$>$0.5\,\Lstar\phantom{0}} &\ch{$L$$>$\Lstar\phantom{1.00}} \\%
\ch{[kpc]} &\ch{[\kms]} &                                  &                                  &                        &                                \\%
\ch{(1)}&\ch{(2)}&\ch{(3)}&\ch{(4)}&\ch{(5)}&\ch{(6)}
}\startdata
$W$$>$50 mA\\
\hline
$<$200 & $<$200 & 4 of 9; 44\% & 4 of 8; 50\% & 4 of 8; 50\% & 1 of 3; 33\% \\
$<$200 & $<$400 & 7 of 9; 77\% & 6 of 8; 75\% & 6 of 8; 75\% & 2 of 3; 66\% \\
$<$200 & $<$1000 & 9 of 9; 100\% & 8 of 8; 100\% & 8 of 8; 100\% & 3 of 3; 100\% \\
\hline
$<$400 & $<$200 & 9 of 22; 40\% & 9 of 16; 56\% & 6 of 11; 54\% & 2 of 5; 40\% \\
$<$400 & $<$400 & 17 of 22; 77\% & 13 of 16; 81\% & 9 of 11; 81\% & 4 of 5; 80\% \\
$<$400 & $<$1000 & 22 of 22; 100\% & 16 of 16; 100\% & 11 of 11; 100\% & 5 of 5; 100\% \\
\hline
$<$1000 & $<$200 & 34 of 135; 25\% & 27 of 95; 28\% & 20 of 62; 32\% & 6 of 30; 20\% \\
$<$1000 & $<$400 & 68 of 135; 50\% & 47 of 95; 49\% & 33 of 62; 53\% & 14 of 30; 46\% \\
$<$1000 & $<$1000 & 101 of 135; 74\% & 72 of 95; 75\% & 48 of 62; 77\% & 24 of 30; 80\% \\
\hline
$<$2000 & $<$200 & 86 of 451; 19\% & 70 of 312; 22\% & 47 of 211; 22\% & 22 of 105; 20\% \\
$<$2000 & $<$400 & 200 of 451; 44\% & 150 of 312; 48\% & 99 of 211; 46\% & 46 of 105; 43\% \\
$<$2000 & $<$1000 & 305 of 451; 67\% & 222 of 312; 71\% & 147 of 211; 69\% & 73 of 105; 69\% \\
\hline
$<$3000 & $<$200 & 152 of 966; 15\% & 112 of 669; 16\% & 79 of 457; 17\% & 35 of 230; 15\% \\
$<$3000 & $<$400 & 360 of 966; 37\% & 262 of 669; 39\% & 177 of 457; 38\% & 80 of 230; 34\% \\
$<$3000 & $<$1000 & 550 of 966; 56\% & 395 of 669; 59\% & 276 of 457; 60\% & 134 of 230; 58\% \\
\hline
\hline\\
$W$$>$300 mA\\
\hline
$<$200 & $<$200 & 5 of 14; 35\% & 5 of 13; 38\% & 4 of 10; 40\% & 2 of 4; 50\% \\
$<$200 & $<$400 & 6 of 14; 42\% & 6 of 13; 46\% & 5 of 10; 50\% & 2 of 4; 50\% \\
$<$200 & $<$1000 & 7 of 14; 50\% & 7 of 13; 53\% & 6 of 10; 60\% & 2 of 4; 50\% \\
\hline
$<$400 & $<$200 & 8 of 33; 24\% & 7 of 25; 28\% & 6 of 17; 35\% & 3 of 9; 33\% \\
$<$400 & $<$400 & 12 of 33; 36\% & 11 of 25; 44\% & 9 of 17; 52\% & 5 of 9; 55\% \\
$<$400 & $<$1000 & 15 of 33; 45\% & 14 of 25; 56\% & 11 of 17; 64\% & 6 of 9; 66\% \\
\hline
$<$1000 & $<$200 & 21 of 174; 12\% & 18 of 121; 14\% & 15 of 80; 18\% & 7 of 43; 16\% \\
$<$1000 & $<$400 & 39 of 174; 22\% & 30 of 121; 24\% & 26 of 80; 32\% & 12 of 43; 27\% \\
$<$1000 & $<$1000 & 77 of 174; 44\% & 56 of 121; 46\% & 42 of 80; 52\% & 22 of 43; 51\% \\
\hline
$<$2000 & $<$200 & 48 of 569; 8\% & 41 of 397; 10\% & 31 of 269; 11\% & 13 of 138; 9\% \\
$<$2000 & $<$400 & 115 of 569; 20\% & 88 of 397; 22\% & 65 of 269; 24\% & 29 of 138; 21\% \\
$<$2000 & $<$1000 & 218 of 569; 38\% & 165 of 397; 41\% & 114 of 269; 42\% & 60 of 138; 43\% \\
\hline
$<$3000 & $<$200 & 97 of 1216; 7\% & 74 of 855; 8\% & 55 of 585; 9\% & 24 of 309; 7\% \\
$<$3000 & $<$400 & 228 of 1216; 18\% & 173 of 855; 20\% & 125 of 585; 21\% & 61 of 309; 19\% \\
$<$3000 & $<$1000 & 403 of 1216; 33\% & 295 of 855; 34\% & 207 of 585; 35\% & 108 of 309; 34\% \\
\hline
\hline\\
\enddata
\tablecomments{%
1: This table gives the number and percentage of galaxies with $v$$<$2500~\kms\
brighter than a given luminosity limit for which it is possible to find a \Lya\
absorber within a given impact parameter and velocity difference and detected
line strength larger than 50~\mA\ (upper half of table) or 300~\mA\ (lower half
of table).
}
\end{deluxetable}

\subsection{\Lya, \Lyb\ and \OVI\ Detection Fraction as Function of Impact Parameter} 
\par In this section we study the fraction of \Lya\ absorbers as function of
impact parameter to the associated galaxy. We can do this because our galaxy
sample is more or less complete, so we can properly count non-detections. Some
previous studies also discussed detection fractions, but only for relatively
bright galaxies and relatively small impact parameters. We first compare the
numbers in these studies to our results, scaling the impact parameters listed in
the other studies to a Hubble constant of $H_o$=71~\kms\
(Sect.~\SSdetfraccompare). We then describe how we construct the detection
fraction histograms as function of impact parameter (Sect.~\SSdetfracconstruct),
including a correction for incompleteness in the \NED\ data, and justifying our
selection criteria. The results are shown in Tables~\TdetfracA\ and \TdetfracB,
as well as Figs.~\Fdethistall\ to \Fdethistgrpcntcen. We discuss them
systematically in Sects.~\SSdetfraclyafield\ (\Lya\ for field galaxies),
\SSdetfraclyagroup\ (\Lya\ for group galaxies), \SSdetfraclyb\ (\Lyb) and
\SSdetfracOVI\ (\OVI).

\subsubsection{Comparison with Previous Studies}
\par Lanzetta et al.\ (1995) reported a 100\% detection fraction for impact
parameters \ip$<$100\Hm~kpc and equivalent width limit \Wlim$>$150~\mA,
decreasing to 66\% for \ip$<$230\Hm~kpc and 11\% at \ip$>$230\Hm~kpc. However,
because of the nature of their survey (a single limiting apparent magnitude over
a large redshift range), it is not completely clear how to compare this to our
sample. If we use a limit of \Rgal$>$11~kpc ($L$$>$0.25\,\Lstar), we find
percentages of 80\%, 50\% and 8\% for the three fractions of lines with
\Wlim$>$150~\mA, i.e.\ the same pattern, but about three quarters as many
detections.
\par Bowen et al.\ (1996) found a 44\% fraction for \ip$<$430\Hm~kpc,
$L$$>$0.5\Lstar\ and $W$$>$300~\mA. Using the same criteria we find 7 detections
for 31 galaxies, or 23\%, which is about half as many. In their subsequent
paper, Bowen et al.\ (2002) claimed a 100\% detection fraction for
\ip$<$285\Hm~kpc, $L$$>$0.5\Lstar\ and $W$$>$45~\mA. With just their sightlines,
we find 4 of 5 (80\%) galaxies are detected, while the complete sample gives
67\% (12 of 18). Clearly, with a relatively low equivalent width limit, most
luminous galaxies with low impact parameter are found to have associated \Lya\
absorption.
\par For the two sightlines toward H\,1821+643 and PG\,1116+215, Tripp et al.\
(1998) found that there was an \Lya\ line within 1000~\kms\ of all 42 galaxies
with \ip$<$600\Hm~kpc in their sample (note that often more than one galaxy is
associated with a particular absorber). The luminosity cutoff of the galaxy
sample of Tripp et al.\ (1998) varied with redshift -- at $z$$<$0.10 (where a
substantial fraction of their absorbers occurs), their limit $B$$<$19
corresponds to about an \Lstar\ galaxy. We find an absorber for 90\% (10 of 11)
of $L$$>$\Lstar\ galaxies with \ip$<$600~kpc and \Dv$<$1000~\kms.
\par In the Impey et al.\ (1999) study, eleven absorbers found in \GHRS\ G140L
spectra (0.8~\AA, or $\sim$150--190~\kms\ resolution) were compared against
galaxies in the Virgo cluster region, with the sample complete down to
$M_B$=$-$16 (0.04\,\Lstar). The 3$\sigma$ detection limit varies between about
30 and 120~\mA. They claimed a detection fraction of 60\% for galaxies with
impact parameter $<$385\Hm~kpc and $L$$>$0.25\Lstar, and 20\% for
\ip$<$700\Hm~kpc and $L$$>$\Lstar. Using the same impact parameter and
luminosity criteria and choosing \Wlim$>$60~\mA, we find fractions of 56\% (22
detections for 39 galaxies) and 41\% (14 detections for 34 galaxies),
respectively.
\par Finally, Chen et al.\ (2001) looked at 47 galaxies with $z$=0.07--0.89 with
\ip$<$180\Hm~kpc, and find associated detections for 34 of these (61\%). For
their detection limit ($\sim$300~\mA) we find 5 detections for 14 galaxies with
\ip$<$180\Hm~kpc, or 36\%, i.e.\ a detection fraction that is about half as
high.

\subsubsection{Constructing Detection Fraction Histograms}
\begin{figure}\plotfiddle{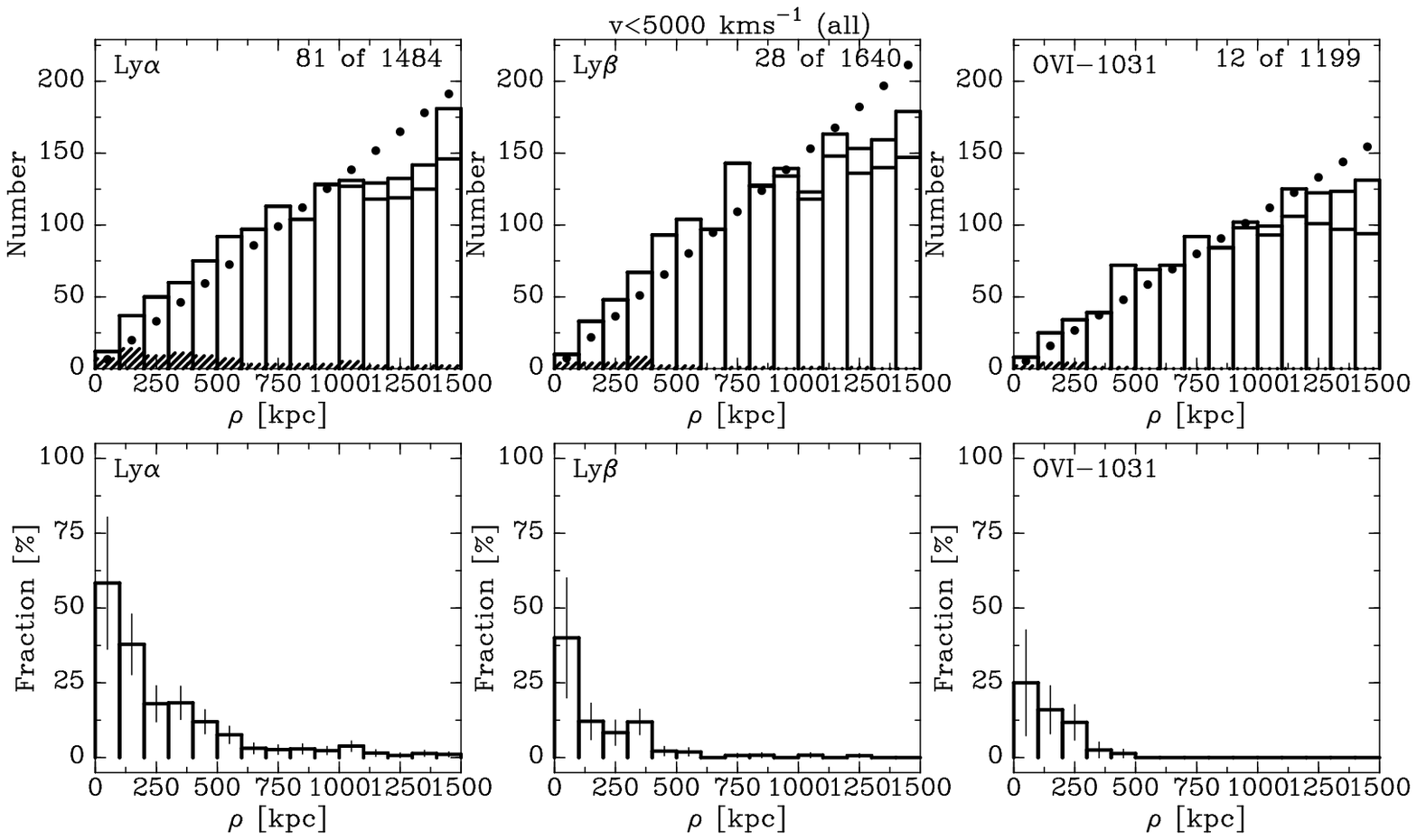}{0in}{0}{450}{300}{0}{-150}\figurenum{13}\caption{\captiondethistall}\end{figure}
\begin{figure}\plotfiddle{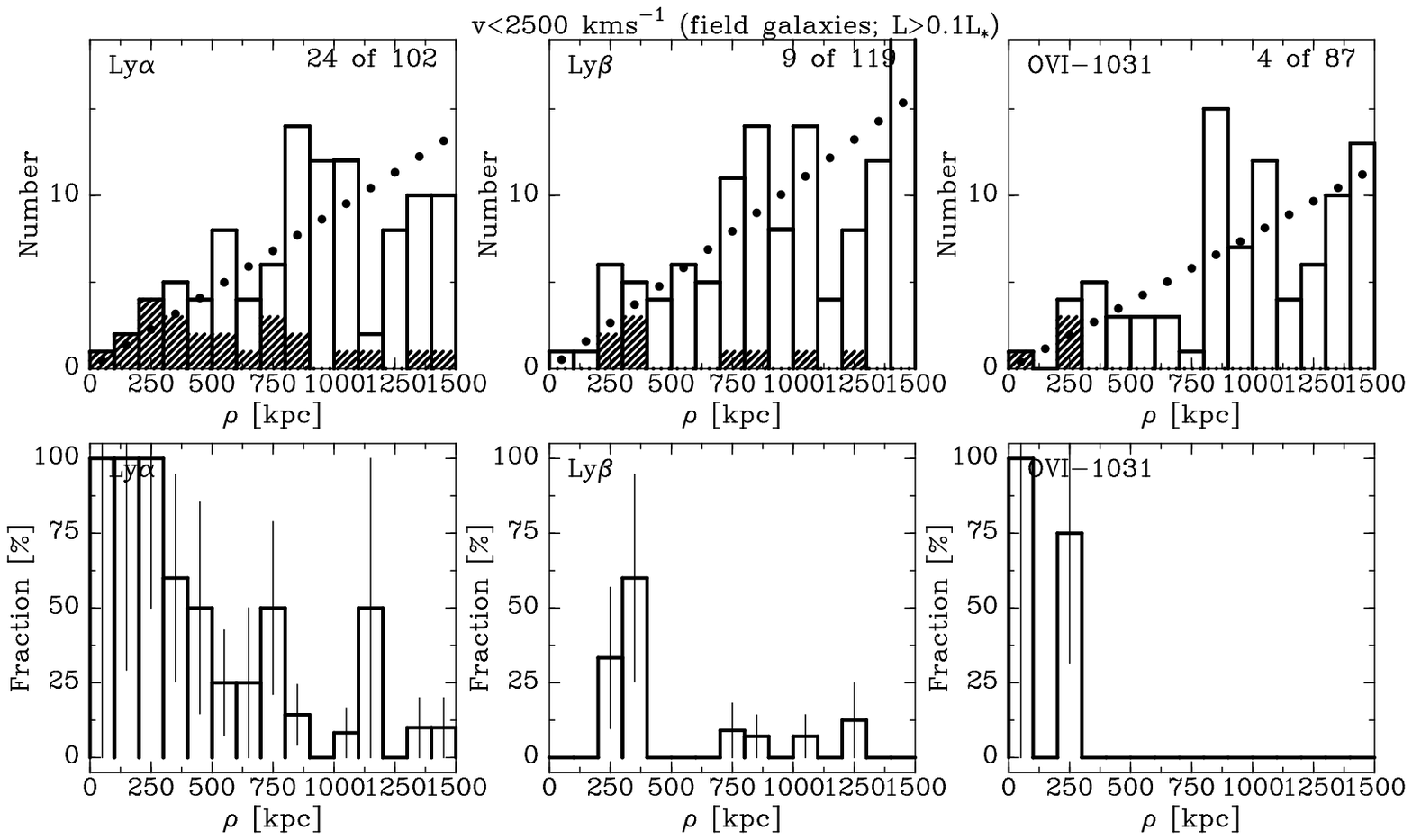}{0in}{0}{450}{300}{0}{-150}\figurenum{14}\caption{\captiondethistfield}\end{figure}
\begin{figure}\plotfiddle{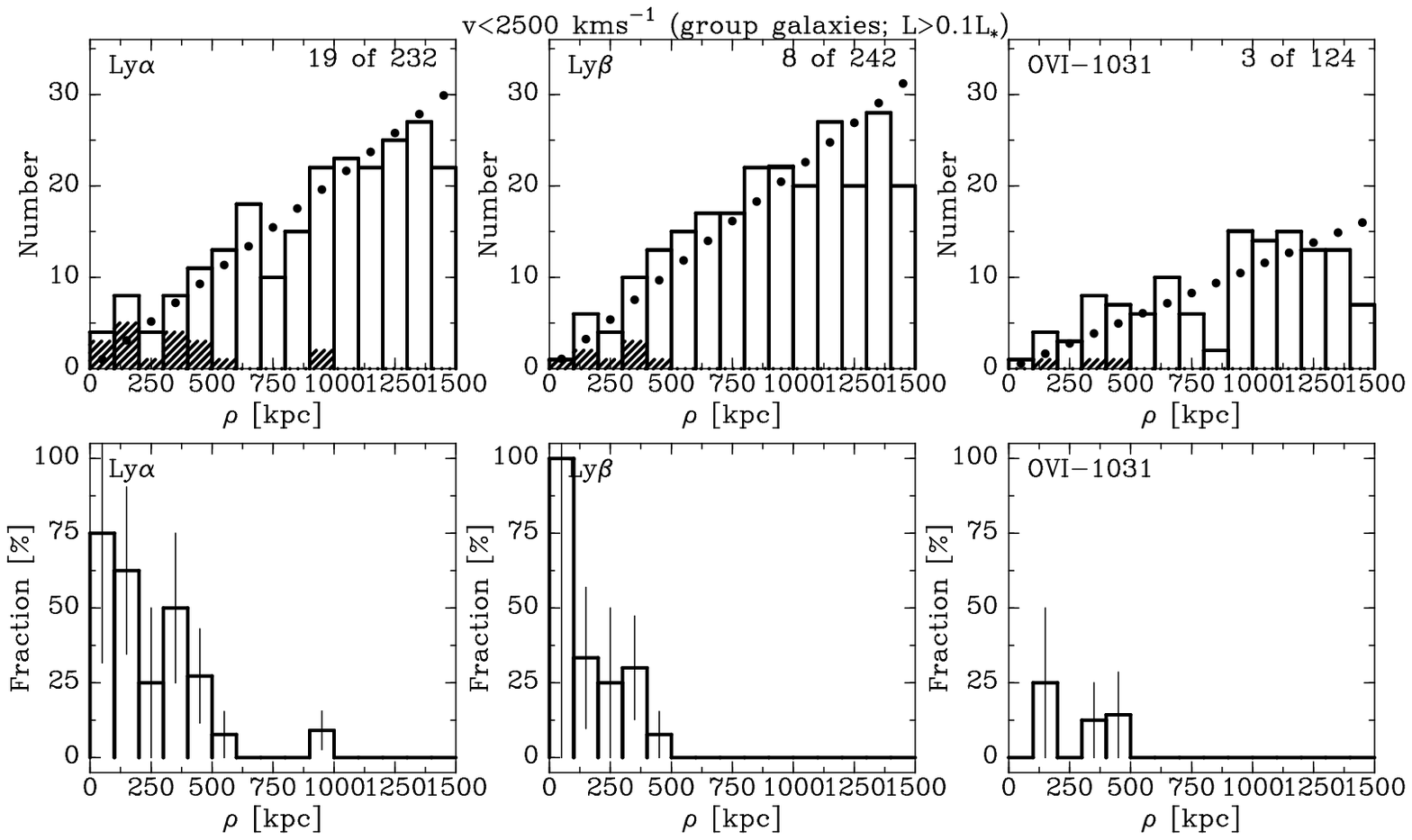}{0in}{0}{450}{300}{0}{-150}\figurenum{15}\caption{\captiondethistgroup}\end{figure}
\begin{figure}\plotfiddle{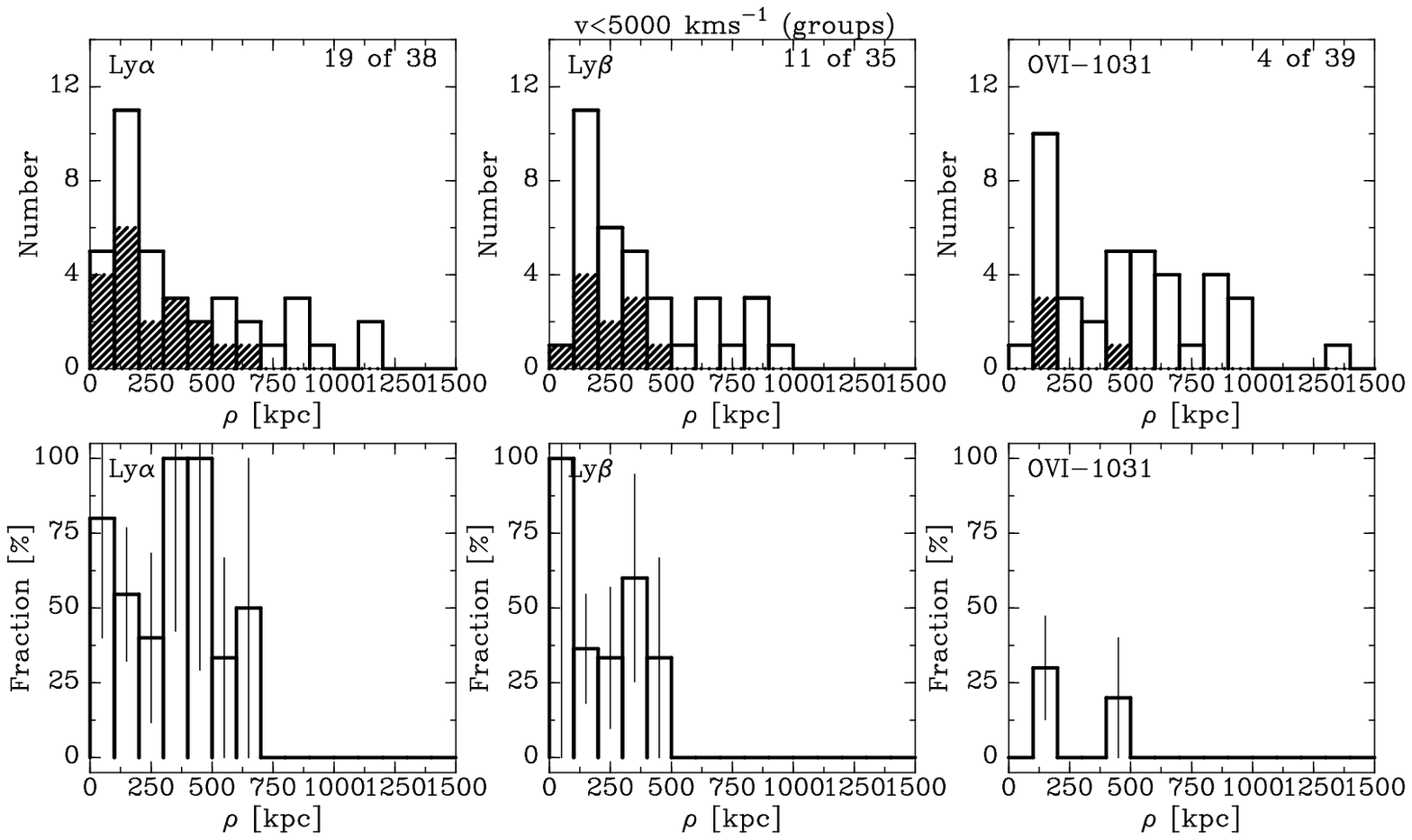}{0in}{0}{450}{300}{0}{-150}\figurenum{16}\caption{\captiondethistgrpcnt}\end{figure}
\begin{figure}\plotfiddle{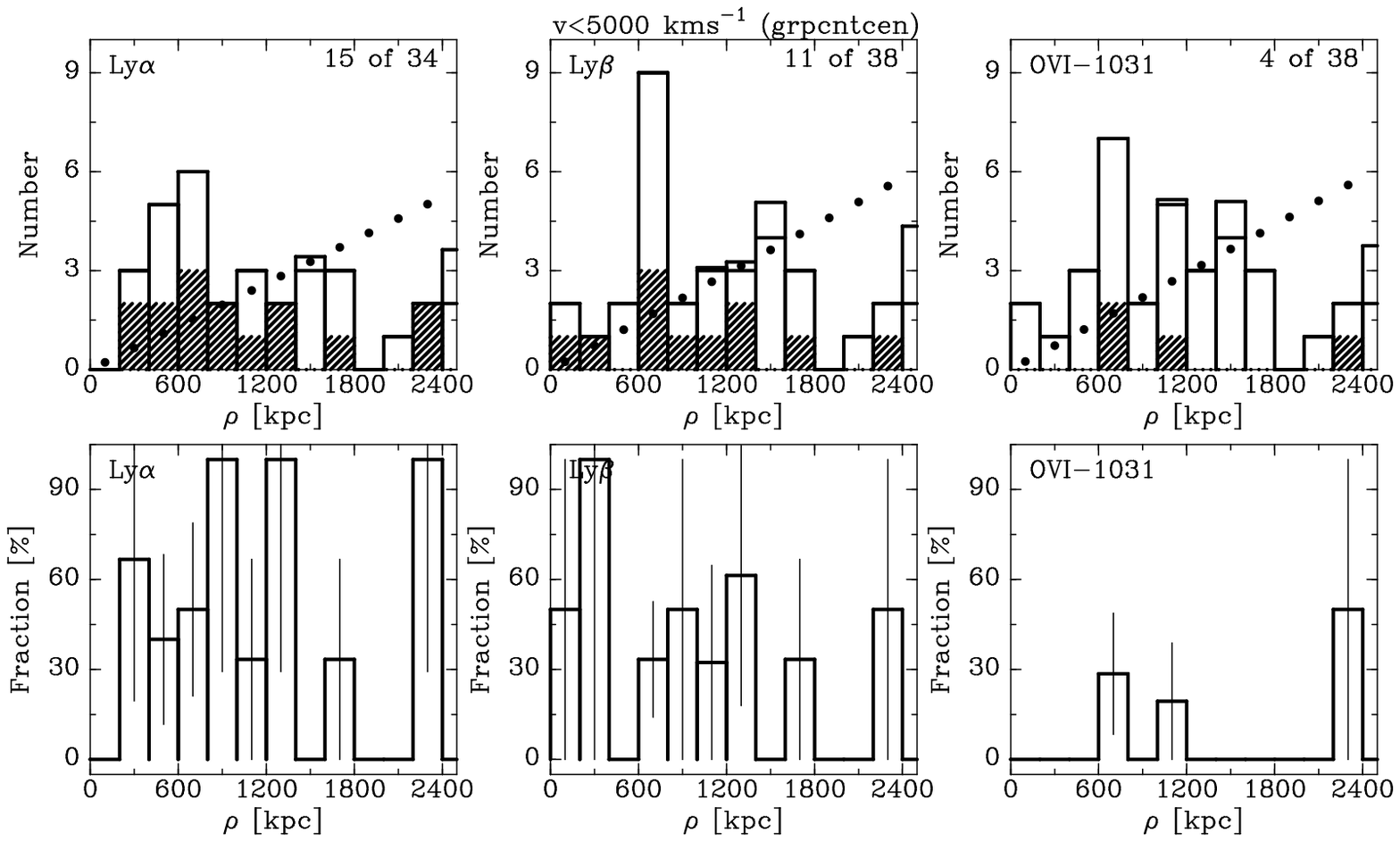}{0in}{0}{450}{300}{0}{-150}\figurenum{17}\caption{\captiondethistgrpcntcen}\end{figure}
\def\rule{------------------------}
\begin{deluxetable}{crrrrrrrrrrrr}
\tablenum{11}
\tablewidth{0pt}
\tabletypesize{\footnotesize}
\tabcolsep=2pt
\tablecolumns{13}
\tablecaption{Detection fraction vs.\ impact parameter$^1$}
\tablehead{%
\ch{\ip}   &\multicolumn{3}{c}{all(\Lya)}        &\multicolumn{3}{c}{field, $v$$<$2500~\kms}    &\multicolumn{3}{c}{groups (\Lya)}      &\multicolumn{3}{c}{all (\OVI)}         \\
           &              &          &           &\multicolumn{3}{c}{(\Lya) $L$$>$0.1\,\Lstar}  &                                       &                                       \\
           &\multicolumn{3}{c}{\rule}            & \multicolumn{3}{c}{\rule}                    &\multicolumn{3}{c}{\rule}              &\multicolumn{3}{c}{\rule}              \\
\ch{[kpc]} &\ch{\ \ \#gal}&\ch{\#det}&\ch{frac.} &\ch{\ \ \ \#gal}&\ch{\#det}&\ch{frac.}        &\ch{\ \ \ \#gal}&\ch{\#det}&\ch{frac.} &\ch{\ \ \ \#gal}&\ch{\#det}&\ch{frac.} \\
\ch{(1)}&\ch{(2)}&\ch{(3)}&\ch{(4)}&\ch{(5)}&\ch{(6)}&\ch{(7)}&\ch{(8)}&\ch{(9)}&\ch{(10)}&\ch{(11)}&\ch{(12)}&\ch{(13)}
}\startdata
0$-$100 &  12 &   7 & 58\%  &   1 &   1 & 100\%  &   5 &   4 & 80\%  &   8 &   2 & 25\%\\
100$-$200 &  37 &  14 & 38\%  &   2 &   2 & 100\%  &  11 &   6 & 55\%  &  25 &   4 & 16\%\\
200$-$300 &  50 &   9 & 18\%  &   4 &   4 & 100\%  &   5 &   2 & 40\%  &  34 &   4 & 12\%\\
300$-$400 &  60 &  11 & 18\%  &   5 &   3 & 60\%  &   3 &   3 & 100\%  &  39 &   1 & 3\%\\
400$-$500 &  75 &   9 & 12\%  &   4 &   2 & 50\%  &   2 &   2 & 100\%  &  72 &   1 & 1\%\\
500$-$600 &  92 &   7 & 8\%  &   8 &   2 & 25\%  &   3 &   1 & 33\%  &  69 &   0 & 0\%\\
600$-$700 &  97 &   3 & 3\%  &   4 &   1 & 25\%  &   2 &   1 & 50\%  &  72 &   0 & 0\%\\
700$-$800 & 113 &   3 & 3\%  &   6 &   3 & 50\%  &   1 &   0 & 0\%  &  92 &   0 & 0\%\\
800$-$900 & 104 &   3 & 3\%  &  14 &   2 & 14\%  &   3 &   0 & 0\%  &  84 &   0 & 0\%\\
900$-$1000 & 128 &   3 & 2\%  &  12 &   0 & 0\%  &   1 &   0 & 0\%  &  98 &   0 & 0\%\\
1000$-$1100 & 131 &   5 & 4\%  &  12 &   1 & 8\%  &   0 &   0 & 0\%  &  93 &   0 & 0\%\\
1100$-$1200 & 129 &   2 & 2\%  &   2 &   1 & 50\%  &   2 &   0 & 0\%  & 106 &   0 & 0\%\\
1200$-$1300 & 132 &   1 & 1\%  &   8 &   0 & 0\%  &   0 &   0 & 0\%  & 101 &   0 & 0\%\\
1300$-$1400 & 141 &   2 & 1\%  &  10 &   1 & 10\%  &   0 &   0 & 0\%  &  97 &   0 & 0\%\\
1400$-$1500 & 180 &   2 & 1\%  &  10 &   1 & 10\%  &   0 &   0 & 0\%  &  94 &   0 & 0\%\\
1500$-$1600 & 166 &   2 & 1\%  &   9 &   0 & 0\%  &   0 &   0 & 0\%  &  92 &   0 & 0\%\\
1600$-$1700 & 200 &   0 & 0\%  &  12 &   0 & 0\%  &   0 &   0 & 0\%  &  99 &   0 & 0\%\\
1700$-$1800 & 180 &   1 & 1\%  &  26 &   0 & 0\%  &   0 &   0 & 0\%  &  95 &   0 & 0\%\\
1800$-$1900 & 185 &   1 & 1\%  &  17 &   0 & 0\%  &   0 &   0 & 0\%  &  93 &   0 & 0\%\\
1900$-$2000 & 202 &   2 & 1\%  &  12 &   0 & 0\%  &   0 &   0 & 0\%  & 126 &   0 & 0\%\\
\enddata
\tablecomments{%
1: For each impact parameter interval given in Col.\ (1), the table gives  the
number of galaxies, the number of galaxies associated with an absorber and the
fraction of absorbers, using four different selection criteria. For the first
group (all(\Lya)) all galaxies and \Lya\ lines with $v$$<$5000~\kms\ are used.
For the second group (field,$v$$<$2500~\kms), only field galaxies and absorbers
with $v$$<$2500~\kms\ are counted; in this context a ``field'' galaxy is one
that was not listed as a group member by Geller \& Huchra (1983) or Garcia
(1993). The third group (groups(\Lya)) counts \Lya\ lines and galaxy groups,
using the group galaxy with the smallest impact parameter. For the fourth group
(all(\OVI)), we count all galaxies and \OVI\ absorbers with $v$$<$5000~\kms. So,
for instance, there are 5 field galaxies with $v$$<$2500~\kms\ and impact
parameter 300--400~kpc, and 3 \Lya\ lines were associated with these galaxies.
The numbers in this table are represented graphically in Figs.~\Fdethistall\ to
\Fdethistgrpcnt.
}
\end{deluxetable}
\begin{deluxetable}{llrrcrrcrrcrrc}
\tablenum{12}
\tablewidth{0pt}
\tabletypesize{\footnotesize}
\tabcolsep=3pt
\tablecolumns{14}
\tablecaption{Detection fraction summary}
\tablehead{%
\ch{Lum. limit} &\ch{\ip}   & \multicolumn{3}{c}{field,\Lya} & \multicolumn{3}{c}{group,\Lya} & \multicolumn{3}{c}{field,\OVI} & \multicolumn{3}{c}{group,\OVI} \\
                &\ch{[kpc]} & \#gal & \#det & frac.          & \#gal & \#det & frac.          & \#gal & \#det & frac.          & \#gal & \#det & frac.          \\
\ch{(1)}&\ch{(2)}&\ch{(3)}&\ch{(4)}&\ch{(5)}&\ch{(6)}&\ch{(7)}&\ch{(8)}&\ch{(9)}&\ch{(10)}&\ch{(11)}&\ch{(12)}&\ch{(13)}&\ch{(14)}
}\startdata
L$>$0.1\,\Lstar$^1$ 
& 0-350 
&   7 &   7 & 100$^{+0}_{-37}$\%  &  21 &  13 & 61$^{+17}_{-17}$\%  &   6 &   4 & 66$^{+33}_{-33}$\%  &  13 &   1 &  7$^{+7}_{-7}$\%  \\
& 350-700 
&  21 &   7 & 33$^{+12}_{-12}$\%  &  45 &   4 &  8$^{+4}_{-4}$\%  &  14 &   1 &  7$^{+7}_{-7}$\%  &  26 &   2 &  7$^{+5}_{-5}$\%  \\
& 700-1500 
&  74 &   9 & 12$^{+4}_{-4}$\%  & 165 &   1 &  0$^{+0}_{-0}$\%  &  68 &   0 &  0$^{+0}_{-0}$\%  &  85 &   0 &  0$^{+0}_{-0}$\%  \\
& 1500-3000 
& 297 &   0 &  0$^{+0}_{-0}$\%  & 566 &   0 &  0$^{+0}_{-0}$\%  & 221 &   0 &  0$^{+0}_{-0}$\%  & 339 &   0 &  0$^{+0}_{-0}$\%  \\
L$>$0.25\,\Lstar$^2$ 
& 0-350 
&  10 &   9 & 90$^{+10}_{-30}$\%  &  25 &  14 & 56$^{+14}_{-14}$\%  &   9 &   7 & 77$^{+22}_{-29}$\%  &  13 &   2 & 15$^{+10}_{-10}$\%  \\
& 350-700 
&  22 &   8 & 36$^{+12}_{-12}$\%  &  40 &   3 &  7$^{+4}_{-4}$\%  &  14 &   0 &  0$^{+0}_{-0}$\%  &  33 &   3 &  9$^{+5}_{-5}$\%  \\
& 700-1500 
&  58 &   6 & 10$^{+4}_{-4}$\%  & 141 &   1 &  0$^{+0}_{-0}$\%  &  64 &   0 &  0$^{+0}_{-0}$\%  &  77 &   0 &  0$^{+0}_{-0}$\%  \\
& 1500-3000 
& 259 &   2 &  0$^{+0}_{-0}$\%  & 463 &   2 &  0$^{+0}_{-0}$\%  & 205 &   0 &  0$^{+0}_{-0}$\%  & 280 &   0 &  0$^{+0}_{-0}$\%  \\
L$>$0.5\,\Lstar$^3$ 
& 0-350 
&   6 &   5 & 83$^{+16}_{-37}$\%  &  18 &  11 & 61$^{+18}_{-18}$\%  &   7 &   5 & 71$^{+28}_{-31}$\%  &   6 &   2 & 33$^{+23}_{-23}$\%  \\
& 350-700 
&  15 &   6 & 40$^{+16}_{-16}$\%  &  29 &   5 & 17$^{+7}_{-7}$\%  &  15 &   0 &  0$^{+0}_{-0}$\%  &  26 &   3 & 11$^{+6}_{-6}$\%  \\
& 700-1500 
&  56 &   9 & 16$^{+5}_{-5}$\%  & 100 &   1 &  1$^{+1}_{-1}$\%  &  53 &   0 &  0$^{+0}_{-0}$\%  &  62 &   0 &  0$^{+0}_{-0}$\%  \\
& 1500-3000 
& 228 &   5 &  2$^{+0}_{-0}$\%  & 367 &   0 &  0$^{+0}_{-0}$\%  & 183 &   0 &  0$^{+0}_{-0}$\%  & 232 &   0 &  0$^{+0}_{-0}$\%  \\
L$>$\Lstar$^3$ 
& 0-350 
&   4 &   3 & 75$^{+25}_{-43}$\%  &   9 &   6 & 66$^{+27}_{-27}$\%  &   6 &   5 & 83$^{+16}_{-37}$\%  &   4 &   1 & 25$^{+25}_{-25}$\%  \\
& 350-700 
&   7 &   2 & 28$^{+20}_{-20}$\%  &  13 &   3 & 23$^{+13}_{-13}$\%  &   8 &   0 &  0$^{+0}_{-0}$\%  &  19 &   3 & 15$^{+9}_{-9}$\%  \\
& 700-1500 
&  35 &   6 & 17$^{+6}_{-6}$\%  &  55 &   0 &  0$^{+0}_{-0}$\%  &  31 &   0 &  0$^{+0}_{-0}$\%  &  36 &   0 &  0$^{+0}_{-0}$\%  \\
& 1500-3000 
& 111 &   4 &  3$^{+1}_{-1}$\%  & 232 &   0 &  0$^{+0}_{-0}$\%  &  92 &   0 &  0$^{+0}_{-0}$\%  & 156 &   0 &  0$^{+0}_{-0}$\%  \\
\enddata
\tablecomments{1: also \vgal$<$2500~\kms; 2: also \vgal$<$3700~\kms; 3: also \vgal$<$5000~\kms.}
\end{deluxetable}
\par Figures \Fdethistall\ through \Fdethistgrpcntcen\ show the number of
galaxies, number of detections and detection fraction as function of impact
parameter, using different sets of criteria, separately for each of \Lya, \Lyb\
and \OVI. In Table~\TdetfracA\ we list the number of galaxies, number of
detections and the detection fraction for the same criteria. We now first
discuss the criteria used to construct these figures, then we discuss the
results. Unless noted otherwise, we only use sightlines for which the limiting
equivalent width is 100~\mA\ or better.
\par In each of these figures, the top three panels show the number of galaxies
in 100~kpc impact parameter bins, with the taller bins corrected for
incompleteness in the \NED\ sample (see below). Hatched regions in the top
panels give the number of detections. The dotted lines show the expected
distributions, calculated as the ratio of the area in the impact parameter bin
to the area in a 2~Mpc radius circle, scaled by the total number of galaxies
with \ip$<$2~Mpc. The bottom panels give the fraction of galaxies with an
associated detection.
\par The correction for \NED\ completeness is necessary because the standard
\NED\ search only allows one to find galaxies within 5\deg\ from a given
direction. This is not a problem for the RC3 part of the galaxy sample.
Therefore, at an impact parameter of $\rho_0$~Mpc, the \NED\ part of the sample
is complete only for galaxies with \vgal$>$813\,$\rho_0$~\kms. Since we only
look at galaxies with \vgal$>$400~\kms, the \NED\ sample {\it is} complete for
impact parameters $<$500~kpc. Therefore, we corrected for the incompleteness at
a given impact parameter $\rho_0$ by scaling the number of additional galaxies
from \NED\ by the ratio of the total number with \ip$<$500~kpc to the number
with \ip$<$500~kpc and \vgal$>$813\,$\rho_0$. Thus, there are two overlapping
histograms in the top panels of Fig.~\Fdethistall, with the bottom one giving
the actual number of galaxies in the sample, and the top histogram showing the
corrected number. The detection fraction is calculated using the corrected data.
\par There still is a small deficit at \ip$>$1500~\kms. This deficit is caused
by a deficit in \NED-only galaxies with \vgal=1000--1100~\kms, which leads to a
slightly lower scaling factor for impact parameters $>$1.3~Mpc. This deficit in
turn is caused by a combination of two factors. First, in the sightlines toward
the Virgo cluster (3C\,273.0, HE\,1228+0131, PG\,1211+143 and PG\,1216+069)
there is a relative deficit of galaxies near 1100~\kms. Second, there are three
sightlines toward galaxy groups with $v$$\sim$1700~\kms\ (MCG+10-16-111,
Mrk\,1383, NGC\,985) leading to a relative increase in the counts near that
velocity.
\par For Fig.~\Fdethistall, we counted every galaxy with systemic velocity
400--5000~\kms\ and impact parameter $<$2~Mpc to any of our 76 sightlines,
independent of brightness, size, completeness of the galaxy survey near the
sightline, or group membership. The first group of columns in Table~\TdetfracA\
gives the corresponding numbers. Clearly, there is a smooth decrease of the
detection fraction with impact parameter.
\par This galaxy sample is inhomogeneous, however. Galaxies in groups strongly
influence the result, since there usually are many galaxies that can be
associated with a single detected line, {\it and} it is likely that the physical
environment in groups differs from that around field galaxies.

\subsubsection{\Lya\ Detection Fraction for Field Galaxies}
\par We now discuss the distribution of the detection fraction of \Lya, \Lyb\
and \OVI\ as function of impact parameter, separately for field and group
galaxies. When combining both kinds, the detection fraction for \Lya\ at
impact parameters $<$400~kpc is 20\% (see Table~\TdetfracA), but this value is
much higher for luminous field galaxies. We find 26 field galaxies with
\ip$<$400~kpc, \vgal$<$5000~\kms\ and $L$$>$0.1\,\Lstar. For 15 of these we
detect an associated \Lya\ line, while for three we have no \Lya\ data, but we
find \Lyb. For two of the galaxies we would be unable to find either the \Lya\
or \Lyb\ line even if it were present, as it is hidden in the \Lya\ line
associated with another galaxy or the Milky Way. Five of the six remaining
non-detections of \Lya\ or \Lyb\ are in noisy spectra, so that the upper limits
are not very significant. The 18 luminous galaxies that are found to have an
associated \Lya\ or \Lyb\ line are:
IC\,4889 (62~kpc from ESO\,185-IG13, \Lyb\ only),
UGC\,8146 (80~kpc from PG\,1259+593),            
UGC\,7697 (139~kpc from Mrk\,771),               
NGC\,3942 (141~kpc from PG\,1149$-$110),         
UGC\,4238 (155~kpc from PG\,0804+761),           
NGC\,1412 (167~kpc from HE\,0340$-$2703),        
Mrk\,412 (196~kpc from 3C\,232),                 
NGC\,6140 (206~kpc from Mrk\,876, 2 lines),      
NGC\,2683 (250~kpc from PG\,0844+349, \Lyb\ only),
UGC\,8849 (274~kpc from PG\,1351+640)            
UGC\,10294 (282~kpc from Mrk\,876),              
NGC\,3104 (296~kpc from PG\,0953+414),           
ESO\,603-G27 (322~kpc from MRC\,2251$-$178),     
UGC\,7625 (339~kpc from HE\,1228+0131),          
CGCG\,291-61 (367~kpc from MCG+10-16-111),
UGC\,4621 (372~kpc from PG\,0844+349, \Lyb\ only; 2 lines),
MCG$-$2-34-6 (391~kpc from PG\,1302$-$102), and
NGC\,7817 (395~kpc from Mrk\,335).
All but three of the \Lya\ lines (the second component toward Mrk\,876, and the
lines toward PG\,0953+414 and PG\,1302$-$102) have equivalent width $>$100~\mA.
\par The only non-detections of \Lya\ for $L$$>$0.1\,\Lstar\ galaxies with
impact parameter $<$400~kpc are associated with NGC\,4939 (104~kpc from
PG\,1302$-$102), UGC\,7226 (362~kpc from Mrk\,205) and UGC\,5922 (391~kpc from
PG\,1049$-$005). However, two of these spectra are relatively noisy (detection
limits 72~\mA\ for NGC\,4939, 31~\mA\ for UGC\,7226, and 75~\mA\ for UGC\,5922).
In two cases the only limits are for \Lyb: UGC\,9452 (278~kpc from Mrk\,477,
$W$$<$39~\mA) and UGC\,5340 (296~kpc from PG\,1001+291, $W$$<$87~\mA). Thus, in
each case it is entirely possible that a better spectrum would reveal a line.
\par For the majority of the associations above, the next nearest field or group
galaxy with velocity within $\pm$500~\kms\ of the \Lya\ line is at significantly
larger impact parameter ($>$493~kpc), although three of the associated galaxies
above do have a dwarf near it with impact parameter similar to that of the main
galaxy. For three of the associations there are two \Lya\ lines, with two
associated galaxies at similar impact parameter, but the second galaxy is either
at \ip$>$400~kpc, is a group galaxy, or is (slightly) smaller than 7.5~kpc.
\par Fig.~\Fdethistfield\ shows the \Lya\ detection fraction for field galaxies
with \vgal$<$2500~\kms\ and $L$$>$0.1\,\Lstar\ (\Rgal$>$7.5~kpc). The second
group of three columns in Table~\TdetfracA\ gives the corresponding numerical
values. This does not include all of the galaxies mentioned above, because some
have \vgal$>$2500~\kms, where the sample is incomplete. There are 24 \Lya\
detections for 178 field galaxies with \ip$<$2~Mpc. The distribution for field
galaxies with \vgal$<$5000~\kms\ and $L$$>$0.5\,\Lstar\ (\Rgal$>$14.6~kpc) is
basically identical, with 24 \Lya\ detections for 137 field galaxies with
\ip$<$2~Mpc.
\par For both the complete sample and the field sample, the detection fraction
for \Lya\ decreases regularly with impact parameter. Table~\TdetfracA\ clearly
shows that in spite of the small number statistics, at all impact parameters the
detection fraction for \Lya\ is higher for the field galaxies than for the full
sample, by a factor two at \ip$<$100~kpc, a factor four at \ip$\sim$500~kpc and
a factor five at \ip$\sim$900~kpc. The effect of the small number of galaxies in
each impact parameter bin is clearly seen for the 1100--1200~kpc bin.  There are
in fact nine galaxies with \Rgal$>$7.5~kpc in this impact parameter bin, but by
accident many are toward sightlines without \Lya\ data, or at velocities where
there are line blends.
\par Table~\TdetfracB\ presents a summary of the detection fractions for \Lya\
and \OVI, separately for field and group galaxies, for three complete samples:
galaxies with $L$$>$0.1\,\Lstar\ (\vgal$<$2500~\kms), $L$$>$0.25\,\Lstar\
(\vgal$<$3700~\kms) and $L$$>$0.5\,\Lstar\ (\vgal$<$5000~\kms). In this table we
include an estimate of the uncertainty in the detection fraction, based on the
square root of the number of detections. This shows that for \Lya\ detections
associated with a complete sample of bright field galaxies the detection
fraction is 85--100\% for impact parameters $<$350~kpc. At larger impact
parameters, the detection fraction decreases to almost 0\% only for
\ip$>$1500~kpc.
\par From the histograms in Fig.~\Fdethistfield\ and the discussion above, we
conclude that {\it the fraction of $L$$>$0.1\,\Lstar\ field galaxies that have
an associated \Lya\ line is 100\%v for impact parameters \ip$<$350~kpc and
decreases monotonically to about 0 at \ip$\sim$1500~kpc.}

\subsubsection{\Lya\ Detection Fraction for Group Galaxies}
\par \fix{There are three ways in which we can compare the detection rate of
field galaxies to that of groups and group galaxies: a) we can count group
galaxies with $L$$>$0.1\,\Lstar, b) we can count groups, using the impact
parameter to the group galaxy nearest the sightlines, c) or we can count groups,
using the impact parameter to the center of the group.}
\par Counting galaxies, Fig.~\Fdethistgroup\ shows that the number of group
galaxies with \vgal$<$2500~\kms\ and $L$$>$0.1\,\Lstar\ (\Rgal$>$7.5~kpc) is
about twice that of the number of field galaxies (382 vs 178) while the number
of detections is similar (19 vs 24). For \vgal$<$5000~\kms\ and
$L$$>$0.5\,\Lstar\ (\Rgal$>$14.6~kpc) there are 18 detections for 246 galaxies,
with the same distribution. Fig.~\Fdethistgroup\ shows that the detection
fraction decreases regularly with impact parameter, just like for field
galaxies. However, compared to the field galaxies (for which the detection rate
was 100\% for \ip$<$350~kpc), the detection rate at \ip$<$350~kpc is 61\% (13 of
21) for $L$$>$0.1\,\Lstar\ and 61\% (11 of 18) for $L$$>$0.5\,\Lstar. These
detections include the ones toward 3C\,232 (NGC\,3067) and Mrk\,205 (NGC\,4319)
where the sightline passes through the disk of the galaxy.
\par \fix{Counting groups in the first way, we determine impact parameters from
the group galaxy with $L$$>$0.1\,\Lstar\ that has the smallest impact parameter
to the sightline. Fig.~\Fdethistgrpcnt\ shows the result for the 38 groups with
$v$$<$5000~\kms\ for which the sightline passes between the group galaxies or
close to the edge of the group (i.e.\ within half a group's radius). There are
another 19 cases where Table~\Tres\ lists a galaxy that is a group member, but
the sightline passes far off ($>$ one group radius) to the side of the group.
For each of the groups we then checked whether there is a detection associated
with a group galaxy or with the group as a whole, which is the case for 19
groups. Because of the  small number statistics we can't really say that there
is a regular decrease of detection fraction with impact parameter to the nearest
group galaxy. For impact parameters $<$350~kpc, 15 of 24 (63\%) of groups are
detected, a fraction that is comparable to the number of bright group galaxies
associated with an absorption line.}
\par \fix{Calculating impact parameters relative to the group center, we find
that it is $>$3~Mpc for 4 of the groups. For the rest, the detection fraction is
consistently about 50\% as all impact parameters (Fig.~\Fdethistgrpcntcen),
although the number of groups in each 200~kpc wide impact parameter interval is
small. Thus, on scales of hundreds of kpc, the gas in groups does not seem to be
more concentrated to the group centers.}
\par From the histograms in Fig.~\Fdethistgroup\ to \Fdethistgrpcntcen\ we
conclude that {\it the covering factor of \Lya\ around bright group galaxies
($L$$>$0.1\,\Lstar) is about 60\% for impact parameters $<$350~kpc} and that
{{\it about 50\% of galaxy groups have associated \Lya\ absorption.}

\subsubsection{\Lyb\ Galaxy Detection Fraction}
\par The histograms in Figs.~\Fdethistall, \Fdethistfield, \Fdethistgroup\ and
\Fdethistgrpcnt\ show that in spite of the small number of detections the galaxy
detection fraction for \Lyb\ has the same distribution as that of \Lya, as it
should, but with about one third of the number of detections. For the 25 \Lya\
lines seen toward $L$$>$0.1\,\Lstar\ galaxies with \ip$<$350~kpc, \Lyb\ is
detected in five cases, all with $W$(\Lya)$>$100~\mA, while a non-detection is
found in another five. Three more \Lya\ lines are sufficiently strong that a
corresponding \Lyb\ absorption is expected to be present, but either there is no
\Lyb\ data, or the \Lyb\ line is blended. Counting these, we find that there
should be 8 \Lyb\ lines accompanying the 25 \Lya\ lines. Thus, the detection
fraction of \Lyb\ is about one third that of \Lya.

\subsubsection{\OVI\ Galaxy Detection Fraction}
\par As can be seen in Fig.~\Fdethistall\ to \Fdethistgrpcnt, the analysis of
the \OVI\ detection fraction is hampered by the small number of positive
detections, so we can only estimate detection fractions with large statistical
uncertainties.
\par We find eight field galaxies for which we can associate an \OVI\ detection
with the galaxy (see Sect.~\SSOVIabs\ for a detailed discussion of each case).
IC\,4889 (62~kpc from ESO\,185-IG13, \Rgal=28.9~kpc),
UGC\,8146 (80~kpc from PG\,1259+593, \Rgal=12.6~kpc),
NGC\,4939 (104~kpc from PG\,1302$-$102, \Rgal=24.6~kpc),
NGC\,6140 (206~kpc from Mrk\,876, \Rgal=27.1~kpc),
NGC\,2683 (250~kpc from PG\,0844+349, \Rgal=23.0~kpc),
UGC\,10294 (282~kpc from Mrk\,876, \Rgal=27.6~kpc),
NGC\,3104 (296~kpc from PG\,0953+414, \Rgal=11.5~kpc), and
ESO\,603-G31 (422~kpc from MRC\,2251$-$178, \Rgal=9.1~kpc).
In addition there are five group galaxies with associated \OVI:
NGC\,247 (125~kpc from Ton\,S180, \Rgal=15.7~kpc),
UGC\,3804 (199~kpc from 1H\,0717+714, \Rgal=22.8~kpc),
NGC\,253 (374~kpc from Ton\,S210, \Rgal=20.7~kpc),
NGC\,5987 (424~kpc from Mrk\,290, \Rgal=51.7~kpc), and
NGC\,954 (562~kpc from HE\,0226$-$4110, \Rgal=33.0~kpc).
\par Of these 13 cases, the detection toward HE\,0226$-$4110 is not included in
the statistics discussed below, as it is at $v$$>$5000~\kms. Finally, the \OVI\
detection at 1008~\kms\ toward 3C\,273.0 is a special case in this regard. The
nearest galaxy (as listed in Table~\Tres) is MCG0-32-16 at 191~kpc, the nearest
$L$$>$0.1\,\Lstar\ galaxy is NGC\,4457 (\Rgal=13.8~kpc) at 469~kpc, while the
nearest $L$$>$0.5\,\Lstar\ galaxy is NGC\,4517 (\Rgal=53.8~kpc) at 662~kpc.
\par Of this set of \OVI\ lines, the ones toward ESO\,185-IG13, PG\,1259+593,
Mrk\,876 and 3C\,273.0 are unambiguously detected. The \OVI\ detections toward
PG\,0844+349, PG\,0953+414, Ton\,S180, Ton\,S210, 1H\,0717+714, Mrk\,290,
HE\,0226$-$4110 and MRC\,2251$-$178 are not unambiguous, in the sense that we
only see one of the two \OVI\ lines, with the other line either blended or too
weak, or the very existence of the feature might be questioned. However, in our
judgement, these features are real, they measure as $>$3$\sigma$ and
intergalactic \OVI\ is in all cases the most likely identification. The least
certain identification is that of the \OVI\ line seen toward PG\,1302$-$102.
There is no corroborating \Lya, \Lyb\ or \OVIb\ and the strength of the feature
is just a little over 3$\sigma$. However, visually the absorption is clear, and
it can be seen in each of the two channels of each of the two \FUSE\
observations of the target.

\par Looking at the \OVI\ detection rate for $L$$>$0.1\,\Lstar\ field galaxies
with \vgal$<$2500~\kms\ in Table~\TdetfracB, we find a 66\% detection rate for
\ip$<$350~\kms, with detections for UGC\,8146, NGC\,6140, NGC\,2683 and
NGC\,3104, as well as non-detections for UGC\,5340 (296~kpc from PG\,1001+291)
and UGC\,7625 (339~kpc from HE\,1228+0131). Note, however, that both
non-detections are for sightlines with relatively low S/N, and all four detected
lines are weaker than the detection limits toward these two sightlines. In any
case, for this sample, the detection rate is 100\% for \ip$<$296~kpc, with just
one (possible) \OVI\ line at \ip$>$296~kpc.

\par When looking at somewhat brighter field galaxies ($L$$>$0.25\,\Lstar,
\vgal$<$3700~\kms) with low impact parameter ($<$350~kpc), we count detections
toward IC\,4889, UGC\,8146, NGC\,4939, NGC\,6140, NGC\,2683, UGC\,10294 and
NGC\,3104, and non-detections for UGC\,7697 (139~kpc from Mrk\,771, and
ESO\,603-G27 (322~kpc from MRC\,2251$-$178), for a 77\% detection fraction.
However, the detection limit toward Mrk\,771 is 94~\mA, while all but one of the
five detections have equivalent width $<$40~\mA; therefore the fact that we do
not see \OVI\ is probably not very significant. Thus, we see \OVI\ in all but
one sightline with impact parameter $<$300~kpc.

\par Finally, for the brightest galaxy sample ($L$$>$\Lstar), we find a 71\%
detection fraction, counting detections associated with IC\,4889, NGC\,4939,
NGC\,6140, NGC\,2683 and UGC\,10294, and non-detections for UGC\,7697 and
ESO\,603-G27. Thus, where for bright field galaxies the detection fraction of
\Lya\ is 85--100\%, for \OVI\ it is about 70\%, although the small number of
sightlines means that there is an uncertainty of about 30\% in this number.
\par The situation is different for group galaxies, as can be seen in
Table~\TdetfracA\ and Fig.~\Fdethistgrpcnt. If we count the number of
intersected groups, we find that the 76 sightlines pass between the galaxies of
39 groups for which it is possible to find detections and non-detections of
\OVI. Only four groups (10\%) yield a detection: GH\,158 toward Mrk\,290, LGG\,4
toward Ton\,S180 and Ton\,S210, LGG\,141 toward 1H\,0717+714 and LGG\,292 toward
3C\,273.0. Using different complete samples with different luminosity limits, we
also find that the detection fraction for group galaxies is low, typically
somewhere between 7\% and 15\%. We need to mention here, however, that for the 53
group galaxies near our sightlines that have \Rgal$>$7.5~kpc ($L$$>$0.1\,\Lstar)
and \ip$<$350~kpc, we find only two detections, but there is no \OVI\ data
available in 15 cases, and in 20 cases the possible \OVI\ line is blended with
interstellar absorption. For the 16 galaxies where \OVI\ could have been (but
was not) detected, only one \Lya\ absorber is found, but for the 15 cases with
no \OVI\ data, there are seven \Lya\ lines. Thus, the \OVI\ detection rate for
group galaxies may be artifically depressed. Nevertheless, it does appear to be
the case that the \OVI\ detection fraction associated with group galaxies is
much lower than that associated with field galaxies, by a factor on the order of
five. It might be as much as a factor ten lower, but the statistics are too
uncertain to support this strongly. Another difference between the field and
group galaxy samples is that we find only one field galaxy that may have
associated \OVI\ at impact parameter $>$300~kpc, but two of the four group
galaxy associations are at \ip$>$300~kpc. Again, the small number of detections
means that this is not a firm conclusion.
\par Stocke et al.\ (2006) used a larger sample of \OVI\ lines (40) and thus
were able to look at the distribution of nearest-neighbor galaxies, similar to
the analysis done for \Lya\ in Sect.~\SSnearest. They found that the median
distance between \OVI\ absorbers and $L$$>$0.1\,\Lstar\ galaxies is 335~kpc, and
almost all \OVI\ absorbers are found within 400~kpc of such galaxies. This
conclusion is confirmed by our results for \OVI. In fact, for all of our \OVI\
detections with $v$$<$5000~\kms\ we can find an $L$$>$0.1\,\Lstar\ galaxy within
450~kpc and \Dv$<$120~\kms.
\par A final way of looking at the relation between \OVI\ absorbers and galaxies
is to find the nearest galaxy above a given luminosity for each absorber. We
find that there is an $L$$>$0.25\,\Lstar\ galaxy within 450~kpc and with
\Dva$<$300~\kms\ for each \OVI\ absorber at $v$$<$5000~\kms, with 9 of 13 having
a galaxy within 300~kpc. An $L$$>$\Lstar\ galaxy can be found within 200~kpc and
300~\kms\ for 4 of the 13 absorbers, within 300~kpc for 7 of the 13, and within
450~kpc for 9 of the 13, with just one case where the nearest such galaxy is at
\ip$>$1~Mpc. Clearly, the \OVI\ absorbers concentrate near luminous galaxies.
\par Two recent papers address the question of a correlation between \OVI\
absorbers and galaxies from the theoretical side. Ganguly et al.\ (2008) used
hydrodynamical simulations from Cen \& Fang (2006) to generate 10,000 synthetic
spectra through these datasets. Identifying \OVI\ absorbers in the spectra and
correlating with the simulated galaxies, they found that 80\% of the \OVI\
absorbers with $W$$>$30~\mA\ lie within 3~Mpc and 1000~\kms\ of an
$L$$>$0.1\,\Lstar\ galaxy, 20\% have impact parameter $<$1~Mpc to such a galaxy,
while just 5\% lie within 500~kpc. This is clearly incompatible with our
findings (and with those of Stocke et al.\ 2006), since we find that 100\% of
the \OVI\ absorbers lie within 120~\kms\ and 450~kpc of a 0.1\,\Lstar\ galaxy.
That is, the observational data show a tight correlation between \OVI\ absorbers
and galaxies, while the interpretation of the simulations would suggest that the
majority originates in the intergalactic medium far from galaxies.
\par Oppenheimer \& Dav\'e (2008) also looked at the relationship between \OVI\
absorption and galaxies. They derived that most \OVI\ is photoionized, and not
directly associated with galaxies, but, they note, ``OVI\ typically is nearest
to $\sim$0.1\,\Lstar\ galaxies''. They also stated that ``the majority of \OVI\
absorbers are between 100--300~kpc from their nearest galactic neighbor'',
although they do not show a plot of an absorber parameter vs.\ impact parameter.
Their results support our conclusion that a luminosity limit of 0.1~\Lstar\ is
the appropriate parameter to study the relationship between \OVI\ absorbers and
galaxies.
\par Finally, we note the evidence for \OVI\ near the Milky Way. Sembach et al.\
(2003) discovered that high-velocity Galactic \OVI\ absorption
($\vert$\vlsr$\vert$$<$400~\kms) is seen in 80\% of high-latitude sightlines
observed with \FUSE. Some of this is associated with high-velocity clouds that
are about 5--10~kpc above the Galactic disk (see Wakker et al.\ 2007, 2008), but
about half of the detections appears to originate much farther away, at
distances of 50--100~kpc (see e.g.\ Fox et al.\ 2005). Thus, in a random
sightline passing within 100~kpc of the Milky Way, there would be a probability
of about 25--50\% to detect an \OVI\ absorber. In our sample of extragalactic
targets, we find 11 cases with impact parameter $<$100~kpc, for 6 of which we
can search for \OVIa\ and/or \OVIb\ with detection limit better than 50~\mA\
(NGC\,4319, 6~kpc from Mrk\,205; NGC\,4291, 51~kpc from Mrk\,205; IC\,4489,
62~kpc from ESO\,185-IG13; [vCS96]\,000254.9+195654.3, 78~kpc from Mrk\,335;
UGC\,8146, 80~kpc from PG\,1259+5930). We find \OVI\ in two cases (ESO\,185-IG13
and PG\,1259+593), just about the expected number.
\par Summarizing the numbers above: we conclude: {\it (1) For impact parameters
$<$350~kpc, the detection rate for \OVI\ is 60--80\% for field galaxies,
10--30\% for group galaxies, and $\sim$10\% for galaxy groups.} {\it (2) Only
one field galaxy with \ip$>$300~kpc may show associated \OVI, but three of the
five \OVI\ lines associated with a bright group galaxy have \ip=300--450~kpc.}
{\it (3) 100\% of the \OVI\ detections at $v$$<$5000~\kms\ can be associated
with a galaxy with $L$$>$0.1\,\Lstar\ at \ip$<$450~kpc, which appears to be
incompatible with a simple interpretation of the results of hydrodynamical
simulations.}

\subsection{A Synthetic Map of the Gaseous Envelope of Galaxies}
\par Here we ask whether the intergalactic gas near galaxies knows about the
direction of rotation of the underlying galaxy. To answer this, it is necessary
to know which side of a galaxy is approaching (relative to the systemic
velocity). Then we can create a map with the plane of the galaxy rotated to be
horizontal and the approaching side on (e.g.) the left. To rotate the galaxies,
we use the position angle given in the RC3, if given. If the RC3 gives no
position angle, we visually align the galaxies, using the Digital Sky Survey
image that can be extracted from \NED. For 68 galaxies it is not possible to
determine a position angle because they are too small or too unstructured. The
literature contains data that allows us to determine the orientation (i.e.\
which is the approaching side) for 44 of the 329 galaxies with \ip$<$1~Mpc
listed in Table~\Tres, with the references given in Note~3. Detections are
associated with 17 of these galaxies.

\begin{figure}\plotfiddle{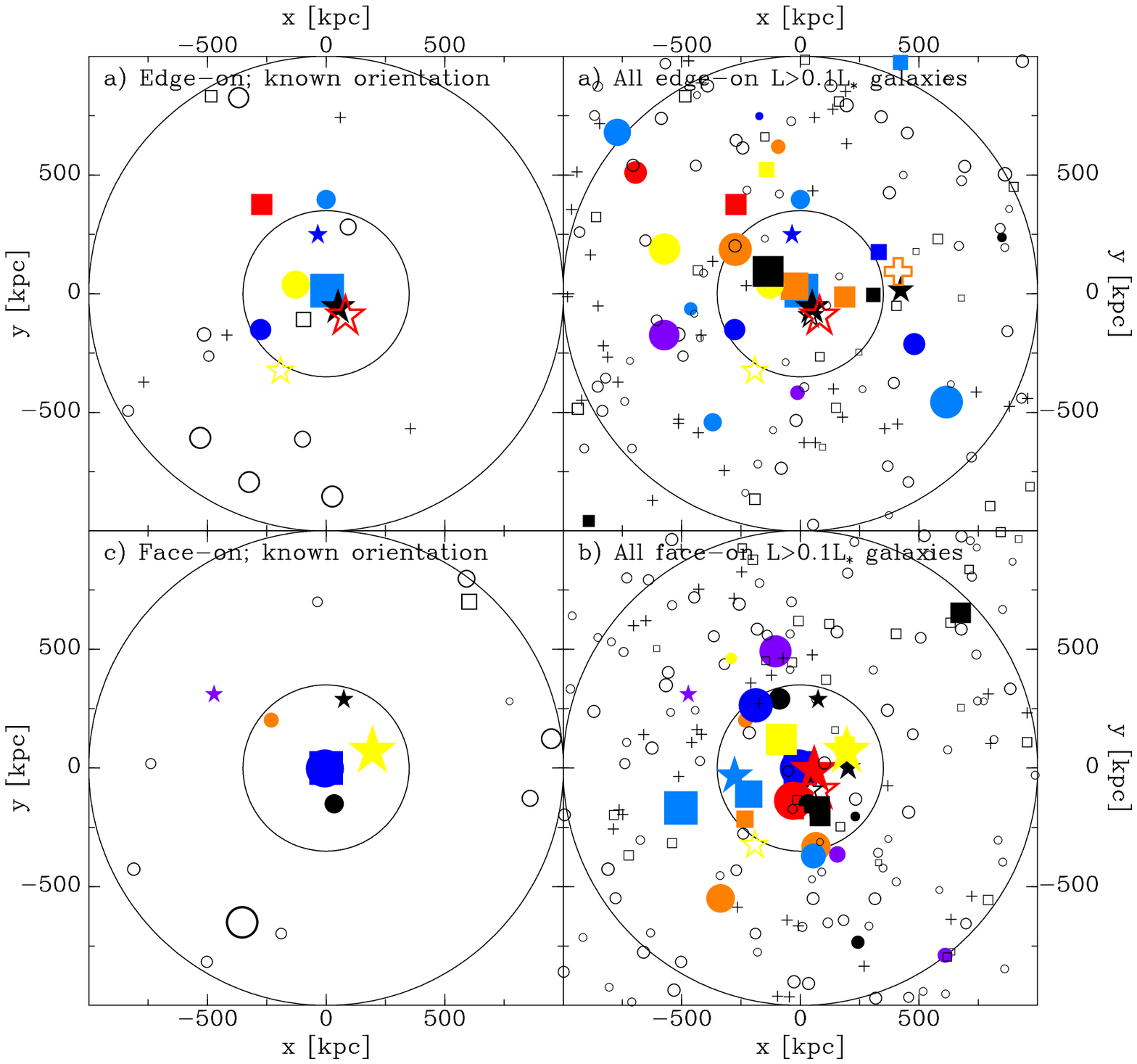}{0in}{0}{350}{350}{40}{-150}\figurenum{18}\caption{\captionhalomap}\end{figure}
\begin{figure}\plotfiddle{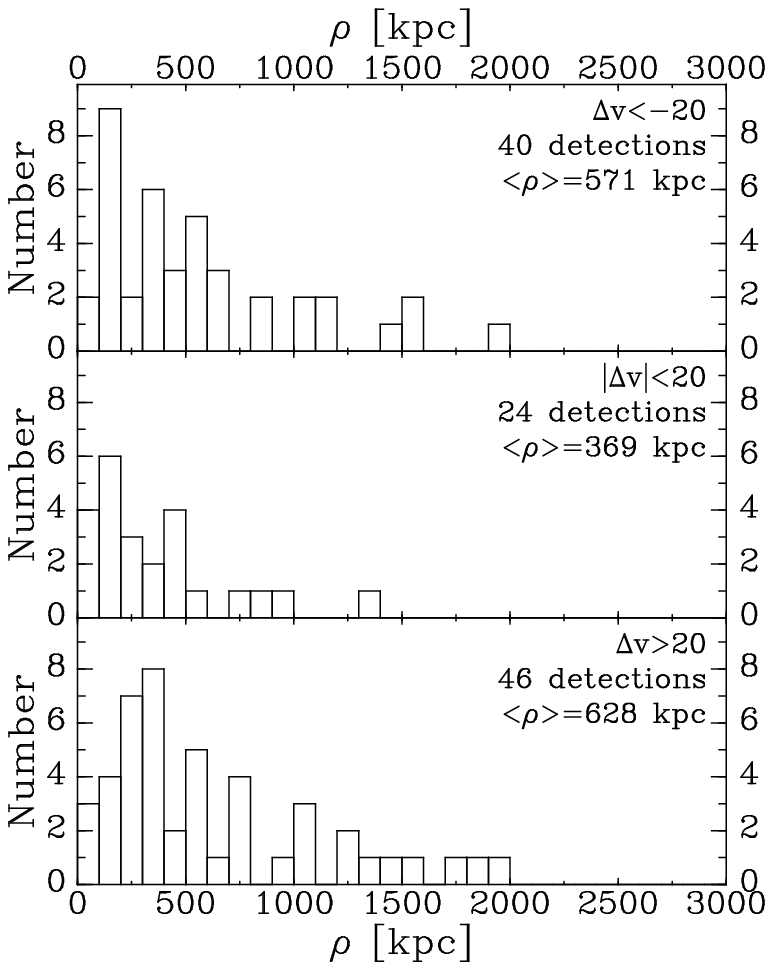}{0in}{0}{350}{400}{40}{-150}\figurenum{19}\caption{\captionimpvdiff}\end{figure}
\par Figure~\Fhalomap\ presents the results. The left panels (a, c) include just
the galaxies with known position angle, known orientation, and (in case of
detections) a clear association between an intergalactic absorber and a galaxy.
In the right panel (b) we include all luminous ($L$$>$0.1\,\Lstar) galaxies with
known position angle and inclination with which we either associate a detection
or find a non-detection with equivalent limit better than 50~\mA. For all of
these the impact parameter is correct, and the galaxies are rotated to have the
major axis horizontal, but if the galaxy's orientation is unknown, the direction
to the AGN sightline might have to be rotated by 180\deg.
\par Figure~\Fhalomap a includes 23 edge-on galaxies (inclination $>$60\deg; 9
with associated detections, 13 with non-detections, one having non-detections
against two different AGNs). Figure~\Fhalomap c includes 17 face-on galaxies
(inclination $<$60\deg; 7 with associated detection, 9 with non-detections, and
one with two non-detections). The colored symbols show the detections. The
symbol shapes encode whether the detection/non-detection is found for just \HI,
just \OVI\ or both (see figure caption for details). The circles have radii of
350~kpc and 1~Mpc.
\par From Fig.~\Fhalomap a, c we can see that at impact parameters $<$400~kpc
most galaxies have an associated intergalactic line, while non-detections are
generally found at larger impact parameters. 
\par Relative to underlying edge-on galaxies, four of the absorbers are
relatively close to the galaxy's plane, and of these one has \Dv$<$0 while lying
on the approaching side (toward PKS\,2155$-$304), one has \Dv$>$0 while lying on
the receding side (toward Ton\,S180), one has the ``wrong'' sign of the velocity
difference (toward Mrk\,771), while the detection toward PG\,1259+593 has small
\Dv. Four other detections occur away from the underlying galaxies' plane and
have both \Dv$<$0 and \Dv$>$0. Thus it appears that in general the absorbers do
not have velocities that would fit the rotation curves of the underlying
galaxies, a conclusion already reached by C\^ot\'e et al.\ (2005), who included
PG\,1259+593 and Ton\,S180 in their sample. However, with such a small sample
more cases are needed to confirm whether this is generally true. Although the
orientations of most of the galaxies in Fig.~\Fhalomap b are uncertain by
180\deg, this figure includes about 15 galaxies with associated detections that
could lie in an extended flat disk, for about 10 of which we do not yet know the
orientation.
\par For the detections associated with more face-on galaxies there also is no
clear pattern to the velocity differences, which range from $-$68 to +64~\kms.
Similarly, when looking at all face-on galaxies (Fig.~\Fhalomap d), there is no
clear pattern to the velocity differences (even though the orientation of most
symbols is uncertain by 180\deg). Both red and blue symbols are seen at almost
any position angle.
\par One systematic pattern is suggested by Fig.~\Fhalomap b and d. We
illustrate it in Fig.~\Fimpvdiff, showing that detections with velocity
differences less than 20~\kms\ (the black symbols) on average occur almost twice
as close to the associated galaxy (average impact parameter 369~kpc) as
detections having $\vert$\Dv$\vert$$>$20~\kms\ (average impact parameter
$\sim$600~kpc). All but one of the 20 associations with \Dv$<$20~\kms\ have
\ip$<$1~Mpc, whereas all but one of the 33 associations at \ip$>$1~Mpc have
\Dv$>$20~\kms. \fix{However, this effect has to remain only a suggestion until
we can obtain more data. A KS-test shows that we can only accept the hypothesis
that the distributions differ with $\sim$70\% confidence, i.e., it is about a
one sigma effect.}
\par From Figs.~\Fhalomap\ and \Fimpvdiff\ we conclude that {\it intergalactic
gas knows about the presence of a nearby galaxy, better matching the galaxies'
velocities the closer in it is, though the gas apparently does not generally
know the direction in which the galaxy rotates.}

\section{Discussion}
\subsection{The Relation Between IGM Absorbers and Galaxies}
\par By combining sightlines analyzed in a number of previous papers aimed at
studying low-redshift intergalactic absorption and its relation to galaxies,
adding data obtained for other studies, including \FUSE\ data for \OVI\ and
concentrating on just the lowest redshift absorbers and galaxies, we have been
able to make progress in understanding the connection between intergalactic gas
and galaxies. The advantage of our approach is that we are not limited to
finding galaxies with small angular separation to an extragalactic sightline,
and in addition our galaxy sample is expected to be complete down to
0.1\,\Lstar\ for $v$$<$2500~\kms. Further, we are able to separately look at
field and group galaxies, as there is a consistently defined catalogue of galaxy
groups for all nearby galaxies.
\par In previous work arguments have been presented that strong \Lya\ absorbers
are physically associated with galaxies (Lanzetta et al.\ 1995; Chen et al.\
1998; Bowen et al.\ 2002; C\^ot\'e et al.\ 2005), probing their 200--300~kpc
radius gaseous halos, while absorbers with larger impact parameters originate in
intergalactic filaments. The main arguments in favor of this interpretation
are (1) the fact that absorbers with \Dv$<$500~\kms\ are almost always found for
sightlines with galaxy impact parameters below a few hundred kpc and (2) a
claimed anti-correlation between impact parameter and \Lya\ equivalent width.
However, the occurrence of \Lya\ absorbers with low \Dv\ at large impact
parameters, the fact that not all authors find a correlation between equivalent
width and impact parameter and the presence of ``void absorbers'' (Penton et
al.\ 2002) suggests a more complicated picture, in which (almost) all absorbers
trace the large-scale structure of intergalactic gas. Some authors even suggest
that absorbers close to galaxies may not be associated with them. Cosmological
simulations suggest that the galaxies reside in the denser regions of filaments
or sheetlike gaseous structures (Dav\'e et al.\ 1999). In these simulations the
density of gas is highest near galaxies and falls off with radius (i.e., the
simulations predict that on average stronger lines occur closer to galaxies),
but even far from galaxies there is enough gas to produce \Lya\ absorption.
\par Our results support the second model, but also explain the results of the
studies that concluded that galaxies have large halos. Supporting the conclusion
that \Lya\ absorbers are related to galaxies we find that the properties of the
\Lya\ absorbers change with impact parameter. (1) The 90th percentile of the
linewidth distribution increases from FWHM$\sim$100~\kms\ at \ip$>$700~kpc to
$\sim$150~\kms\ at \ip=100~kpc (Sect.~\SSlinewidth), although there is a large
spread at any impact parameter, while the 10th percentile (widths of
$\sim$50~\kms) is independent of impact parameter. (2) The 90th percentile of
the equivalent width distribution increases from $\sim$100~\mA\ at 1~Mpc to
about 500~\mA\ at \ip=100~kpc, while no weak lines ($<$100~\mA) occur at
\ip$<$150~kpc (Sect.~\SSewvsb). (3) For impact parameters $<$350~kpc to galaxies
brighter than 0.1\,\Lstar, 100\% of field and 61\% of group galaxies have
associated (\Dv$<$400~\kms) \Lya\ absorption (Sects.~\SSdetfraclyafield,
\SSdetfraclyagroup). (4) \OVI\ absorption is only found within 500~kpc of
luminous ($L$$>$0.25\,\Lstar) galaxies (Sect.~\SSdetfracOVI). Thus, on average,
stronger and wider \Lya\ lines occur near galaxies, while almost all luminous
galaxies have an associated \Lya\ absorber.
\par On the other hand, there is evidence supporting the conclusion that \Lya\
absorbers originate in intergalactic filaments. (1) About half of the \Lya\
lines originate more than 400~kpc away from the nearest galaxy
(Sects.~\SSassoccount, \SSnearest, \SSassoc). (2) A substantial fraction
($\sim$20\%) of absorbers occurs far from ($>$3~Mpc) the nearest \Lstar\ galaxy,
although few occur far from the nearest 0.1\,\Lstar\ galaxy -- in fact for
$v$$<$2500~\kms\ we find an $L$$>$0.1\,\Lstar\ galaxy within 1.5~Mpc of every
absorber (Sect.~\SSvoid). (3) About 10\% of strong ($>$200~\mA) \Lya\ lines
occur far ($>$1~Mpc) from the nearest galaxy (Sect.~\SSewvsb). (4) The fraction
of galaxies that have an associated \Lya\ line decreases regularly with impact
parameter, i.e., there is no break in the distribution of detection fraction
(Sects.~\SSdetfrac, \SShalomap).
\par The picture that is most consistent with the absorber properties summarized
above is one in which the intergalactic gas filaments are densest near galaxies,
with area covering factor near 100\% within about 300~kpc, but these
concentrations merge smoothly into a more tenuous intergalactic medium that
connects the galaxies. This picture works when all galaxies brighter than about
0.1\,\Lstar\ are considered and 50\% of the \Lya\ lines originate within 400~kpc
of such a galaxy, 75\% within 1~Mpc. If only galaxies with $L$$>$\Lstar\ were to
be taken into account, a different picture would emerge, one in which a little
more than half the galaxies have associated gas within 300~kpc, but most (80\%)
of the \Lya\ absorptions originate far (\ip$>$400~kpc) from \Lstar\ galaxies.

\subsection{\OVI\ Absorbers and Thermal Properties of the Gas Near Galaxies}
\par Although we have far fewer \OVI\ detections than were discussed by Tripp et
al.\ (2008) and Danforth \& Shull (2008) we find similar values for
$dN$(\OVI)/$dz$ and a basically identical linewidth distribution
(Sects.~\SSlinewidth, \SSdndz). This leads us to the conclusion that the
intergalactic \OVI\ lines seen by Tripp et al.\ (2008) at redshifts 0.2--0.5 are
likely to have the same relation to galaxies as the ones we find in the nearby
universe. All 14 \OVI\ absorbers in our sample originate within 560~kpc from an
$L$$>$0.1\,\Lstar\ galaxy, 13 of which originate near an $L$$>$0.25~\Lstar\
galaxy, 9 (65$\pm$20\%) occur within 560~kpc of an $L$$>$\Lstar\ galaxy, while 6
(40$\pm$20\%) lie within 300~kpc of an $L$$>$\Lstar\ galaxy (Sects.~\SSOVIabs,
\SSdetfracOVI). Thus, we predict that searches for galaxies near higher 
redshift \OVI\ absorbers will turn up an \Lstar\ galaxy within 500~kpc about
half the time. At $z$=0.08 this requires a limiting magnitude of about 18.2,
while at $z$=0.25 \Lstar\ corresponds to $m$$\sim$20.8. We conclude that a
proper analysis of the relation between \OVI\ absorbers and galaxies requires
galaxy surveys that are about 3 times (2.5 magnitudes) deeper than that of Tripp
et al.\ (1998), in order to locate all galaxies brighter than 0.1\,\Lstar.
\par Three lines of evidence combine to suggest that the denser intergalactic
gas near galaxies also is hotter, a property that is predicted by cosmological
evolution models (Dav\'e et al.\ 2001) and is due to the heating of the gas by
infall. (1) \OVI\ absorption is only found at impact parameters $<$500~kpc from
luminous ($L$$>$0.25\,\Lstar) galaxies. Even though for many absorbers the \OVI\
appears to be generated by photoionization, a contribution from collisional
ionization is suggested in about half of the cases (Sects.~\SSOVIabs,
\SSdetfracOVI). (2) At impact parameters below 700~kpc the maximum linewidth of
\Lya\ lines increases with decreasing impact parameter (Sect.~\SSwidthimp). (3)
The fraction of wide \Lya\ lines is the largest at $z$=0 (Sect.~\SSbevol). By
itself, item (1) could be explained by the fact that all detected \OVI\ lines
have smaller equivalent widths than the \HI\ lines in the same system, so for
the weaker \HI\ lines at large impact parameters it would be easier to miss any
accompanying \OVI. Item (2) might be explained by increasing turbulence in the
gas near galaxies \fix{or by velocity gradients introduced by tidal stretching},
but the linewidths are much larger than would be expected from turbulence or
tidal effects; an explanation that we still need to fully exclude is that the
broad lines are multi-component absorbers. Item (3) could be explained if the
gas has higher turbulence near galaxies and over time the average impact
parameter has decreased. However, in combination with the modeling, these three
items are most consistent with thermal evolution of infalling intergalactic gas.

\subsection{The Baryon Content of the IGM}
\par Penton et al.\ (2002) and Danforth \& Shull (2008) combined the count of
the number of \Lya\ absorbers per unit redshift, the measured column density
distribution of \Lya\ lines with photoionization modeling to derive an estimate
of the fraction of baryons in the photoionized \Lya\ forest. They found a
fraction of 29$\pm$4\%. Using a different sample of sightlines, Lehner et al.\
(2007) derived a similar value. Since the fraction of baryons inside galaxies is
estimated to be $\sim$8\% (Fukugita \& Peebles 2004), the photoionized \Lya\
forest appears to contain about 3--4 times more baryons than the galaxies. We
note that this estimate depends on the assumption that it is the general
extragalactic radiation field that does the ionizing. The radiation field will
be stronger near galaxies, but for an \Lstar\ galaxy like the Milky Way, the
extragalactic field dominates outside radii of 50--150~kpc (see Fig.~9 in Fox et
al.\ 2005), which is where almost all \Lya\ forest lines originate. Thus, using
the extragalactic radiation field to derive the ionization correction is
justified, except possibly for the 10\% of the absorbers with \ip$<$150~kpc.
\par With 8\% of the baryons inside galaxies, and 30\% in the \Lya\ forest, the
remaining $\sim$60\% of the baryons is suspected to be in the form of hotter
($T$$>$\dex5~K) gas, although the observational data supporting this is sparse.
For instance, Nicastro et al.\ (2005) claimed to have detected the \dex6~K IGM,
but their conclusion is disputed by Kaastra et al.\ (2006) and Rasmussen et al.\
(2007). Tripp \& Savage (2000) estimated that about 10\% of the IGM may be
traced by \OVI\ absorbers, similar to the number found by Danforth \& Shull
(2008) in their much larger survey. Lehner et al.\ (2007) concluded that
10--20\% of the baryons show themselves as broad \Lya\ lines.
\par We find that for 50\% of the \Lya\ absorbers at $v$$<$2500~\kms\ there is
an $L$$>$0.1~\Lstar\ galaxy within \ip=370~kpc and \Dv$<$400~\kms\ (see
Sect.~\SSassoccount). Conversely, 77\% of $L$$>$0.1~\Lstar\ galaxies at
$v$$<$2500~\kms\ have a \Lya\ absorber with \ip$<$400~kpc and \Dv$<$400~\kms\
(see Table~\Tgalaxycount). Thus, it appears that most galaxies have extended
envelopes (halos) of associated photoionized and warm collisionally ionized gas
that has been detected via narrow and broad \Lya\ absorption.
\par To summarize, 8\% of the baryons is found inside galaxies, 30\% is in the
photoionized \Lya\ forest, 10--20\% may be collisionally ionized and seen as
broad \Lya\ lines, and finally, 50\% of the \Lya\ lines originate within 400~kpc
of a luminous galaxy. Therefore, we conclude that the gas out to 400~kpc from
$L$$>$0.1~\Lstar\ galaxies seen in \Lya\ absorption represents 20--25\% of the
baryons, i.e., there are at least 3 times as many baryons in the gaseous
envelopes (``halos'') of the galaxies than there are inside the galaxies. The
true baryonic content of these extended halos may be larger by another factor
$\sim$2--3, since the \Lya\ and \OVI\ observations are not sensitive to the hot
($T$$>$\dex6~K) phase of the intergalactic gas predicted by the hydrodynamicl
structure formation simulations. However, the baryonic content of the gas now
detected in these extended structures already greatly exceeds the baryonic
content of galaxies and these baryons likely play an important role in galaxy
evolution.
\par Since the critical density to have a closed universe ($\Omega$=1) is
9.2\tdex{-30}\,h$^2_{71}$\,g\,\cmm3, and since baryons represent 4.6\% of this
(Fukugita \& Peebles 2004), the average baryon density is
4.2\tdex{-31}\,g\,\cmm3. We previously found (Sect.~\SSnearest) that on average
the nearest neighbor of an \Lstar\ galaxy is at 2.1~Mpc, that of a 0.5~\Lstar\
galaxy at 1.65~Mpc and that of a 0.1\,\Lstar\ galaxy at 940~kpc (these values
differ from the medians (1.5, 1.2 and 0.6~Mpc, respectively) because the
distribution of nearest-neigbour distances is skewed). Thus, the average
0.1\,\Lstar\ galaxy has 4.8\tdex{10}\,\Msun\ of intergalactic baryons associated
with it, every 0.5\,\Lstar\ has 2.7\tdex{11}\,\Msun\ and every \Lstar\ galaxy
5.4\tdex{11}\,\Msun. Within 400~kpc the mass of intergalactic baryons associated
with each kind of galaxy is about half this amount.
\par Several lines of evidence support the contention that galaxies have been
and are still accreting new material. Sancisi et al.\ (2008) reviewed this and
infer that the observations show a visible (i.e.\ seen in the form of neutral
hydrogen) accretion rate for $L$$>$0.5\,\Lstar\ on the order of 0.2~\Msunpyr.
Theoretically, a rate on the order of 1~\Msunpyr\ is expected (e.g.\ Chiappini
et al.\ 2001), so Sancisi et al.\ (2008) concluded that a large fraction of the
accreting gas is not neutral. With a rate of 1~\Msunpyr, in 10~Gyr a 0.5~\Lstar\
galaxy would accrete about \dex{10}\,\Msun. This is a small fraction of the
total amount of intergalactic material associated with it, but a substantial
fraction of the material that currently lies within 400~kpc.

\par Whether or not one should call the intergalactic gas near the galaxies a
``halo'' or ``corona'' turns out to be a vague question. Certainly, the gas near
galaxies knows about their presence. The gas is affected by them, and probably
falling in and heating up. \fix{However, the increase in maximum linewidth with
decreasing impact parameter may have alternative explanations, such as an
increase in the amount of tidal material. Also, there may be alternative sources
of heating, \fix{such as from the mechanical energy deposited by galaxy
outflows.} In any case, }these ``halos'' have no firm boundary. They merge
smoothly into the more tenuous intergalactic filaments connecting galaxies. The
term ``corona'' might be appropriate for the hotter gas, if further studies
could show that the hot gas is more concentrated near galaxies. It is possible
that the notion that galaxies have ``halos'' can be kept if the intergalactic
gas generally is kinematically related to the underlying galaxies. However, our
first attempt at addressing this possibility suggests that the intergalactic
lines do {\it not} know about the rotation of the underlying galaxies, which
supports the notion that the intergalactic absorption lines originate in
filaments connecting galaxies. Yet, absorption is seen much more frequently near
galaxies than away from them, at rates approaching 100\% for luminous galaxies.
This implies that the galaxies are surrounded by gaseous envelopes.

\section{Conclusions}
\par We have analyzed intergalactic absorption lines in \HST\ and \FUSE\ spectra
of 76 AGNs, searching for detections and non-detections of absorption near the
velocity of nearby (\vgal$<$5000~\kms) galaxies near those sightlines. Compared
to previous studies of this subject, ours has several advantages -- (1) the
galaxy sample is much larger, (2) at \vgal$<$2500~\kms\ the galaxy sample is
basically complete down to 0.1\,\Lstar, (3) we can separate the sample into
group and field galaxies, (4) for each galaxy we can record non-detections as
well as associated detections, (5) we can compare \OVI\ with \HI. Previous
studies (Morris et al.\ 1993; Lanzetta et al.\ 1995; Tripp et al.\ 1998;, Impey
et al.\ 1999;, Chen et al.\ 2001; Bowen et al.\ 2002; Penton et al.\ 2002;
C\^ot\'e et al.\ 2005; Aracil et al.\ 2006; Prochaska et al.\ 2006) used smaller
and/or less complete samples. Although none of these papers combined all our
analyses, they pointed to many of the same conclusions. With our study we are
also able to reconcile some apparent contradictions between the conclusions
reached in these papers. We now summarize the conclusions, which are also
highlighted at the end of each subsection in Sects.~\Sabsresults,
\Sassocresults\ and \Sgalresults.

\par (1 -- see Sect.~\SSmeasure). We have analyzed 52 \HST\ and 63 \FUSE\
spectra of 76 AGN (both QSOs and Seyfert galaxies), and identify a total of 133
intergalactic absorber systems at recession velocities $<$6000~\kms. We measure
115 \Lya, 40 \Lyb, 13 \OVIa, and 5 \OVIb\ lines. Of these systems, 45 are
presented for the first time (including 29 \Lya, 36 \Lyb, 8 \OVIa\ and all 5
\OVIb\ lines). On the other hand, we do not confirm 20 previously published
\Lya\ lines or 6 previously claimed \OVIa\ lines.

\par (2 -- see Sect.~\SSOVIabs). The properties of the \OVI\ absorbers in our
sample generally match those of the larger well-studied sample of Tripp et al.\
(2008). Since we find that all our absorbers originate within 550~kpc of an
$L$$>$0.1\,\Lstar\ galaxy, we suggest that this is generally true for
intergalactic \OVI\ at low redshift. \fix{For eight of the fourteen \OVI\
systems we can make some attempt at explaining their origin. In three cases
photoionization appears to be the more likely description, while for the other
five collisional ionization appears needed to explain the absorber properties.}

\par (3 -- see Sects.~\SSlinewidth\ and \SSdndz). The distributions of
\Lya/\OVI/ linewidths, $dN$(\Lya)/$dz$ and $dN$(\OVI)/$dz$ at $z$=0.01 are all
similar to the distributions found from studies at redshifts 0--0.5, suggesting
that the relationship between \OVI\ absorbers and galaxies is the same at
$z$$\sim$0 as at $z$$\sim$0 to 0.5.

\par (4 -- see Sect.~\SSbevol). The fraction of broad \Lya\ lines is higher in
the nearby universe ($z$$<$0.017) than at higher redshifts, with 55\% of the
lines having $b$$>$40~\kms, compared to 30\% at $<$$z$$>$=0.25 (lookback time
2.5~Gyr) and 20\% at $z$$\sim$2 (7.5~Gyr ago).

\par (5 -- see Sect.~\SSassoccount). For the great majority (96\%) of
intergalactic absorbers a galaxy (of any luminosity) can be found with impact
parameter \ip$<$3~Mpc and velocity difference \Dv$<$400~\kms. For a large
fraction (75\%) there is a galaxy within 400~\kms\ and within 1~Mpc, and most of
these are brighter than 0.1\,\Lstar. A bright ($L$$>$\Lstar) galaxy is found
within 400~kpc and 200~\kms\ for just 17\% of the \Lya/\Lyb\ lines.
Table~\Tassoccount\ summarizes the fraction of absorbers for which it is
possible to find a galaxy of a given brightness within some impact parameter and
velocity difference. This is the completement of Table~\Tgalaxycount, which
gives the fraction of galaxies having an associated \Lya\ absorber.

\par (6 -- see Sect.~\SSnearest). We analyze the distribution of the nearest
neighbor galaxy to an absorber and that of the nearest neighbor galaxy to
another galaxy. We find that the median distance between group galaxies brighter
than $L$$>$0.1\,\Lstar\ is 500~kpc, while the median distance between field
galaxies brighter than $L$$>$0.1\,\Lstar\ is 1.5~Mpc. For absorbers, the
nearest-neighbor group galaxy is at a distance of 350~kpc, while in the field
the median nearest-neighbor galaxy is at 600~kpc. Therefore, we conclude that
most \Lya\ absorbers are associated with galaxies, both in the field and in
groups. Previous similar analyses reached the opposite conclusion because they
mixed field and group galaxies and only included \Lstar\ galaxies -- the median
separaton between \Lstar\ galaxies is 1.1~Mpc, while the median absorber-\Lstar\
galaxy separation is 1.0~Mpc.

\par (7 -- see Sect.~\SSassoc). For a little more than half of the intergalactic
absorbers there is just one galaxy that is likely to be associated with it. For
a quarter there either are two equally likely galaxies or there are two
absorption lines and two likely galaxies. For about one eighth of the lines the
absorption occurs in a group of galaxies and no unambiguous choice can be made.
In the remaining 7 cases the association between an absorber and a galaxy is
ambiguous, with two or more galaxies equally likely candidates.

\par (8 -- see Sect.~\SSvoid). We conclude that it depends on the luminosity
limit and completeness of the galaxy sample whether an absorber occurs far from
the nearest galaxy and can be called a ``void absorber''. Penton et al.\ (2002)
found a fraction of 22$\pm$8\% of absorbers occuring more than 3~Mpc from an
\Lstar\ galaxy, while we find 17$\pm$4\% using these criteria for our full
\vgal$<$5000~\kms\ sample. However, an $L$$>$0.1\,\Lstar\ galaxy is found within
1.5~Mpc of each absorber with $v$$<$2500~\kms, where our galaxy sample is
complete down to that luminosity limit. Further, just 7\% of the absorbers with
$v$$<$2500~\kms\ occur more than 3~Mpc from an \Lstar\ galaxy.

\par (9 -- see Sect.~\SSvdiffhist). For unambiguous associations, the difference
in velocity between the intergalactic absorption and the galaxy's systemic
velocity ranges from $-$118 to 147~\kms, with a dispersion of 60~\kms. For all
associations the range is $-$443 to +349~\kms, although
$\vert$\Dv$\vert$$>$300~\kms\ for just four of the associations that we make
(see Sect.~\SSassoc).

\par (10 -- see Sect.~\SSwidthimp). At impact parameters less than about 500~kpc
the width of the \Lya\ lines increases with decreasing impact parameter. This is
unlikely to be caused by kinematical broadening due to projection effects. We
cannot completely exclude that turbulent motions are responsible, but the most
likely explanation is that there is an increase in the temperature of the gas
within a few hundred kpc of galaxies. Such heating inside gravitational wells is
predicted in hydrodynamical simulations of structure formation in the universe
(Cen \& Ostriker 1999; Dav\'e et al.\ 2001; Cen \& Fang 2006).

\par (11 - see Sect.~\SSewvsb). Previous studies of the relation between \Lya\
equivalent width and impact parameter have led to conflicting conclusions. Some
authors concluded that there is a strong anti-correlation, others concluded that
there is none. We find that some of the confusion is due to the usage of log-log
plots as well as to the fact that different ranges in impact parameter are
included in different studies. We find that at any impact parameter there is a
wide range in equivalent widths. However, the strength of the strongest line
does anti-correlate with impact parameter, becoming progressively higher at
lower impact parameter. We find that 80\% of strong ($W$$>$300~\mA) lines occur
within 600~kpc of a galaxy, while 70\% originate within 350~kpc. On the other
hand, weak lines only occur far from galaxies, and the weaker the line, the
larger the minimum impact parameter; specifically all \Lya\ lines with
$W$$<$25~\mA\ have \ip$>$0.75/2.5~Mpc to the nearest 0.1/1.0~\Lstar\ galaxy,
while all \Lya\ lines with $W$$<$50~\mA\ have \ip$>$0.3/1.0~Mpc. Thus, we
conclude that there are patterns in the relation between \Lya\ equivalent width
and impact parameter, but there is no simple one-to-one correspondence.

\par (12 -- see Sect.~\SSgaldens). We conclude that total \Lya\ equivalent width
in windows 500--1000~\kms\ wide around detections does not directly correlate
with the density of galaxies in cylinders of radii of 500--2000~kpc and velocity
depths of 500--1000~\kms. However, we do find that for a given density of
galaxies the largest equivalent width found does change with density, such that
at a density of 1 galaxy per Mpc$^3$ the equivalent width in a 500~\kms\ range
can reach 3~\AA, whereas at a density of 0.01 galaxy per Mpc$^3$ the maximum
equivalent width in a 500~\kms\ range is 300~\mA.

\par (13 -- see Sect.~\SSgalaxycount). For 100\% of $L$$>$0.1~\Lstar\ galaxies
it is possible to find a $>$50~\mA\ \Lya\ line with impact parameter $<$400~kpc
and velocity difference \Dv$<$1000~\kms, while the percentage is 80\% for
\Dv$<$400~\kms. Strong lines ($>$300~\mA) are found within 400~kpc and 400~\kms\
for about 50\% of the galaxies. At impact parameters $<$1~Mpc strong lines with
\Dv$<$400~\kms\ are found for about 25\% of the $L$$>$0.1\,\Lstar\ galaxies,
while a line with $W$$>$50~\mA\ is seen for 50\% of such galaxies.
Table~\Tgalaxycount\ summarizes the fraction of galaxies for which it is
possible to find an absorber of a given equivalent width within some impact
parameter and velocity difference. This is the completement of
Table~\Tassoccount.

\par (14 -- see Sects.~\SSdetfraclyafield\ and \SSdetfraclyagroup). Using the
individually determined associations listed in Table~\Tres, we find that 100\%
(7 of 7) of field and 60\% (13 of 21) of group galaxies brighter than
0.1~\Lstar\ and with velocity $<$2500~\kms\ have an associated \Lya\ absorber at
impact parameter $<$350~kpc. The fraction of galaxies with associated absorbers
decreases monotonically to about 0 at \ip$\sim$1500~kpc. Similarly, about 50\%
of galaxy groups are found to have associated \Lya\ absorption.

\par (15 - see Sect.~\SSdetfracOVI). At impact parameters $<$350~kpc associated
\OVI\ is detected for about 67\% (4 of 6) of field galaxies with
\vgal$<$2500~\kms\ brighter than 0.1~\Lstar. Only 8\% (1 of 13) of such group
galaxies are found to have associated \OVI. Only one field galaxy has associated
\OVI\ at \ip$>$300~kpc, but the impact parameter for three of the five \OVI\
lines associated with a bright group galaxy is 300--450~kpc.

\par (16 -- see Sect.~\SShalomap). Our sample includes four edge-on galaxies for
which we know which side is approaching/receding and which have an intergalactic
absorber with impact parameter $<$350~kpc that lies near the extended plane of
the galaxies. We do not find a correlation between the difference in velocity
between the absorber and the galaxy, suggesting that the intergalactic gas does
not know about the rotation of the underlying galaxy. However, we need to
confirm this disconnect using more examples.

\par (17) We can combine conclusions (2), (3) and (15) to imply that many
intergalactic \OVI\ lines originate in photoionized gas within 500~kpc of bright
($L$$>$0.1~\Lstar) galaxies. As is the case in larger and more redshifted
samples of \OVI\ absorbers, there are cases in which the \HI\ and \OVI\ lines
have similar widths, but also cases where broader \HI\ lines are seen. It is not
clear yet, however, whether these broad lines are single- or multi-component
absorbers.

\par (18) Conclusions (3), (4) and (10) suggest that there is an increase in the
temperature of the gas within a few hundred kpc of galaxies, \fix{although we
cannot clearly exclude alternative explanations such as an increase in the
amount line broadening caused by an increase in the amount of tidal material
with high velocity gradients.} However, heating inside gravitational wells is
predicted in hydrodynamical simulations of structure formation in the universe
(Cen \& Ostriker 1999; Dav\'e et al.\ 2001; Cen \& Fang 2006).

\par (19) Combining conclusions (5)--(9) implies that intergalactic \Lya/\Lyb\
and \OVI\ absorbers are associated with galaxies. For the parts of parameter
space where our sample is complete, we show that an $L$$>$0.1\,\Lstar\ galaxy
can be found within 1.5~Mpc of each absorber, while each $L$$>$0.1\,\Lstar\
field galaxy has an associated absorber within 350~kpc. About half of the
$L$$>$0.1\,\Lstar\ group galaxies have an associated absorber within 350~kpc,
while associated \Lya\ absorption is found for about 50\% of the galaxy groups.

\par (20) Conclusions (11)--(14) show that weak lines are seen at all impact
parameters, but strong lines above a given equivalent width limit only occur
below a given impact parameter. Similarly, the strongest lines occur where the
galaxy density is highest, although weak lines also occur in high-density
regions. This suggests that denser patches of intergalactic gas are more often
found closer to galaxies than at large impact parameter.

\par (21) All the arguments summarized above can be reconciled if galaxies have
gaseous envelopes (``halos'') that are several hundred kpc in radius, smoothly
connecting to intergalactic filaments. These halos consist of intergalactic gas
that is in the process of falling in toward the galaxies and possibly heating up
as it falls, but the gas has not yet taken on the kinematics of those galaxies.
The baryonic content of this photoionized and warm collisionally ionized gas
located within 400~kpc of galaxies exceeds by a factor $\sim$2--4 the baryonic
content of the galaxies. This gas likely plays a crucial role in the evolution
of galaxies.

\acknowledgements
The data in this paper were obtained with the NASA-CNES-CSA {\it Far Ultraviolet
Spectroscopic Explore}, FUSE, operated for NASA by The Johns Hopkins University
under NASA contract NAS5-32985 and with the NASA ESA {\it Hubble Space
Telescope}, at the Space Telescope Science Institute, which is operated by the
Association of Universities for Research in Astronomy, Inc.\ under NASA contract
NAS5-26555. Spectra were retieved from the Multimission Archive (MAST) at STScI.
The study made use of the NASA/IPAC Extragalactic Database (NED), which is
operated by the Jet Propulsion Laboratory, California Institute of Technology,
under contract with NASA. Over the course of this study, BPW was supported by
NASA grants NNG04GA39G, NNG06GG39G (FUSE), GO-00754.01-A (STScI) and NNX07AH42G
(ADP). BDS was supported by grants NAS5-31248 (NASA/FUSE) and MSN111587 from the
University of Colorado (NASA Cosmic Origins Spectrograph).

\newpage
\bigskip\bigskip
Appendix
\par Here we present notes on individual sightlines. We refer to Fig.~\Fassoc\
and Table~\Tres\ for even more details. The notes often give the parameters of a
galaxy near the line of sight, in the format e.g., (\vrb=925, 13.4, 140). These
three numbers give the systemic velocity ($v$) in \kms, diameter at 25th
magnitude surface brightness ($D$) in kpc, and impact parameter (\ip) in kpc.

{\it 1H\,0419$-$577. --}
Although there is just a \FUSE\ spectrum of this target, and it has low S/N
($\sim$5), it is included in the sample because the impact parameter to
NGC\,1574 (\vrb=925, 13.4, 140) is low. A strong \Lyb\ and a possible \Lyg\ line
are seen at 1112~\kms. There is a hint of \OVIa\ absorption, but only at the
2$\sigma$ level. The confirming \OVIb\ line is hidden in geocoronal \OI*
emission, and the orbital-night-only data have no signal. Table~\Tres\ lists
NGC\,1574 as the associated galaxy because it is the northernmost galaxy in the
LGG\,112 (\vavg=1095~\kms) group, and 1H\,0419$-$577 lies only 140~kpc from it
(see filled square in Fig.~\Fassoc(1)). However, the other group galaxies also
have fairly low impact parameters, between 442 and 627~kpc, while a small galaxy
(LSBG\,F157$-$081; \vrb=1215, 3.7, 70) is somewhat closer. We choose to list
NGC\,1574 as the associated galaxy, because it has smaller \ip/\Rgal\ (10) than
LSBG\,F157$-$081 (19), but we could also have interpreted the absorption as a
general group detection.
\par Two more galaxy groups lie near the sightline, and like LGG\,112 these
groups are quite well-defined. Non-detections are listed for LGG\,119
(\vgal=1095~kpc, nearest galaxy ESO\,118-G34 at 700~kpc) and LGG\,114
(\vgal=1481~kpc, nearest galaxy APMBGC157+016+068 at 317~kpc). A final
non-detection entry is given Table~\Tres\ for NGC\,1533 (\vrb=790, 17.7, 396),
as it does not belong to any of the groups.

{\it 1H\,0707$-$495. --}
The feature listed as \Lyb\ at 1302~\kms\ is associated with ESO\,207-G09
(\vrb=1029, 8.0, 482), but needs confirmation with an observation of \Lya. The
association is relatively unambiguous (see Fig.~\Fassoc(2)). A galaxy with
unknown velocity (ESO207-G31) could have \ip$\sim$370~kpc, if its velocity is
similar to that of the \Lyb\ line.

{\it 1H\,0717+714. --}
This sightline passes close to several galaxies in the LGG\,141 group
(\vavg=3001~\kms), with UGC\,3804 being the closest at 199~kpc (filled square in
Fig.~\Fassoc(3)). UGC\,3921, UGC\,3940 and IC\,2184 have velocities similar to
those of the group galaxies (2475, 2462 and 3605~\kms, respectively), but lie
outside the group on the sky, and are thus listed separately in Table~\Tres.
\par No \Lya\ data exist for this sightline, making the analysis more difficult.
However, there is a feature that is likely to be \Lyb\ at 2888~\kms. On the
other hand, it is possible that this feature is a weak \CII\ line at
\vlsr=$-$200~\kms\ associated with the nearby ($<$1\deg) high-velocity cloud
complex~A. No 21-cm \HI\ emission is detected at this velocity in the direction
of 1H\,0717+714, but if we interpret the absorption as \CII, its strength
suggests $N$(\HI)$\sim$4\tdex{17}~\cmm2, which is below the 21-cm detection
limit. An interpretation as \CII\ is not supported by the higher Lyman lines, as
there is no evidence for a component at $-$200~\kms. Further, complex~A has
velocities of $\sim$$-$160~\kms\ nearest 1H\,0717+714, while velocities of
$\sim$$-$200~\kms\ are seen only 3\deg\ or more away.
\par Further arguing in favor of interpreting the feature as \Lyb\ is the clear
66\E15\E9~\mA\ absorption feature at 2914~\kms\ on the \OVIa\ velocity scale. In
the combined orbital day plus orbital night data this is blended with geocoronal
\OI* emission, but it is very clear in the orbital-night-only data. This feature
could be \Lyb\ at 4750~\kms, but there are no known galaxies near that velocity.
It is more likely that it is redshifted \OVIa\ matching the \Lyb\ line. The only
problem with this interpretation is that the corresponding \OVIb\ line is not
clearly visible. Considering the errors on the probable \OVIa\ feature, the
equivalent width of the other \OVI\ line is expected to be between 25 and
40~\mA. Since the detection limit is 23~\mA, the apparent non-detection is not
too problematic, but data with higher S/N ratio are sorely needed (they were
approved, but not executed before \FUSE\ was decommissioned).

{\it 3C\,232. --}
With Mrk\,205, this is one of two targets with good data that lies behind the
disk of a nearby galaxy, NGC\,3067 (\vrb=1476, 17.0, 14) in the case of 3C\,232.
A very strong \Lya\ line ($N$(\HI)=8\tdex{19}~\cmm2) is detected in the \GHRS\
spectra of 3C\,232, as are many other ions (\OI, \CII, \CIV, \MgI, \MgII, \AlII,
\SiII, \SiIII, \SiIV, \AlII, \FeII). These lines were analyzed in detail by
Tumlinson et al.\ (1999) and Keeney et al.\ (2005). Unfortunately, the FUV flux
of 3C\,232 is too low (0.5\tdex{-14}~\fu) to obtain a useful \FUSE\ spectrum.
\par In addition to NGC\,3067, there are 16 other galaxies in the GH\,50
(\vavg=1442~\kms) group with impact parameter between 427 and 1000~Mpc. If any
any of these have associated absorption, it will be hidden by the strong line
originating in NGC\,3067.
\par There are three galaxies with sufficiently different velocity to warrant
separate entries in Table~\Tres: UGC\,5272 (\vrb=520, 6.5, 359), UGC\,5340
(\vrb=503, 8.3, 660) and Mrk\,412 (\vrb=4479, 11.7, 196). These are not
discussed by Tumlinson et al.\ (1999) and Keeney et al.\ (2005). Any \Lya\
absorption near 520~\kms\ is obscured by the Galactic \Lya\ line, however. There
is a clear feature that is probably \Lya\ at 4526~\kms, and that can be
associated with Mrk\,412, which is the only known galaxy with similar velocity;
see filled symbol in Fig.~\Fassoc(5).

{\it 3C\,249.1. --}
One possible 3$\sigma$ \Lya\ line is found at 1861~\kms. There are four galaxies
with similar velocity and impact parameter near this velocity: UGC\,5854
(\vrb=1808, 7.7, 505), UGC\,5841 (\vrb=1766, 12.2, 538), NGC\,3329 (\vrb=1812,
14.3, 543) and UGC\,5814 (\vrb=1881, 13.3, 641); see Fig.~\Fassoc(6). In
Table~\Tres\ the absorption line is associated with the largest of these three
(NGC\,3329), but this is an arbitrary choice. For the other galaxies near this
velocity non-detections are listed.
\par The velocity of the galaxy with the smallest angular distance (UGC\,6065)
has not yet been measured. It has an impact parameter of 238\,(\vgal/4000)~kpc,
and diameter of 25\,(\vgal/4000)~kpc. So, if this galaxy had a velocity like
that of the \Lya\ line at 1861~\kms, its diameter would be similar to those of
the other five galaxies near this velocity, but it would have an impact
parameter of only $\sim$107~kpc, and it would be listed as the galaxy associated
with the \Lya\ line.

{\it 3C\,263. --}
As many as 17 galaxies with velocity near about 1100~\kms\ lie within 1~Mpc of
this sightline. For five of these the ratio \ip/\Rgal\ is less than 125, and
these are the ones listed in Table~\Tres: NGC\,3682 (\vrb=1532, 11.8, 626),
UGC\,6448 (\vrb=991, 5.1, 643, UGC\,6390 (\vrb=1008, 10.3, 634), NGC\,3945
(\vrb=1220, 15.6, 950) and UGC\,6534 (\vrb=1273, 15.2, 955). No absorption is
seen near their velocities.

{\it 3C\,273.0. --}
In this spectrum there are four \Lya\ lines with $v$$<$6000~\kms. Two (at 1010
and 1580~\kms) are strong, two (at 2160, 2274~\kms) are weak. This sightline was
analyzed in great detail by Sembach et al.\ (2001) and Tripp et al.\ (2002), and
it was included in the sample of Penton et al.\ (2000). The equivalent widths
and velocities all agree between these papers and Table~\Tres. The two strong
lines are associated with the Virgo cluster and assigning it to any individual
galaxy (if any) would be very ambiguous. The LGG catalogue splits the galaxies
near this sightline into two groups -- LGG\,287 (\vavg=1655~\kms) and LGG\,292
(\vavg=938~\kms), see Fig.~\Fassoc(7)/(8). The impact parameters (191 and
311~kpc) listed in Table~\Tres\ correspond to the nearest galaxy that has a
velocity within \E200~\kms\ of the \Lya\ detection. The probable detection of
\OVI\ at 1008~\kms\ was previously reported in three other papers. However, the
three papers disagree on the value of the equivalent width. Sembach et al.\
(2001) reported 26\E10~\mA, Danforth et al.\ (2006) gave 35\E6~\mA, Tripp et
al.\ (2008) listed 31\E7~\mA, while we find 21\E3\E7~\mA. Our value is smaller
than any of the others because we correct for the 5~\mA\ contribution of \H2\
L(6-0) P(4)\lm1035.181 line, unlike the other authors. Taking this into account,
we agree with Sembach et al.\ (2001).
\par Sembach et al.\ (2006) also reported an \OVIb\ absorption to go with the
1580~\kms\ \Lya\ line. However, the detailed modeling of the \H2\ lines shows
that this is actually Galactic \H2\ L(5-0) R(3)\lm1041.158, as can be seen in
Fig.~\Fspectra.
\par For the two weak components, there are two galaxies with \vgal\ between
2000 and 2500~\kms. UGC\,7625 (\vrb=2234, 9.6, 771) has a large impact
parameter, while 2MASX\,J122815.85+024202.5 (\vrb=2286, 5.0, 429) is closer to
the AGN. Both galaxies have almost the same ratio of \ip/\Rgal. In Table~\Tres\
the \Lya\ component at 2274 is (somewhat arbitrarily) associated with
2MASX\,J122815.85+024202.5, and the one at 2160~\kms\ with UGC\,7625. This
results in \Dv=$-$12 and +74~\kms\ instead of $-$126 and $-$40~\kms,
respectively.

{\it 3C\,351.0. --}
As there is a Lyman limit system at $z$=0.22 toward this sightline, there is no
flux below 1112~\AA, and the \Lyb\ and \OVI\ lines cannot be checked. Good \Lya\
data exist, however. The intrinsic absorption system at $z$=0.3721 was studied
by Yuan et al.\ (2002).
\par There are 12 galaxies with \vgal\ between 3012 and 3855~\kms\ within 1~Mpc
of this target (see Fig.~\Fassoc(10)). Just four of these (NGC\,6292, NGC\,6306,
NGC\,6307, NGC\,6310) are included in the RC3. Within 5 degrees (3~Mpc) of
3C\,351.0 there are tens of galaxies whose velocities cluster around 3300~\kms,
with a range from 2500 to 4000~\kms. Garcia (1993) did not define these galaxies
as a group, probably because not all radial velocities had been measured at the
time. However, there is clearly a group in this part of the sky, and we identify
it as LGG179A in Table~\Tres, as its right ascension lies between that of LGG179
and LGG180.
\par Two strong \Lya\ features are found near these velocities, at 3598 and
3465~\kms. The velocities and impact parameters of the galaxies are such that
Table~\Tres\ associates the detection at 3598~\kms\ with Mrk\,892 (\vrb=3617,
10.2, 170, smaller filled square in Fig.~\Fassoc(10)). The one at 3465~\kms\ is
listed as associated with NGC\,6292 (\vrb=3411, 22.1, 314), which has the
smallest velocity difference (see larger filled square in Fig.~\Fassoc(10)). 
NGC\,6310 (\vrb=3386, 28.5, 402, large open square with plus next to larger
black square in Fig.~\Fassoc(10)) has the same \ip/\Rgal\ ratio as NGC\,6292,
but a non-detection is listed because of the larger impact parameter and larger
velocity difference with the absorption. Of the other nine galaxies, six have
velocity between 3012 and 3350~\kms, and three lie between 3736 and 3855~\kms.
All are fairly large ($>$7~kpc) and a non-detection is given in Table~\Tres\ for
each of these.
\par Two galaxies with different velocity have small impact parameters and are
included separately in Table~\Tres: UGC\,10770 (\vrb=1108, 5.9, 531) and
SDSS\,J170349.45+601806.1 (\vrb=5183, 6.6, 594). \Lya\ absorption at 5175~\kms\
is associated with the SDSS galaxy (not shown in Fig.~\Fassoc\ because
$v$$>$5000~\kms).

{\it ESO\,141-G55. --}
Several galaxies have impact parameters between 400~kpc and 1~Mpc, concentrated
around three velocities. IC\,4824 and ESO\,141-G42 have \vgal=953 and 935~\kms,
and lie close together at 769 and 790~kpc impact parameter. IC\,4826 and
IC\,4819 have similar velocity (1925 and 1841~\kms), and similar impact
parameter (865 and 868~kpc), but lie in opposite directions from ESO\,141-G55,
so separate entries are given in Table~\Tres. This is also the case for
ESO\,141-G51 (\vrb=3497, 14.2, 410) and IC\,4843/ESO\,141-G46 (\vgal=3975 and
4079~\kms, \ip=662 and 878~kpc). No \Lya\ or \OVI\ absorption is seen in the
spectrum of ESO\,141-G55. Penton et al.\ (2000) listed 1.5$\sigma$ (12\E8~\mA)
features near 8449 and 9078~\kms. However, these are likely to be spurious.

{\it ESO\,185-IG13. --}
This nearby galaxy (\vgal=5600~\kms) is the only AGN in the sample with
\vgal$<$7000~\kms. In spite of this and in spite of its low S/N (4.1) \FUSE\
spectrum it is included because the impact parameter to IC\,4889
(\vgal=2528~\kms) is just 62~kpc and very strong \Lyb, \Lyg, \CIII\ and \OVI\
are seen. Follow-up \FUSE\ observations were approved twice, but never executed.
Non-detections are listed for three other galaxies: IC\,4888,
2MASXJ\,194221.91$-$550627.5 and ES0\,185-G03. The last two have large impact
parameter (672 and 985~kpc) and \vgal$\sim$3000~\kms, and only $\sim$90~\mA\
\OVIa\ upper limits are given. Like IC\,4889, IC\,4888 has low impact parameter
(123~kpc), and almost the same velocity, but it's diameter is one-third that of
IC\,4889.

{\it ESO\,438-G09. --}
Only a \STIS-G140M spectrum is available for this target. It lies behind the
LGG\,230 group, with nine galaxies clustering around 1425~\kms\ (see
Fig.~\Fassoc(12)). ESO\,438-G05 is the nearest, with an impact parameter of
173~kpc. A very strong \Lya\ line is seen at 1426~\kms, which is associated with
the group in Table~\Tres. Bowen et al.\ (2002) previously listed this line, and
associated it with UGCA\,226 (\vrb=1500, 19.9, 178). However, there are several
other nearby group galaxies: ESO\,437-G05 (\vrb=1507, 19.2, 173), ESO\,438-G12
(\vrb=1322, 6.8, 245) and ESO\,438-G10 (\vrb=1487, 9.6, 248). There is no reason
to prefer associating the absorption with any particular one of these.
\par An additional \Lya\ line at 2215~\kms\ (also first listed by Bowen et al.\
2002) is not considered as originating in the group, because of the large
velocity difference and because the group galaxies have velocities ranging from
1230 to 1517~\kms. Instead it is listed under 2MASX\,J111343.40$-$274328.8
(\vrb=2100, 4.8, 577~kpc), which is the only candidate galaxy with
\ip$<$900~kpc.

{\it Fairall\,9. --}
Although this sightline passes relatively close to the LGG\,19 group
(\vavg=5035~\kms), the two nearest galaxies in that group (NGC\,484,
ESO\,113-G35) have large impact parameters (735, 799~kpc), and no absorption is
seen. Penton et al.\ (2000) reported on this sightline, and only list absorbers
at $v$$>$5000~\kms.

{\it H\,1821+643. --}
This sightline lies at relatively low galactic latitude (27\deg), and Galactic
extinction clearly influences the number of bright galaxies that are visible
near it. \FUNNY{Sweep that dust}. Only four small galaxies with
\vgal$<$6000~\kms\ are known within 1.5~Mpc of the sightline, and two of these
have \ip/\Rgal$>$125. That leaves NGC\,6690 (\vrb=488, 8.7, 858), which is
included in Table~\Tres. Penton et al.\ (2000) list no absorption lines below
6000~\kms, mostly because their GHRS spectrum does not extend below 4500~\kms.
The \STIS-E140M spectrum shows two low redshift \Lya\ absorbers, at 2836 and
4084~\kms\ for which the nearest galaxies with similar velocities have
\ip$\sim$2.5~Mpc. The line at 2836~\kms\ is the weakest line in our sample
(17\E5~\mA). In addition, there is a \Lya\ line at 5253~\kms, but no galaxies
with similar velocity are known with impact parameter $<$3~Mpc.
\par Low-redshift \OVI\ absorbers (at $z$=0.22497) were first discovered by
Savage et al.\ (1998) and Tripp et al.\ (2000) in this sightline.

{\it HE\,0226$-$4110. --}
Lehner al.\ (2006) analyzed the IGM absorption in this sightline. They listed
the system at 5235~\kms, which includes \Lya\ and both \OVI\ lines. However, the
\OVIa\ line is contaminated by interstellar \H2\ L(4-0) R(1)\lm1049.960.
Nevertheless, the very good \H2\ model for this sightline (see Wakker 2006)
shows that not all of this feature can be \H2, and about half is likely to be
\OVI, centered at 5240~\kms. Unfortunately, the corresponding \OVIb\ line is
contaminated by \OVI\lm787.711 at $z$=0.3406 (see Lehner et al.\ 2006). Aiding
the interpretation is the possible detection of weak \CIII\ (23\E9~\mA) and
\CIV\ (39\E11~\mA) absorption. The lines can be associated with NGC\,954
(\vgal=5353~\kms), which has impact parameter 562~kpc, and which belongs to the
LGG\,62 group. If this interpretation is correct, it is the \OVI\ detection with
the largest impact parameter. We note that the galaxy survey in this region of
sky is relatively good: the RC3 include three galaxies with \ip$<$1.5~Mpc, and
\NED\ lists four more, including three low-surface brightness galaxies.
\par There may be a weak ($\sim$2$\sigma$) \Lya\ line at 1413~\kms, near the
velocity of NGC\,986A. This feature was not considered significant by Lehner et
al.\ (2006), and we also list it as a non-detection in Table~\Tres.
\par On the other hand, Danforth \& Shull (2008) claimed an absorption line at
3642~\kms\ that was not found by Lehner et al.\ (2006). The nearest galaxy
within 400~\kms\ has an impact parameter of 1.5~Mpc. We measured this feature as
22\E10~\mA, and decided that it is a noise fluctuation, just like some other
similar-looking features near it.

{\it HE\,0340$-$2703. --}
The \STIS-G140M spectrum of this target shows intrinsic \Lyg, \Lyd, \Lye,
\SVI\llm933.378, 944.523 ($z$=0.2830), as well as \Lyb, \OVIa\ and \OVIb\ at
$z$=0.1655. In addition to these lines there are three features that are
probably \Lya\ at 1361, 1785 and 4100~\kms, although there is a chance that they
are \Lyb\ or \Lyg\ at higher redshift.
\par Assuming that these are \Lya, we associate the feature at 1785~\kms\ with
NGC\,1412 (\vrb=1790, 12.3, 167), although ESO\,483-G32 (\vrb=1756, 9.8, 152)
(for which we list a non-detection) is just as a likely a candidate (see the
adjacent filled and open circles in Fig.~\Fassoc(17)). Several other galaxies
with \Dva$<$400~\kms\ and \ip=1~Mpc can be found, but as Fig.~\Fassoc(17)\
shows, NGC\,1412 or ESO\,483-G32 is the most likely associated galaxy.
Table~\Tres\ lists a separate non-detection for 6dF\,J0342278$-$260243
(\vrb=1738, 3.2, 325), the one remaining galaxy with \ip/\Rgal$<$125.
\par For the line at 4100~\kms\ there are two candidates with large \ip\ (shown
by a large filled circle and a smaller open overlapping circle in
Fig.~\Fassoc(18)): 2MASX\,J034134.24$-$27491.87 (\vrb=4125, 10.9, 913) and
ESO\,419-G03 (\vrb=4109, 26.9, 942). We choose the largest of these as the
associated galaxy in Table~\Tres, giving a non-detection for the other.
\par The \Lya\ line at 1361~\kms\ can be associated with the large galaxy
NGC\,1398 (\vrb=1407, 29.8, 244). The sightline passes near the LGG\,97 group,
whose center lies about 5\deg\ ($\sim$1.6~Mpc) to the west of the QSO. The 32
group members defined by Garcia (1993) include two galaxies with \ip$<$1~Mpc:
NGC\,1371 (\vrb=1471, 30.6, 838) and NGC\,1385 (\vrb=1493, 18,4, 861). There are
another five galaxies with \ip$<$1~Mpc to the west of the QSO which were not
included in the group, but which have a velocity within 100~\kms\ of that of the
group galaxies: NGC\,1398 (\vgal=1407~\kms, \ip=244~kpc), ESO\,482-G46
(\vgal=1525~\kms, \ip=543~kpc), ESO\,482-G39 (\vgal=1381~\kms, \ip=588~kpc),
ESO\,482-G11 (\vgal=1595~\kms, \ip=646~kpc), and ESO\,482-G06 (\vgal=1535,
\ip=705~kpc). \FUNNY{This is too complicated.}

{\it HE\,1029$-$1401. --}
There are high S/N ($\sim$28) \STIS-G140M data for this bright target, but no
\FUSE\ observation was done. Three \Lya\ lines are found, at 2004, 2457 and
4567~\kms, as reported by Penton et al.\ (2004). The middle one of these can be
associated with the galaxy 6dF\,J1033307$-$144736 (\vrb=2475, 10.1, 427). For
the other two the nearest galaxies have impact parameter $>$1~Mpc:
MCG$-$2-27-1 (\vrb=2028, 13.9, 1065) and MCG$-$2-27-9 (\vrb=4529, 377, 1035)
(triangles in Fig.~\Fassoc(19)/(21)).

{\it HE\,1143$-$1810. --}
Several galaxies have impact parameter $<$1~Mpc, concentrating around three
velocities. First there are ESO\,571-G18 (\vrb=1391, 7.5 368), [KKS2000]25
(\vrb=1227, 7.4, 437) and NGC\,3887 (\vrb=1209, 20.2, 596). A second group is
formed by ESO\,572-G06 (\vrb=1737, 9.9, 811), ESO\,572-G09 (\vrb=1737, 12.9,
884~kpc), and ESO\,572-G07 (\vrb=1466, 9.3, 976) which are part of the LGG\,263
group. Finally, there is ESO\,571-G16 (\vrb=3637, 26.8, 851). There is no \Lyb\
detected, so we list a non-detection for each of these galaxies in Table~\Tres.
There is no data allowing us to search for \Lya\ lines.

{\it HE\,1228+0131. --}
This is a sightline with low S/N \STIS-E140M and \FUSE\ data ($\sim$5), that
passes through the Virgo cluster. At velocities below 2000~\kms, galaxies in two
groups lie near it. For \vgal=700--1200~\kms\ the nearest galaxy is part of
LGG\,292: MCG0-32-16 with \vrb=1105, 6.3, 131. Although five more small galaxies
have smaller impact parameter, Table~\Tres\ lists NGC\,4517 (\vrb=1121, 53.8,
383), because it is so large. No absorption is found in the 700--1200~\kms\
velocity range.
\par There are 36 galaxies in the velocity range 1250--1850~\kms, which spans
the velocities of the galaxies in the LGG\,287 group (\vavg=1655~\kms). Two
large galaxies are among the five with the lowest impact parameters: NGC\,4536
(\vrb=1804, 32.9, 338) and NGC\,4517A (\vrb=1530, 32.3, 466). No other large
galaxy has impact parameter $<$700~kpc (see Fig.~\Fassoc(22)). The two
absorption lines at 1482 and 1700~\kms\ are listed as generic group absorption
in Table~\Tres. Both were previously reported by Penton et al.\ (2000). These
authors actually listed three lines, at 1666, 1745 and 1860~\kms, but the
spectrum is too noisy to support splitting the strong \Lya\ line into separate
components at 1666 and 1745~\kms, and the claimed 1860~\kms\ detection may well
be noise.  The latter system has a very strong \Lya\ line, and \Lyg\ through at
least \Lyz\ are seen, as are \CIII\ and \CI, but not \OVI, although that may be
because of the low S/N ratio of the data. The \Lyb\ line of this system is
contaminated both by Milky Way \OVIa\ and by intrinsic \SVI\ absorption.
\par Rosenberg et al.\ (2003) discussed this sightline, analyzing the metal-line
absorption and estimating absorber-sizes. They also considered the galaxies with
the lowest impact parameters and concluded that it is more likely that they are
associated with gaseous filaments than with individual galaxies.
\par There is a third \Lya\ line at 2306~\kms, which can be associated with
UGC\,7625 (\vrb=2234, 9.6, 339), one of eight galaxies with \ip$<$1~Mpc and
\vgal=2100--2500~kms, but one of only two with \ip$<$550~kpc). \Lyb\ is also
seen for this system. Formally, the dwarf [ISI96]1228+0116 (\ip=182~kpc) is the
nearest galaxy.

{\it HS\,0624+6907. --}
Two galaxies are known near this relatively high extinction sightline: UGC\,3580
(\vrb=1201, 18.0, 729; to the north) and UGC\,3403 (\vrb=1264, 12.6 932; to the
west). No detections are found below 6000~\kms, although the \STIS-E140M data
are relatively noisy.
\par Aracil et al.\ (2006) make a more extensive study of the spectrum of this
target, listing just one detection with $v$$<$6000~\kms. However, we do not
confirm that there is a line at 5262~\kms, for which Aracil et al.\ (2006)
quoted an equivalent width of 41\E10~\mA.

{\it HS\,1543+5921. --}
This object is a $z$=0.807 QSO lying directly behind the dwarf galaxy
SBS\,1543+593, which is a member of the GH152 (=LGG402) group (\vavg=2828~\kms).
Several papers have been written on this pair, including Bowen et al.\ (2001)
and Bowen et al.\ (2005), who reported on the absorption lines. Chengalur \&
Kanekar (2002) and Rosenberg et al.\ (2006) showed an \HI\ map of SBS\,1543+593.
With \ip=0.3~kpc, it is the closest coincidence in our sample.

{\it IRAS\,09149$-$6206. --}
Near this low galactic latitude sightline lie nine galaxies with velocities
between 1900 and 3200~\kms. For seven of these, the possible associated \Lyb\
absorption is hidden by the Milky Way \CII\lm1036.337 or intrinsic
\CIII\lm977.020 absorption. Limits on \OVIa\ can be set for five galaxies, but
in two cases they require orbital-night-only data. No detections are found.

{\it IRAS\,F22456$-$5125. --}
This sightline shows eight intrinsic systems, whose \Lyd\ and \Lye\ lines fall
between 1025 and 1045~\AA. Fortunately, they are not located near the velocities
of the only two known galaxies with \ip$<$1~Mpc: ESO\,238-G05 (\vrb=706, 5.9,
616) and ESO\,149-G03 (\vrb=594, 3.3, 895). No absorption is seen, and the
\ip/\Rgal\ ratio is high enough only for ESO\,238-G05 to list it in Table~\Tres.

{\it MCG+10-16-111. --}
This is a relatively low redshift (\vgal=8124~\kms) Seyfert galaxy that lies in
the direction of several galaxy groups. Thus, \NED\ includes 94 galaxies with
\vgal\ between 400 and 6000~\kms\ and impact parameter $<$1~Mpc (26 of which are
listed in the RC3). For the discussion below, these are divided into several
groupings, based on their group membership, velocity, and position relative to
MCG+10-16-111. With nine \Lya\ lines at $v$$<$5500~\kms, this sightline is the
most complicated in our sample. Bowen et al.\ (2002) originally observed the AGN
and discussed the detections. They also included an image that shows several of
the galaxies near MCG+10-16-111. However, our interpretation of the associations
between the \Lya\ absorptions and these galaxies differs in some details from
that presented by Bowen et al.\ (2002).
\par Five of the galaxies have \vgal=400--900~\kms. One of these is the large
galaxy NGC\,3556 (\vrb=695, 25.8, 462). This is the galaxy most likely
associated with the \Lya\ line at 942~\kms\ (see first panel for MCG+10-16-111,
Fig.~\Fassoc(25)). We further list a non-detection for UGC\,6249 (\vrb=1058,
8.1, 620), the only other galaxy with $v$$\sim$1000 and \ip/\Rgal$<$125. We
choose to associate the detection with NGC\,3556 even though the difference in
velocity between with the absorber is smaller for UGC\,6249, because NGC\,3556
is much larger and has much smaller impact parameter.
\par Twelve galaxies with velocities between 1000 and 1500~\kms\ are members of
the LGG\,244 group (\vavg=1230~\kms) and have \ip$<$1~Mpc. The sightline does
not pass between the galaxies of this group, however; the group galaxy with the
smallest impact parameter is CGCG\,291-76 (\ip=616~kpc). There is no \Lya\ line
with \vgal\ in the velocity range spanned by the galaxies in LGG\,244, so a
non-detection is listed for this group in Table~\Tres.
\par The sightline also goes through the LGG\,234 group (\vavg=1692~\kms), which
has thirty galaxies with \ip$<$1~Mpc. The ones included in the RC3 are shown by
filled squares in the LGG\,234 panel of Fig.~\Fassoc(26). Most of these group
galaxies are small. Bowen et al.\ (2002) associated the strong \Lya\ line seen
at 1654~\kms\ with NGC\,3619 (\vrb=1542, 21.2, 145, larger of the two filled
squares at \ip$<$200~kpc), which is the nearest large group galaxy. There are
also seven dwarfs (\Rgal$<$5~kpc) near NGC\,3619 with impact parameters ranging
from 94 to 248~kpc, as well as UGC\,6304 (\vrb=1762, 11.3, 163, smaller filled
square with \ip$<$200~kpc). NGC\,3619 and UGC\,6304 lie in opposite directions
from MCG+10-16-111, and have velocity differences of 112 and $-$108~\kms\ with
the absorption lines, so there is no real reason to preferentially associate
either galaxy with the absorption line. In Table~\Tres\ the \Lya\ line at
1654~\kms\ is listed as generally associated with the LGG\,234 group.
\par The two absorption lines at 2022 and 2136~\kms\ are listed under NGC\,3625
(\vrb=1966, 18.1, 190) and NGC\,3613 (\vrb=2054, 35.3, 41). Both galaxies are
members of the LGG\,232 group. It is likely that at least one of the two lines
is associated with NGC\,3613, which has the fourth smallest impact parameter for
any galaxy in our sample. It is quite possible that both lines originate in the
NGC\,3613 halo, but there is no a-priori reason to exclude associating one with
NGC\,3625. Other, smaller group members lie at 203~kpc (UGC\,6344) and 506~kpc
(NGC\,3669). In addition there are many dwarfs with impact parameter $<$1~Mpc.
\par In addition to the three galaxy groups listed above, there are several more
galaxies with velocities between 2500 and 7000~\kms. Table~\Tres\ includes
UGC\,6335 (\vrb=2927, 20.6, 957), CGCG\,291-61 (\vrb=3188, 14.4, 367), and
MCG+10-16-118 (\vrb=5357, 16.6, 208). The last two of these can be associated
with \Lya\ lines at 3113 and 5363~\kms, while the other one yields a
non-detection. Figure~\Fassoc\ includes a panel for the 3113~\kms\ \Lya\ line
(Fig.~\Fassoc(28)), but not for the one at 5363~\kms, as it has $v$$>$5000~\kms.
\par Finally, there are three more features that might be \Lya\ at 3541, 3792
and 4043~\kms, even though there are no galaxies known with such velocities
within 2.5~Mpc. These are 3--5$\sigma$ detections, but they seem secure, unless
they are blueshifted absorption lines intrinsic to MCG+10-16-111
($v$=8124~\kms).

{\it MRC\,2251$-$178. --}
This target is surrounded by three galaxies with \vgal$\sim$3270~\kms, lying
toward the south (ESO\,603-G27, \vrb=3267, 15.6, 322), north (ESO\,603-IG23,
\vrb=3282, 12.5, 412), and east (MCG$-$3-58-13, \vrb=3271, 10.0, 846). A
two-component \Lya\ line is present at 3212 and 3046~\kms, previously listed by
Penton et al.\ (2004). In Table~\Tres\ these are listed as associated with the
nearest two galaxies, while an upper limit is given for MCG$-$3-58-13. The two
associations are shown in a single panel in Fig.~\Fassoc(33), since it is not
obvious which line should go with which galaxy. Danforth et al.\ (2006) claimed
a 29\E16~\mA\ \OVI\ absorber at 3205~\kms, but we cannot confirm this, instead
setting an upper limit of 27~\mA. As can be seen from Fig.~\Fspectra, the wiggle
in the spectrum that Danforth et al.\ (2006) probably identified as \OVIa\ is
more likely to be due to the \FUSE\ detector flaw. We cannot actually check
this, since Danforth et al.\ (2006) do not show their version of the data.
\par ESO\,603-G31 (\vrb=2271, 9.1, 422) has an associated \Lya\ line at
2265~\kms. This \Lya\ line was listed as two detections by Penton et al.\
(2004), though the rms is not high enough to justify splitting this absorption
line. There is also a 3$\sigma$ feature that can be interpreted as \OVIb\ at
2283~\kms. Unfortunately, the corresponding \OVIa\ line is hidden by geocoronal
\OI* emission, which is present even in the orbital-night only data. This
feature cannot be \Lyb\ at higher velocity, as there is no corresponding \Lya\
line, so \OVI\ is the most likely interpretation. Danforth et al.\ (2006) did
not list this feature.
\par Finally, there is a feature (also listed by Penton et al.\ 2004) that is
best interpreted as a weak \Lya\ line at 4371~\kms. The nearest galaxy with
similar velocity (NGC\,7381, \vgal=4521~\kms, see Fig.~\Fassoc(34)) has high
impact parameter: 2470~kpc.

{\it Mrk\,9. --}
Several galaxies in the group LGG\,143 (\vavg=3420~\kms) lie within 1~Mpc of
this sightline: UGC\,3943 (\vrb=3527, 24.8, 422), UGC\,3897 (\vrb=3529, 19.4,
877), UGC\,3885 (\vrb=3809, 15.3, 904) and UGC\,3855 (\vrb=3167, 29.2, 950). No
absorption is identified, although no \Lya\ data is available and \Lyb\ at the
velocities of these galaxies would be hidden by Galactic \H2. Absorption is also
absent near 1092~\kms, the velocity of UGC\,4121 (\ip=895~kpc).

{\it Mrk\,106. --}
This sightline passes within a few hundred kpc of the group GH\,44 (\vavg=619).
Table~\Tres\ includes UGC\,4879 (\vrb=600, 2.8, 266), and NGC\,2841 (\vrb=638,
13.9, 453), which is the nearest large galaxy in the group. The RC3 includes
many other small (\Rgal$<$5~kpc) group galaxies with \ip$<$1~Mpc. Unfortunately,
any possible \Lyb\ absorption is hidden in Galactic \OI* emission, which is
still present in the orbital-night-only data, while any \OVI\ is hidden by what
appears to be \Lyb\ at 2407~\kms.
\par If there are any \HI, \OVI\ and \CIII\ absorption lines associated with
CGCG\,265$-$14 (\vrb=3334, 13.1, 772) or UGC\,4984 (\vrb=3386, 15.5, 881), they
are invisible, as they are all hidden by Galactic lines.
\par A feature is seen at 1033.94~\AA\ that is best interpreted as intergalactic
\Lyb\ at 2407~\kms, although it could possibly be \OVIa\ at 586~\kms. A spectrum
containing 1220~\AA\ is needed to resolve this issue. UGC\,4800
(\vgal=2433~\kms) has similar velocity, and is included in Table~\Tres, even
though its impact parameter is as large as 1030~kpc.

{\it Mrk\,110. --}
There is a \FUSE\ spectrum for this target, but it has an S/N ratio of about 1,
so it is not useful. The \STIS-G140M spectrum shows two clear and one possible
features. The clearest feature is alnost certainly \Lya\ at 3579~\kms. The
nearest galaxy with similar velocity is UGC\,4984 (\vgal=3386~\kms), but it has
a very large impact parameter (\ip=1975~kpc). UGC\,4934 could have smaller
impact parameter ($\sim$1400~kpc), if its velocity were like that of the
absorber. There is also a feature at the wavelength where the corresponding
\SiIII\ line is expected. However, this is more likely to be another \Lya\ line
at 1297~\kms. The nearest galaxy with known velocity within 200~\kms\ is
UGC\,5354, which has impact parameter 1700~kpc.
\par There is one further possible feature, which could be a broad, but shallow
line at 2247~\kms. However, it is only 2.5$\sigma$ and we do not consider it
to be real.
\par There is no detectable absorption near 638~\kms, the velocity of NGC\,2841,
which has an impact parameter of only 144~kpc. This was previously reported by
C\^ot\'e et al.\ (2005). However, the detection limit for this line is only
108~\mA, because of the damping wings of the Galactic \Lya\ absorption.

{\it Mrk\,205. --}
Mrk\,205 lies behind the disk of the nearby galaxy NGC\,4319 (\vgal=1357~\kms).
A strong \Lya\ line is detected at 1289~\kms. By itself the \Lya\ spectrum
suggests two components, but this must to be due to some hot pixels near
1260~\kms, since the \Lyb\ absorption shows a single saturated component. There
is also absorption in many other high- and low-ionization species, including
\CIV, \SiIV, \CIII, \SiIII, \NII, \SIII, \OI, \SiII, \CII, \NI, \AlII\ and
\FeII, as well as \H2. \OVI\ and \NV\ are not detected, however, although the
\OVIa\ line would be hidden by Galactic \CII, and the \OVIb\ line is near
geocoronal \OI* emission, so the 37~\mA\ upper limit is derived from the
night-only data. Bowen \& Blades (1993) originally reported on the \MgII\
absorption from NGC\,4319, but the full \STIS-E140M and \FUSE\ spectra of this
sightline have not yet been analyzed in detail.
\par NGC\,4319 is not the only galaxy with low impact parameter to Mrk\,205, as
can be seen from the open square in Fig.~\Fassoc(38). For instance, three other
large galaxies and two dwarfs have impact parameters $<$200~kpc: NGC\,4291
(\vgal=1757~\kms) is at 51~kpc, NGC\,4386 at 136~kpc, NGC\,4363 at 148~kpc,
Mailyan\,68 at 71~kpc and CGCG\,352-27 at 78~kpc. Based on their velocities, the
large galaxies with impact parameter $<$1~Mpc can be divided into two groups:
four galaxies with \vgal\ between 1187 and 1357~\kms\ and ten galaxies with
\vgal=1570 to 1980~\kms. The latter are all members of the GH\,107 group, so a
second entry for Mrk\,205 is given in Table~\Tres\ for the velocity of
NGC\,4291, the group member with the lowest impact parameter. This velocity is
sufficiently different from that of the \Lya\ lines that any line associated
with NGC\,4291 would appear in the velocity range $\sim$1500 to 2000~\kms.
Unfortunately, the only upper limits that can be set are for \Lya\ and \OVIa.
\par A third entry (an upper limit) for Mrk\,205 is given in Table~\Tres\ for
UGC\,7226 (\vgal=2267~\kms), which is not part of the GH\,107 group.

{\it Mrk\,279. --}
This is a sightline with very high S/N -- $\sim$20 at $\lambda$$<$1000, $\sim$40
for 1000--1180~\AA, and $\sim$25 at $\lambda$$>$1200~\AA. There are three large
galaxies with \vgal$<$6000~\kms, but all have \ip$>$600~kpc: UGC\,8737
(1873~\kms), NGC\,5832 (453~\kms), FGC\,1680 (3865~\kms). No \Lya\ or \OVI\ are
detected at \vgal$<$6000~\kms. Penton et al.\ (2000) listed six significant and
two $<$3$\sigma$ detections at \vgal=5000--8000~\kms, based on a \GHRS\
spectrum. However, \STIS-E140M data show that the claimed detections at 6372,
6445 and 6925~\kms\ are much more likely to be intrinsic \SiIII\ lines, as there
are clear \Lya\ and \CIII\ absorptions at the corresponding velocities. Their
5631~\kms\ \Lya\ line is really \NV\ in complex~C (see Fox et al.\ 2004), while
the 5246 and 5486~\kms\ features are not seen in the \STIS\ spectrum. This
leaves just one confirmed \Lya, at 7779~\kms.

{\it Mrk\,290. --}
This is a sightline with five groups of entries in Table~\Tres. The galaxies
with similar velocities usually lie in different directions, while in general
their velocities differ by a few hundred \kms. Therefore, all galaxies are
listed separately in Table~\Tres.
\par To the north of Mrk\,290 lie several galaxies with \vgal$\sim$700~\kms,
which are in the LGG\,396 group. NGC\,5963 (\vrb=656, 12.3, 307) is the nearest.
It is associated with a \Lyb\ line seen at 720~\kms\ in the orbital-night-only
data (see filled square in Fig.~\Fassoc(39)). Four other large group galaxies
with similar velocity are listed in Table~\Tres, though all have impact
parameters $>$734~kpc (NGC\,5907, NGC\,5879, NGC\,5866B, NGC\,5866; see open
squares with plusses in Fig.~\Fassoc(39)). Eight dwarfs (not listed in
Table~\Tres) have impact parameters between 307 and 550~kpc, but the association
of the \Lyb\ line with NGC\,5963 is comparatively secure.
\par There is no \Lya\ data for the velocity of NGC\,5981 (\vrb=1764, 22.5,
725), while \Lyb, and \OVI\ are hidden by Galactic \OVI\ and \H2, respectively,
\par This sightline also passes within 0\fdg5 of the group GH\,158 (also known
as LGG\,402), which has \vavg=2882~\kms. In the RC3 the nearest of these group
galaxies is NGC\,5987 at 424~kpc, which has a very large diameter
(\Rgal=52~kpc). \NED\ includes nine more small galaxies lying next to the group
having velocities between 2900 and 3350~\kms, and impact parameters $<$1~Mpc.
2MASX\,J153514.22+573052.9, CGCG\,297-17 and SDSS\,J153733.02+583446.7 are large
enough (\ip/\Rgal$<$80) to merit separate entries in Table~\Tres. Pisano et al.\
(2004) obtained VLA and DRAO \HI\ 21-cm data for the field around Mrk\,290, but
failed to detect \HI\ emission from any of the group galaxies. They also
discussed a \FUSE\ spectrum of Mrk\,290, based on the first 13~ks observation.
After this paper was published, another 92~ks of data was taken toward Mrk\,290,
which turned out to be almost three times brighter as before during the longest
(54~ks) individual exposure. With these much improved data, there appears to be
\OVI\ at 3073~\kms, seen in both the \OVIa\ and the \OVIb\ lines, although the
\OVIb\ line only shows up as a wing in Galactic \ArI. The equivalent width given
in Table~\Tres\ is double the value measured using only the positive-velocity
half of the line. Unfortunately, the corresponding \Lyb\ line is hidden in
Galactic \CII, while there is no data for \Lya. This \OVI\ feature thus remains
rather uncertain, but it is listed under the entry for NGC\,5987 (\ip=424~kpc).
\par The velocity of SDSS\,J153802.76+573018.3 (\vgal=3525~\kms) is similar to
that of the GH\,158 galaxies, but sufficiently different that this galaxy is
listed separately.
\par The \GHRS\ spectrum of this target only allows a search for \Lya\ at
velocities above 4000~\kms. One weak line is found at 4638~\kms, which was
included by Penton et al.\ (2000). The nearest known galaxies with similar
velocity are NGC\,5971 (\vrb=4306, 29.5, 1586, triangle in Fig.~\Fassoc(41)) and
SBS\,1533+574B (\vrb=4287, 9.1, 721, open circle in Fig.~\Fassoc(41)).

{\it Mrk\,335. --}
There are three \Lya\ detections at $v$$<$6000~\kms\ toward this target. The two
at 1954 and 2286~\kms\ were previously reported by Penton et al.\ (2000), based
on a \GHRS-G140M spectrum. We observed this sightline with the \STIS-E140M,
finding a similar equivalent width for the feature at 1954~\kms\ (229\E12~\mA\
versus 229\E30~\mA\ in Penton et al.), while for the other feature we find
114\E17~\mA, whereas Penton et al.\ (2000) quoted 81\E26~\mA. The stronger \Lya\
also shows a weak \Lyb\ absorption, adjacent to the strong Galactic L(6-0) R(4)
\H2\ line at 1032.520~\AA. The 2$\sigma$ feature at 4268~\kms\ that Penton et
al.\ (2000) listed is not confirmed by the E140M spectrum, however.
\par The correspondence between \Lya\ absorptions and galaxies is complex for
this sightline. The only simple association is for the component at 2286~\kms,
which can be associated with NGC\,7817 (\vrb=2308, 30.4, 395, Fig.~\Fassoc(44)).
\par For the (stronger) feature at 1954~\kms, the existence of better galaxy
data than is usually the case makes the choice of association more complicated.
On the other hand, as can be seen in Fig.~\Fassoc(43), there are no large
galaxies with low impact parameter near this velocity. This feature might be
associated with NGC\,7817 (see large plussed circle in Fig.~\Fassoc(43)),
although a \Dv\ of 355~\kms\ would be much higher than is normally found. The
only other non-dwarf galaxies with similar velocity within 1~Mpc are
ESDO\,F538-2 (\vrb=2175, 7.2, 320) and NGC\,7798 (\vrb=2404, 12.4, 915). For
most sightlines this is where the search would stop, and we would associate the
absorption with ESDO\,F538-2, even though \Dv=219~\kms. However, van Gorkom et
al.\ (1996) did a deep search for \HI\ clouds near Mrk\,335, and found the tiny
(\Rgal=0.7~kpc) dwarf [vCS96]\,000254.9+195654.3 at a velocity of 1950~\kms\ and
with impact parameter of only 78~kpc (see tiny black circle in
Fig.~\Fassoc(43)). As \ip/\Rgal$\sim$110 for this galaxy, only a factor two
larger than the value of 43 for ESO\,F538-2, but \Dv=7~\kms, Table~\Tres\ lists
the absorption at 1954~\kms\ as being associated with
[vCS96]\,000254.9+195654.3, while non-detections are listed for ESDO\,F538-2 and
NGC\,7798.
\par In the velocity range 713 to 1108~\kms\ there are 17 known galaxies with
impact parameter $<$1~Mpc, 10 of which are included in the RC3. However, the two
largest (NGC\,100 and NGC\,7814) have large impact parameter (814~kpc). For all
other galaxies the ratio \ip/\Rgal\ is larger than 90, and all have
\Rgal$<$7~kpc. The smallest impact parameter (170~kpc) is for ESDO\,F538-1, a
galaxy with \Rgal=1.1~kpc. UGC\,47 (\Rgal=3.3~kpc) has the next smallest \ip\
(447~kpc). No absorption feature is seen in the velocity range 700 to 1150~\kms,
but there is a weak feature at 1308~\kms\ (lying outside the spectral range
analyzed by Penton et al.\ 2000). However, there are no known galaxies within
1~Mpc with \vgal\ between 1150 and 1970~\kms. In Table~\Tres\ the galaxy listed
for this feature is UGC\,12893 (\vrb=1108, 6.0, 697), which has the smallest
velocity difference. However, this is rather ambiguous assignment, as can be
seen in Fig.~\Fassoc(42).
\par Finally, UGC\,44 is a large (\Rgal=21.5~kpc) galaxy with \vgal=5936~\kms,
but there is no \Lya\ near that velocity, although there is a strong feature at
6280~\kms. Because of the relatively small velocity difference between this
feature and the velocity of Mrk\,335 itself (\vgal=7730~\kms), it is not listed
in Table~\Tres.

{\it Mrk\,421. --}
A weak but clear \Lya\ line is detected at 3007~\kms\ in this sightline. This
detection was first reported by Penton et al.\ (2000). Savage et al.\ (2005a)
showed that there is no \OVI\ absorption at this velocity. They also studied in
detail the galaxies in the field near this sightline, showing that there are no
known galaxies with similar velocity within 1~Mpc, and just two dwarfs within
3.5~Mpc, one of which is HS\,1059+3934 (\vrb=3274, 3.8, 1041).

{\it Mrk\,477. --}
Although the S/N ratio of this spectrum near \OVIa\ is about 10, the continuum
of this target shows strong fluctuations that are probably due to intrinsic
emission lines, making the continuum fit uncertain. Furthermore, the SiC data
are too noisy to be useful, and there is no \Lya\ data. The only measurements
that generally can be made are $\sim$40~\mA\ upper limits on \Lyb\ and
\OVIb.
\par The galaxies near this sightline come in three groups. At velocities
between 750 and 950~\kms\ there are many dwarfs and one substantial galaxy with
large impact parameter (NGC\,5866, \vrb=769, 16.7, 930). The nearest dwarf
(SDSS\,J144303.81+535457.5, \vrb=838, 1.5, 138) has \ip/\Rgal\ $<$100, so it is
included in Table~\Tres. Six galaxies lie between 2056 and 2181~\kms\ of which
UGC\,9452 (\vrb=2173, 17.1, 278), NGC\,5687 (\vrb=2119, 23.1, 743) and
KUG\,1437+524 (\vrb=2181, 9.6, 743) are included in Table~\Tres. Finally, two of
the three galaxies near 3300~\kms\ are in Table~\Tres: NGC\,5751 (\vrb=3242,
21.6, 419) and SBS\,1436+529 (\vrb=3389, 10.0, 761).

{\it Mrk\,478. --}
A strong \Lya\ line is seen at 1573~\kms\ toward Mrk\,478, which was included by
Penton et al.\ (2004), although their reported equivalent width is smaller than
our measurement (194\E31~\mA\ vs 254\E14~\mA). Based on the equivalent and
velocity width of the \Lya\ line, an $\sim$100~\mA\ \Lyb\ line is expected.
There indeed seems to be an 78~\mA\ feature at the right place, except that it
would just be a 2$\sigma$ detection. The nearest large galaxy is NGC\,5727
(\vrb=1491, 16.1, 645). As can be seen from the filled circle in
Fig.~\Fassoc(46), this is not the only possible galaxy that can be associated
with the absorption line. However, it is by far the largest galaxy. Two of the
smaller galaxies have \ip/\Rgal$<$125 and are listed separately in Table~\Tres:
UGC\,9519 (\vrb=1692, 5.3, 624) and UGC\,9562 (\vrb=1253, 6.4, 748). However,
SDSSJ144003.48+340559.6 (\vrb=1492, 3.1, 605) is not included, as it is a dwarf
close to NGC\,5727.
\par A final galaxy for which a non-detection is listed in Table~\Tres\ is
UGC\,9540 (\vrb=802, 4.3, 400). Two more similar sized galaxies with similar
velocities are not listed because they have \ip/\Rgal$>$125.

{\it Mrk\,501. --}
There are two galaxies with similar velocity and impact parameter that may be
associated with the \Lya\ line at 4593~\kms: CGCG\,225-6 (\vrb=4648, 10.7, 517)
and NGC\,6257 (\vrb=4692, 18.0, 521). The first of these lies to the north of
Mrk\,501, the second to the south. In Table~\Tres\ the \Lya\ line is listed
under the larger galaxy, NGC\,6257, while a non-detection is listed for
CGCG\,225-6. This choice is arbitrary, but it allows us to account for having
one detection for two galaxies. Penton et al.\ (2000) also reported this
absorption line. Four more galaxies have \ip$<$1~Mpc: UGC\,10625 (\vrb=2048,
11.7, 700), NGC\,6239 (\vrb=922, 11.3, 807), NGC\,6255 (\vrb=918, 15.2, 854) and
NGC\,6907 (\vrb=850, 12.2, 895). Separate non-detection entries are given each
of these.

{\it Mrk\,509. --}
A strong \Lya\ line is detected at 2545~\kms, previously reported by Penton et
al.\ (2000). The \FUSE\ spectrum shows the corresponding \Lyb\ line. The nearest
galaxy with known velocity between 2000 and 3000~\kms\ is MCG$-$1-54-3
(\vgal=2231~\kms) at \ip=4834~kpc, so this seems to be a clear case of an
absorber in a void. However, there are five galaxies with unknown velocity that
potentially could have impact parameter between 1400 and 2500~kpc, {\it if}
their velocity is similar to that of the absorber: [RC3]A2052$-$1056,
[RC3]A2030$-$0838, MCG$-$1-52-4, MCG$-$2-52-17 and MCG$-$2-53-13. On the other
hand, the likelihood that this is actually the case is low.

{\it Mrk\,586. --}
Absorption is absent for NGC\,851 (\vrb=3111, 11.9, 967). On the other hand,
what may be \Lyb\ is detected at 1464~\kms. No \Lya\ data is available to
confirm this identification, and the redshift of Mrk\,586 is sufficiently high
(0.1553) that it could be a higher-redshift ionic line, although the \FUSE\
spectrum does not show evidence for a redshifted system. If it is \Lyb, the
nearest galaxy with similar velocity has impact parameter $>$1~Mpc (NGC\,864,
\vgal=1560~\kms, \ip=1240~kpc).

{\it Mrk\,734. --}
This is one of a few sightlines that passes near the center of a galaxy group
with many galaxies, namely GH\,78. Geller \& Huchra (1983) include nine galaxies
in this group, eight of which have velocities ranging from 628 to 1191~\kms,
with one standing out at 1726~\kms. Without this last one, the average velocity
of the group galaxies is 921~\kms. The galaxy with the lowest impact parameter
to Mrk\,734 is NGC\,3666 (\vrb=1062, 12.7, 133), while the other seven original
group galaxies have \ip=230 to 603~kpc. \NED\ includes another 13 small galaxies
inside the GH\,78 velocity range that can be considered part of the group.
\par Although the \FUSE\ spectrum of this target has relatively low S/N ratio
($\sim$5.5), two \Lyb\ absorption lines are clearly detected, at 478 and
757~\kms. \Lyg\ may also be present at 745~\kms. There is a 2$\sigma$ feature
that may be \OVI\ at $\sim$720~\kms, but the S/N ratio of the spectrum is
insufficient to confirm it. The two \Lya\ lines are listed as generic group
detections in Table~\Tres.
\par There are four other galaxies near Mrk\,734 that are not in the GH\,78
group for which non-detections are listed separately: 2MASX\,J112139.77+112924.2
(\vrb=5944, 6.2, 388),
SDSS\,J111938.66+112643.3 (\vrb=3054, 5.3, 500), and IC\,2763 (\vrb=1574, 10.4,
617).

{\it Mrk\,771. --}
This sightline is included in the sample because the \STIS-G140M data are good,
although the \FUSE\ spectrum has low S/N (4.6).
\par There are three similar-strength \Lya\ absorptions, at 1184, 1891 and
2557~\kms. All three were reported by Penton et al.\ (2004). The 2557~\kms\ line
was also reported by C\^ot\'e et al.\ (2005). It can clearly be associated with
UGC\,7697 (\vrb=2536, 24.6, 139; see C\^ot\'e et al.\ 2005), which is the only
galaxy with velocity between 2000 and 3000~\kms\ and impact parameter $<$1~Mpc
(filled circle in Fig.~\Fassoc(53)).
\par There are two galaxies with \vgal\ between 1700 and 2100~\kms\ (see
Fig.~\Fassoc(52)). The small galaxy KUG\,1229+207 (\vrb=1921, 6.5, 184) has
small impact parameter, while the large galaxy NGC\,4450 (\vrb=1956, 23.4, 851)
is more distant. Table~\Tres\ lists the \Lya\ absorption as associated with
KUG\,1229+207.
\par The most difficult to associate is the 1184~\kms\ absorption. There are
five large galaxies in the LGG\,289 group that have \vgal\ within 250~\kms\ from
this detection (\vgal=1027--1403~\kms, open symbols in Fig.~\Fassoc(51)). They
have \ip=502, 666, 807, 846 and 962~kpc. The nearest galaxy (with
$\Delta$\vgal=259~\kms), however, is IC\,3436 (\vrb=925, 1.8, 183), but this is
a small dwarf not in the LGG\,289 group. Table~\Tres\ lists the absorption line
as associated with NGC\,4561 (the \ip=502~kpc group galaxy), but this is one of
the most uncertain associations in the table.

{\it Mrk\,817. --}
This sightline passes through the edge of the GH\,144 group (\vavg=1967~\kms),
with the closest large galaxy being UGC\,9391 (\vrb=1921, 14.5, 308). Three
other large galaxies are listed in the RC3 with velocities between 1900 and
2300~\kms\ and \ip$<$1~Mpc (UGC\,9477, NGC\,5667 and NGC\,5678). \NED\ gives
another six smaller such galaxies, two of which have impact parameter
$<$308~kpc. The nearest is the dwarf PWWF\,J1437+5905 (\vrb=2233, 1.0, 191)
which was identified by Pisano et al.\ (2004) in a VLA \HI\ map of the field
around Mrk\,817. A somewhat larger dwarf (SDSS\,J143903.89+544717.6, \vrb=2134,
4.0, 202) has similar impact parameter.
\par We find four \Lya\ features in this sightline, at 1922, 2085, 4670 and
5081~\kms\ (confirming the entries in the list of Penton et al.\ (2000)). The
first two can be associated with the GH\,144 group, and are listed under the
entries for SDSS\,J143903.89+544717.6 and UGC\,9391 (see Fig.~\Fassoc(54)/(55)),
with velocities of 1922 and 2085~\kms. We find accompanying \Lyb\ absorption at
2081~\kms.
\par For the feature at 4670~\kms\ we cannot find any galaxies with similar
velocity at impact parameter $<$5~Mpc. This feature is therefore listed as
intergalactic in Table~\Tres.
\par The feature at 5081~\kms\ has extremely strong \OVI\ associated with it,
and it is likely to be gas expanding away from Mrk\,817 at 4500~\kms.

{\it Mrk\,876. --}
There are three \Lya\ features at $v$$<$6000~\kms\ in this sightline. The strong
one at 3481~\kms\ is associated with UGC\,10294 (\vrb=3516, 27.6, 282;
Fig.~\Fassoc(58)). The corresponding \Lyb\ line contaminates Galactic \OVIb,
which is indicated by the fact that the $N_a(v)$ profiles of the two Galactic
\OVI\ lines do not match (see Wakker et al.\ 2003). \Lyg\ may be seen at the
2$\sigma$ level. A 4$\sigma$ (18\E4\E7~\mA) feature is also present where
redshifted \OVIa\ is expected, but the corresponding \OVIb\ line is hidden by
\H2\ absorption. Danforth et al.\ (2006) gave an upper limit of 16~\mA\ for
associated \OVI\ absorption. There are only two other galaxies near this
velocity: NGC\,6135, and UGC\,10376, but they have impact parameters of 630 and
799~kpc. Note that the RC3 gives a velocity of 822~\kms\ for UGC\,10376, but
according to Schneider et al.\ (1992) it is 3246~\kms.
\par The other two \Lya\ lines, at 936 and 1109~\kms, appear to be associated
with NGC\,6140 (\vrb=910, 27.1, 206) the only large galaxy with velocity between
700 and 1375~\kms\ within 1~Mpc (see Fig.~\Fassoc(57)). The four dwarf galaxies
in this velocity range all have much larger impact parameters: UGC\,10369
(\vrb=998, 5.9, 493), UGC\,10194 (\vrb=870, 6.9, 573), CGCG319-39 (\vrb=933,
2.9, 763) and IC\,1218 (\vrb=1109, 4.8, 767). Non-detections are listed in
Table~\Tres\ for the first of these  two galaxies, as they have \ip/\Rgal$<$125.
\par The line at 1109~\kms\ is unusually narrow ($b$=14~\kms, FWHM=23~\kms), but
its identification as \Lya\ is fairly secure, since there are no other known
intergalactic absorber systems that would produce an ionic line at its
wavelength.
\par Using a G140M \STIS\ spectrum (30~\kms\ resolution) C\^ot\'e et al.\ (2005)
reported a strength of 390~\mA\ at 935~\kms\ for the \Lya\ absorber, Shull et
al.\ (2000) gave 324\E52~\mA\ at 958~\kms, whereas we find 476\E14~\mA\ at
936~\kms\ from our \STIS-E140M data. The \Lyb\ line corresponding to the \Lya\
at 936~\kms\ is strongly contaminated by two-component \H2\ L(6-0)
R(3)\lm1028.985 absorption. However, the S/N ratio in this sightline is so high
(33) that a good (two-component) \H2\ model can be made (see Wakker 2006), which
fits all uncontaminated \H2\ lines extremely well. This shows that the feature
contains 79\E6~\mA\ of \Lyb, though the systematic error is large (33~\mA).
Shull et al.\ (2000) reported 110\E50~\mA\ for this line, but that clearly
includes the \H2\ absorption.
\par In addition, there is \OVIa\ absorption, centered at 945~\kms. This line is
contaminated by \H2\ L(6-0) P(4) at 1035.7830~\AA, but the very good \H2\ model
does not account for all the absorption near this wavelength, unlike what is the
case for all other $J$=4 \H2\ lines. Danforth et al.\ (2006) reported this \OVI\
absorption, but their \H2\ model is not as precise. Consequentially, they listed
an equivalent width that is too high (26\E7~\mA). We measure it as 17\E4\E8~\mA.

{\it Mrk\,926. --}
This is the only sightline in the sample with only one known galaxy with
\vgal=400--6000~\kms\ and impact parameter $<$1~Mpc: the dwarf
SDSS\,J230556.27$-$100257.0 (\vrb=2304, 2.6, 697). Because \ip/$D$$>$125 this
galaxy is not listed in Table~\Tres. There are also no \Lya\ lines with
$v$$<$5000~\kms, as was previously noted by Penton et al.\ (2004). The \FUSE\
spectrum has low S/N and thus contributes no useful pathlength to the \Lyb\ and
\OVI\ search.

{\it Mrk\,1095. --}
Few galaxies lie near this target. UGC\,3303 (\vrb=522, 6.4, 571) has low
velocity, but the only line for which a limit can be set is \OVIa. UGC\,3258
(\vrb=2821, 8.6, 999) has large impact parameter, and no absorption is found. The
only probable \Lya\ feature is found at 4048~\kms, and this was listed by Penton
et al.\ (2000). There are some galaxies near 4000~\kms, but they have large
impact parameter. The smallest are 834 and 836~kpc for two tiny galaxies
([OHG88]\,0510-0037, [OHG88]\,0510-0036, \Rgal=1.8~kpc). The nearest large
galaxy with \Dva$<$300~\kms\ is UGC\,3262 (\vgal=4285~\kms) at 1306~kpc.

{\it Mrk\,1383. --}
This sightline is one of the few that passes between galaxies in a galaxy group
(LGG\,386 or GH\,145). The group has \vavg=1701~\kms, and 32 galaxies have
impact parameter $<$1~Mpc. No absorption is seen near this velocity, however.
The closest of nine group galaxies in the RC3 is UGC\,9348, with \ip=622~kpc.
\NED\ includes seven small (\Rgal$<$3~kpc) galaxies with smaller impact
parameter (as low as 236~kpc for 2dFGRS\,N413Z236), as well as the larger galaxy
SDSS\,J143229.08+001734.4 (\Rgal=15.4~kpc) at 602~kpc. Another 13 galaxies found
only in \NED\ have impact parameter $>$622~kpc.

{\it Mrk\,1513. --}
Like Penton et al.\ (2004), we find no \Lya\ absorption below 5000~\kms. The
only galaxy with impact parameter less than 1~Mpc is UGC\,11782 at 412~kpc. This
lack of galaxies is only partly the result of the target's low galactic latitude
-- there are relatively few galaxies in this region of the sky, which is in the
direction diametrically opposite to the Virgo cluster.

{\it MS\,0700.7+6338. --}
Like VII\,Zw\,118, this sightline passes through the LGG\,140 group
(\vavg=4404~\kms). The closest group galaxy is UGC\,3660 (\vrb=4262, 31.0, 356).
Strong \Lyb\ and probable \CIII\ absorption is seen at 4322~\kms, but \OVI\ is
absent. In Table~\Tres\ this system is listed as generally associated with the
group. There is one other galaxy within 1~Mpc that is not in the group
(UGC\,3685, \vrb=1797, 25.9, 941). However, near this velocity only \OVIb\ is
not blended with interstellar \OVI, \OI\ or \H2.

{\it NGC\,985. --}
NGC\,985 passes through the southern outskirs of the LGG\,71 galaxy group, which
has \vavg=1406~\kms. Forty eight group galaxies have impact parameter less than
1~Mpc (see Fig.~\Fassoc(61)). Eleven of these are included in the RC3, and eight
of these have diameter $>$10~kpc. Their velocities range from 1145 to 1781~\kms,
although the velocity range for the large (\Rgal$>$10~kpc) galaxies is smaller
(1241--1534~\kms). Of these, NGC\,988 (\vgal=1504~\kms) lies much closer than
other group galaxies (\ip=175~kpc, vs $>$394~kpc for the rest, see large open
square in Fig.~\Fassoc(61)). No \Lya\ line is within the velocity range spanned
by the velocities of the group galaxies, so Table~\Tres\ lists a generic group
non-detection.
\par There are two galaxies with higher velocity: DDO\,23 (\vrb=2110, 15.3, 993)
and Mrk\,1042 (\vrb=2133, 3.1, 996). Two weak (3.5 and 3.2$\sigma$), broad
absorption lines may be present at velocities of 1924 and 2183~\kms. These were
previously reported by Bowen et al.\ (2002). Both are rather shallow (10\%
depth) and the one at 1924~\kms\ may not be real. It is hard to justify an
association with NGC\,988 or the LGG\,71 group, as proposed by Bowen et al.\
(2002). NGC\,988 has low impact parameter, but the velocity difference is large
(420~\kms). This is much larger than in any other case with impact parameter
$<$1~Mpc, for which we can always find a galaxy within 320~\kms. The group
galaxy with the velocity closest to that of the absorber is called
NGC\,1052-[PBF2005]GC47 (\vgal=1781~\kms) in \NED\, but it is very small. The
nearest (in velocity) substantial group galaxy is NGC\,991 (\vrb=1534, 12.9,
475), for which \Dv=390 and 649~\kms. The velocities of the absorptions resemble
those of DDO\,23 better (differences 186 and 78~\kms), but the impact parameter
to this galaxy is large. In Table~\Tres\ both features are listed under DDO\,23,
but this is by no means a clear-cut association and may be the most debatable
one in the table.

{\it PG\,0804+761. --}
Penton et al.\ (2004) and C\^ot\'e et al.\ (2005) separately observed this
sightline with the \STIS-G140M. grating. Both reported the \Lya\ line at
1537~\kms, though there is disagreement about the equivalent width. Integrating
from 1420 to 1680~\kms, we find 114\E14~\mA. C\^ot\'e et al.\ (2005) gave
260~\mA, while Penton et al.\ (2004) split the line into two components, with a
total equivalent width of 139\E35~\mA. The evidence for a two-component
absorption line is weak, but if it is a single component, the line would have a
large, but not extreme width (178~\kms, see Sect.~\SSlinewidth). This \Lya\ line
is associated with UGC\,4238 (\vrb=1544, 16.3, 155; filled circle in
Fig.~\Fassoc(63)). Two other galaxies with similar velocity are included in
Table~\Tres, but they have much larger impact parameter (NGC\,4466, \vrb=1416,
8.4, 839) and NGC\,2591 (\vrb=1323, 18.5, 907).
\par A second \Lya\ line is seen at 1144~\kms. Penton et al.\ (2004) also
reported this feature. At the wavelength where the corresponding \OVIa\
absorption is expected there is a 47~\mA\ feature. Danforth et al.\ (2006)
listed this as a 36\E10~\mA\ \OVIa\ absorber. However, it is much more likely
that this is Galactic \CII\ at $-$140~\kms, associated with HVC complex~A, whose
edge lies about 1 degree away (see Wakker 2001). The strength of the \CII\ line
would typically correspond to a total hydrogen column density of about
2\tdex{17}\,\cmm2. Weak \HI\ absorption, on the order of a few \dex{16}\,\cmm2\
is seen in the\ FUSE\ spectrum, so this gas appears to be mostly ionized, as is
expected at these column densities. An association between this component and a
galaxy is somewhat uncertain, as can be seen from the symbols in
Fig.~\Fassoc(62). It could be associated with UGC\,4238 (\vrb=1544, 16.3,
155~kpc), but \Dv\ is too large (400~\kms). For the same reason, we do not
associate the \Lya\ with UGC\,4527 (\vrb=721, 4.7, 438), since \Dv=423~\kms.
Instead, in Table~\Tres\ UGC\,3909 (\vrb=945, 11.3, 829, \Dv=199~\kms) is listed
as the associated galaxy, in spite of the large impact parameter. Within the
range implied by all other associations. UGC\,3909 is preferred over NGC\,2591
(\vrb=1323, 18.5, 907), because it has slightly smaller impact parameter.
Furthermore, there are four galaxies with unknown velocity that could have
smaller impact parameter (UGC\,4360, UGC\,4413 at \ip$\sim$560~kpc, UGC\,4563,
UGC\,4194 at \ip$\sim$800~kpc). This is therefore one of the most uncertain
associations in Table~\Tres.
\par The weak \Lya\ at 2282~\kms\ (28\E7~\mA) was not reported by Penton et al.\
(2004), but it can be seen in their spectra. In Table~\Tres\ it is associated
with UGC\,4202 (\vrb=2296, 12.0, 875), which is not listed in the RC3.
\par At 5549~\kms\ there is a very strong \Lya, with accompanying \Lyb, though
no \OVI. No galaxy is known within 3~Mpc with \vgal\ between 5000 and 6000~\kms.
This may be because no deep searches were made in this direction.

{\it PG\,0838+770. --}
A reasonably good \FUSE\ spectrum exists for this target, which passes within
10~kpc of UGC\,4527 (\vgal=721~\kms, \Rgal=4.7~kpc), a small irregular galaxy. A
strong \Lyb\ absorption line is seen centered at 716~\kms, but \OVI\ is absent.
\par The sightline also passes through the LGG\,165 group, which has 13 galaxies
with velocities between 1257 and 1544~\kms\ within 1~Mpc. NGC\,2591 has the
lowest impact parameter (444~kpc). No associated absorption is seen. UGC\,4238
(\vrb=1544, 16.3, 803) has similar velocity as the group galaxies, but it is not
a group member, and it lies in the opposite direction seen from PG\,0838+770. It
is therefore listed separately.
\par UGC\,4623 (\vgal=2885~\kms) is the only known galaxy with
\vgal$>$2000~\kms, and has impact parameter 467~kpc, but no absorption is seen
near this velocity.

{\it PG\,0844+349. --}
No high-resolution \GHRS\ or \STIS\ spectrum exists for this sightline. The
\FUSE\ spectrum shows three features just redward of \OVIa. Two of these are
interpreted as \Lyb\ at 2260 and 2326~\kms\ and are associated with UGC\,4621
(\vrb=2306, 10.3, 372), one of the few cases that we associate two lines with a
single galaxy. Four galaxies other than UGC\,4621 have similar velocities (see
Fig.~\Fassoc(67)): HS\,0846+3522 (\vrb=2481, 0.6, 387),
SDSS\,J084619.14+351858.2 (\vrb=2368, 3.8, 391), CG222 (\vrb=2429, 1.1, 411),
and KUG\,0847+350 (\vrb=2354, 4.0, 421), but all are dwarfs with larger impact
parameter. A fourth galaxy with similar velocity (UGC\,4660; \vrb=2203, 11.6,
826) is listed as a non-detection in Table~\Tres.
\par The third feature just redward of \OVIa\ seems to be \OVI\ at 365~\kms\
that can be associated with NGC\,2683 (\vrb=410, 23.0, 250). This is supported
by the apparent \Lyb\ line detected at 351~\kms, as well as by extra (though
blended) absorption at the wavelength where \OVIb\ would be. Three additional
galaxies in Table~\Tres\ have \vgal$\sim$400~\kms\ (see Fig.~\Fassoc(66)):
[KK98]69 (\vrb=463, 6.4, 228), UGC\,4787 (\vrb=552, 6.6, 816) and UGC\,4704
(\vrb=596, 13.2, 967).

{\it PG\,0953+414. --}
Toward this sightline the RC3 includes just two galaxies with impact parameter
less than 1~Mpc -- NGC\,3104 (\vrb=612, 11.5, 296) and NGC\,3184 (\vrb=593,
23.4, 758). There is a weak feature in the damping wing of Galactic \Lya\ that
is probably \Lya\ at 621~\kms\ (see Fig.~\Fassoc(68)). In addition, a 5$\sigma$
(39$\pm$8$\pm$7~\mA) feature at 637~\kms\ seems to be \OVI, as there are no
higher redshift systems that produce absorption at that wavelength.
Unfortunately, the corresponding \OVIb\ line is only measureable in the lower
S/N night-only data; the upper limit (22~\mA) is compatible with the detection
of the \OVIa\ line, however. \par \NED\ lists twenty more galaxies with impact
parameter below 1~Mpc near this sightline, most of which are small. Of the 13
with velocity $\sim$600~\kms, 12 are KUG and SDSS galaxies with impact parameter
$>$600~kpc (i.e.\ much larger than the 296~kpc to NGC\,3104), and with
\ip/\Rgal$>$125, so none is listed in Table~\Tres. However, the table does give
upper limits on \Lya\ lines for two relatively small galaxies with
\ip/\Rgal$<$125 (KUG\,0956+420 and Mrk\,1427 at 332 and 390~kpc, respectively).
\par Also in \NED\ is KUG\,0952+418 (\vrb=4695, 9.2, 449), whose velocity is
similar to that of three closely-spaced \Lya\ absorption lines, at 4670, 4807
and 4961~\kms. These three absorption lines were previously listed by Savage et
al.\ (2002). At impact parameters larger than 1~Mpc, there are many galaxies
with \vgal$\sim$2000~\kms, as well as a few with \vgal$\sim$4800~\kms, including
some in the group GH\,49 at 2.2~Mpc. Finally, UGC\,5290 (\vrb=5030, 20.8, 1292)
lies about 1~Mpc away. No galaxies are known within 3~Mpc with \vgal\ between
2800 and 4600~\kms, however. The most probable explanation for these \Lya\
absorbers thus seems to be an intergalactic filament that is associated with the
galaxies near 4800~\kms. Table~\Tres\ lists all three as associated with
KUG\,0952+418, but a deeper search for faint galaxies may turn up others.

{\it PG\,1001+291. --}
The situation for PG1001+291 is complicated. The \STIS-E140M spectrum is of
reasonably good quality (S/N$\sim$8), but the \FUSE\ spectrum only has
S/N$\sim$4, making the upper limits to \Lyb\ and the \OVI\ lines not very
significant, and resulting in effectively useless \CIII\ data. There are several
galaxies with \vgal$\sim$500~\kms, with UGC\,5427 being small (\Rgal=3.6~kpc),
but having an impact parameter of only 84~kpc. A feature is seen in the \Lya\
line at 487~\kms, but it is close to where the Galactic line completely
saturates, so the line is noisy. Two other small galaxies have similar velocity:
UGC\,5340 (\vrb=503, 8.3, 296) and UGC\,5272 (\vrb=520, 6.5, 731). Because of
the small impact parameter, Table~\Tres\ lists the \Lya\ line at 487~\kms\ as
associated with UGC\,5427, as well as non-detections for UGC\,5340 and
UGC\,5227.
\par An additional \Lya\ feature is found at 1069~\kms, but the nearest galaxies
with similar velocities have large \ip\ and are small (Fig.~\Fassoc(71)):
MCG+5-24-11 (=UGCA\,201, \vrb=1363, 3.5, 188, \ip/\Rgal=54), UGC\,5464
(\vrb=1011, 7.2, 337, \ip/\Rgal=47) and UGC\,5478 (\vrb=1378, 11.2, 686).
Table~\Tres\ chooses to associate the 1069~\kms\ absorption with UGC\,5464
(\Dv=$-$58~\kms), rather than with MCG+5-24-11 (\Dv=294~\kms), even though the
latter has smaller impact parameter. In contrast, Bowen et al.\ (1996)
associated the 1069~\kms\ feature with MCG+5-24-11.
\par In addition to these lines, there is a \Lya\ line at 4602~\kms. The nearest
galaxy with similar velocity is UGC\,5461 at \ip=1249~kpc.

{\it PG\,1011$-$040. --}
The RC3 includes three galaxies near this sightline: IC\,600 (\vrb=1309, 15.1,
417), MCG$-$1-26-12 (\vrb=662, 10.4, 750), and NGC\,3115 (\vrb=658, 26.6, 900).
Two additional large galaxies can be found in \NED: LCRS\,B101019.9$-$032413
(\vrb=3395, 9.8, 680) and 2MASX\,J101213.26$-$040226.2 (\vrb=5619, 20.2, 852).
No absorption features are found in any line near the velocities of these
galaxies. It is possible that \Lya\ absorption is present, but there are no
\Lya\ data. \FUNNY{Time for \HST time.} No previous paper has reported results
for this sightline.

{\it PG\,1049$-$005. --}
Non-detections are listed for seven galaxies near this sightline: UGC\,5922
(\vrb=1846, 9.2, 391), IC\,653 (\vrb=5538, 45.2, 438), CGCG\,10-41 (\vrb=1810,
4.5, 475), UGC\,6011 (\vrb=5547, 33.5, 685), UGC\,5943 (\vrb=4544, 22.4, 689),
NGC\,3365 (\vrb=986, 23.2, 943) and NGC\,3521 (\vrb=805, 49.2, 962). The S/N
ratio of the \STIS-G140M data is relatively low, however (7.2), and the
detection limit is at best 75~\mA. C\^ot\'e et al.\ (2005) claimed an 80~\mA\
line at 5538~\kms, but this is (a) near the limit of detection and (b) more
likely to be Galactic \NV\lm1238.821 There is a hint of an absorption line near
2253~\kms, but it measures as 135\E105~\mA, and this is probably not
significant.

{\it PG\,1116+215. --}
This sightline has high S/N \FUSE\ and \STIS-E140M data (S/N$\sim$25 and
$\sim$10, respectively). Both Penton et al.\ (2004) and Sembach et al.\ (2004)
reported the \Lya\ lines at 1479 and 4884~\kms, with equivalent widths that are
within 1$\sigma$ of the values in Table~\Tres. The feature at 1479~\kms\ can be
associated with UGC\,6258 (\vrb=1454, 14.4, 543~kpc; see Fig.~\Fassoc(73)).
Danforth et al.\ (2006) listed a corresponding \OVIb\ feature. However, this can
be shown to be Ly$\xi$ at $z$=0.138, at which redshift there is a Lyman-limit
system in which 19 Lyman lines, from \Lya\ to Ly$\tau$, can be identified.
\par There is another \Lya\ absorber at 4884~\kms. The only galaxy with
\ip$<$3000 and \Dva$<$400~\kms\ is NGC\,3649 (\vrb=4979, 26.9, 1742; triangle in
Fig.~\Fassoc(74)). There is a feature at the position of \Lyb\ that corresponds
to this \Lya\ absorption, even though it is much narrower. The most likely
explanation is that the \Lya\ is actually a two-component absorber. Danforth et
al.\ (2006) list an upper limit for \OVI\ at this velocity, but any \OVI\
absorption would be blended with Ly$\iota$ at $z$=0.138 and their upper limit
does not take this into account.

{\it PG\,1149$-$110. --}
There are eight galaxies at various velocities that lie close to this target.
Unfortunately, the \STIS-G140M spectrum has a relatively low S/N ratio of 5.2,
and there is an unidentified emission feature near 1232~\AA. Nevertheless, the
following galaxies are listed in Table~\Tres: NGC\,3942 (\vrb=3696, 232., 141),
MCG$-$2-30-33 (\vrb=1273, 5.1, 442), NGC\,3892 (\vrb=1697, 24.3, 529),
MCG$-$2-30-39 (\vrb=1483, 13.6, 541), LCRS\,B115151.0$-$113904 (\vrb=3009, 20.4,
640), LCSB\,S1630P (\vrb=1971, 11.2, 642), PGC\,37027 (\vrb=2379, 14.5, 812),
and Mrk\,1309 (\vrb=1715, 13.2, 909). \Lya\ absorption lines are seen at 1665
and 3728~\kms, and they are associated with NGC\,3942 (Fig.~\Fassoc(75)) and
NGC\,3892 (Fig.~\Fassoc(76)), respectively. These lines were also reported by
Bowen et al.\ (2005). However, Bowen et al.\ gave an equivalent width of
1100\E30~\mA\ for the feature at 1665~\kms, which is clearly too strong, with
too low an error. We find 437\E69~\mA.

{\it PG\,1211+143. --}
Together with 3C\,273.0 and HE\,1228+0131 this is one of three sightlines
passing through the Virgo cluster. That means that many galaxies have impact
parameter $<$1~Mpc. A histogram of the velocities of these galaxies shows three
groups: \vgal$<$400~\kms, \vgal=450--1600~\kms, and \vgal=1650--2550~\kms.
Together, the RC3 and \NED\ include 55 galaxies in the first group, 91 in the
second and 44 in the third. Many (but not all) of these galaxies were assigned
to one of the LGG groups by Garcia (1993). The sightline passes between the
galaxies of the LGG\,285 and LGG\,289 groups. So, Table~\Tres\ gives these two
entries for PG\,1211+143, with the impact parameter that of the nearest group
galaxy with \Rgal$>$9~kpc. LGG\,289 (\vavg=1282~\kms) includes IC\,3077
(\vrb=1411, 9.3, 215), as well as NGC\,4206 (\vrb=702, 40.5, 418) and three more
large galaxies with \ip$>$700~kpc. LGG\,285 (\vavg=1282~\kms) includes IC\,3061
(\vgal=2361~\kms) at \ip=121~kpc, NGC\,4189 (\vgal=2113~\kms) at 413~kpc and
seven more galaxies with \ip=420--1000~kpc.
\par Absorption near these velocities is only seen at 2110~\kms, with equivalent
width 104\E8~\mA. It is listed under the LGG\,285 group in Table~\Tres\ (at
\ip$>$121~kpc). This line was previously reported by Penton et al.\ (2004),
although they gave $v$=2130~\kms, and equivalent width 186\e19~\mA. This
discrepancy may partly be because of differences in the continuum placement, but
also because Penton et al.\ (2004) included the $<$1$\sigma$ noisy wings in
the velocities over which they integrated. However, we only get 125~\mA\ if we
do that.
\par Danforth et al.\ (2006) mysteriously listed a 45\E14~\mA\ \OVIb\ feature at
2130~\kms. However, as can be seen in Fig.~\Fspectra, there clearly is nothing
significant within 200~\kms\ of this velocity. We derive an upper limit of
19~\mA. Since Danforth et al.\ (2006) did not show their data, it is not
possible to determine which feature they had in mind.
\par Penton et al.\ (2004) also listed the two lines at 4932 and 5015~\kms, with
equivalent widths of 189\E46 and 154\E40~\mA. We find 165\E7\E4 and
231\E7\E4~\mA, respectively. Since the higher-velocity line is clearly stronger
than the lower-velocity one (see Fig.~\Fspectra), the Penton et al.\ (2004)
result cannot be correct. The nearest galaxy with known velocity within
$\pm$400~\kms\ of these absorption lines is CGCG\,69-129 (\vrb=4987, 10.0,
1919), which is the one listed in Table~\Tres. Potentially, IC\,3073 has much
smaller impact parameter (670~kpc) if its velocity were $\sim$5000~\kms.
\par Tumlinson et al.\ (2005) analyzed two absorption systems at 19400 and
15300~\kms, which have many \HI\ lines, as well as \OVI, \CIII\ and \SiIII.

{\it PG\,1216+069. --}
This sightline passes within 1~Mpc from a large number of galaxies (265 are
known in \NED), including galaxies in six LGG groups -- LGG\,288
(\vavg=505~\kms), LGG\,292 (\vavg=938~\kms), LGG\,289 (\vavg=1282~\kms),
LGG\,287 (\vavg=1655), LGG\,278 (\vavg=2078~\kms) and LGG\,281
(\vavg=2473~\kms). There is a \STIS-E140M spectrum with relatively low S/N (9
per 6.5~\kms\ resolution element), and a low S/N ($\sim$5) \FUSE\ spectrum,
which makes the analysis difficult. Nevertheless, we can discern three \Lya\
absorption lines, at 1106, 1443 and 1895~\kms. The third of these is a damped
\Lya\ system, which was discussed in detail by Tripp et al.\ (2005). Bowen et
al.\ (1996) first identified the damped \Lya\ absorber in a low S/N \GHRS\
spectrum, listing a velocity of 1650~\kms.
\par In Table~\Tres\ we associate the absorption at 1106~\kms\ with LGG\,292
(nearest galaxy NGC\,4241 at \ip=104~kpc; Fig.~\Fassoc(79)), that at 1443~\kms\
with LGG\,289 (nearest galaxy UGC\,7423 at \ip=264~kpc; Fig.~\Fassoc(80)), and
that at 1895~\kms\ with LGG\,278 (nearest galaxy NGC\,4223 at \ip=283~kpc;
Fig.~\Fassoc(81)), although the first two of these are certainly open for
discussion.
\par In addition to the six galaxy groups at \vgal$<$2500~\kms, the sightline
passes within 1~Mpc from six galaxies that have \vgal\ between 2500 and
5600~\kms: NGC\,4257 (\vrb=2756, 16.0, 689), NGC\,4246 (\vrb=3725, 40.1, 644),
NGC\,4247 (\vrb=3810, 18.8, 742), NGC\,4296 (\vrb=4227, 25.4, 596), IC\,771
(\vgal=5477, 26.9, 841), and IC\,3136 (\vrb=5594, 28.8, 684). There are two
\Lya\ lines, at 3774 and 3808~\kms, which, because of its low impact parameter
are both associated with SDSS\,J121903.72+063343.0 (\vrb=3833, 5.1, 103,
\ip/\Rgal=20.2). Non-detections are listed for NGC\,4246 (\vrb=3725, 40.1, 644,
\ip/\Rgal=16) and NGC\,4247 (\vrb=3838, 18.8, 742, \ip/\Rgal=39), even though
these are much larger galaxies and NGC\,4246 has the lowest \ip/\Rgal\ ratio.
The associations are a bit ambiguous, however.

{\it PG\,1259+593. --}
PG\,1259+593 is one of the few sightlines with both high S/N \FUSE\ and high S/N
\STIS-E140M data. Richter et al.\ (2004) analyzed all IGM lines in detail,
listing the \Lya\ and \Lyb\ detections at 678 and 2275~\kms, although they gave
velocities of 686 and 2278~\kms. Their listed equivalent widths are similar to
ours within 1$\sigma$ for the 2275~\kms\ component, but they differ for the
678~\kms\ component. The difference in the \Lya\ equivalent width (231\E9~\mA\
in Table~\Tres\ vs 190\E24 in Richter et al.\ (2004)) is probably caused by the
fact that Richter et al.\ fitted a polynomial to the continuum around this line,
while we model the Galactic damped \Lya\ profile. In the case of \Lyb\
($<$15~\mA\ vs 23\E6~\mA\ in Richter et al.\ 2004) the difference is due to the
different manner by which the \H2\ line is removed. C\^ot\'e et al.\ (2005) also
presented a plot of the \STIS-G140M spectrum of the \Lya\ line, measuring it as
330\E80~\mA. Our reduction of this spectrum gives 245\E70~\mA, more in line with
the equivalent width found from the E140M data.
\par An apparently double \OVI\ feature is seen, centered at 627~\kms,
surrounded by \Lyb\ at 2269~\kms\ and \Lyd\ at $z$=0.0894. The two apparent
features are centered at 627 and 689~\kms.  Even though this feature is offset
in velocity from \Lya\ by $-$50~\kms, the identification of the 627~\kms\
feature as \OVI\ is fairly secure, because a) there are no high redshift systems
that produce absorption at this wavelength, and b) there is a corresponding
feature for \OVIb\ that is half the strength. The other feature does not have
such a counterpart. We measure 22\E5~\mA\ for the 627~\kms\ absorption, and
14\E5~\mA\ for the 689~\kms\ absorption. We conclude that there is \OVIa\ at
627~\kms, while the 689~\kms\ absorption is considered not significant.
\par These features were reported in three previous papers. Richter et al.\
(2004) gave an equivalent width of 63\E22~\mA, centered at 690~\kms, but
integrated from 580 to 710~\kms. However, this would not only include the
probable \OVI, but also the \Lyd\ line at $z$=0.0894; further we find an
equivalent width of 50\E10~\mA\ when integrating over this velocity range.
Danforth et al.\ (2006) listed 14\E9~\mA\ at 687~\kms; apparently only measuring
the least significant of the three features. Tripp et al.\ (2008) gave
40\E5~\mA\ at 630~\kms. Using their integration range of 590 to 730~\kms\ we
find 36\E8~\mA, though a fit would give a central velocity of 644~\kms. Clearly,
different authors do not agree about the detailed interpretation and measurement
of these two features, but they all agree that \OVIa\ is present.
\par The impact parameter of UGC\,8146 (\vgal=669~\kms) is only 80~kpc, so the
\Lya\ and \OVI\ absorptions can be confidently associated with that galaxy (see
Fig.~\Fassoc(83)). Two other small galaxies have similar velocity, but are not
included in Table~\Tres\ because \ip/\Rgal$>$100. NGC\,4964 (\vrb=755, 4.4, 672)
has large impact parameter, while SDSS\,J130206.46+584142.9 (\vrb=623, 1.8, 72),
is a dwarf companion of UGC\,8146.
\par A second \Lya\ absorber is visible at 2275~\kms. There are several large
galaxies with diameter $>$10~kpc and \vgal$\sim$2500~\kms\ within 1~Mpc
(Fig.~\Fassoc(84)): UGC\,8046 (\vrb=2572, 11.8, 584), UGC\,8040 (\vrb=2522,
16.8, 595) and NGC\,4814 (\vrb=2513, 35.0, 694). The feature is listed as
associated with the edge-on galaxy UGC\,8040, which is the largest of the two
galaxies with \ip$\sim$590~kpc. Associating it with NGC\,4814 instead would have
been justifiable, however. Six more dwarfs with similar velocity and \ip$<$1~Mpc
are listed in \NED. \NED\ also lists SDSS\,J125926.78+591735.0 (\vgal=2867, 6.5,
255), which is included separately in Table~\Tres.
\par Finally, a third \Lya\ feature at $v$$<$5000~\kms\ is visible at 4501~\kms.
For this velocity neither the RC3 nor \NED\ lists any galaxy closer than
MCG+10-19-23 at impact parameter 1.8~Mpc, so this feature is classified as
intergalactic in Table~\Tres.

{\it PG\,1302$-$102. --}
The \STIS-E140M spectrum of this sightline shows several absorption lines
between 1216 and 1224~\AA, but all but two of these can be identified as \Lyb\
at $z$=0.188 to 0.192, or \Lyg\ at $z$=0.254. The remaining lines are probably
\Lya\ at 1045 and 1316~\kms. The latter identification is secure, as \Lyb\ and
\CIII\ absorption are also seen at this velocity, while the first is uncertain.
Danforth et al.\ (2006) reached the same interpretation for these lines. In the
RC3 and \NED\ there are 22 galaxies with impact parameter $<$1~Mpc near this
sightline. Separate entries are given in Table~\Tres\ for the seventeen of these
for which the ratio \ip/\Rgal$<$125. For all except two these are upper limits,
which is unsurprising since all but three of the galaxies have impact parameter
$>$650~kpc.
\par The table includes the following galaxies. Toward the north lie NGC\,4939
(\vrb=3111, 24.6, 104), DDO\,163 (\vrb=1123, 11.0, 930), NGC\,4818 (\vrb=1065,,
24.0, 987), and NGC\,4948A (\vrb=1553, 10.5, 991). Toward the east is
MCG$-$2-34-6 (\vrb=1213, 18.2, 391), while toward the south lies NGC\,5068
(\vrb=672, 10.9, 995). In a westerly direction lies the group LGG\,307, of which
NGC\,4920 (\vrb=1336, 6.8, 473) is the nearest. \FUNNY{Is this too much detail.}
Other listed group members are MCG$-$1-33-60 (\vrb=1487, 20.1, 830),
MCG$-$2-33-85 (\vrb=1582, 11.6, 863), UGCA\,312 (\vrb=1307, 8.6, 900), and
UGCA\,308 (\vrb=1322, 9.4, 986). Also toward the west are MCG$-$2-33-75
(\vrb=1247, 6.8, 658), MCG$-$2-33-95 (\vrb=2753, 17.4, 657), while in a
southwesternly directly lie MCG$-$2-33-97 (\vrb=2705, 14.9, 782), NGC\,4933
(\vrb=2965, 34.9, 820), UGCA\,307 (\vrb=824, 9.1, 882), and NGC\,4802
(\vrb=1013, 12.8, 901).
\par As Fig.~\Fassoc(86)\ shows, two galaxies have considerably lower impact
parameter than the rest. These are the two that we associate the two \Lya\ lines
with. The line at 1045 ~\kms\ is associated with MCG$-$2-34-6, while the one at
1316~\kms\ is listed under NGC\,4920. This assignment is chosen because it
minimizes the velocity differences between absorption and galaxy (\Dv=$-$168,
20~\kms\ for this choice, instead of \Dv=$-$103, +291~\kms\ for the
alternative). MCG$-$1-33-83 potentially has smaller impact parameter (308~kpc),
but its velocity is unknown.
\par The association of the \Lya\ line at 3909~\kms\ with NGC\,4939 is much less
ambiguous, as can be seen in Fig.~\Fassoc(87). NGC\,4939 is the only galaxy near
this velocity with low impact parameter.

{\it PG\,1341+258. --}
Bowen et al.\ (2002) claimed a 120\E20~\mA\ detection of \Lya\ at 1425~\kms, but
we do not see this line, and instead derive a 30~\mA\ upper limit around this
velocity. From the plot in the Bowen et al.\ (2002) paper it looks like there
may be a feature, which we measure to be at 1454~\kms, but with equivalent width
20\E12~\mA, i.e.\ not significant. The only other measurement from this
sightline is a non-detection for CGCG\,132-10 (\vrb=3188, 6.9, 825).

{\it PG\,1351+640. --}
In this spectrum absorption is seen at 1771 and 1447~\kms. The corresponding
\Lyb\ lines are hidden by Galactic \OVI\ and intrinsic \Lyd\ absorption,
respectively. The \OVIa\ lines are hidden by strong \H2\ lines, while an upper
limit can be derived for the \OVIb\ line.
\par There are five galaxies with \vgal\ between 1247 and 1971~\kms\ and with
\ip$<$500~kpc: UGC\,8894 (\vrb=1771, 12.9, 274), UGCA\,375 (\vrb=1763, 7.3, 299;
Mrk\,277 in the RC3, lying in a direction on the sky opposite to UGC\,8894),
SDSS\,J134800.10+633120.8 (\vrb=1665, 2.5, 284), and SDSS\,J134711.13+625006.3
(\vrb=1496, 2.5, 482). Associating the \Lya\ at 1771~\kms\ with UGC\,8894 seems
an obvious choice (see Fig.~\Fassoc(89)). However, associating the 1447~\kms\
absorption is more tricky. In Table~\Tres\ it is listed with UGCA\,375
(\Dv=316~\kms, see Fig.~\Fassoc(88)), but it is easy to argue that it should
also be associated with UGC\,8894 (\Dv=324~\kms) or even with
SDSS\,J134711.13+625006.3 or SDSS\,J134800.10+6733120.8 (\Dv=49~\kms\ and
218~\kms, open circles). However, \ip/\Rgal=192 and 113 for these galaxies,
which would be larger than is the case for any other association. Obviously, any
association choice is dubious.
\par Penton et al.\ (2004) have this sightline in their sample, but surprisingly
they did not list the two 3.5$\sigma$ features at 1771 and 1447~\kms, even
though they can be seen in their figure of the spectrum and even though they
sometimes listed 2$\sigma$ features that do not seem to be real.

{\it PG\,1444+407. --}
This sightline has reasonable quality \STIS-E140M data (S/N$\sim$9), but
low-quality \FUSE\ data (S/N$\sim$4). Therefore, the upper limits for \Lyb\ and
\OVI\ are not very significant, especially for velocities where only
orbital-night data can be used. For \Lya\ the limits are about 50~\mA. No \Lya\
appears associated with UGC\,9497 (\vrb=633, 4.0, 445), nor with
SDSS\,J145001.59+402142.4 (\vrb=4814, 8.0, 653). However, \Lya\ is clearly
present at 5638~\kms, and is associated with UGC\,9502 (\vrb=5672, 27.8, 637).
\par A line also appears at 2630~\kms, which in Table~\Tres\ is listed under
SDSS\,J145045.59+413742.1 (\vrb=2582, 4.2, 897; Fig.~\Fassoc(90)), in spite of
the large impact parameter. Several other galaxies with similar velocity are
present with impact parameters of 1200 to 1800~kpc. A potential alternative
candidate is UGC\,9495, whose velocity is unknown. If it is similar to the
velocity of the detected \Lya\ line at 2630~\kms, its impact parameter would be
551~kpc, and its diameter 12.9~kpc, in which case we would associate it with the
detection.

{\it PG\,1553+113. --}
There are few galaxies in the part of the sky near PG\,1553+113. Only one can be
found in the RC3 or \NED\ with impact parameter $<$1~Mpc: UGC\,10014 (\vrb=1121,
7.1, 916).

{\it PG\,1626+554. --}
There are just two galaxies with low impact parameter near this target:
NGC\,6182 (\vrb=5138, 38.1, 338) and NGC\,6143 (\vrb=1595, 6.8, 403). No \Lya\
data exist, and no \Lyb\ absorption is seen.

{\it PHL\,1811. --}
PHL\,1811 lies in a part of the sky where there are few nearby galaxies. In
fact, only two galaxies with \ip$<$1~Mpc can be found in \NED:
SDSS\,J214451.59$-$084537.6 (\vrb=1276, 0.2, 674) and
SDSS\,J215446.45$-$084616.9 (\vrb=5498, 12.0, 787), which is listed in
Table~\Tres. A very narrow feature (FWHM 14~\kms) can be seen at 5402~\kms,
which may be \Lya. The sightline contains a Lyman limit system at $z$=0.08093
which was discussed in detail by Jenkins et al.\ (2003, 2005).
\par Jenkins et al.\ (2003) listed possible \Lya\ lines at 3537 and 5202~\kms.
However these are actually \OVIab\ at $z$=0.1919, which is confirmed by the fact
that these lines have the same velocity structure, as well as by the presence of
a corresponding (weak) \Lya\ absorption. Danforth et al.\ (2006) gave a lower
limit of 63~\mA\ for the \OVI\ line corresponding to the erroneously claimed
3537~\kms\ \Lya. This is not only inappropriate, since the line is redshifted
\OVIa, but also mysterious, since there is a strong redshifted ($z$=0.0735)
\Lyg\ line at the corresponding wavelength.

{\it PKS\,0405$-$12. --}
Prochaska et al.\ (2004) and Williger et al.\ (2006) analyzed the \FUSE\ and
\STIS-E140M spectra of this sightline in detail. Williger et al.\ (2006) listed
seven \Lya\ features at $v$$<$5000~\kms. Four of those are less than 1.5$\sigma$
and probably not real. Lehner et al.\ (2007) reidentified the features and
concluded that the absorptions that are believable are at 1220.605, 1227.359,
1230.136, 1233.808 and 1236.105~\AA. The first two of these are best interpreted
as \OVI\ redshifted to $z$=0.1828, especially since a strong \HI\ absorption
system (\Lya\ to \Lyz) is seen at this redshift. The feature at 1230.136~\AA\ is
probably \Lya\ at 3574~\kms. The nearest galaxy with \vgal$\sim$3500~\kms\ is
2MASX\,J040607.61$-$102327.2 at \ip=1.5~Mpc. The feature at 1236.105~\AA\ is
listed as intergalactic \Lya\ at $v$=5035~\kms\ -- no galaxies with similar
velocity are known within 3~Mpc. The feature at 1233.808~\AA\ was listed as a
3$\sigma$ detection by Lehner et al.\ (2007), but we decided that it is not
significant.

{\it PKS\,0558$-$504. --}
Near this sightline lie ESO\,205-G34, NGC\,2104 and NGC\,2101, all of which have
impact parameter $>$500~kpc and velocity $\sim$1100~\kms. In addition, there is
NGC\,2152, for which no velocity has been determined, but which potentially has
a small impact parameter: \ip=172$\times$(\vgal/2000)~kpc. Finally, there is
ESO\,205-G07 at \vgal=2000~\kms. No \Lya\ data exist for PKS\,0558$-$504, and no
\Lyb\ absorption with $v$$<$5000~\kms\ is seen.

{\it PKS\,2005$-$489. --}
For this sightline Penton et al.\ (2004) listed three \Lya\ lines at velocities
below 5100~\kms. Two of these are strong lines centered at 4973 and 5071~\kms,
and for both \Lyb\ absorption is also detected. The higher-velocity \Lyb\ line
is contaminated by the \FUSE\ detector flaw near 1043~\AA, and its measured
equivalent width is too large; however, we can't correct for this. The nearest
galaxies with similar velocity are ESO\,233-G37 (\vrb=4950, 26.4, 599) and
2MASX\,J200943.13$-$481105.2 (\vrb=5116, 16.6, 787). The feature that Penton et
al.\ (2004) identified as a third \Lya\ at 2752~\kms\ is more likely to be
\SiIII\ absorption at a velocity of 5073~\kms. It has an equivalent width of
29\E4~\mA. In this system \Lya, \Lyb\ and \CIII\ are detected at velocities of
5071, 5079 and 5065~\kms, and the \CIII\ line is rather strong (see
Table~\Tion). Photoionization modeling using {\it CLOUDY} (Ferland et al.\ 1998)
shows that the column densities of \HI, \CIII\ and \SiIII\ can be explained with
a 2.8~kpc thick feature having log\,$n$=$-$4.0~\cmm3, and metallicity 0.1 times
solar. Running a {\it CLOUDY} model with just the \HI\ and \CIII\ column
densities leads to a prediction for the \SiIII\ equivalent width on the order of
25~\mA, for gas densities $\sim$\dex{-4}\,\cmm3.
\par The sightline passes through the group LGG\,430, which is defined by 12
galaxies that have \vavg=2955\E213~\kms, with velocities ranging from 2504 to
3200~\kms. Seven of the defining galaxies have impact parameter $<$1~Mpc to
PKS\,2005$-$489. The RC3 lists another 6 galaxies within the group's defining
velocity range, and \NED\ includes another 10, for a total of 23 galaxies with
\vgal\ in the range 2504 to 3200~\kms\ and \ip$<$1~Mpc. Since the feature at
1226.916~\AA\ is most likely \SiIII\ at 5073~\kms, there apparently is no \Lya\
absorption associated with the group.

{\it PKS\,2155$-$304. --}
Penton et al.\ (2000) listed eight \Lya\ lines at $v$=2600--5700~\kms\ in this
sightline, based on a \GHRS\ spectrum (20~\kms\ resolution), although the
features at 2632, 2785, 4031, 4709, and 5618~\kms\ are all less than
2.5$\sigma$. Although Penton et al.\ (2000) listed them as having ``significance
level'' $>$4, none of these features are confirmed in the \STIS-E140M spectrum
of PKS\,2155$-$304, with upper limits that are three times better than those
derived from the \GHRS\ spectrum. Only the feature at 5618~\kms\ is confirmed in
the E140M data, but now has a velocity of 5673~\kms.
\par The three features that Penton et al.\ (2000) listed at velocities of 4951,
5013 and 5119~\kms\ must have been mismeasured by them. Based on the \GHRS\ data
they gave equivalent widths of 64, 82 and 218~\mA. We can reproduce these values
by summing over the following velocity ranges: 4850--4970~\kms, 4970--5040~\kms\
and 5040--5230~\kms. As can be seen in Fig.~\Fspectra, the first of these would
then cover half of the most-negative-velocity component, plus some velocities
where no absorption is seen in the E140M data; although there appeared to be a
hint of absorption in the G140M data. The second of the Penton et al.\ (2000)
components would cover the other half of the leftmost absorber. The third
component covers a velocity range that spans an asymmetrical absorption. None of
these three components seem to be justifiable as-is. Even if the equivalent
widths given by Penton et al.\ (2000) were based on gaussian fits, it is hard to
see how they were derived.
\par We instead measure three different components in the \STIS-E140M spectrum
(which shows the same structure as the \GHRS\ data), at velocities of 4990, 5101
and 5164~\kms. We also find \Lyb\ absorption with similar component structure,
although the absorption at the highest velocities is confused with the \FUSE\
detector flaw near 1043~\AA. We associate these absorbers with ESO\,466-G32, a
galaxy at \vgal=5153~\kms\ with impact parameter 306~kpc, which is not included
in the RC3, but listed in \NED.
\par The sightline also passes through the edge of the LGG\,450 group
(\vavg=2601~\kms). The RC3 includes two group galaxies (ESO\,466-G36 and
NGC\,7163), while \NED\ lists five more large (ESO\,466-G29, 2dF\,GRSS407Z162,
MCG$-$5-51-2, ESO\,466-G43 and 2dF\,GRSS408Z175) as well as three more small
galaxies with velocity between 2380 and 2875~\kms. However, no absorption is
seen anywhere in this velocity range down to a limit of $\sim$30~\mA\ (in spite
of the 2.5$\sigma$ features claimed by Penton et al.\ (2000)).

{\it RX\,J0048.3+3941. --}
For this sightline the impact parameter with M\,31 is just 22~kpc. High-negative
velocity \OVI\ absorption ($-$300~\kms) is present, as it is in many sightlines
in the southern sky (see Wakker et al.\ 2003). A few other galaxies have impact
parameter $<$1~Mpc. Five are included in the RC3, but four are small and have
velocity similar to UGC\,655 (\vrb=829, 6.9, 602~kpc), the largest and the only
one listed in Table~\Tres. \NED\ lists two more galaxies (UGC\,578 and
CGCG\,535-25), having velocities of 1471 and 4001~\kms\ and impact parameters of
499 and 810~kpc. No intergalactic lines are seen near any of these velocities,
although in most cases the possible intergalactic line is blended with
interstellar atomic or molecular absorption.

{\it RX\,J0100.4$-$5113. --}
Near this target lies ESO\,151-G19 (\vrb=1386, 6.5, 902). There may be a
2$\sigma$ associated \Lya\ feature in the \STIS\ spectrum, at 1111~\kms, but
this is not a convincing detection and it is not listed in Table~\Tres. There is
another possible (3$\sigma$) \Lya\ feature at 4874~\kms; the nearest galaxy is
ESO\,195-G17 (\vrb=1189, 18.6, 1189), although a galaxy with unknown velocity
(ESO\,195-G24) would have impact parameter $\sim$800~kpc if it were at
\vgal$\sim$4700~\kms.

{\it RX\,J1830.3+7312. --}
As Bowen et al.\ (2002) reported, there are five \Lya\ lines seen toward this
target in the \STIS-G140M spectrum. The two absorptions at 1968 and 1549~\kms\
can be ssociated with two galaxies lying close to the sightline: NGC\,6645A
(\vrb=1558, 18.2, 308; Fig.~\Fassoc(95)) and NGC\,6654 (\vrb=1821, 18.2, 186;
Fig.~\Fassoc(96)). There are five more galaxies with similar velocity and
\ip$<$1~Mpc: MCG+12-17+27 (\vrb=1404, 5.2, 262), UGC\,11331 (\vrb=1554, 10.5,
296), CGC\,G340-51 (\vrb=1469, 4.3, 337), NGC\,6643 (\vrb=1489, 26.4, 638), and
UGC\,11193 (\vrb=1489, 8.9, 985). As Fig.~\Fassoc(96)\ shows, the association
between NGC\,6654 and the 1968~\kms\ \Lya\ absorber is easy to justify. For the
other absorber, the large galaxy NGC\,6654A is preferred over the smaller ones
near it. These galaxies were not included as a group in the LGG list of Garcia 
(1993), because some redshifts were missing at the time. To account for this, a
new group has been defined, using the galaxies above, excepting UGC\,11193. The
right ascension of this group would place it between LGG\,420 and LGG\,421, so
it is identified as LGG\,420A in Table~\Tres.
\par For a third \Lya\ feature, at 2383~\kms, the nearest galaxies are
UGC\,11295 and UGC\,11382, which have impact parameters of 1307 and 1332~kpc
(see triangle in Fig.~\Fassoc(97)). Finally, there are lines at 4260 and
4770~\kms, but the only galaxy near that velocity with \ip$<$3000~kpc is
UGC\,11334 (\vrb=4582, 37.8, 1022~kpc; Fig.~\Fassoc(98)/(99)).

{\it Ton\,S180. --}
Two galaxies in the nearest galaxy group, the Sculptor Group, have low impact
parameter: NGC\,247 (\vrb=159, 15.7, 125) and NGC\,253 (\vrb=251, 20.7, 166;
closed and open symbol in Fig.~\Fassoc(100)). There is a clear \OVI\ pair at
260~\kms, with an accompanying 2.5$\sigma$ detection of a \CIII\ line. \Lyb\ is
confused with geocoronal \OI* emission, but seems absent, while any \Lya\ is
completely hidden in the Galactic \Lya\ absorption. The \OVI\ velocity lies
below the nominal survey limit of 400~\kms, but no positive-velocity Galactic
high-velocity clouds are known in this part of the sky, so an association with
NGC\,247 or NGC\,253 is likely. Three other Sculptor group galaxies (NGC\,45,
NGC\,300 and NGC\,24) have larger impact parameter (437, 534 and 876~kpc) and
similar velocity (468, 142 and 554~\kms); no features are seen near those
velocities.
\par Three weak but clear \Lya\ absorbers are found between 1500 and 5000~\kms,
all of which were listed by Penton et al.\ (2004). In Table~\Tres\ the absorber
at 2792~\kms\ is associated with the relatively small galaxy
2MASX\,J005700.66$-$232044.2 (\vrb=2657, 4.9, 560; Fig.~\Fassoc(102)), which is
not in the RC3. The absorber at 1939~\kms\ can be associated with either
ESO\,541-G05 (\vrb=1958, 8.3, 774; included in the RC3) or ESO\,474-G45
(\vrb=1863, 7.2, 777; not in the RC3), both of which are small, but lie in
opposite directions from Ton\,S180 (see Fig.~\Fassoc(101)). Table~\Tres\
includes both galaxies, but the \Lya\ is listed under ESO541-G05, while an upper
limit is given for ESO\,474-G45. The third \Lya\ feature is at 3001~\kms, but
the nearest galaxy with similar velocity is ESO\,474-G25 (\vgal=2850~\kms;
Fig.~\Fassoc(103)), which has \ip=1.5~Mpc.
\par Finally, there is a strong \Lya\ and \Lyb\ line at 5519~\kms, already
reported by Penton et al.\ (2004). The nearest galaxy with similar velocity is
2MASX\,J010208.03$-$224559.7 (\vrb=5611, 16.6, 1547).

{\it Ton\,S210. --}
This sightline passes through the nearby Sculptor group (LGG\,4). The galaxy
with the smallest impact parameter is NGC\,253 (374~kpc). Formally, its velocity
of 251~\kms\ excludes it from our sample. However, there is an \OVI\ absorption
line at 288~\kms, which is very unlikely to be related to the Milky Way as
Ton\,S210 lies near the South Galactic Pole. This feature is also unlikely to be
related to the Magellanic Stream, as the Stream has velocities of $\sim$0~\kms\
in this part of the sky. It is also unlikely that this is a high redshift line,
as there are no strong intergalactic absorption-line systems in this sightline.
The most likely association is with NGC\,253, or possibly with the Local Group.
Unfortunately, the possible corresponding \Lya\ absorption is hidden in the
Galactic \Lya\ line, and \Lyb\ absorption is limited to $<$20~\mA, which still
means \Lya\ could be a strong as 100~\mA\ (see Table~\Tres).
\par We also give non-detections for NGC\,45 (\vrb=468, 6.5, 702), NGC\,613
(\vrb=1475, 27,4, 872) and ESO\,413-G02 (\vrb=5588, 17.9, 942).

{\it VII\,Zw\,118. --}
This is one of two targets passing through the (large) group LGG\,140 (the other
sightline being MS\,0700.7+6338). The nearest group galaxy is UGC\,3648
(\vrb=4530, 25.9, 620), with UGC\,3642 (\vrb=4498, 28.2, 738), and UGC\,3660
(\vrb=4252, 31.0, 845) also within 1~Mpc. UGC\,3648 was not listed as a group
galaxy by Garcia (1993), but only because its velocity was not yet known at the
time. A weak \Lya\ absorption at 4613~\kms\ is seen near the group's velocity,
and it is listed as associated with UGC\,3648 in Table~\Tres\ (see
Fig.~\Fassoc(107)).
\par A much stronger \Lya\ line is found at 2438~\kms, which can be associated
with UGC\,3748 (\vrb=2479, 12.2, 753; Fig.~\Fassoc(106)). \Lyb\ is also seen in
this system. We find 70\E8~\mA\ for \Lyb, while Shull et al.\ (2000) reported
110\E50~\mA. Penton et al.\ (2004) listed both features, but split both of them
into two components. Based on just the noisy line structure there is little
evidence to support this, except that single lines would have relatively large
(though not extraordinaly so) FWHM (148 and 109~\kms), and that the \Lyb\ line
has an FWHM of only 63~\kms. The 2438~\kms\ \Lya\ line also may be slightly
asymmetric.
\par A weak (3$\sigma$) \Lya\ feature is also seen at 1697~\kms. The nearest
galaxy with similar velocity is UGC\,3685 (\vgal=1797~\kms), at 1.4~Mpc
(Fig.~\Fassoc(105)). However, a galaxy with unknown velocity, UGCA\,133, might
have \ip=800~kpc if its velocity were about 1700~\kms.

\newpage
\bigskip\bigskip
Appendix B
\par We include a table of all galaxies in our sample that have velocity less
than 7000 km/s and impact parameter less than 2 Mpc to one of the 76
extragalactic sightlines. A sample is included, but the full table can be found
in the on-line version of this paper. It is important to note that this table in
inhomogeneous. Near some sightlines deep searches for dwarf galaxies exist, near
others they don't. The table should therefore not be used for statistical
analyses for faint galaxies, except for galaxies with diameter $>$7.5~kpc.
\par The values in the "Note" column give extra information. Notes starting with
``GH'' or ``LGG' indicate the galaxy is a member of a GH or LGG group; this is
(followed by the distance that would have been derived if the galaxy's velocity
were used directly. Other notes can be ``ASSUME'', when the galaxy's velocity is
unknown, ``ASSUME-CHECKED'' when the galaxy's velocity remained unknown after we
checked NED, or ``MINDIST'', when we assumed the minimum distance of 2.9~Mpc to
estimate the impact parameter. A plain number gives the velocity distance for
galaxies for which a better distance estimate was instead taken from the
literature.

\begin{deluxetable}{lllrrrrrrrl}
\tablenum{13}
\tablewidth{0pt}
\tabletypesize{\scriptsize}
\tabcolsep=3pt
\tablecolumns{11}
\tablecaption{Galaxy-absorber coincidences$^1$}
\tablehead{%
\ch{AGN} &\ch{Galaxy} &\ch{type} & \ch{$l$}    & \ch{$b$}    & \ch{v$_{\rm hel}$} & \ch{D$_{\rm gal}$} & \ch{R$_{\rm gal}$} & \ch{$\alpha$} & \ch{$\rho$} & Note \\
         &            &          & \ch{[\deg]} & \ch{[\deg]} & \ch{[\kms]}        & \ch{[Mpc]}         & \ch{[kpc]}         & \ch{[\deg]}   & \ch{[kpc]}         \\
\ch{(1)}&\ch{(2)}&\ch{(3)}&\ch{(4)}&\ch{(5)}&\ch{(6)}&\ch{(7)}&\ch{(8)}&\ch{(9)}&\ch{(10)}&\ch{(11)}
}\startdata
1H0419$-$577   &   LSBGF157$-$081           &   ...... & 267.22 & $$-$$41.78 &  1215 &  15.4 &  3.7 &  0.27 &   72 & NED \\
1H0419$-$577   &   IC2039                 &   .L..0*P& 266.36 & $$-$$44.58 &   250 &   2.9 &  0.8 &  2.62 &  132 & MINDIST \\
1H0419$-$577   &   NGC1574                &   .LAS$-$*.& 266.89 & $$-$$42.58 &   925 &  13.6 & 13.4 &  0.59 &  140 & LGG112 \\
1H0419$-$577   &   HIPASSJ0423$-$56         &   ...... & 265.95 & $$-$$42.60 &  1345 &  17.2 &  0.0 &  0.97 &  291 & NED \\
1H0419$-$577   &   APMBGC157+016+068      &   ...... & 265.87 & $$-$$42.67 &  1350 &  17.3 &  4.7 &  1.05 &  317 & NED \\
1H0419$-$577   &   IC2038                 &   .S..7P*& 266.34 & $$-$$44.60 &   712 &   8.1 &  4.1 &  2.64 &  373 & 0 \\
1H0419$-$577   &   ESO118$-$G19             &   ...... & 268.72 & $$-$$42.59 &  1239 &  15.7 &  3.7 &  1.40 &  384 & NED \\
1H0419$-$577   &   NGC1533                &   .LB.$-$..& 266.45 & $$-$$44.43 &   790 &   9.2 &  7.4 &  2.46 &  396 & 0 \\
1H0419$-$577   &   NGC1533:[RMF2004]2     &   ...... & 266.52 & $$-$$44.36 &   831 &   9.8 &  0.0 &  2.38 &  406 & NED \\
1H0419$-$577   &   NGC1533:[RMF2004]1     &   ...... & 266.53 & $$-$$44.36 &   846 &  10.0 &  0.0 &  2.38 &  415 & NED \\
\enddata
\tablecomments{%
1: Column (1) gives the name of the AGN target, Col.\ (2) the galaxies near the
AGN sightline. Column (3) is the galaxy type, taken from the RC3 (de Vaucouleurs
et al.\ 1991). Columns (4) and (5) give the Galactic longitude and latitude of
the galaxy. Column (6), (7), and (8) are the  galaxy's heliocentric velocity,
estimated distance and estimated diameter (with values of 0.0 meaning the
diameter is unknown). Columns (9) and (10) are the angular and line-of-sight
separation between target and galaxy (in degrees and kpc). Column (11) gives a
note: group membership (followed by the distance that would have been derived if
the galaxy's velocity was used directly), whether the galaxy's velocity is
unknown (note=ASSUMEV), or unknown and checked in NED (note=ASSUME-CHECKED), or
whether the minimum distance (2.9 Mpc) was assumed (note=MINDIST). A plain
number gives the velocity distance for galaxies for which a better distance
estimate was instead taken from the literature.
}
\end{deluxetable}

\end{document}